\documentclass[12pt,oneside,italian,english]{book}
\usepackage{ae,aecompl}

\usepackage[T1]{fontenc}
\usepackage[utf8]{luainputenc}
\usepackage[paperwidth=22cm,paperheight=28cm]{geometry}
\usepackage{fancyhdr}
\usepackage{float}
\usepackage{mathtools}
\usepackage{amsthm}
\usepackage{amsmath}
\usepackage{amssymb}
\usepackage{accents}
\usepackage{stmaryrd}
\usepackage{graphicx}
\usepackage{epstopdf}
\usepackage{esint}

\makeatletter

\newcommand{\lyxmathsym}[1]{\ifmmode\begingroup\def\b@ld{bold}
  \text{\ifx\math@version\b@ld\bfseries\fi#1}\endgroup\else#1\fi}

\newcommand{\docedilla}[2]{\underaccent{#1\mathchar'30}{#2}}
\newcommand{\cedilla}[1]{\mathpalette\docedilla{#1}}

\let\SF@@footnote\footnote
\def\footnote{\ifx\protect\@typeset@protect
    \expandafter\SF@@footnote
  \else
    \expandafter\SF@gobble@opt
  \fi
}
\expandafter\def\csname SF@gobble@opt \endcsname{\@ifnextchar[
  \SF@gobble@twobracket
  \@gobble
}
\edef\SF@gobble@opt{\noexpand\protect
  \expandafter\noexpand\csname SF@gobble@opt \endcsname}
\def\SF@gobble@twobracket[#1]#2{}

\newenvironment{lyxlist}[1]
{\begin{list}{}
{\settowidth{\labelwidth}{#1}
 \setlength{\leftmargin}{\labelwidth}
 \addtolength{\leftmargin}{\labelsep}
 }}
{\end{list}}
\theoremstyle{plain}
\newtheorem{thm}{\protect\theoremname}
  \theoremstyle{remark}
  \newtheorem{rem}[thm]{\protect\remarkname}
  \theoremstyle{remark}
  \newtheorem*{rem*}{\protect\remarkname}

\usepackage{hyperref}
\hypersetup{
colorlinks=true,%
linkcolor=blue,%
linktocpage=true,%
pageanchor=true
}

\usepackage{lipsum}
\newcommand\abstractname{Abstract}  
\makeatletter
\if@titlepage
  \newenvironment{abstract}{%
      \titlepage
      \null\vfil
      \@beginparpenalty\@lowpenalty
      \begin{center}%
        \bfseries \abstractname
        \@endparpenalty\@M
      \end{center}}%
     {\par\vfil\null\endtitlepage}
\else
  \newenvironment{abstract}{%
      \if@twocolumn
        \section*{\abstractname}%
      \else
        \small
        \begin{center}%
          {\bfseries \abstractname\vspace{-.5em}\vspace{\z@}}%
        \end{center}%
        \quotation
      \fi}
      {\if@twocolumn\else\endquotation\fi}
\fi
\makeatother

\makeatother

\usepackage{babel}
  \addto\captionsenglish{\renewcommand{\remarkname}{Remark}}
  \addto\captionsenglish{\renewcommand{\theoremname}{Theorem}}
  \addto\captionsitalian{\renewcommand{\remarkname}{Osservazione}}
  \addto\captionsitalian{\renewcommand{\theoremname}{Teorema}}
  \providecommand{\remarkname}{Remark}
\providecommand{\theoremname}{Theorem}

\begin{document}
\begin{titlepage} 
\begin{center} {{\Large{\textsc{Universit\`a degli Studi di Salerno}}}} \rule[0.1cm]{15.0cm}{0.1mm} 
\rule[0.5cm]{15.0cm}{0.6mm} {\small{\bf Corso di Laurea in Fisica \\ Tesi di Laurea Magistrale }} 
\end{center} 
\vspace{15mm} 

\begin{figure}[!] 
\centering 
\includegraphics[scale=0.40]{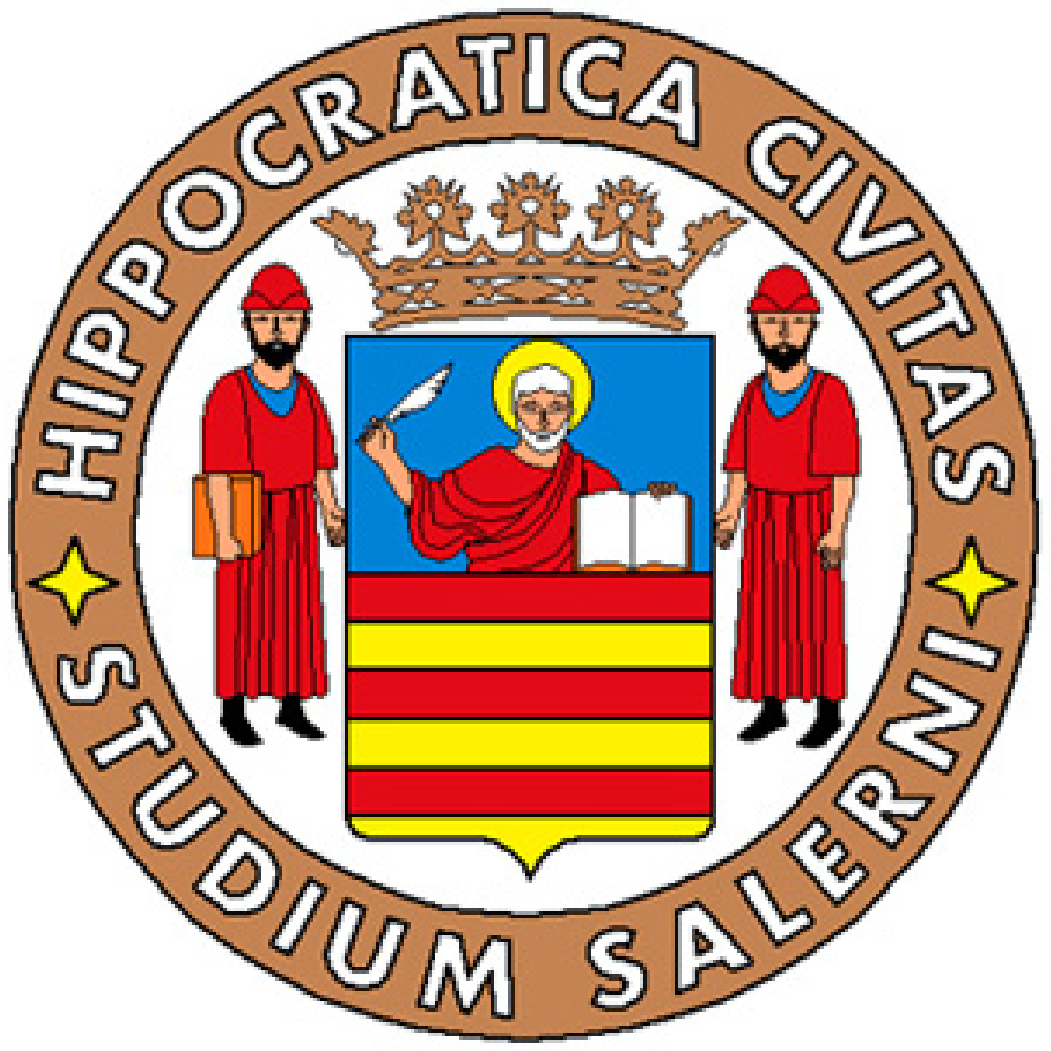} 
\end{figure}

\begin{center} {\Large{\bf Ghost and singularity free theories of gravity}}\\ 
\end{center} 
\vspace{40mm} 
\par 
\noindent 
\begin{minipage}[t]{0.47\textwidth} 
{\large{\bf Candidate:\\ 
Luca Buoninfante}} 
\vspace{4mm}
{\large{\bf \\Registration Number:\\ 
0522600065}} 
\end{minipage} 
\hfill 
\begin{minipage}[t]{0.47\textwidth} \raggedleft 
{\large{\bf Supervisors:\\ 
Prof. Gaetano Lambiase}}
\vspace{0.2mm}
{\large{\bf \\Dr. Anupam Mazumdar}}
\vspace{4mm}
{\large{\bf \\Co-Supervisor:\\ 
Prof. Massimo Blasone}} 
\end{minipage} 
\vspace{20mm} 
\begin{center}{\large{\bf Academic Year 2015/2016}}\\
\end{center} 
\end{titlepage}

\newpage
\thispagestyle{empty}

$ $

\newpage

\begin{minipage}{\textwidth}

\title{Ghost and singularity free theories of gravity}

\author{L. Buoninfante}

\date{{Dipartimento di Fisica ``E.R. Caianiello”
Università di Salerno, I-84084 Fisciano (SA), Italy.} }

\maketitle
\vspace{0.5cm}

\begin{abstract}
Albert Einstein's General Relativity (GR) from 1916 has become the widely accepted theory of gravity and been tested observationally to a very high precision at different scales of energy and distance. At the same time, there still remain important questions to resolve. At the classical level cosmological and black hole singularities are examples of problems which let us notice that GR is incomplete at short distances (high energy). Furthermore, at the quantum level GR is not ultraviolet (UV) complete, namely it is not perturbatively renormalizable. Most of the work try to solve these problems modifying GR by considering finite higher order derivative terms. Fourth Derivative Gravity, for example, turns out to be renormalizable, but at the same time it introduces ghost. To avoid both UV divergence and presence of ghost one could consider sets of infinite higher derivative terms that can be expressed in the form of entire functions satisfying the special property do not introduce new poles other than GR graviton one. By making a special choice for these entire functions, one could show that such a theory describes a gravity that, at least in the linear regime, can avoid both the presence of ghost and classical singularities (both black hole and cosmological singularities).\\ 
In this master's thesis we review some of these aspects regarding gravitational interaction, focusing more on the classical level. Most of the calculations are done in detail and an extended treatment of the formalism of the spin projector operators is presented.
\end{abstract}

\end{minipage}

\newpage
\thispagestyle{empty}

$ $

\selectlanguage{italian}%
\thispagestyle{empty}

\selectlanguage{english}%

\date{$ $}

\newpage
\thispagestyle{empty}

\tableofcontents{}

\chapter*{Convention and notation\addcontentsline{toc}{chapter}{Convention and notation} 
\markboth{Convention and notation}{CONVENTION AND NOTATION}}

\section*{Natural units}

We are going to list all the conventions and the notations we shall
use in this thesis. We shall also follow Ref. {[}\ref{-M.-Schwartz,}{]}.\\
In relativistic quantum field theory, it is standard to set
\[
c=2.998\times10^{8}\, m\, s^{-1}=1
\]
which turns meters into seconds and
\[
\hbar=\frac{h}{2\pi}=1.054572\times10^{-34}\, J\, s=1
\]
which turns joules into inverse seconds.\\
The use $\hbar=1=c$ units (natural units) can simplify particle physics
notation considerably. Since one typically deals with particles that
are both relativistic and quantum mechanical, a lot of $\hbar's$
and $c'$s will encumber the equations if natural units are not adopted.
This makes all quantities have dimensions of energy (or mass, using
$E=mc^{2})$ to some power. Quantities with positive mass dimension,
(e.g. momentum $p$) can be thought of as energies and quantities
with negative mass dimension (e.g. position $x$) can be thought of
as lengths.\\
Some examples are:
\[
[dx]=[x]=[t]=M^{-1},
\]
\[
[\partial_{\mu}]=[p_{\mu}]=[k_{\mu}]=M^{1},
\]
\[
[\mathrm{velocity}]=\frac{[x]}{[t]}=M^{0}.
\]
Thus
\[
[dx^{4}]=M^{-4}.
\]
The action is a dimensionless quantity
\[
[S]=[\int d^{4}x\mathcal{L}]=M^{0},
\]
and, consequently, it implies that 
\[
[\mathcal{L}]=M^{4}.
\]
For example, a free scalar field has Lagrangian%
\footnote{To be more rigorous we should call $\mathcal{L}$ Lagrangian density.
While the Lagrangian is defined as $L=\int d^{3}x\mathcal{L}.$ However
in this thesis we shall just use the word Lagrangian.%
} $\mathcal{L}=\frac{1}{2}\partial_{\mu}\phi\partial^{\mu}\phi$ so
\[
[\phi]=M^{1}
\]
and so on. In general bosons, whose kinetic terms have two derivatives,
have mass dimension $1$ and fermions, whose kinetic terms have one
derivative, have mass dimension $\frac{3}{2}.$ \\
Since $\hbar=1$ then $p_{\mu}=k_{\mu},$ so in this thesis we will
use the four-wave vector to indicate the four-momentum.

\section*{$2\pi-$factors}

Often people get confused because they don't know whether consider
or not the factors of $2\pi.$ The origin of all $2\pi'$s is the
relation
\[
\delta(x)=\int_{-\infty}^{+\infty}dpe^{\pm2\pi ikx}.
\]
This identity holds with either sign. To remove the $2\pi$ from the
exponent, we can rescale either $x$ or $p.$ Since position generally
is not an angular coordinate, it makes sense to rescale $p.$ Then
\[
2\pi\delta(x)=\int_{-\infty}^{+\infty}dpe^{\pm ikx}.
\]
Our convention for Fourier transforms is that the momentum space integrals
have $\frac{1}{2\pi}$ factors while position space integrals have
no $2\pi$ factors:
\[
f(x)=\int\frac{d^{4}p}{(2\pi)^{4}}\tilde{f}(p)e^{\pm ikx},\,\,\,\,\tilde{f}(p)=\int d^{4}xf(x)e^{\pm ikx}
\]
Since we adopt the convention that the Fourier transform of the partial
derivative $\partial_{\mu}$ is $ik_{\mu},$ 
\[
\partial_{\mu}f(x)=ik_{\mu}f(x)),
\]
we choose the ``$+$'' sign for $f(x)$ and the ``$-$'' sign
for $\tilde{f}(p).$

\section*{Metric signature}

In the Minkowski space-time the metric has signature 
\[
\eta_{\mu\nu}=\begin{pmatrix}+1\\
 & -1\\
 &  & -1\\
 &  &  & -1
\end{pmatrix}.
\]
Because of this convention one has $k^{2}=k_{0}^{2}-\bar{k}^{2}=m^{2}>0.$
The alternative choice, $\eta_{\mu\nu}=\mathrm{diag(-1,+1,+1,+1)},$
would make $k^{2}<0.$ Note that with the symbol ``$\,\,\bar{}\,\,"$
we indicate the three-spatial vectors. Then, the norm of a three-vector
$\bar{a}$ is indicated by $|\bar{a}|.$ \\
The sign of terms in Lagrangians is set so that they have positive
energy density. In fact, given the Lagrangian $\mathcal{L}=\mathcal{L}_{kin}-V,$
we know that the potential energy $V$ is positive in a stable system.
For example, for a scalar field the mass term $\frac{1}{2}m^{2}\phi^{2}$
gives positive energy, so $V=\frac{1}{2}m^{2}\phi^{2}$ and $\mathcal{L}=\mathcal{L}_{kin}-\frac{1}{2}m^{2}\phi^{2}.$
The kinetic term sign can then be chosen to obtain the correct dispersion
relation. Thus, since $\boxempty=\partial_{\mu}\partial^{\mu}\rightarrow-k^{2}$
in  momentum space and $k^{2}=m^{2}$ on-shell, the field equations
of motion is $(\boxempty+m^{2})\phi=0.$%
\footnote{If we consider the Fourier transform of this field equation we obtain
$(-k^{2}+m^{2})\phi(k)=0,$ that implies $k^{2}=m^{2},$ i.e. the
dispersion relation is satisfied. %
} Therefore we have%
\footnote{To obtain the second form for the Lagrangian we have integrated by
parts and neglected the surface terms.%
}
\[
\mathcal{L}=-\frac{1}{2}\phi(\boxempty+m^{2})\phi=\frac{1}{2}\partial_{\mu}\phi\partial^{\mu}\phi-\frac{1}{2}m^{2}\phi^{2}.
\]
Since the propagator is defined as the inverse of the operator $-(\boxempty+m^{2}),$
its form also depends on the convention sign:
\[
\mathcal{P}(x-y)=\int\frac{d^{4}k}{(2\pi)^{4}}e^{i(x-y)}\frac{1}{k^{2}-m^{2}+i\epsilon}.
\]
$ $\\
For other kind of Lagrangians, like photon and graviton ones, the
signs are always chosen in a way to give positive defined energy and
consistency with dispersion relation. \\
\\
In all the equations we shall employ the modern summation convention
where contracted indices can be raised or lowered without ambiguity:
\[
a\cdot b=a_{\mu}b^{\mu}=a^{\mu}b_{\mu}=\eta_{\mu\nu}a^{\mu}b^{\nu}=\eta^{\mu\nu}a_{\mu}b_{\nu}.
\]

\section*{Indices}

In this thesis  we shall use the index $0$ for the temporal coordinate,
and the other indices $1,2,3$ for the spatial coordinates. \\
Then, latin indices $i,j,k,l$ and so on generally run over three
spatial coordinate labels, usually, $1,2,3$ or $x,y.z.$ Greek indices
$\mu,\nu,\rho,\sigma$ and so on generally run over the four coordinate
labels in a general coordinate system.\\
Note that in this thesis we shall frequently suppress the indices,
especially when we work with the spin projector operators. Thus, for
instance, $\mathcal{P}_{\mu\nu\rho\sigma}^{2}$ will be just written
as $\mathcal{P}_{\mu\nu\rho\sigma}^{2},$ and in the same way also
in the formulas that contain the spin projector operators there will
be a suppression of the indices.

$ $

The indices are lowered and raised by the metric tensor $g_{\mu\nu}(x)$,
that in the Minkowski space-time is represented by $\eta_{\mu\nu}.$
\\
The adopted conventions for the geometric objects are given by:\\
Riemann tensor:

\[
\mathcal{R_{\,\mu\lambda\nu}^{\alpha}}=\partial_{\lambda}\Gamma_{\,\mu\nu}^{\alpha}-\partial_{\nu}\Gamma_{\,\lambda\mu}^{\alpha}+\Gamma_{\,\lambda\rho}^{\alpha}\Gamma_{\,\mu\nu}^{\rho}-\Gamma_{\,\nu\rho}^{\alpha}\Gamma_{\,\lambda\mu}^{\rho};
\]
Ricci tensor, $\mathcal{R}_{\mu\nu}=\mathcal{R}_{\,\mu\alpha\nu}^{\alpha}=g^{\alpha\rho}\mathcal{R}_{\alpha\mu\rho\nu}:$
\[
\mathcal{R}_{\mu\nu}=\partial_{\alpha}\Gamma_{\,\mu\nu}^{\alpha}-\partial_{\nu}\Gamma_{\,\mu\alpha}^{\alpha}+\Gamma_{\,\mu\nu}^{\alpha}\Gamma_{\,\alpha\beta}^{\beta}-\Gamma_{\,\mu\beta}^{\alpha}\Gamma_{\,\nu\alpha}^{\beta};
\]
Then, we also have the curvature scalar $\mathcal{R}=\mathcal{R}_{\,\mu}^{\mu}=g^{\mu\nu}\mathcal{R}_{\mu\nu}.$
\\
By lowering the upper index with the metric tensor we can obtain the
completely covariant Riemann tensor:
\[
\mathcal{R}_{\mu\nu\lambda\sigma}=\frac{1}{2}\left(\partial_{\nu}\partial_{\lambda}g_{\mu\sigma}+\mathcal{\partial_{\mu}\partial_{\sigma}}g_{\nu\lambda}-\mathcal{\partial_{\sigma}\partial_{\nu}}g_{\mu\lambda}-\partial_{\mu}\partial_{\lambda}g_{\nu\sigma}\right)+g_{\alpha\beta}\left(\Gamma_{\,\nu\lambda}^{\alpha}\Gamma_{\,\mu\sigma}^{\beta}-\Gamma_{\,\sigma\nu}^{\alpha}\Gamma_{\,\mu\lambda}^{\beta}\right).
\]
It is worth to introduce the linearized forms for Riemann tensor,
Ricci tensor and scalar curvature as we shall frequently use them.
By performing the following perturbation around Minkowski metric,
\[
g_{\mu\nu}(x)=\eta_{\mu\nu}+h_{\mu\nu}(x),
\]
the curvature tensors become 
\[
\begin{array}{rl}
\mathcal{R}_{\mu\nu\lambda\sigma}= & \frac{1}{2}\left(\partial_{\nu}\partial_{\lambda}h_{\mu\sigma}+\mathcal{\partial_{\mu}\partial_{\sigma}}h_{\nu\lambda}-\mathcal{\partial_{\sigma}\partial_{\nu}}h_{\mu\lambda}-\partial_{\mu}\partial_{\lambda}h_{\nu\sigma}\right),\\
\\
\mathcal{R_{\mu\nu}}= & \frac{1}{2}\left(\partial_{\rho}\partial_{\nu}h_{\mu}^{\rho}+\partial_{\rho}\partial_{\mu}h_{\nu}^{\rho}-\partial_{\mu}\partial_{\nu}h-\boxempty h_{\mu\nu}\right),\\
\\
\mathcal{R}= & \partial_{\mu}\partial_{\nu}h^{\mu\nu}-\boxempty h.
\end{array}
\]
$ $\\
Let us introduce a notation for the expressions containing either
symmetric or antisymmetric terms. The indices enclosed in parentheses
or brackets satisfy, respectively, the properties of symmetry or antisymmtery
defined by the following rules:
\[
T_{(\mu\nu)}\equiv\frac{1}{2}\left(T_{\mu\nu}+T_{\nu\mu}\right),\,\,\,\,\, T_{[\mu\nu]}\equiv\frac{1}{2}\left(T_{\mu\nu}-T_{\nu\mu}\right).
\]
$ $\\
Finally, let us discuss on the coupling constants appearing in the
Einstein equations. The usual form of the field equation for General
Relativity is given by
\[
G_{\mu\nu}\equiv\mathcal{R}_{\mu\nu}-\frac{1}{2}g_{\mu\nu}\mathcal{R}=\kappa\tau_{\mu\nu},
\]
where $G_{\mu\nu}$ is the Einstein tensor and $\tau_{\mu\nu}$ the
energy-momentum tensor. \\
It is important to explicit the form of the constant $\kappa$ in
natural units. Its expression in SI units is well known, and is given
by $\kappa=\frac{8\pi G}{c^{4}},$ where the value of the Newton constant
is $G=6.67\times10^{-8}\, g^{-1}\, cm^{3}\, s^{-2}.$ In natural units,
since $c=1,$ one has $\kappa=8\pi G.$ Often it is useful to display
the Planck mass in the gravitational field equations. Indeed, the
Planck mass is defined as 
\[
m_{P}\coloneqq\sqrt{\frac{\hbar c}{G}}\simeq1.2\times10^{19}\, GeV/c^{2}
\]
and in natural units $G=\frac{1}{M_{P}^{2}}.$ To get rid of the $2\pi$
factor is useful to introduce the reduced Planck mass that is defined
as
\[
M_{P}\coloneqq\sqrt{\frac{\hbar c}{8\pi G}}\simeq2.4\times10^{18}\, GeV/c^{2}=4.3\times10^{-9}\, kg.
\]
In this way the coupling constant reads as 
\[
\kappa=\frac{1}{M_{P}^{2}},
\]
and the gravitational field equations turn out to be expressed in
terms of the reduced Planck mass.

\section*{Acronyms }

GR: General Relativity%
\footnote{In this thesis we shall frequently use this acronym. Some authors
use the expression General Relativity also to refer to modified Einstein-Hilbert
action because the most of the fundamental principles of Einstein's
theory are still valid. Anyway, in this thesis every time we use the
expression General Relativity (or its acronym) we mean Einstein's
GR, whose action is the Einstein-Hilbert action.%
}.

H-E: Einstein-Hilbert.

ED: ElectroDynamics.

IDG: Infinite Derivative theories of Gravity.

UV: UltraViolet.

\chapter*{Introduction\addcontentsline{toc}{chapter}{Introduction} 
\markboth{Introduction}{INTRODUCTION}}

\renewcommand{\theequation}
{\arabic{equation}}
\setcounter{equation}{0}

Albert Einstein's General Relativity (GR) from 1916 has become the
widely accepted theory of gravity and been tested observationally
to a very high precision. Although a vast amount of observational data {[}\ref{-M.W.-Clifford,}{]} have made GR a remarkable theory, there still remain fundamental questions to resolve. At the classical level, cosmological and black hole
singularities are examples of problems which
let us suspect that the theory is incomplete at short distances (high energy). Furthermore, at the quantum level GR is not UV complete,
namely it turns out to be perturbatively non-renormalizable. \\
One can easily do a power counting to understand whether GR, i.e.
Einstein-Hilbert action, can be renormalizable {[}\ref{-R.-Percacci,}{]}.
The superficial degree of divergence of a loop integral in GR turns out to be%
\footnote{Note that the superficial degree of divergence in any dimension $d$
is given by $D=dL-2I+2V.$ %
} 
\begin{equation}
D=4L-2I+2V,
\end{equation}
where $L$ is the number of loops, $I$ is the number of internal propagators and  $V$ is the number of vertices. Using the well known topological relation
\begin{equation}
L=1+I-V,
\end{equation}
we get
\begin{equation}
D=2L+2.
\end{equation}
Thus, as the number of loops increases the superficial degree of divergence increases too, making the theory of GR perturbatively non-renormalizable.\\
In 1972, ’t Hooft and Veltman {[}\ref{-G.-=002019t}{]} calculated
the one-loop effective action of Einstein’s theory. They found that
gravity coupled to a scalar field is non-renormalizable, but also
showed how to introduce counter-terms to make pure GR finite at one-loop.
In the following years the non-renormalizability of gravity coupled
to various types of matter was also established. The crucial result
was only obtained several years later by Goroff and Sagnotti {[}\ref{-M.H.-Goroff}{]}
and van de Ven {[}\ref{-A.-E. M}{]}, who showed the existence of
a divergent term cubic in curvature in GR action at two loops. \\
All these works suggested that the perturbative treatment of Einstein’s
theory as a quantum field theory, either on its own or coupled to
generic matter fields, leads to the appearance of divergences that
spoil the predictivity of the theory. There were subsequently several
attempts that tried to resolve this problem. Most of them emerge in
the context of quantum field theory, while other attempts are based
on different principles. Below we list different approaches that physicists
used to follow and are still following {[}\ref{-R.-Percacci,}{]}.
\begin{itemize}
\item First, one could change the gravitational action, so the field equations.
Examples of this approach are $f(\mathcal{R})$ theories, Fourth Derivative
Gravity, Infinite Derivative Gravity (IDG) and so on. In 1977, Stelle
proved that a theory containing four-derivative terms in the Lagrangian
(i.e. terms quadratic in curvature) is perturbatively renormalizable
{[}\ref{-Stelle...renormaliz}{]}. Unfortunately it also appeared
that this kind of Lagrangian leading to a renormalizable theory contain
propagating ghosts%
\footnote{See Appendix $C$ for a discussion on ``good'' and ``bad'' ghosts.%
}. These ghosts, would be physical particles and hence would violate
the unitarity. On the other hand, a Lagrangian that do not contain
ghosts turns out non-renormalizable. Thus, at the perturbative level,
it seems to be present a problem of incompatibility between unitarity
and renormalizability.
\item A second attempt was based on the introduction of new particles and
new symmetries, creating a new theory of gravity. So far the most
important examples in this class are supergravity theories (SUGRA),
whose pioneers are Freedman, Ferrara and van Nieuwenhuizen {[}\ref{-D.Z.-Freedman,}{]}.
Supersymmetric theories are very special because the balance of bosonic
and fermionic degrees of freedom leads to cancellation of divergences
in loop diagrams and indeed even the simplest SUGRAs do not have the
two-loop divergence that is present in GR. Besides the improved quantum
behavior, these theories have other kind of either theoretical and
experimental difficulties that thwarted this hope. 
\item A third possibility is that the non-renormalizability is an intrinsic
pathology of the perturbative approach, and not of gravity itself.
There have been more than one way of implementing this idea. The Hamiltonian
approach to quantum gravity can be viewed as falling in this broad
category. Examples of this subapproach are Geometrodynamics and Loop
Quantum Gravity. There was also the covariant formalism, in which
most work has been based, more or less explicitly, on the Feynman
“sum over histories” approach. Misner was one of the pioneer {[}\ref{-C.-W.misner}{]}.
Different versions of the gravitational functional integral were developed,
like the Euclidean version and the lattice approach. Then, there was
the non-perturbative way out of the issue of the UV divergences that
is known as “non-perturbative renormalizability” and originates from
the work of Wilson on the renormalization group. 
\item A very popular attempt that doesn't follow the principles of quantum
field theory is given by String Theory {[}\ref{-M.-Green,}{]}. This
is the main approach to construct an unifying quantum framework of
all interaction. The quantum aspect of the gravitational field only
emerges in a certain limit in which the different interactions can
be distinguished from each other. All particles have their origin
in excitations of fundamental strings. The fundamental scale is given
by the string length; it is supposed to be of the order of the Planck
length. 
\end{itemize}
$ $\\
This thesis is focused on the first category of attempts. Most of
the work on modifying GR has focused upon studying finite higher order
derivative gravity and, as we have already mentioned, an example is
Fourth Order Derivative Theory of Gravity by Stelle which is quadratic in curvatures.
In 1977, Stelle considered the following action%
\footnote{Such modified gravitational action has been also studied in Ref. {[}\ref{-G.-Lambiase,}{]},{[}\ref{-S.-Capozziello,}{]},{[}\ref{-G.-Lambiase,-1}{]},{[}\ref{-S.Capozziello,-G.}{]}
and {[}\ref{-S.-Calchi}{]}.%
}$^{,}$%
\footnote{Note that the square Riemann tensor doesn't appear in Stelle action
because of the existence of the so called Euler topological invariant.
In fact, the following relation holds:
\[
\mathcal{R}^{\mu\nu\rho\sigma}\mathcal{R}_{\mu\nu\rho\sigma}-4\mathcal{R}^{\mu\nu}\mathcal{R}_{\mu\nu}+\mathcal{R}^{2}=\nabla_{\mu}K^{\mu},
\]
 where the total derivative gives a zero contribute in the action.
Thus, the Riemann tensor can be rewritten in terms of the Ricci tensor
and curvature scalar, unless than a total derivative.%
}
\begin{equation}
S=\int d^{4}x\sqrt{-g}\left(\alpha\mathcal{R}+\beta\mathcal{R}^{2}+\gamma\mathcal{R}^{\mu\nu}\mathcal{R}_{\mu\nu}\right)
\end{equation}
and he proved that the theory is renormalizable for appropriate values
of the coupling constants. Unfortunately, precisely for these values
of the coupling constants the theory exhibits a bad behavior. It has a negative
energy propagating degree of freedom that causes instability around the Minkowski vacuum and violation of unitary in the quantum regime. The spin-$2$ component of the
UV modified graviton propagator is roughly given by
\begin{equation}
\Pi=\Pi_{GR}-\frac{\mathcal{P}^{2}}{k^{2}-m^{2}},
\end{equation}
and it shows the presence of the so called Weyl ghost in the spin-$2$
component%
\footnote{It is a ghost because of the presence of the minus sign that comes
from a negative kinetic energy in the Lagrangian. In Appendix $C$
more details on ghost and unitarity are discussed. %
} that violates stability and unitarity conditions. Thus, in the UV
regime the special form of modified Stelle propagator makes the loop integral appearing in the Feynman diagrams convergent at $1$-loop and beyond but unfortunately this costs the presence of a massive spin-$2$ ghost.\\
Instead, as for $f(\mathcal{R})$ we have the opposite situation.
In fact, the theory is free-ghost but at the same time is non-renormalizable.
It seems that one is lead to conclude that there is incompatibility
between renormalizability and unitarity.\\
In 1989 Kuz'min {[}\ref{-kuzmin}{]} and in 1997 Tomboulis {[}\ref{-E.-T.}{]} noticed that if one considers a non-polynomial Lagrangian containing an infinite series in higher derivatives gauge theories and theories of gravity can be made perturbatively super-renormalizable. Following this direction, in 2006 the authors Biswas, Mazumdar and Siegel {[}\ref{-T.-Biswas,bouncing}{]}
argued that the absence of propagating ghosts and the renormalizability of the theory
can $\mathit{only}$ be realized if one considers an infinite number
of derivative terms, and in particular it could be done by making use of the exponential function,
that are allowed by the condition of general covariance {[}\ref{-T.-Biswas, conroy}{]}.
In such a modified gravity they also argued that the theory could
be asymptotically free. The infinite derivative action considered
in {[}\ref{-T.-Biswas,bouncing}{]} includes cosmological
non-singular bouncing solutions, i.e. solutions that avoid the presence of Big
Bang and Big Crunch. Based on the action introduced in {[}\ref{-T.-Biswas,bouncing}{]} other progress was made. Cosmological perturbation
analysis were performed in Ref. {[}\ref{-Biswas-T, cosmological pert}{]}
which makes the bouncing model more robust,
and tells us about some possible connection to inflationary cosmology.
All these results suggest that it might be possible to make gravitational
interaction weaker both at short distances and
at early times, in a consistent way. \\
Indeed, in 2012 Biswas, Gerwick, Koivisto and Mazumdar {[}\ref{-T.-Biswas,prl}{]}
noticed that, by including an infinite number of derivative terms, the theory could be asymptotically free in the UV regime preserving general covariance and without violating fundamental physical principles, such as unitarity. The action considered by the authors in {[}\ref{-T.-Biswas,prl}{]}
is given by%
\footnote{See Ref. {[}\ref{-kuzmin}{]}, {[}\ref{-E.-T.}{]}, {[}\ref{-T.-Biswas,bouncing}{]}, {[}\ref{-moffat}{]}  and {[}\ref{-modesto}{]} for previous works in which the same idea to introduce infinitely many higher derivatives appears. See also {[}\ref{-anselmi}{]} for a different interesting way to proceed. %
}$^{,}$%
\footnote{See also {[}\ref{-shapmod1}{]}, {[}\ref{-shapmod2}{]}, {[}\ref{-S.-Capozziello}{]}, {[}\ref{-S.-Capozziello-1}{]},
{[}\ref{-S.-Capozziello-2}{]}, {[}\ref{-S.-Capozziello,-1}{]} and
{[}\ref{-G.-K.}{]}.%
}$^{,}$%
\footnote{To be more precise we should write $\mathcal{F}_{i}(\frac{\boxempty}{M^{2}})$
to have a dimensionless argument ($[\boxempty]=[M]^{2}),$ but for
simplicity we shall always write just $\mathcal{F}_{i}(\boxempty),$
implying that a squared mass is present also in the denominator.%
}
\begin{equation}
S=\int\sqrt{-g}\left(-\mathcal{R}+\mathcal{R}\mathcal{F}_{1}(\boxempty)\mathcal{R}+\mathcal{R}^{\mu\nu}\mathcal{F}_{2}(\boxempty)\mathcal{R}_{\mu\nu}+\mathcal{R}^{\mu\nu\rho\sigma}\mathcal{F}_{3}(\boxempty)\mathcal{R}_{\mu\nu\rho\sigma}\right),\label{eq:int6}
\end{equation}
where the $\mathcal{F}_{i}(\boxempty)'$s are functions of the D'Alambertian
operator, $\boxempty=g^{\mu\nu}\nabla_{\mu}\nabla_{\nu},$ and contain
an infinite set of derivatives:
\begin{equation}
\mathcal{F}_{i}(\boxempty)=\sum_{n=0}^{\infty}f_{i,n}\boxempty^{n},\,\,\,\,\, i=1,2,3.
\end{equation}
One requires that the $\mathcal{F}_{i}(\boxempty)'$s are analytic
at $\boxempty=0$ so that one can recover GR in the infrared regime.
Theories described by the action (\ref{eq:int6}) are called $\mathit{Infinite}$
$\mathit{Derivative}$ $\mathit{theories}$ $\mathit{of}$ $\mathit{Gravity}$
(IDG). \\
It is also interesting the fact that such infinite higher derivative actions
appear in non-perturbative string theories.\\
Making an appropriate choice for the functions $\mathcal{F}_{i}(\boxempty),$
the authors in Ref. {[}\ref{-T.-Biswas,prl}{]}
\footnote{See also Ref. {[}\ref{-modesto}{]}.%
} obtained the following modified propagator {[}\ref{-Biswas-T,non local}{]}
\begin{equation}
\Pi=\frac{1}{a(k^{2})}\Pi_{GR},
\end{equation}
where
\begin{equation}
a(k^{2})=e^{\frac{k^{2}}{M^{2}}},
\end{equation}
whose expression in coordinate space is given by
\begin{equation}
a(\boxempty)=e^{-\frac{\boxempty}{M^{2}}}.
\end{equation}
Thus, the GR propagator is just modified by the multiplicative factor
$a(k^{2}).$ Since $a(k^{2})$ has no zeros on the complex plane%
\footnote{Typical functions with this kind of characteristic can be written
as
\[
a(\boxempty)=e^{-\gamma(\boxempty)},
\]
 where $\gamma(\boxempty)$ is an analytic function of $\boxempty.$
Physically, as it has already said for the functions $\mathcal{F}_{i}(\boxempty),$
the analyticity property of $\gamma(\boxempty)$ is required to recover
GR in the infrared limit. It is then easy to see that for any polynomial
$\gamma(\boxempty),$ as long as the highest power has positive coefficient,
the propagator will be even more convergent than the exponential case
{[}\ref{-T.-Biswas,}{]}, {[}\ref{-T.-Biswas,prl}{]}.%
}, this modification doesn't introduce any new pole, in fact the only
present pole is the massless transverse and traceless graviton. The
parameter $M$ corresponds to the scale at which the modification
becomes important. Since this model is non-local because of the presence
of an infinite set of higher derivatives, $M$ could be the scale
at which non-local effects emerge. \\
So far, good and promising results have been obtained in the linearized regime.
In fact, the authors in Ref. {[}\ref{-T.-Biswas,prl}{]} have
found that this models describe a singularity-free gravity, although
their result only holds for mini black holes with a mass much smaller than the
Planck mass%
\footnote{The result holds in the Newtonian approximation where the gravitational
potential is very weak.%
}. Moreover, as we already mentioned, the theory also admits periodic
cosmological solutions with bounce showing a scenario in which the singularity issue of the standard cosmology could be solved. \\
However, to say something more about classical singularities one should
study the full theory in a generic curved background. In Ref. {[}\ref{-T.-Biswas, conroy}{]}
the authors have obtained the full field equations for the most general
IDG action quadratic in the curvatures we wrote above and also
verified the consistency with the already known results in the linear
regime. One of the main aim of these IDG theories is to study astrophysical
black holes in the framework of the full (i.e. not linearized) theory
and try to understand whether singularities are present, but so far
no new meaningful results have been obtained.\\
$ $\\
As for the quantum level, the current results from quantum loop computations
in IDG theories seem very promising. In particular it seems that the presence of the exponential functions containing infinite derivatives could make convergent the loop integrals in the Feynman diagrams, giving a strong clue for the renormalizability of the theory. So far, just a toy
scalar model has been considered to face the problem of renormalizability by the authors in Ref. {[}\ref{-S.-Talaganis,}{]}.
They got a modified superficial degree of divergence due to
the presence of the exponential contribution; the relevant term is given by
\begin{equation}
D=1-L.
\end{equation}
Thus, if the number of loops is such that $L\geq2,$ the superficial
degree of divergence, i.e. the power momenta in the integrals, becomes negative and the loop amplitudes are
superficially convergent. Unlike what happens in GR,
it seems that IDG theories can be perturbatively renormalizable. \\
For a toy scalar model it has been found that the 2-point
function is divergent at one loop but, by adding appropriate counter terms, it can be made renormalized and the UV behavior of all other
$1$-loop diagrams as well as the 2-loop, 2-point function show the same renormalizable behavior. \\
Modesto in Ref. {[}\ref{-modesto}{]} (see also {[}\ref{-modesto2}{]}) reconsiders the theory of Tomboulis {[}\ref{-E.-T.}{]} and states that by introducing special entire functions the theory of gravitational interaction can be made renormalizable at one loop and also at higher loops. In this way, since only a finite number of diagrams diverges in the UV limit the theory should be super-renormalizable.

$ $

\textbf{Organization of this thesis}\\
\textbf{$ $}\\
The goal of this thesis is to reach a satisfactory understanding of
$\mathit{ghost}$ $\mathit{and}$ $\mathit{singularity}$ $\mathit{free}$
$\mathit{theories}$ $\mathit{of}$ $\mathit{gravity}$ in the context
of IDG models. Our study is mostly focused on the classical aspects,
although something is said about the quantum level.\\
The work is structured as follows: 
\begin{lyxlist}{00.00.0000}
\item [{Chapter$\,\,1:$}] We shall study the theory of photon field to
worm up before dealing with the theory of graviton field. By starting
from the photon Lagrangian we determine the main results, like the
field equations, the counting of the degrees of freedom for both on-shell
and off-shell photon and the photon propagator. We work by using the
formalism of the spin projector operator by which the spin components
of the photon field are more explicit. We also determine a set of
polarization vectors in terms of which we rewrite the photon propagator.
In the last subsection we perform the unitarity analysis verifying
whether the theory contains ghosts.
\item [{Chapter$\,\,2:$}] Once warming up with the theory of the photon
field, we can easily approach the theory of the graviton field that
corresponds to the linearized Einstein gravity. This chapter has the
same structure of the previous one. Also here we make extensive use
of the formalism of the spin projector operators. This formalism turns
out to be very useful either to determine the graviton propagator
and to distinguish the different graviton spin components. We shall
see that the physical part of the propagator (saturated propagator)
has a scalar component other than the spin-$2$ component. We also
determine a set of polarization tensors in terms of which we rewrite
the graviton propagator. In the last subsection we perform also for
the graviton Lagrangian the unitarity analysis showing that ``bad''
ghosts are absent, but there is the presence of a ``good'' ghost,
corresponding to the scalar graviton component, that is fundamental
either to ensure the unitarity of the theory and to give the exact
number of propagating mode components for the graviton, i.e. the two
traceless and transverse ones.
\item [{Chapter$\,\,3:$}] We  introduce the most general quadratic action
of gravity. We consider the linear regime and determine field equations
and propagator always by using the spin projector operators. Without
specifying the form of the coefficients that appear in the theory
we cannot say anything about either how many degrees of freedom propagates
and unitarity. Indeed at the end of the chapter we make three special
choices for the coefficients and we obtain GR, $f(\mathcal{R})$ theory
and conformally invariant gravity as subclasses. 
\item [{Chapter$\,\,4:$}] The starting point is the linear quadratic action
obtained in the previous chapter. We are going to make a special choice
for the coefficients to obtain a free-ghost theory of gravity. We
also notice that the theory is singularity-free in the weak approximation
(Newtonian approximation) and we are able to state that mini black
holes don't have any singularities. Indeed, we obtain a modified non-singular
Newtonian potential that gives us the well known Newtonian potential
in the infrared limit. At the end of the chapter we discuss about
the parameter $M$ coming from the exponential factor, trying to understand
its physical meaning and to obtain some constraints.
\item [{Conclusions:}] We summarize what we have done in this thesis and
also give an overview of what physicists are working on.
\item [{Appendix$\,\, A:$}] We discuss on the concepts of on-shell and
off-shell particles.
\item [{Appendix$\,\, B:$}] We treat in details the vector and tensor
decompositions in terms of the spin projector operators. First we
introduce the group representations and then the associated spin projector
decompositions.
\item [{Appendix$\,\, C:$}] We introduce the concept of unitarity. We
then give the definition of ghost field and discuss the difference
between ``good'' and ``bad'' ghost in connection with the violation
of the unitarity. We also give a prescription to verify whether ghosts
violate the unitarity conditions, i.e whether bad ghosts are present.
In the last section we discuss Fourth Derivative Gravity as an application.
\item [{Appendix$\,\, D:$}] We present the table of the Clebsch-Gordan
coefficients because it turns to be useful to determine the graviton
polarization tensors by starting from the photon polarization vectors.
\end{lyxlist}
\renewcommand{\theequation}
{\arabic{chapter}.\arabic{equation}}

\chapter{Vector field: photon}

In this first chapter we are going to treat the theory of the vector
field that describes the propagation of the electromagnetic wave (photon)
in electrodynamics (ED). It will turns out to be a good exercise to
warm up before discussing the linearized GR, i.e. the theory of the
symmetric two-rank tensor, that we shall treat in the next chapter.
We have organized both chapters in the same way, but, of course, for
GR case the work will be harder as we have to deal with a tensor field
and not with a simple vector field. In particular, for both chapters,
we shall use the $\mathit{spin}$ $\mathit{projectors}$ $\mathit{formalism}$
that will be very useful to calculate the propagator and understand
which are the physical spin components {[}\ref{-P.-Van}{]}, {[}\ref{-E.J.-Rivers,}{]},
{[}\ref{-J.-Helayel}{]}.

\section{Photon Lagrangian}

The free real massless vector field is described by the Lagrangian
\begin{equation}
\mathcal{L}_{V}=-\frac{1}{4}(\partial_{\mu}A_{\nu}-\partial_{\nu}A_{\mu})(\partial^{\mu}A^{\nu}-\partial^{\nu}A^{\mu}),\label{eq:1}
\end{equation}
where $A^{\mu}$ is a four-vector. We can rewrite $\mathcal{L}_{V}$
as a quadratic form in the following way
\begin{equation}
\mathcal{L}_{V}=\frac{1}{2}A_{\mu}\mathcal{O}^{\mu\nu}A_{\nu},\label{eq:2}
\end{equation}
where the symmetric operator $\mathcal{O}^{\mu\nu}$ is given by
\begin{equation}
\mathcal{O}^{\mu\nu}\coloneqq\boxempty\left(\eta^{\mu\nu}-\frac{\partial^{\mu}\partial^{\nu}}{\boxempty}\right).\label{eq:14oper}
\end{equation}
The Euler-Lagrange equations for $A^{\mu}$ are given by
\[
\partial_{\mu}\frac{\partial\mathcal{L}_{V}}{\partial(\partial_{\mu}A_{\nu})}=\frac{\partial\mathcal{L}_{V}}{\partial A_{\nu}},
\]
and since ${\displaystyle \partial_{\mu}\frac{\partial\mathcal{L}_{V}}{\partial(\partial_{\mu}A_{\nu})}=}\left(\boxempty\eta_{\mu\nu}-\partial_{\mu}\partial_{\nu}\right)A^{\nu}$
and ${\displaystyle \frac{\partial\mathcal{L}_{V}}{\partial A_{\nu}}}=0$
we obtain the field equations 
\begin{equation}
\left(\boxempty\eta_{\mu\nu}-\partial_{\mu}\partial_{\nu}\right)A^{\nu}=0.\label{eq:3}
\end{equation}
We can decompose (see Appendix $B)$ every four-vector in terms of
spin-$1$ and spin-$0$ components under the rotation group $SO(3),$
i.e. $A^{\mu}\in\mathbf{0}\oplus\mathbf{1},$ by introducing a complete
set of projectors:
\begin{equation}
\{\theta,\omega\}:\,\,\,\,\,\begin{cases}
{\displaystyle \theta_{\mu\nu}\coloneqq\eta_{\mu\nu}-\frac{\partial_{\mu}\partial_{\nu}}{\boxempty}}\\
 & ,\\
{\displaystyle \omega_{\mu\nu}\coloneqq\frac{\partial_{\mu}\partial_{\nu}}{\boxempty}}
\end{cases}\label{eq:4}
\end{equation}
that in momentum space becomes
\begin{equation}
\theta_{\mu\nu}=\eta_{\mu\nu}-\frac{k_{\mu}k_{\nu}}{k^{2}},\,\,\,\,\,\omega_{\mu\nu}=\frac{k_{\mu}k_{\nu}}{k^{2}}.\label{eq:4momentum}
\end{equation}
 It is easy to show that the following properties hold
\begin{equation}
\begin{array}{rl}
 & \theta+\omega=\mathbb{I}\Leftrightarrow\theta_{\mu\nu}+\omega_{\mu\nu}=\eta_{\mu\nu}\\
\\
 & \theta^{2}=\theta,\,\,\,\omega^{2}=\omega,\,\,\,\theta\omega=0\\
\\
\Leftrightarrow & \theta_{\mu\nu}\theta_{\rho}^{\nu}=\theta_{\mu\rho},\,\,\,\omega_{\mu\nu}\omega_{\rho}^{\nu}=\omega_{\mu\rho},\,\,\,\theta_{\mu\nu}\omega_{\rho}^{\nu}=0,
\end{array}\label{eq:5}
\end{equation}
so we can decompose the four-vector $A^{\mu}$ as 
\begin{equation}
A_{\mu}=\theta_{\mu}^{\nu}A_{\nu}+\omega_{\mu}^{\nu}A_{\nu}.
\end{equation}
This special decomposition corresponds to that in which $A^{\mu}$
decomposes in transverse and longitudinal components. In fact, if
$k^{\mu}$ is the $4\textrm{-}$momentum associated to the electromagnetic
wave (or photon) we can immediately see that
\begin{equation}
k^{\mu}\theta_{\mu\nu}=0,\,\,\,\,\, k^{\mu}\omega_{\mu\nu}=k_{\nu};
\end{equation}
hence $\theta$ and $\omega$ project along the transverse and longitudinal
components respectively. \\
Furthermore, we can also verify that the transverse component has
spin-$1$ and the longitudinal one spin-$0$ by calculating the trace
of the two projectors:
\begin{equation}
\begin{array}{rl}
\eta^{\mu\nu}\theta_{\mu\nu}=3=2(1)+1\,\,(\mathrm{spin}\textrm{-}1),\\
\\
\eta^{\mu\nu}\omega_{\mu\nu}=1=2(0)+1\,\,(\mathrm{spin}\textrm{-}0);
\end{array}
\end{equation}
By means this formalism of the spin projector operators in the space
of four-vectors. one can make more clear which are the spin components
of the vector field and, also, make the discussion more elegant.\\
The field equations (\ref{eq:3}) can be easily recast in terms of
the projectors:
\begin{equation}
\boxempty\theta_{\mu\nu}A^{\nu}=0.\label{eq:6}
\end{equation}
In momentum space the last equations become 
\begin{equation}
-k^{2}\theta_{\mu\nu}A^{\nu}=0\Rightarrow k^{2}=0\Rightarrow E^{2}=|\bar{k}|^{2},
\end{equation}
namely the vector field $A^{\mu}$ is such that only the spin-$1$
massless component propagates.\\
Note that the equation (\ref{eq:6}) tells us that the transverse
and longitudinal components decouple by means the field equations.
If we want to speak in terms of spin components, we can say that only
the spin-$1$ component propagates. In the next section we will see
how to keep only the two physical degrees of freedom by imposing a
gauge. \\
To conclude this section we shall introduce the gauge transformation
under which the Lagrangian (\ref{eq:1}) is invariant.\\
$ $

\textbf{Gauge invariance of photon Lagrangian}

$ $\\
Let us observe that the Lagrangian (\ref{eq:1}), and the field equations
(\ref{eq:3}), are invariant under gauge transformations
\begin{equation}
A_{\mu}\rightarrow A'_{\mu}=A_{\mu}+\partial_{\mu}\alpha,\label{eq:7}
\end{equation}
where $\alpha$ is any differentiable function. In fact, by rewriting
the field equations (\ref{eq:3}) in terms of the transformed vector
field $A'_{\mu},$ we obtain
\begin{equation}
\begin{array}{rl}
 & \left(\boxempty\eta_{\mu\nu}-\partial_{\mu}\partial_{\nu}\right)A^{'\nu}=0.\\
\\
\Leftrightarrow & \left(\boxempty\eta_{\mu\nu}-\partial_{\mu}\partial_{\nu}\right)A^{\nu}+\left(\boxempty\eta_{\mu\nu}-\partial_{\mu}\partial_{\nu}\right)\partial^{\nu}\alpha=0,
\end{array}
\end{equation}
and 
\begin{equation}
\left(\boxempty\eta_{\mu\nu}-\partial_{\mu}\partial_{\nu}\right)\partial^{\nu}\alpha=0\Rightarrow\left(\boxempty\eta_{\mu\nu}-\partial_{\mu}\partial_{\nu}\right)A^{\nu}=0,
\end{equation}
i.e. the field equations don't change under the gauge transformation
(\ref{eq:7}). This invariance preserves the masslessness of the field.

\section{Photon degrees of freedom}

Now we want to determine the number of degrees of freedom of a vector
field. We shall see that by using field equations and gauge invariance
we can get rid of the unphysical degrees of freedom. A the end we
shall see that an on-shell photon has only two degrees of freedom,
instead an off-shell photon%
\footnote{For a discussion on on-shell and off-shell photons see Appendix $A.$
Note also that as synonyms of on-shell and off-shell we shall also
use real and virtual respectively. %
} three degrees of freedom.

\subsection{On-shell photon}

Let us start working with on-shell photon.\\
First of all let us expand $A_{\mu}(k)$ in the basis of four-vectors
$\left\{ k^{\mu},\tilde{k}^{\mu},\varepsilon_{1}^{\mu},\varepsilon_{2}^{\mu}\right\} :$
\begin{equation}
\begin{array}{l}
k^{\mu}\equiv(k^{0},\bar{k}),\,\,\,\,\,\tilde{k}^{\mu}\equiv(\tilde{k}^{0},-\bar{k}),\,\,\,\,\,\varepsilon_{i}^{\mu}\equiv(0,\bar{\varepsilon}_{i}),\\
\\
k^{\mu}\varepsilon_{i,\mu}=0=\tilde{k}^{\mu}\varepsilon_{i,\mu},\,\,\,\,\,\,\,\varepsilon_{i}^{\mu}\varepsilon_{j,\mu}=-\bar{\varepsilon}_{i}\cdot\bar{\varepsilon}_{j}=-\delta_{ij},
\end{array}\,\,\,\,\, i=1,2,\label{eq:8}
\end{equation}
thus
\begin{equation}
A_{\mu}(k)=ak_{\mu}+b\tilde{k}_{\mu}+c^{i}\varepsilon_{i,\mu}\label{eq:9}
\end{equation}
By substituting (\ref{eq:9}) in the field equations (\ref{eq:3})
written in  momentum space, we obtain
\[
\begin{array}{rl}
 & \left(k^{2}A_{\mu}(k)-k_{\mu}k^{\nu}A_{\nu}(k)\right)=0\\
\\
\Leftrightarrow & ak^{2}k_{\mu}+bk^{2}\tilde{k}_{\mu}+c^{i}k^{2}\varepsilon_{i,\mu}-ak_{\mu}k^{2}-bk_{\mu}k\cdot\tilde{k}-c^{i}k_{\mu}k^{\nu}\varepsilon_{i,\nu}=0
\end{array}
\]
and by using the orthonormality relations (\ref{eq:8}) one has 
\[
bk^{2}\tilde{k}_{\mu}+c^{i}k^{2}\varepsilon_{i,\mu}-bk_{\mu}k\cdot\tilde{k}=0.
\]
Then
\begin{equation}
\begin{array}{rl}
 & k\cdot\tilde{k}=\eta_{\mu\nu}k^{\mu}\tilde{k}^{\nu}=(k^{0})^{2}+(\bar{k})^{2}\neq0\\
\\
\Rightarrow & bk^{2}\tilde{k}_{\mu}+c^{i}k^{2}\varepsilon_{i,\mu}-bk_{\mu}\left((k^{0})^{2}+(\bar{k})^{2}\right)=0.
\end{array}\label{eq:10}
\end{equation}
If we consider $\mu=0$ we have
\begin{equation}
\begin{array}{rl}
 & bk^{2}\tilde{k}_{\mu}-bk_{\mu}\left((k^{0})^{2}+(\bar{k})^{2}\right)=0\\
\\
\Leftrightarrow & -2b|\bar{k}|^{2}k^{0}=0\Rightarrow b=0;
\end{array}
\end{equation}
the field equations allow to get rid of one degree of freedom, so
now we have $4-1=3$ degrees of freedom. \\
If $b=0$ then (\ref{eq:10}) becomes $k^{2}c^{i}\varepsilon_{i,\mu}=0,$
and its spatial part gives us 
\begin{equation}
k^{2}c^{i}\bar{\varepsilon}_{i}=0\Rightarrow k^{2}c^{i}=0\Leftrightarrow k^{2}=0\,\vee\, c^{i}=0.\label{eq:10sol}
\end{equation}
Hence we have two possible solutions that satisfy the last equation.\\
If $k^{2}=0$ one has 
\begin{equation}
A_{\mu}(k)=ak_{\mu}+c^{i}\varepsilon_{i,\mu},
\end{equation}
i.e. $A^{\mu}$ describes a massless particle. \\
While, if $c=0$ one has
\begin{equation}
A_{\mu}(k)=ak_{\mu}.
\end{equation}
Now, by choosing a gauge, we will see that only one of them is physically
admissible. \\
The gauge transformation (\ref{eq:7}) in  momentum space is 
\begin{equation}
A'_{\mu}(k)=A_{\mu}(k)+ik_{\mu}\alpha(k),
\end{equation}
and in the basis (\ref{eq:8}) we have chosen becomes
\[
\begin{array}{rl}
 & \left(a'k_{\mu}+c'^{i}\varepsilon_{i,\mu}\right)=\left(ak_{\mu}+c^{i}\varepsilon_{i,\mu}\right)+ik_{\mu}\alpha(k)\\
\\
\Leftrightarrow & c'^{i}=c^{i},\,\,\,\,\, a'(k)=a(k)+i\alpha(k).
\end{array}
\]
By choosing $\alpha(k)=-\frac{1}{i}a(k)$ we can eliminate the coefficient
$a$ in (\ref{eq:9}), so we get rid of another unphysical degree
of freedom. At this point, we can immediately notice that if we choose
$c^{i}=0$ as the solution of the equation (\ref{eq:10sol}) we will
obtain $A_{\mu}=0,$ but this is not a physical solution. Thus the
solution of the equation (\ref{eq:10sol}) is $k^{2}=0$ that tells
us again that the photon is massless. Hence ,by means the constraint
of the field equations and the freedom of choosing a gauge, we obtain
that for an on-shell photon the vector field $A_{\mu}$assumes the
following form: 
\begin{equation}
a=0=b\Rightarrow A_{\mu}(k)=c^{i}\varepsilon_{i,\mu},\,\,\, k^{2}=0.\label{eq:11}
\end{equation}
In conclusion we got rid of the unphysical degrees of freedom, keeping
only the $\mathit{two}$ physical one, whose information is held in
the $\mathit{two}$ coefficients $c^{1}$ and $c^{2}.$

\subsection{Off-shell photon}

The above discussion holds just for on-shell gauge field. As for off-shell
photon we can't impose the field equations, so we can eliminate just
one degree of freedom via gauge symmetry. It means that we cannot
eliminate the coefficient $b,$ that will turns out to be different
from zero giving the third degree of freedom in the counting. Hence,
we are able to state that an off-shell photon has $\mathit{three}$
degrees of freedom. \\
\\
We can summarize the two cases with the following expression:%
\footnote{More generally, if we are in $D-$dimensions we have:
\[
D\textrm{-}\mathrm{dimension:}\,\,\,\begin{cases}
\mathrm{off\textrm{-}shell}: & (D-1)\,\,\mathrm{d.o.f.}\\
\mathrm{on\textrm{-}shell}: & (D-2)\,\,\mathrm{d.o.f.}
\end{cases};
\]
as particular case we can see that in $D=2$
\[
2\textrm{-}\mathrm{dimension:}\,\,\,\begin{cases}
\mathrm{off\textrm{-}shell}: & 1\,\,\mathrm{d.o.f.}\\
\mathrm{on\textrm{-}shell}: & 0\,\,\mathrm{d.o.f.}
\end{cases},
\]
namely on-shell photons don't exist; then in $D=3$
\[
3\textrm{-}\mathrm{dimension:}\,\,\,\begin{cases}
\mathrm{off\textrm{-}shell}: & 2\,\,\mathrm{d.o.f.}\\
\mathrm{on\textrm{-}shell}: & 1\,\,\mathrm{d.o.f.}
\end{cases}.
\]
}

\begin{equation}
4\textrm{-}\mathrm{dimension:}\,\,\,\begin{cases}
\mathrm{off\textrm{-}shell}: & 3\,\,\mathrm{d.o.f.}\\
\mathrm{on\textrm{-}shell}: & 2\,\,\mathrm{d.o.f.}
\end{cases}.\label{eq:12}
\end{equation}

\section{Photon propagator}

Now our aim is to obtain the photon propagator; we can do it by working
with the spin projector operators. In general given a Lagrangian written
as a quadratic form in terms of an operator $\mathcal{O},$ the propagator
is defined as the inverse operator $\mathcal{O}^{-1}.$ A generic
operator $\mathcal{O}$ acting in the four-vectors space can be expanded
in the basis $\left\{ \theta,\omega\right\} :$
\begin{equation}
\mathcal{O}=a\theta+b\omega;
\end{equation}
 one can say the same for its inverse
\begin{equation}
\mathcal{O}^{-1}=c\theta+d\omega.
\end{equation}
In general the coefficients $a,b,c,d$ can be complex numbers.\\
Imposing that $\mathcal{O}\mathcal{O}^{-1}=\mathbb{I},$ or equivalently
$\mathcal{O}^{\mu\rho}\mathcal{O}_{\rho\nu}^{-1}=\delta_{\nu}^{\mu},$
we can obtain the propagator once we know the form of operator $\mathcal{O};$
in fact:
\[
\left(a\theta+b\omega\right)\left(c\theta+d\omega\right)=\mathbb{I}
\]
\[
\Leftrightarrow ac\theta+bd\omega=\mathbb{I}\Leftrightarrow c=\frac{1}{a},\,\,\, d=\frac{1}{b}
\]
\begin{equation}
\Rightarrow\mathcal{O}^{-1}=\frac{1}{a}\theta+\frac{1}{b}\omega.
\end{equation}
In the case of the Lagrangian (\ref{eq:1}) we have $\mathcal{O}=-k^{2}\theta,$
i.e. $a=-k^{2}$ and $b=0.$ We notice that, since $b=0$ we cannot
directly invert the operator $\mathcal{O}.$ Thus the operator $\mathcal{O}$
we have defined for the Lagrangian (\ref{eq:1}) is not invertible%
\footnote{Physically we can interpret this result saying that the fact that
$b=0$ implies that the spin-$0$ component (longitudinal component)
doesn't propagate, so it won't appear in the physical part of the
propagator. %
}.\\
We encounter the same problem also starting from the field equations.
In fact, by considering the presence of a source $J^{\mu}$ we have
to add the term $-A_{\mu}J^{\mu}$ to the Lagrangian and the field
equations become
\begin{equation}
\left(\boxempty\eta_{\mu\nu}-\partial_{\mu}\partial_{\nu}\right)A^{\nu}=J_{\mu}\Leftrightarrow\boxempty\theta_{\mu\nu}A^{\nu}=(\theta_{\mu\nu}+\omega_{\mu\nu})J^{\nu}\Leftrightarrow\mathcal{O}_{\mu\nu}A^{\nu}=(\theta_{\mu\nu}+\omega_{\mu\nu})J^{\nu}.\label{eq:13}
\end{equation}
Again to obtain the propagator we have to invert the same operator
$\mathcal{O}$ and in principle we could do it by acting on both members
with spin projection operators%
\footnote{We are suppressing the indices for simplicity. %
}: 
\begin{equation}
\theta\rightarrow-k^{2}\theta A=\theta J,\label{eq:14}
\end{equation}
\begin{equation}
\omega\rightarrow0A=\omega J\Rightarrow\omega J=0.\label{eq:15}
\end{equation}
We can notice that it's impossible to invert both equations, indeed
from (\ref{eq:14}) we can obtain the transverse component $\theta A=-\frac{\theta}{k^{2}}J,$
but we are not able to do the same with (\ref{eq:15}) because we
can't invert the zero at the first member. Hence we have found that
the spin-$0$ projection is undetermined and it means that there is
a gauge freedom and concurrently a restriction on the source, i.e.
the equation (\ref{eq:15}), $\omega J=0.$%
\footnote{It is worth observing the connection between undetermined spin component
and gauge freedom. In the case of ED, from (\ref{eq:2}), we can see
that the Lagrangian is composed only by the spin-$1$ component
\begin{equation}
\mathcal{L}_{V}=\frac{1}{2}A_{\mu}\boxempty\theta^{\mu\nu}A_{\nu}.\label{eq:16}
\end{equation}
One can notice the Lagrangian in the last equation is invariant under
spin-$0$ transformation, $\delta A_{\mu}\sim\omega_{\mu\nu}A^{\nu},$
in fact $\theta\omega=0;$ and we also know that there is a restriction
on the source in terms of the spin-$0$ component, $\omega J=0.$
It means that for the Lagrangian (\ref{eq:16}) there is a gauge symmetry
that corresponds to the gauge invariance under transformations $\delta A_{\mu}=\partial_{\mu}\alpha.$
The arbitrary function $\alpha$ is the scalar associated to the spin-$0$
symmetry. %
}

$ $\\
This mathematical obstacle can be overcome by adding a gauge fixing
term to the Lagrangian (\ref{eq:1}). Moreover, we know that in any
gauge theory the choice a gauge is needed to get rid of because of
the spurious degrees of freedom. Thus we have seen that, already at
classical level, we need a gauge fixing term because of both mathematical
and physical reasons.\\
Although we have this problem of inversion for the operator $\mathcal{O},$
we can always obtain the physical part of the propagator. Indeed,
the propagator always contains a gauge dependent part that is not
physical and a gauge independent part that is physical, namely the
part of the propagator appearing when we want to calculate, for example,
the scattering amplitudes. The physical part of the propagator is
often called $saturated$ $propagator,$ or sandwiched propagator,
because it corresponds to the sandwich of the propagator between two
conserved currents%
\footnote{In general, given a propagator $\mathcal{P}$ and two conserved currents
$J_{1},$ $J_{2},$ the saturated propagator is given by the sandwich
\begin{equation}
J_{1}\mathcal{P}J_{2}.
\end{equation}
Note that we are not writing the indices neither for the propagator
$\mathcal{P}$ nor for the conserved currents $J_{1},$ $J_{2}$ for
simplicity, but in general $\mathcal{P}$ can have two indices, for
example two for photon case ($\mathcal{D}_{\mu\nu})$ and four indices
for graviton case ($\Pi_{\mu\nu\rho\sigma}),$ as we shall see below. %
}.\\
\\
Now, first we are going to invert the operator $\mathcal{O}$ by introducing
a gauge fixing term so we can obtain the propagator that, of course,
will turn out to be gauge dependent. Secondly we are going to determine
the saturated propagator, that doesn't need the introduction of a
gauge fixing term.

\subsection{Gauge fixing term for photon Lagrangian}

By introducing the Lorenz gauge fixing term $-\frac{1}{2\alpha}\left(\partial_{\mu}A^{\mu}\right)^{2},$
the total Lagrangian is
\begin{equation}
\begin{array}{rl}
\tilde{\mathcal{L}}{}_{V}= & \mathcal{L}_{V}-\frac{1}{2\alpha}\left(\partial_{\mu}A^{\mu}\right)^{2}\\
\\
= & \frac{1}{2}A_{\mu}\tilde{\mathcal{O}}^{\mu\nu}A_{\nu},
\end{array}
\end{equation}
 where $\tilde{\mathcal{O}}^{\mu\nu}\coloneqq\boxempty\eta^{\mu\nu}+\left(\frac{1}{\alpha}-1\right)\partial^{\mu}\partial^{\nu},$
or in terms of the spin projector operators in momentum space 
\begin{equation}
\tilde{\mathcal{O}}^{\mu\nu}=-k^{2}\left(\theta^{\mu\nu}+\frac{1}{\alpha}\omega^{\mu\nu}\right).\label{eq:45oper}
\end{equation}
We notice that (\ref{eq:45oper}) corresponds to the operator (\ref{eq:14oper})
plus an additive term given by the gauge fixing term. Now we can invert
the operator in (\ref{eq:2}) once we go into momentum space, in fact
$b=\frac{1}{\alpha}\neq0,$ i.e. the longitudinal component is present
too. Thus, the photon propagator in a generic gauge is%
\footnote{Note that often the propagator is also defined as the vacuum expectation
value of the time ordered product:
\[
\left\langle T\left\{ A_{\mu}(-k)A_{\nu}(k)\right\} \right\rangle =i\mathcal{D}_{\mu\nu}(k)=-\frac{i}{k^{2}}\left(\theta_{\mu\nu}+\alpha\omega_{\mu\nu}\right).
\]
The expression $T\left\{ A_{\mu}(-k)A_{\nu}(k)\right\} $ is the Fourier
transform of the time ordered product that in coordinate space is
defined as
\[
T\left\{ A_{\mu}(x)A_{\nu}(y)\right\} \coloneqq A_{\mu}(x)A_{\nu}(y)\Theta(x_{0}-y_{0})+A_{\nu}(y)A_{\mu}(x)\Theta(y_{0}-x_{0}),
\]
where the function $\Theta(x_{0}-y_{0})$ is equal to $1$ if $x_{0}>y_{0},$
and to $0$ otherwise.%
}
\begin{equation}
\mathcal{D}^{\mu\nu}(k)\equiv\tilde{\mathcal{O}}^{-1\mu\nu}=-\frac{1}{k^{2}}\left(\theta^{\mu\nu}+\alpha\omega^{\mu\nu}\right)=-\frac{1}{k^{2}}\left[\eta_{\mu\nu}+\left(\alpha-1\right)\frac{k_{\mu}k_{\nu}}{k^{2}}\right].\label{eq:17-1}
\end{equation}
Particular nomenclatures associated with $\alpha$ are
\begin{equation}
\begin{cases}
\alpha=1: & \mathrm{Feynman\,\mathrm{gauge}}\\
\alpha=0: & \mathrm{Landau\,\mathrm{gauge}}
\end{cases};
\end{equation}
as we have already observed the physics is unaffected by the value
of $\alpha.$ \\
\\
Note that the propagator in (\ref{eq:17-1}) contains a part that
depends on the coefficient $\alpha,$ i.e. a gauge dependent part.
If we consider the sandwich between two currents $J(-K)$ and $J(k)$
we can already notice that, because of the conservation law 
\begin{equation}
\partial^{\mu}J_{\mu}=0\Leftrightarrow k^{\mu}J_{\mu}=0,\label{eq:17-2}
\end{equation}
the saturated propagator is gauge independent:
\begin{equation}
J^{\mu}(-k)\mathcal{D}_{\mu\nu}(k)J^{\nu}(k)=J^{\mu}(-k)\frac{-\eta_{\mu\nu}}{k^{2}}J^{\nu}(k).
\end{equation}
This last equation will be the result of the next calculation.\\

\subsection{Saturated photon propagator}

Consider again the field equations (\ref{eq:13}) in  momentum space
and add to the both members the term $-k^{2}\omega_{\mu\nu}A^{\nu}:$
\begin{equation}
-k^{2}\left(\theta_{\mu\nu}+\omega_{\mu\nu}\right)A^{\nu}(k)=J_{\mu}(k)-k^{2}\omega_{\mu\nu}A^{\nu}(k).
\end{equation}
Now by multiplying with $\theta_{\rho}^{\mu}$ we obtain
\[
-k^{2}\left(\theta_{\rho}^{\mu}\theta_{\mu\nu}+\theta_{\rho}^{\mu}\omega_{\mu\nu}\right)A^{\nu}(k)=\theta_{\rho}^{\mu}J_{\mu}(k)-k^{2}\theta_{\rho}^{\mu}\omega_{\mu\nu}A^{\nu}(k),
\]
and, since $\theta_{\rho}^{\mu}\theta_{\mu\nu}=\theta_{\rho\nu}$
and $\theta_{\rho}^{\mu}\omega_{\mu\nu}=0,$ we have
\begin{equation}
-k^{2}\theta_{\mu\nu}A^{\nu}(k)=\theta_{\mu\nu}J^{\nu}(k)\Leftrightarrow\theta_{\mu\nu}A^{\nu}(k)=-\frac{1}{k^{2}}\theta_{\mu\nu}J^{\nu}(k)\label{eq:17}
\end{equation}
Then by multiplying with $\omega_{\rho}^{\mu}$ we obtain 
\begin{equation}
\begin{array}{rl}
 & -k^{2}\left(\omega_{\rho}^{\mu}\theta_{\mu\nu}+\omega_{\rho}^{\mu}\omega_{\mu\nu}\right)A^{\nu}(k)=\omega_{\rho}^{\mu}J_{\mu}(k)-k^{2}\omega_{\rho}^{\mu}\omega_{\mu\nu}A^{\nu}(k)\\
\\
\Leftrightarrow & -k^{2}\omega_{\mu\nu}A^{\nu}(k)=\omega_{\mu\nu}J^{\nu}(k)-k^{2}\omega_{\mu\nu}A^{\nu}(k).
\end{array}\label{eq:18}
\end{equation}
The equation (\ref{eq:15}) says that $\omega_{\mu\nu}J^{\mu}=0,$
so by multiplying (\ref{eq:18}) with $J^{\mu}(-k)$ we obtain
\begin{equation}
J^{\mu}(-k)\omega_{\mu\nu}A^{\nu}(k)=-\frac{1}{k^{2}}J^{\mu}(-k)\omega_{\mu\nu}J^{\nu}(k),
\end{equation}
and doing the same for (\ref{eq:17}) one has
\begin{equation}
J^{\mu}(-k)\theta_{\mu\nu}A^{\nu}(k)=-\frac{1}{k^{2}}J^{\mu}(-k)\theta_{\mu\nu}J^{\nu}(k).
\end{equation}
By combining the last two equations we obtain 
\begin{equation}
J^{\mu}(k)\left(\theta_{\mu\nu}+\omega_{\mu\nu}\right)A^{\nu}(k)=J^{\mu}(-k)\frac{-1}{k^{2}}\left(\theta_{\mu\nu}+\omega_{\mu\nu}\right)J^{\nu}(k)=J^{\mu}(-k)\frac{-\theta_{\mu\nu}}{k^{2}}J^{\nu}(k),
\end{equation}
or, without writing the indices,
\begin{equation}
J(-k)\left(\theta+\omega\right)A(k)=J(-k)\frac{-1}{k^{2}}\left(\theta+\omega\right)J(k)=J(-k)\frac{-\theta}{k^{2}}J(k).
\end{equation}
We can notice that we have calculated the saturated propagator%
\footnote{We have seen that the saturated propagator, that corresponds to the
physical part (gauge-independent), is invertible. Note that we cannot
say that $-\frac{\theta}{k^{2}}$ is the propagator or the inverse
of the operator $\mathcal{O}:$ in fact $\mathcal{O}\theta=-k^{2}\theta\frac{-1}{k^{2}}\theta=\theta\neq\mathbb{I}.$
But we can say that $J(-k)\mathcal{O}(k)J(k)$ is invertible and the
inverse operator turns out to be $J(-k)\frac{-\theta}{k^{2}}J(k).$ %
}:
\begin{equation}
J^{\mu}(-k)\mathcal{D}_{\mu\nu}(k)J^{\nu}(k)=J^{\mu}(-k)\frac{-\theta_{\mu\nu}}{k^{2}}J^{\nu}(k);\label{eq:satprqed}
\end{equation}
hence the quantity on the right side between the two conserved currents
is the physical part of the photon propagator. We can notice that
a virtual (off-shell) photon has only the spin-$1$ component, namely
the two transverse components and the longitudinal one. \\

\section{Photon propagator and polarization sums}

\subsection{Polarization vectors%
\footnote{We shall also follow Ref. {[}\ref{-Greiner-book}{]}.%
}}

In the section $1.2$ we have seen that a virtual photon is a spin-$1$
particle, namely it has three components; instead a real photon has
only two components, the longitudinal part is absent, and we showed
that the difference between off-shell and on-shell is that for the
latter we can also impose the field equations in addition to the gauge
condition. By rewriting the saturated propagator in terms of the polarization
vectors of the photon we can see explicitly that the two transverse
and the longitudinal components for the off-shell photon and only
the two transverse for the on-shell one are present.\\
\\
To construct the set of polarization vectors we have to specify if
we are considering either massless photon $(m=0)$ or massive photon
$(m\neq0),$ {[}\ref{-Greiner-book}{]}. Since the longitudinal component
of a massive photon can be also chosen for the longitudinal component
of a massless off-shell photon, we will study both massive and massless
cases. In both cases the set of polarization vectors $\left\{ \epsilon_{(0)},\epsilon_{(1)},\epsilon_{(2)},\epsilon_{(3)}\right\} $
has to form a $4$-dimensional orthonormal and complete basis satisfying
\begin{equation}
\epsilon_{(\lambda),\mu}\epsilon_{(\lambda')}^{\mu}=\eta_{\lambda\lambda'}\,\,\,\,(\mathrm{orthonormality)}\label{eq:25}
\end{equation}
and $ $
\begin{equation}
\sum_{\lambda=0}^{3}\eta_{\lambda\lambda}\epsilon_{(\lambda),\mu}\epsilon_{(\lambda),\nu}=\eta_{\mu\nu}\,\,\,\,(\mathrm{completeness).}\label{eq:26}
\end{equation}

We shall start with the construction of the set in the case of massive
photon.

$ $

\textbf{Massive photon}

$ $\\
We choose a frame of reference in which the plane wave has spatial
momentum $\bar{k}.$ Now we choose two space-like $\mathit{transverse}$
$\mathit{polarization}$ $\mathit{vectors}$ 
\begin{equation}
\epsilon_{(1)}^{\mu}\equiv\left(0,\bar{\epsilon}_{(1)}\right),\,\,\,\,\epsilon_{(2)}^{\mu}\equiv\left(0,\bar{\epsilon}_{(2)}\right)\label{eq:27}
\end{equation}
 imposing the conditions
\begin{equation}
\bar{\epsilon}_{(1)}\cdot\bar{k}=0=\bar{\epsilon}_{(2)}\cdot\bar{k}
\end{equation}
 and 
\begin{equation}
\bar{\epsilon}_{(i)}\cdot\bar{\epsilon}_{(j)}=\delta_{ij}.
\end{equation}
The third polarization vector is chosen such that its spatial component
points in the direction of the momentum $\bar{k},$ that is normalized
according to (\ref{eq:25}). We will adopt the further condition that
the four-vector $\epsilon_{(3)}^{\mu}$ is orthogonal to the four-momentum
$k^{\mu},$
\begin{equation}
k^{\mu}\epsilon_{(3),\mu}=0.
\end{equation}
Taking this equation and the normalized condition (\ref{eq:25}) for
$\lambda=3=\lambda',$ we found the components of the thus constructed
$\mathit{longitudinal}$ $\mathit{polarization}$ $\mathit{vector}:$
\begin{equation}
\epsilon_{(3)}^{\mu}\equiv\left(\frac{|\bar{k}|}{m},\frac{\bar{k}}{|\bar{k}|}\frac{k^{0}}{m}\right).\label{eq:28}
\end{equation}
The normalization condition $\epsilon_{(3)}^{2}=-1$ is satisfied
since $\epsilon_{(3)}^{2}=\frac{\bar{k}^{2}}{m^{2}}-\frac{\bar{k}^{2}}{\bar{k}^{2}}\frac{(k^{0})^{2}}{m^{2}}=\frac{\bar{k}^{2}}{m^{2}}-\frac{(k^{0})^{2}}{m^{2}}=-1.$
It is worth noting that the longitudinal polarization vector (\ref{eq:28})
is not well defined in the case of massless photon because we have
the mass $m$ at the denominator. This problem will be addressed below.\\
To complete the vector basis in Minkowski space we need to introduce
a fourth $\mathit{time\textrm{-}like}$ $\mathit{polarization}$ $\mathit{vector}$
with index $\lambda=0.$ We can simply use the $4\textrm{-}$momentum
$k^{\mu},$ namely
\begin{equation}
\epsilon_{(0)}^{\mu}\coloneqq\frac{k^{\mu}}{m},\label{eq:29}
\end{equation}
where the factor $\frac{1}{m}$ ensures the normalization condition
according to (\ref{eq:25}). Also, it is obvious that the four-vector
in (\ref{eq:29}) is orthogonal to the three space-like polarization
vectors. Let us write down the $4$-dimensional scalar product of
our set of polarization vectors with the momentum vector:
\begin{equation}
k^{\mu}\epsilon_{(0),\mu}=k,\,\,\,\, k^{\mu}\epsilon_{(\lambda),\mu}=0,\,\,\,\lambda=1,2,3.
\end{equation}
One can also check that the completeness relation (\ref{eq:26}) is
satisfied for the set we have just constructed.\\
We know that the Lorenz condition $k\cdot\epsilon=0$ has to hold%
\footnote{For a massive photon the condition $k\cdot\epsilon=0$ is a consistency
relation that holds once we impose the field equations and the current
conservation.%
} but the vector polarization $\epsilon_{(0)}$ doesn't satisfy it.
We have only three physical polarization vectors: in fact for a massive
photon we have three degrees of freedom. A virtual massless photon,
as we have already showed in (\ref{eq:12}) with $D=4,$ has three
degrees of freedom too, and we shall see that for it we can choose
the same longitudinal component, with the only difference that we
cannot use the mass $m$ as vector component (see below). \\
$ $

\textbf{Massless photon}

$ $\\
To construct the polarization states of the massless photon we can
begin as in the massive case and introduce two $\mathit{transverse}$
$\mathit{polarization}$ $\mathit{vectors},$ $\lambda=1,2,$ as in
(\ref{eq:27}). However, now the momentum $k^{\mu}$ can no longer
be used as a basis vector: it cannot be normalized to $1$ since the
dispersion relation now reads $k^{2}=0.$ In addition, as we have
already noticed above, the longitudinal polarization vector (\ref{eq:28})
is not defined for $k^{2}=0.$ In the massless case is impossible
to construct a third polarization vector which is normalizable and
at the same time such that the scalar product with $k^{\mu}$ is zero.\\
To avoid this problem we arbitrarily define a time-like unit vector
and choose it as $\mathit{time\textrm{\textrm{-}}like}$ $\mathit{polarization}$
$\mathit{vector},$ which in the chosen Lorentz frame simply is given
by
\begin{equation}
\epsilon_{(0)}^{\mu}\coloneqq n^{\mu}\equiv\left(1,0,0,0\right),\,\,\,\,\, n^{2}=+1.\label{eq:30}
\end{equation}
The $\mathit{longitudinal}$ $\mathit{polarization}$ $\mathit{vector}$
can be then written in covariant form as%
\footnote{Since we are also writing $k^{2}$ we are considering the more general
off-shell case. To obtain the on-shell third component we have just
to impose $k^{2}=0.$ %
} 
\begin{equation}
\epsilon_{(3)}^{\mu}\coloneqq\frac{k^{\mu}-n^{\mu}(n\cdot k)}{\sqrt{(n\cdot k)^{2}-k^{2}}}.\label{eq:31}
\end{equation}
This vector indeed has the correct normalization
\begin{equation}
\epsilon_{(3)}^{\mu}\epsilon_{(3),\mu}=\frac{k^{\mu}k_{\mu}-2(n\cdot k)^{2}+n^{2}(n\cdot k)}{(n\cdot k)^{2}-k^{2}}=-1
\end{equation}
We can easily verify that in this special Lorentz frame, where $n^{0}=1$
and $n^{2}=+1,$ the longitudinal polarization vector becomes
\begin{equation}
\epsilon_{(3)}^{\mu}\equiv\left(0,\frac{\bar{k}}{|\bar{k}|}\right).
\end{equation}
The $4$-dimensional scalar product of the basis vectors and the momentum
vector reads
\begin{equation}
k\cdot\epsilon_{(1)}=0=k\cdot\epsilon_{(2)},\,\,\, k\cdot\epsilon_{(0)}=-k\cdot\epsilon_{(3)}=k\cdot n,
\end{equation}
which, of course, is valid in any frame of reference. \\
One can easily verify that the set we have just constructed satisfies
the orthonormality and completeness relations (\ref{eq:25}) and (\ref{eq:26}).\\
\\
Finally one also show that if we choose the spatial momentum $\bar{k}$
along the third direction in the Minkowski space, $\bar{k}=|\bar{k}|\hat{z},$
the set of polarization vectors reduces in the simple form 
\begin{equation}
\epsilon_{(0)}\equiv\begin{pmatrix}1\\
0\\
0\\
0
\end{pmatrix},\,\,\,\epsilon_{(1)}\equiv\begin{pmatrix}0\\
1\\
0\\
0
\end{pmatrix},\,\,\,\epsilon_{(2)}\equiv\begin{pmatrix}0\\
0\\
1\\
0
\end{pmatrix},\,\,\,\epsilon_{(3)}\equiv\begin{pmatrix}0\\
0\\
0\\
1
\end{pmatrix}.
\end{equation}
From this set we can easily go to a new set of polarization vectors
whose transverse polarizations describe states with helicity $j_{z}=+1$
and $j_{z}=-1.$ They can be easily introduced in the following way
\begin{equation}
\begin{cases}
\epsilon_{{\scriptscriptstyle (1,+1)}}^{\mu}=\frac{1}{\sqrt{2}}\left(\epsilon_{(1)}^{\mu}+i\epsilon_{(2)}\right)\\
\\
\epsilon_{{\scriptscriptstyle (1,-1)}}^{\mu}=\frac{1}{\sqrt{2}}\left(\epsilon_{(1)}^{\mu}-i\epsilon_{(2)}\right)
\end{cases}.
\end{equation}
We can check that they correspond to the two helicity states by acting
with the rotation matrix around the third axis
\begin{equation}
R_{\nu}^{(z)\mu}(\vartheta)=\begin{pmatrix}1 & 0 & 0 & 0\\
0 & \cos\vartheta & \sin\vartheta & 0\\
0 & -\sin\vartheta & \cos\vartheta & 0\\
0 & 0 & 0 & 1
\end{pmatrix}.\label{eq:75 rotation}
\end{equation}
In fact we obtain
\begin{equation}
\begin{array}{rl}
R_{\nu}^{(z)\mu}(\vartheta)\epsilon_{{\scriptscriptstyle (1,+1)}}^{\nu}= & \begin{pmatrix}1 & 0 & 0 & 0\\
0 & \cos\vartheta & \sin\vartheta & 0\\
0 & -\sin\vartheta & \cos\vartheta & 0\\
0 & 0 & 0 & 1
\end{pmatrix}\begin{pmatrix}0\\
1\\
i\\
0
\end{pmatrix}=\begin{pmatrix}0\\
\cos\vartheta+i\sin\vartheta\\
-\sin\vartheta+i\cos\vartheta\\
0
\end{pmatrix}\\
= & e^{i\vartheta}\begin{pmatrix}0\\
1\\
i\\
0
\end{pmatrix}=e^{i\vartheta}\epsilon_{{\scriptscriptstyle (1,+1)}}^{\mu};
\end{array}
\end{equation}
and
\begin{equation}
\begin{array}{rl}
R_{\nu}^{(z)\mu}(\vartheta)\epsilon_{{\scriptscriptstyle (1,-1)}}^{\nu}= & \begin{pmatrix}1 & 0 & 0 & 0\\
0 & \cos\vartheta & \sin\vartheta & 0\\
0 & -\sin\vartheta & \cos\vartheta & 0\\
0 & 0 & 0 & 1
\end{pmatrix}\begin{pmatrix}0\\
1\\
-i\\
0
\end{pmatrix}=\begin{pmatrix}0\\
\cos\vartheta+i\sin\vartheta\\
-\sin\vartheta-i\cos\vartheta\\
0
\end{pmatrix}\\
= & e^{-i\vartheta}\begin{pmatrix}0\\
1\\
-i\\
0
\end{pmatrix}=e^{-i\vartheta}\epsilon_{{\scriptscriptstyle (1,-1)}}^{\mu}.
\end{array}
\end{equation}
Finally let us introduce the three polarization vectors for a off-shell
massless photon. We learned, from (\ref{eq:12}) with $D=4,$ that
a virtual massless photon has three degrees of freedom that correspond
to the three component of a spin-$1$ vector, indeed by studying the
photon propagator we saw that only the spin-$1$ component, $j=1,$
is present (see (\ref{eq:satprqed})). The two transverse polarization
vectors $\epsilon_{{\scriptscriptstyle (1,+1)}}^{\mu}$ and $\epsilon_{{\scriptscriptstyle (1,-1)}}^{\mu}$
correspond to the $j_{z}=+1$ and $j_{z}=-1$ helicity spin-$1$ components.
To complete the spin-$1$ components it remains to define the longitudinal
one with helicity $j_{z}=0,$ namely such that $R_{\nu}^{(z)\mu}(\vartheta)\epsilon_{{\scriptscriptstyle (1,0)}}^{\nu}=\epsilon_{{\scriptscriptstyle (1,0)}}^{\mu}$
(scalar component). We can define as longitudinal polarization the
same polarization vector used for the massive photon, with the only
difference that we cannot use $m$ as part of the components$:$
\begin{equation}
\epsilon_{{\scriptscriptstyle (1,0)}}\equiv\frac{1}{k}\begin{pmatrix}k^{3}\\
0\\
0\\
k^{0}
\end{pmatrix},\,\,\,\, R_{\nu}^{(z)\mu}(\vartheta)\epsilon_{{\scriptscriptstyle (1,0)}}^{\nu}=\epsilon_{{\scriptscriptstyle (1,0)}}^{\mu}
\end{equation}
Hence, the three polarization vectors for an off-shell massless photon
are
\begin{equation}
\epsilon_{{\scriptscriptstyle (1,+1)}}\equiv\frac{1}{\sqrt{2}}\begin{pmatrix}0\\
1\\
i\\
0
\end{pmatrix},\,\,\,\,\epsilon_{{\scriptscriptstyle (1,-1)}}\equiv\frac{1}{\sqrt{2}}\begin{pmatrix}0\\
1\\
-i\\
0
\end{pmatrix},\,\,\,\,\epsilon_{{\scriptscriptstyle (1,0)}}\equiv\frac{1}{k}\begin{pmatrix}k^{3}\\
0\\
0\\
k^{0}
\end{pmatrix}.\label{eq:31-1}
\end{equation}

\subsection{Photon propagator in terms of polarization vectors}

Now we can come back to what we anticipated at the beginning of the
subsection $1.4.1.$ Our aim is to rewrite the saturated propagator
in terms of the polarization vectors: we shall see that for a real
(on-shell) photon only the two transverse polarization vectors $\epsilon_{{\scriptscriptstyle (1,+1)}}$
and $\epsilon_{{\scriptscriptstyle (1,-1)}}$ are present; instead
for a virtual (off-shell) photon also the longitudinal polarization
vector $\epsilon_{{\scriptscriptstyle (1,0)}}$ is present. Hence
we have another confirmation that the saturated propagator reads the
interaction of a spin-$1$ particle with two conserved currents%
\footnote{In fact in (\ref{eq:satprqed}) one realized that the physical part
of the photon propagator has just the spin-$1$ component $\theta;$
while the spin-$0$ component $\omega$ is absent.%
}. Let us start studying the off-shell case.

$ $

\textbf{Off-shell photon}

$ $\\
Let us consider the saturated photon propagator in (\ref{eq:satprqed})
or (\ref{eq:19}): 
\begin{equation}
\begin{array}{rl}
J^{\mu}(-k)\mathcal{D}_{\mu\nu}(k)J^{\nu}(k)= & {\displaystyle -\frac{1}{k^{2}}}J^{\mu}(-k)\eta_{\mu\nu}J^{\nu}(k)\\
\\
= & -{\displaystyle \frac{1}{k^{2}}}\left[-J^{0}(-k)J^{0}(k)+J^{1}(-k)J^{1}(k)\right.\\
\\
 & \left.+J^{2}(-k)J^{2}(k)+J^{3}(-k)J^{3}(k)\right]
\end{array}\label{eq:32}
\end{equation}
and a virtual photon of four-momentum $k^{\mu}\equiv\left(k^{0},0,0,k^{3}\right).$
Current conservation implies that
\begin{equation}
k^{0}J^{0}=k^{3}J^{3}.\label{eq:33}
\end{equation}
Note that%
\footnote{We are considering the complex conjugation because now the polarization
vectors have also complex components.%
}

\begin{equation}
\begin{array}{rl}
J^{\mu}(-k)\left({\displaystyle \sum_{{\scriptscriptstyle j_{z}=+1,-1}}\epsilon_{{\scriptscriptstyle (1,j_{z})},\mu}\epsilon_{{\scriptscriptstyle (1,j_{z}),\nu}}^{*}}\right)J^{\nu}(k)= & J^{\mu}(-k)\epsilon_{{\scriptscriptstyle (1,+1)},\mu}\epsilon_{{\scriptscriptstyle (1,+1),\nu}}^{*}J^{\nu}(k)\\
\\
 & +J^{\mu}(-k)\epsilon_{{\scriptscriptstyle (1,-1)},\mu}\epsilon_{{\scriptscriptstyle (1,-1),\nu}}^{*}J^{\nu}(k)\\
\\
= & {\displaystyle \frac{1}{2}\left[J^{1}(-k)+iJ^{2}(-k)\right]}\left[J^{1}(k)-iJ^{2}(k)\right]\\
\\
 & +{\displaystyle \frac{1}{2}\left[J^{1}(-k)-iJ^{2}(-k)\right]}\left[J^{1}(k)+iJ^{2}(k)\right]\\
\\
= & J^{1}(-k)J^{1}(k)+J^{2}(-k)J^{2}(k),
\end{array}\label{eq:35}
\end{equation}
and also that
\begin{equation}
\begin{array}{rl}
J^{\mu}(-k)\left(\epsilon_{{\scriptscriptstyle (1,0)},\mu}\epsilon_{{\scriptscriptstyle (1,0),\nu}}^{*}\right)J^{\nu}(k)= & {\displaystyle \frac{1}{k^{2}}}\left[J^{0}(-k)k_{3}+J^{3}(-k)k_{0}\right]\left[J^{0}(k)k_{3}+J^{3}(k)k_{0}\right]\\
\\
= & {\displaystyle \frac{1}{k^{2}}}\left[-J^{0}(-k)k^{3}+J^{3}(-k)k^{0}\right]\left[-J^{0}(k)k^{3}+J^{3}(k)k^{0}\right]\\
\\
= & {\displaystyle \frac{1}{k^{2}}}\left[J^{0}(-k)J^{0}(-k)(k^{3})^{2}-J^{0}(-k)J^{3}(-k)k^{0}k^{3}\right.\\
\\
 & \left.-J^{3}(-k)J^{0}(-k)k^{0}k^{3}+J^{3}(-k)J^{3}(-k)(k^{0})^{2}\right].
\end{array}\label{eq:37}
\end{equation}
Then, since the conservation relation (\ref{eq:33}) holds, we obtain
\begin{equation}
J^{\mu}(-k)\left(\epsilon_{{\scriptscriptstyle (1,0),\mu}_{,}}\epsilon_{{\scriptscriptstyle (1,0),\nu}}^{*}\right)J^{\nu}(k)=-J^{0}(-k)J^{0}(k)+J^{3}(-k)J^{3}(k)\label{eq:38}
\end{equation}
Hence, from the relations (\ref{eq:33}), (\ref{eq:35}) and (\ref{eq:38}),
we deduce that in the off-shell case the saturated propagator can
be rewritten as
\begin{equation}
J^{\mu}(-k)\mathcal{D}_{\mu\nu}(k)J^{\nu}(k)=-\frac{1}{k^{2}}J^{\mu}(-k)\left(\sum_{{\scriptscriptstyle j_{z}=+1,-1},{\scriptscriptstyle 0}}\epsilon_{{\scriptscriptstyle (1,j_{z})},\mu}\epsilon_{{\scriptscriptstyle (1,j_{z}),\nu}}^{*}\right)J^{\nu}(k).\label{eq:39}
\end{equation}
As we have already seen  in the section $1.2,$ we have had another
confirmation that a virtual photon has $\mathit{three}$ degrees of
freedom. 

$ $

\textbf{On-shell photon}

$ $\\
As for on-shell photon $(k^{0}=k^{3})$ the conservation law (\ref{eq:33})
becomes 
\begin{equation}
J^{0}=J^{3},
\end{equation}
and so equation (\ref{eq:32}) reduces to
\begin{equation}
J^{\mu}(-k)\mathcal{D}_{\mu\nu}(k)J^{\nu}(k)=-\frac{1}{k^{2}}\left[J^{1}(-k)J^{1}(k)+J^{2}(-k)J^{2}(k)\right].\label{eq:34}
\end{equation}
Hence, from (\ref{eq:35}) the saturated propagator for on-shell photon
can be rewritten as
\begin{equation}
J^{\mu}(-k)\mathcal{D}_{\mu\nu}(k)J^{\nu}(k)=-\frac{1}{k^{2}}J^{\mu}(-k)\left(\sum_{{\scriptscriptstyle j_{z}=+1,-1}}\epsilon_{{\scriptscriptstyle (1,j_{z})},\mu}\epsilon_{{\scriptscriptstyle (1,j_{z}),\nu}}^{*}\right)J^{\nu}(k),\label{eq:36}
\end{equation}
and this expression explicitly shows that an on-shell photon has only
$\mathit{two}$ degrees of freedom, i.e. the $\mathit{two}$ transverse
components, as we have already shown in the section $1.2.$ \\

\section{Ghosts and unitarity analysis in Electrodynamics}

In the Appendix $C$ we show a method by which we can verify whether
ghosts%
\footnote{In this case we mean `` bad'' ghost, i.e. ghosts whose presence
violates the unitarity of the theory. Instead the ``good'' ghosts
are ghosts whose presence is necessarily required to preserve the
unitarity (see Appendix $C)$.%
} and tachyons are absent, and, so, whether the theory preserves the
unitarity. The method states that to verify whether ghosts and tachyons
are absent in a given Lagrangian, one has to require that the propagator
has only first order poles at $k^{2}-m^{2}=0$ with $\mathit{real}$
masses $m$ (no tachyons) and with $\mathit{positive}$ residues (no
ghosts) {[}\ref{-P.-Van}{]}, {[}\ref{-A.-Accioly}{]}. Therefore,
to verify that the presence of ghosts doesn't violate the unitarity,
we couple the propagator to external conserved currents, $J^{\mu},$
compatible with the symmetry of the theory, and afterward we verify
the positivity of the residue of the current-current amplitude%
\footnote{Note that the positivity of the imaginary part of the amplitude residue
is just a $\mathit{necessary}$ condition to ensure the unitarity
condition, it is not a $\mathit{sufficient}$ condition.%
}. \\
In ED $m=0,$ since we are considering massless photons, so we know
that tachyons are absent. Now let us consider the following tree level
amplitude%
\footnote{We have used the reality condition $J(-k)=J^{*}(k).$ %
}{[}\ref{-J.-Helayel}{]}:

\begin{equation}
\begin{array}{rl}
\mathcal{A}= & J^{*\mu}(k)\left\langle T\left(A_{\mu}(-k)A_{\nu}(k)\right)\right\rangle J^{\nu}(k)=iJ^{\mu}(-k)\mathcal{D}_{\mu\nu}(k)J^{\nu}(k)\\
\\
= & -{\displaystyle iJ^{\mu}(-k)\frac{1}{k^{2}}\left[\eta_{\mu\nu}+\left(\alpha-1\right)\frac{k_{\mu}k_{\nu}}{k^{2}}\right]J^{\nu}(k)}\\
\\
= & {\displaystyle \frac{-i}{k^{2}}\eta_{\mu\nu}J^{\mu}(-k)J^{\nu}(k).}
\end{array}\label{eq:19}
\end{equation}
We are going to study the residue of the amplitude to check whether
the unitarity is preserved and at same time ghosts are absent.\\
Hence let us calculate the residue of the amplitude (\ref{eq:19})
at $k^{2}=0:$%
\footnote{See Ref. {[}\ref{-M.-R. spiegel}{]} to see how to calculate the residues
of complex functions, or any other books on complex analysis. %
} 
\begin{equation}
\begin{array}{rl}
Res_{k^{2}=0}\left\{ \mathcal{A}\right\} = & {\displaystyle Res_{k^{2}=0}\left\{ \frac{-i}{k^{2}}\eta_{\mu\nu}J^{\mu}(-k)J^{\nu}(k)\right\} }\\
\\
= & {\displaystyle \lim_{{\scriptstyle k^{2}\rightarrow0}}k^{2}\left(\frac{-i}{k^{2}}\eta_{\mu\nu}J^{\mu}(-k)J^{\nu}(k)\right)=-i\eta_{\mu\nu}J^{\mu}(-k)J^{\nu}(k)}.
\end{array}\label{eq:22}
\end{equation}
Since we want to verify the positivity of the imaginary part of the
residue in $k^{2}=0,$ we can valuate the current conservation for
$k^{0}=k^{3}:$\\
\begin{equation}
0=k^{\mu}J_{\mu}=k^{0}J_{0}+k^{3}J_{3}=k^{0}\left(J_{0}+J_{3}\right)=0\underset{{\scriptscriptstyle {\scriptstyle k^{0}\neq0}}}{\Rightarrow}J_{0}=-J_{3}=J^{3};\label{eq:23}
\end{equation}
thus by substituting in the residue (\ref{eq:22}) we find
\begin{equation}
\begin{array}{rl}
Res_{k^{2}=0}{\displaystyle \left\{ \frac{-i}{k^{2}}\eta_{\mu\nu}J^{\mu*}(k)J^{\nu}(k)\right\} }= & -iJ^{0*}J^{0}+iJ^{1*}J^{1}+iJ^{2*}J^{2}+iJ^{3*}J^{3}\\
\\
= & i\left(|J^{1}|^{2}+|J^{2}|^{2}\right)\Rightarrow\mathrm{Im}Res_{k^{2}=0}\left\{ \mathcal{A}\right\} >0.
\end{array}\label{eq:24}
\end{equation}
Finally we have showed that the imaginary part of the residue is positive,
so ghosts are absent and the unitarity is preserved.

\chapter{Symmetric two-rank tensor field: graviton}

In the previous section we have studied the ED case where the main
character was a vector field $A^{\mu}.$ Now our aim is to study the
same topics but in the context of linearized GR where the main physical
quantity is a symmetric two-rank tensor, $h_{\mu\nu}.$ \\
\\
In my opinion the most elegant approach to work on linearized GR is
the geometrical interpretation which gives us the physical meaning
of the theory. This geometrical perspective is based on the fact that
$g_{\mu\nu}=\eta_{\mu\nu}+h_{\mu\nu}$ is the metric of the space-time,
where $h_{\mu\nu}$ represents a perturbation of Minkowski background.
In this approach, the gravitational interaction is described by geometric
tools such as the equation of geodesic deviation, curvature tensors,
and so, it is interpreted as deformation of space-time. This deformation
is associated to the propagation of gravitational waves which are
determined by examining how $h_{\mu\nu}$ contributes to the curvature
of the background space-time. This first way to proceed is also called
$\mathit{top}$ $\mathit{down}$ $\mathit{approach}.$ \\
General Relativity can also be seen as a classical%
\footnote{With the word ``classical'' we mean ``not quantum''.%
} field theory in its linearized form, and from this point of view
one can directly introduce the concept of quantization. In this approach
we can apply all standard field-theoretical methods; we shall treat
linearized gravity as a classical field theory of the symmetric field
$h_{\mu\nu}$ living in a $\mathit{flat}$ space-time with Minkowski
metric $\eta_{\mu\nu}.$ In this way we forget that $h_{\mu\nu}$
has an interpretation in terms of a space-time metric, and instead
we treat it as any other field living in Minkowski space-time. This
second way to proceed is also called $\mathit{bottom}$ $\mathit{up}$
$\mathit{approach}.$ \\
The geometrical and the field-theoretical approaches are complementary;
some aspects of gravitational waves physics can be better understood
from the former pro, some from the latter, and together they give
us a deeper overall understanding. \\
In this Chapter we shall start from the point of view of geometrical
approach and then by linearizing the theory we shall proceed using
tools of classical field theory. Afterwords we will be able to discuss
linearized gravity from the point of view of quantum field theory,
and we will obtain the graviton propagator. We will be able to interpret
the graviton, the particle associated to the gravitational wave, as
a particle mediator of gravitational interaction.\\
In our approach we will make use of the $\mathit{spin}$ $\mathit{projector}$
$\mathit{operators},$ by which we can better understand the number
and which degrees of freedom propagate in General Relativity. \\
In our approach we will take inspiration from%
\footnote{See also Ref. {[}\ref{-A.-Accioly, rey}{]} and {[}\ref{-B.-Pereira-Dias,}{]}
for examples in which the formalism of the spin projector operators
is used.%
} {[}\ref{-P.-Van}{]}, {[}\ref{-J.-Helayel}{]}.

\section{Graviton Lagrangian}

Our starting point is the Lagrangian for any symmetric two-rank tensor
field. We can obtain it in more ways: for example we can consider
all the possible invariants quadratic in the tensor field $h_{\mu\nu}$
and by imposing the field equations we can find the value of the coefficients
for each terms. We shall proceed in a different way, i.e. we want
to linearize starting from the geometrical approach to GR. \\
Once we have perturbed the metric
\begin{equation}
g_{\mu\nu}=\eta_{\mu\nu}+h_{\mu\nu},\,\,\,\,\,\, g^{\mu\nu}=\eta^{\mu\nu}-h^{\mu\nu},
\end{equation}
we can obtain the Lagrangian by considering the quadratic part in
$h_{\mu\nu}$ of Hilbert-Einstein action%
\footnote{Let us note that the coupling constant $\kappa=\frac{1}{M_{p}^{2}}$
doesn't appear in H-E action (\ref{eq:95HE}). According to our convention,
the coupling constant is introduced when the interaction term with
a matter source is considered (see below). %
} 
\begin{equation}
S_{HE}=-\int d^{4}x\sqrt{-g}\mathcal{R}.\label{eq:95HE}
\end{equation}
By performing the perturbation around Minkowski background one has
\begin{equation}
S_{HE}(g_{\mu\nu})=S_{HE}(\eta_{\mu\nu}+\delta g_{\mu\nu})=S_{HE}(\eta_{\mu\nu})+\frac{\delta S_{HE}}{\delta g^{\mu\nu}}\delta g^{\mu\nu}+\mathcal{O}((\delta g)^{2}),
\end{equation}
where we are neglecting cubic terms in the perturbation $h_{\mu\nu}.$
\\
Note that, since $\mathcal{R}(\eta_{\mu\nu})=0,$ then also $S_{HE}(\eta_{\mu\nu})=0;$
moreover terms linear in the perturbation $\delta g^{\mu\nu}$ do
not appear. Thus, the linearized H-E action, quadratic in $\delta g^{\mu\nu}$
is given by
\begin{equation}
S_{HE}=\frac{\delta S_{HE}}{\delta g^{\mu\nu}}\delta g^{\mu\nu}=-\int d^{4}x(\delta g^{\mu\nu})\left(\mathcal{R}_{\mu\nu}-\frac{1}{2}\eta_{\mu\nu}\mathcal{R}\right)\label{eq:97HE}
\end{equation}
Once we note that $\delta g^{\mu\nu}=-h^{\mu\nu},$ by using the linearized
forms of the Riemann tensor, Ricci tensor and scalar tensor:
\begin{equation}
\begin{array}{rl}
\mathcal{R}_{\mu\nu\lambda\sigma}= & {\displaystyle \frac{1}{2}}\left(\partial_{\nu}\partial_{\lambda}h_{\mu\sigma}+\mathcal{\partial_{\mu}\partial_{\sigma}}h_{\nu\lambda}-\mathcal{\partial_{\sigma}\partial_{\nu}}h_{\mu\lambda}-\partial_{\mu}\partial_{\lambda}h_{\nu\sigma}\right),\\
\\
\mathcal{R_{\mu\nu}}=g^{\alpha\rho}\mathcal{R}_{\alpha\mu\rho\nu}= & {\displaystyle \frac{1}{2}}\left(\partial_{\rho}\partial_{\nu}h_{\mu}^{\rho}+\partial_{\rho}\partial_{\mu}h_{\nu}^{\rho}-\partial_{\mu}\partial_{\nu}h-\boxempty h_{\mu\nu}\right),\\
\\
\mathcal{R}= & \partial_{\mu}\partial_{\nu}h^{\mu\nu}-\boxempty h,
\end{array}\label{eq:97linearized curv}
\end{equation}
the perturbed action in (\ref{eq:97HE}) becomes 
\begin{equation}
\begin{array}{rl}
S_{HE}= & -{\displaystyle \int}d^{4}x(-h^{\mu\nu})\left(\mathcal{R}_{\mu\nu}-\frac{1}{2}\eta_{\mu\nu}\mathcal{R}\right)\\
\\
= & \int d^{4}xh^{\mu\nu}\left[\frac{1}{2}\left(\partial_{\rho}\partial_{\nu}h_{\mu}^{\rho}+\partial_{\rho}\partial_{\mu}h_{\nu}^{\rho}-\partial_{\mu}\partial_{\nu}h-\boxempty h_{\mu\nu}\right)\right.\\
\\
 & \left.-\frac{1}{2}\eta_{\mu\nu}\left(\partial_{\alpha}\partial_{\beta}h^{\alpha\beta}-\boxempty h\right)\right]\\
\\
= & \int d^{4}x\left(h_{\sigma}^{\mu}\partial^{\sigma}\partial^{\nu}h_{\mu\nu}-h\partial^{\mu}\partial^{\nu}h_{\mu\nu}-\frac{1}{2}h_{\mu\nu}\boxempty h^{\mu\nu}+\frac{1}{2}h\boxempty h\right)\\
\\
\equiv & \int d^{4}x\mathcal{L}_{HE},
\end{array}\label{eq:40.1}
\end{equation}
thus the Lagrangian for any symmetric two-rank tensor is%
\footnote{We started from the geometrical point of view to find $\mathcal{L}_{HE},$
and in this case $h_{\mu\nu}$ is interpreted as the metric perturbation;
but from the point of view of field theory it is only seen as a generic
field and we can't say that it is related to any metrics at this level.%
} 
\begin{equation}
\mathcal{L}_{HE}\coloneqq h_{\sigma}^{\mu}\partial^{\sigma}\partial^{\nu}h_{\mu\nu}-h\partial^{\mu}\partial^{\nu}h_{\mu\nu}-\frac{1}{2}h_{\mu\nu}\boxempty h^{\mu\nu}+\frac{1}{2}h\boxempty h.\label{eq:40}
\end{equation}
By raising and lowering the indices with the metric tensor $\eta_{\mu\nu},$
we can rewrite the Lagrangian (\ref{eq:40}) in the following way:

\begin{equation}
\begin{array}{rl}
\mathcal{L}_{HE}= & h_{\mu\nu}\left(\partial^{\nu}\partial^{\sigma}\eta^{\mu\rho}\right)h_{\rho\sigma}-h_{\mu\nu}\left(\partial^{\mu}\partial^{\nu}\eta^{\rho\sigma}\right)h_{\rho\sigma}\\
\\
 & -\frac{1}{2}h_{\mu\nu}\left(\eta^{\mu\rho}\eta^{\nu\sigma}\boxempty\right)h_{\rho\sigma}+\frac{1}{2}h_{\mu\nu}\left(\eta^{\mu\nu}\eta^{\rho\sigma}\boxempty\right)\\
\\
= & \frac{1}{2}h_{\mu\nu}\left[2\partial^{\nu}\partial^{\sigma}\eta^{\mu\rho}-2\partial^{\mu}\partial^{\nu}\eta^{\rho\sigma}-\eta^{\mu\rho}\eta^{\nu\sigma}\boxempty+\eta^{\mu\nu}\eta^{\rho\sigma}\boxempty\right]h_{\rho\sigma}\\
\\
= & \frac{1}{2}h_{\mu\nu}\left[-\left(\frac{1}{2}\eta^{\mu\rho}\eta^{\nu\sigma}+\frac{1}{2}\eta^{\mu\sigma}\eta^{\nu\rho}-\eta^{\mu\nu}\eta^{\rho\sigma}\right)\boxempty-\eta^{\mu\nu}\partial^{\rho}\partial^{\sigma}-\eta^{\rho\sigma}\partial^{\mu}\partial^{\nu}\right.\\
\\
 & \left.+\frac{1}{2}\left(\eta^{\nu\rho}\partial^{\mu}\partial^{\sigma}+\eta^{\nu\sigma}\partial^{\mu}\partial^{\rho}+\eta^{\mu\rho}\partial^{\nu}\partial^{\sigma}+\eta^{\mu\sigma}\partial^{\nu}\partial^{\rho}\right)\right]h_{\rho\sigma}.
\end{array}\label{eq:100HElagr}
\end{equation}

Thus, the Lagrangian (\ref{eq:40}) can be recast as

\begin{equation}
\mathcal{L}_{HE}=\frac{1}{2}h_{\mu\nu}\mathcal{O}^{\mu\nu\rho\sigma}h_{\rho\sigma},\label{eq:41}
\end{equation}
 where $\mathcal{O}^{\mu\nu\rho\sigma}$ is given by 
\begin{equation}
\begin{array}{rl}
\mathcal{O}^{\mu\nu\rho\sigma}\coloneqq & -\left(\frac{1}{2}\eta^{\mu\rho}\eta^{\nu\sigma}+\frac{1}{2}\eta^{\mu\sigma}\eta^{\nu\rho}-\eta^{\mu\nu}\eta^{\rho\sigma}\right)\boxempty-\eta^{\mu\nu}\partial^{\rho}\partial^{\sigma}-\eta^{\rho\sigma}\partial^{\mu}\partial^{\nu}\\
\\
 & +\frac{1}{2}\left(\eta^{\nu\rho}\partial^{\mu}\partial^{\sigma}+\eta^{\nu\sigma}\partial^{\mu}\partial^{\rho}+\eta^{\mu\rho}\partial^{\nu}\partial^{\sigma}+\eta^{\mu\sigma}\partial^{\nu}\partial^{\rho}\right),
\end{array}\label{eq:42}
\end{equation}
and satisfies the symmetries 
\begin{equation}
\mathcal{O}^{\mu\nu\rho\sigma}=\mathcal{O}^{\nu\mu\rho\sigma}=\mathcal{O}^{\mu\nu\sigma\rho}=\mathcal{O}^{\rho\sigma\mu\nu}.\label{eq:43}
\end{equation}
By varying the linearized action, we can obtain the Euler-Lagrange
equations%
\footnote{From the geometrical point of view we can obtain the same field equations
by linearizing Einstein equations.%
} for the symmetric two-rank tensor $h_{\mu\nu}.$ In order to do this,
it is more convenient to rewrite the Lagrangian (\ref{eq:40}) only
in terms of the first derivatives of $h_{\mu\nu}$ by means integration
by parts:
\begin{equation}
\mathcal{L}_{HE}=-\partial_{\rho}h_{\alpha}^{\rho}\partial_{\beta}h^{\alpha\beta}+\partial_{\alpha}h\partial_{\beta}h^{\alpha\beta}+\frac{1}{2}\partial_{\rho}h^{\alpha\beta}\partial^{\rho}h_{\alpha\beta}-\frac{1}{2}\partial_{\rho}h\partial^{\rho}h.\label{eq:104L}
\end{equation}
 The field equations are given by 
\begin{equation}
\partial_{\sigma}\frac{\partial\mathcal{L}_{HE}}{\partial(\partial_{\sigma}h_{\mu\nu})}=\frac{\partial\mathcal{L}_{HE}}{\partial h_{\mu\nu}},\label{eq:44}
\end{equation}
 thus by computing the derivatives with respect to $h_{\mu\nu}$ and
$\partial_{\sigma}h_{\mu\nu}$ of the Lagrangian in (\ref{eq:104L})
we obtain
\begin{equation}
\frac{\partial\mathcal{L}_{HE}}{\partial h_{\mu\nu}}=0\label{eq:45}
\end{equation}
and 
\[
\begin{array}{rl}
{\displaystyle \frac{\partial\mathcal{L}_{HE}}{\partial(\partial_{\sigma}h_{\mu\nu})}}= & {\displaystyle -\eta^{\mu\sigma}\partial_{\rho}h^{\nu\rho}-\eta^{\nu\sigma}\partial_{\rho}h^{\rho\mu}+\partial^{\sigma}h^{\mu\nu}}\\
\\
 & {\displaystyle +\eta^{\mu\nu}\partial_{\rho}h^{\sigma\rho}+\eta^{\nu\sigma}\partial^{\mu}h-\eta^{\mu\nu}\partial^{\sigma}h}
\end{array}
\]
\begin{equation}
\begin{array}{rl}
\Rightarrow{\displaystyle \partial_{\rho}\frac{\partial\mathcal{L}_{HE}}{\partial(\partial_{\rho}h_{\mu\nu})}}= & {\displaystyle -\partial^{\mu}\partial_{\rho}h^{\nu\rho}-\partial^{\nu}\partial_{\rho}h^{\rho\mu}+\boxempty h^{\mu\nu}}\\
\\
 & {\displaystyle +\eta^{\mu\nu}\partial_{\rho}\partial_{\sigma}h^{\rho\sigma}+\partial^{\mu}\partial^{\nu}h-\eta^{\mu\nu}\boxempty h.}
\end{array}\label{eq:46}
\end{equation}
 Hence, substituting (\ref{eq:45}) and (\ref{eq:46}) into (\ref{eq:44})
we obtain the field equations in the vacuum:
\begin{equation}
\partial^{\mu}\partial_{\rho}h^{\nu\rho}+\partial^{\nu}\partial_{\rho}h^{\rho\mu}-\boxempty h^{\mu\nu}+\eta^{\mu\nu}\partial_{\sigma}\partial_{\rho}h^{\sigma\rho}-\partial^{\mu}\partial^{\nu}h+\eta^{\mu\nu}\boxempty h=0,\label{eq:47}
\end{equation}
 or equivalently, raising and lowering the indices with $\eta_{\mu\nu},$
\begin{equation}
\left(\eta_{\mu\rho}\eta_{\nu\sigma}\boxempty+\eta_{\rho\sigma}\partial_{\mu}\partial_{\nu}-\eta_{\mu\sigma}\partial_{\nu}\partial_{\rho}-\eta_{\nu\sigma}\partial_{\mu}\partial_{\rho}+\eta_{\mu\nu}\partial_{\sigma}\partial_{\rho}-\eta_{\mu\nu}\eta_{\rho\sigma}\boxempty\right)h^{\rho\sigma}=0.\label{eq:48}
\end{equation}
Now multiply for $\eta^{\mu\nu}$ one gets
\begin{equation}
\begin{array}{rl}
 & \left(2\boxempty\eta_{\rho\sigma}-2\partial_{\rho}\partial_{\sigma}\right)h^{\rho\sigma}+4\left(\partial_{\rho}\partial_{\sigma}-\boxempty\eta_{\rho\sigma}\right)=0\\
\\
\Leftrightarrow & \left(\partial_{\rho}\partial_{\sigma}-\boxempty\eta_{\rho\sigma}\right)h^{\rho\sigma}=0.
\end{array}\label{eq:49}
\end{equation}
Since (\ref{eq:49}) holds, the field equations in the vacuum assume
a simplified form%
\footnote{From the geometrical point of view it means that $\mathcal{R}=0$
in the vacuum. In fact the linearized form of the Ricci scalar is
$\mathcal{R}=\partial_{\rho}\partial_{\sigma}h^{\rho\sigma}-\boxempty h=\left(\partial_{\rho}\partial_{\sigma}-\eta_{\rho\sigma}\boxempty\right)h^{\rho\sigma}.$%
}
\begin{equation}
\left(\eta_{\mu\rho}\eta_{\nu\sigma}\boxempty+\eta_{\rho\sigma}\partial_{\mu}\partial_{\nu}-\eta_{\mu\sigma}\partial_{\nu}\partial_{\rho}-\eta_{\nu\sigma}\partial_{\mu}\partial_{\rho}\right)h^{\rho\sigma}=0\label{eq:50}
\end{equation}
\\
In presence of a source $\tau^{\mu\nu}$ we have to add the term $-\kappa h_{\alpha\beta}\tau^{\alpha\beta}$%
\footnote{We have the minus sign because starting from GR in a generic background
we have: $S=S_{HE}+S_{m},$ where $S_{HE}$ was already examined in
(\ref{eq:40.1}), and the matter part whose variation with respect
to the metric is $\delta S_{m}$=$\kappa\int d^{4}x\delta g^{\mu\nu}\tau_{\mu\nu}=\kappa\int d^{4}x(-g^{\mu\alpha}g^{\nu\beta}\delta g_{\alpha\beta})\tau_{\mu\nu},$
and since $\delta g_{\alpha\beta}=h_{\alpha\beta},$ from the point
of view of the field theory approach we consider $-\kappa h_{\alpha\beta}\tau^{\alpha\beta}.$ %
} to the Lagrangian $\mathcal{L}_{HE},$ and the field equations become
\begin{equation}
\boxempty h^{\mu\nu}-\partial^{\mu}\partial_{\rho}h^{\nu\rho}-\partial^{\nu}\partial_{\rho}h^{\rho\mu}+\eta^{\mu\nu}\partial_{\sigma}\partial_{\rho}h^{\sigma\rho}+\partial^{\mu}\partial^{\nu}h-\eta^{\mu\nu}\boxempty h=-\kappa\tau^{\mu\nu},\label{eq:51}
\end{equation}
 since 
\begin{equation}
\frac{\partial\mathcal{L}_{HE}}{\partial h_{\mu\nu}}=-\kappa\tau^{\mu\nu}.
\end{equation}
Also in this case, in analogy to the case of vector fields, we can
introduce the spin projector operators in the space of the symmetric
two-rank tensors. We can decompose (see Appendix $B)$ a symmetric
two-rank tensor in terms of spin-$2,$ spin-$1$ and two spin-$0$
components under the rotation group $SO(3),$ i.e. $h_{\mu\nu}\in\mathbf{0}\oplus\mathbf{0}\oplus\mathbf{1}\oplus\mathbf{2},$
by introducing the following set of operators%
\footnote{In Appendix $B$ we discuss more in detail the basis of spin projector
operators, taking into account also the possibility to have antisymmetric
operators. %
} 
\begin{equation}
\begin{array}{l}
\mathcal{P}_{\mu\nu\rho\sigma}^{2}={\displaystyle \frac{1}{2}\left(\theta_{\mu\rho}\theta_{\nu\sigma}+\theta_{\mu\sigma}\theta_{\nu\rho}\right)-\frac{1}{3}\theta_{\mu\nu}\theta_{\rho\sigma},}\\
\\
\mathcal{P}_{\mu\nu\rho\sigma}^{1}={\displaystyle \frac{1}{2}}\left(\theta_{\mu\rho}\omega_{\nu\sigma}+\theta_{\mu\sigma}\omega_{\nu\rho}+\theta_{\nu\rho}\omega_{\mu\sigma}+\theta_{\nu\sigma}\omega_{\mu\rho}\right),\\
\\
\mathcal{P}_{s,\,\mu\nu\rho\sigma}^{0}={\displaystyle \frac{1}{3}}\theta_{\mu\nu}\theta_{\rho\sigma},\,\,\,\,\,\,\,\,\,\,\mathcal{P}_{w,\,\mu\nu\rho\sigma}^{0}=\omega_{\mu\nu}\omega_{\rho\sigma},\\
\\
\mathcal{P}_{sw,\,\mu\nu\rho\sigma}^{0}={\displaystyle \frac{1}{\sqrt{3}}\theta_{\mu\nu}\omega_{\rho\sigma},\,\,\,\,\,\,\,\mathcal{P}_{ws,\,\mu\nu\rho\sigma}^{0}=\frac{1}{\sqrt{3}}}\omega_{\mu\nu}\theta_{\rho\sigma};\\
\\
\end{array}\label{eq:52}
\end{equation}
where the projectors $\theta_{\mu\nu}$ and $\omega_{\mu\nu}$ have
been already defined in the previous Chapter (see eqs. (\ref{eq:4})-(\ref{eq:5})).
It is worth recalling their expressions in momentum space:
\[
\theta_{\mu\nu}=\eta_{\mu\nu}-\omega_{\mu\nu},\,\,\,\,\,\,\omega_{\mu\nu}=\frac{k_{\mu}k_{\nu}}{k^{2}}.
\]
The operators defined in (\ref{eq:52}) satisfy the following orthogonality
relations 
\begin{equation}
\begin{array}{lll}
\mathcal{P}_{a}^{i}\mathcal{P}_{b}^{j}=\delta_{ij}\delta_{ab}\mathcal{P}_{a}^{j}, &  & \mathcal{P}_{ab}^{0}\mathcal{P}_{c}^{i}=\delta_{i0}\delta_{bc}\mathcal{P}_{ab}^{i},\\
\\
\mathcal{P}_{ab}^{0}\mathcal{P}_{cd}^{0}=\delta_{ad}\delta_{bc}\mathcal{P}_{a}^{0}, &  & \mathcal{P}_{c}^{i}\mathcal{P}_{ab}^{0}=\delta_{i0}\delta_{ac}\mathcal{P}_{ab}^{0},
\end{array}\label{eq:53}
\end{equation}
where $i,j=2,1,0$ and $a,b,c,d=s,w,\mathrm{absent,}$%
\footnote{Note that the spin projector operators $\mathcal{P}^{2}$ and $\mathcal{P}^{1}$
do not have the lower indices, so it can happen that $a,b,c,d$ are
absent. %
} and also the completeness property%
\footnote{We shall often suppress the indices for simplicity. %
} 
\begin{equation}
\mathcal{P}^{2}+\mathcal{P}^{1}+\mathcal{P}_{s}^{0}+\mathcal{P}_{w}^{0}=\mathbb{I}\Leftrightarrow\left(\mathcal{P}^{2}+\mathcal{P}^{1}+\mathcal{P}_{s}^{0}+\mathcal{P}_{w}^{0}\right)_{\mu\nu\rho\sigma}=\frac{1}{2}\left(\eta_{\mu\rho}\eta_{\nu\sigma}+\eta_{\nu\rho}\eta_{\mu\sigma}\right).\label{eq:54}
\end{equation}
As we have done for the projectors $\theta$ and $\omega$ in the
previous chapter, we can show what is the associated spin for each
spin projector operators in (\ref{eq:52}) (see Appendix $B):$\\
\begin{equation}
\begin{array}{ll}
\eta^{\mu\rho}\eta^{\nu\sigma}\mathcal{P}_{\mu\nu\rho\sigma}^{2}=5=2(2)+1 & (\mathrm{spin\textrm{-}2),}\\
\\
\eta^{\mu\rho}\eta^{\nu\sigma}\mathcal{P}_{\mu\nu\rho\sigma}^{1}=3=2(1)+1 & (\mathrm{spin\textrm{-}1),}\\
\\
\eta^{\mu\rho}\eta^{\nu\sigma}\mathcal{P}_{s,\,\mu\nu\rho\sigma}^{0}=1=2(0)+1 & (\mathrm{spin\textrm{-}0),}\\
\\
\eta^{\mu\rho}\eta^{\nu\sigma}\mathcal{P}_{w,\,\mu\nu\rho\sigma}^{0}=1=2(0)+1 & (\mathrm{spin\textrm{-}0).}
\end{array}\label{eq:55}
\end{equation}
Moreover one can easily show that the following relations hold: 
\begin{equation}
\begin{array}{ll}
\eta^{\mu\nu}\mathcal{P}_{\mu\nu\rho\sigma}^{2}=0=\eta^{\mu\nu}\mathcal{P}_{\mu\nu\rho\sigma}^{2} & (\mathrm{traceless),}\\
\\
k^{\mu}\mathcal{P}_{\mu\nu\rho\sigma}^{2}=0=k^{\mu}\mathcal{P}_{\mu\nu\rho\sigma}^{1} & \mathrm{(transverse)}.
\end{array}\label{eq:56}
\end{equation}

Note that we have introduced six operators of that form the basis
\begin{equation}
\left\{ \mathcal{P}^{2},\mathcal{P}^{1},\mathcal{P}_{s}^{0},\mathcal{P}_{w}^{0},\mathcal{P}_{sw}^{0},\mathcal{P}_{ws}^{0}\right\} 
\end{equation}
in terms of which the symmetric four-rank tensor $\mathcal{O}^{\mu\nu\rho\sigma}$
can be expanded%
\footnote{By the expression ``a symmetric four-rank tensor'' we mean the operator
$\mathcal{O}^{\mu\nu\rho\sigma}$ that appear in a given parity-invariant
Lagrangian, like the operator (\ref{eq:42}). See Appendix $B.2$
for more details.%
}. At the same time, from the properties (\ref{eq:53})-(\ref{eq:54})
we notice that four out of six form a complete set of spin projector
operators, 
\begin{equation}
\left\{ \mathcal{P}^{2},\mathcal{P}^{1},\mathcal{P}_{s}^{0},\mathcal{P}_{w}^{0}\right\} ,
\end{equation}
in terms of which a symmetric two-rank tensor can be decomposed in
one spin-$2,$ one spin-$1$ and two spin-$0$ components. The operators
$\mathcal{P}_{sw}^{0}$ and $\mathcal{P}_{ws}^{0}$ are not projectors
as we can see from the relations (\ref{eq:53}), but are necessary
to close the algebra and form a basis of symmetric four-rank tensors
in terms of which the operator space of the gravitational field equations
can be spanned. They can potentially mix the two scalar multiplets
$s$ and $w$ (See Appendix $B$ for more details).

This basis of projectors represents six field degrees of
freedom. The other four fields in a symmetric tensor field, as
usual, represent the gauge (unphysical) degrees of freedom. $\mathcal{P}^{2}$
and $\mathcal{P}^{1}$ represent transverse and traceless spin-$2$
and spin-$1$ degrees, accounting for four degrees of freedom, while
$\mathcal{P}_{s}^{0}$ and $\mathcal{P}_{w}^{0}$ represent the spin-$0$
scalar multiplets. In addition we need also to
consider the transition operators $\mathcal{P}_{sw}^{0}$ and $\mathcal{P}_{ws}^{0}$ which are
not projectors as we can see from (\ref{eq:53}), but are necessary
to close the algebra and form a basis; they can mix the
two scalar multiplets.\\
\\
In terms of the spin projector operators $h_{\mu\nu}$ decomposes
as
\begin{equation}
h^{\mu\nu}=\mathcal{P}_{\rho\sigma}^{2\mu\nu}h^{\rho\sigma}+\mathcal{P}_{\rho\sigma}^{1\mu\nu}h^{\rho\sigma}+\mathcal{P}_{s,\,\rho\sigma}^{0\mu\nu}h^{\rho\sigma}+\mathcal{P}_{w,\,\rho\sigma}^{0\mu\nu}h^{\rho\sigma}.
\end{equation}
In momentum space, taking into account the presence of the source,
the field equations (\ref{eq:48}) assume the form
\begin{equation}
\begin{array}{rl}
 & {\displaystyle \left(\eta_{\mu\rho}\eta_{\nu\sigma}+\eta_{\rho\sigma}\frac{k_{\mu}k_{\nu}}{k^{2}}-\eta_{\mu\sigma}\frac{k_{\nu}k_{\rho}}{k^{2}}-\eta_{\nu\sigma}\frac{k_{\mu}k_{\rho}}{k^{2}}+\eta_{\mu\nu}\frac{k_{\rho}k_{\sigma}}{k^{2}}-\eta_{\mu\nu}\eta_{\rho\sigma}\right)h^{\rho\sigma}=\frac{\kappa}{k^{2}}\tau_{\mu\nu}}\\
\\
\Leftrightarrow & \Bigl(\eta_{\mu\rho}\eta_{\nu\sigma}+\eta_{\rho\sigma}\omega_{\mu\nu}-\eta_{\mu\sigma}\omega_{\nu\rho}-\eta_{\nu\sigma}\omega_{\mu\rho}+\eta_{\mu\nu}\omega_{\rho\sigma}-\eta_{\mu\nu}\eta_{\rho\sigma}\Bigr)h^{\rho\sigma}={\displaystyle \frac{\kappa}{k^{2}}\tau_{\mu\nu}}.
\end{array}\label{eq:57}
\end{equation}
This equation can be expressed in terms of the spin projector operators.
In fact by manipulating (\ref{eq:57}) we obtain
\begin{equation}
\begin{array}{rl}
\left[{\displaystyle \frac{1}{2}}\left(\eta_{\mu\rho}\eta_{\nu\sigma}+\eta_{\mu\sigma}\eta_{\nu\rho}\right)+\left(\eta_{\rho\sigma}\omega_{\mu\nu}+\eta_{\mu\nu}\omega_{\sigma\rho}\right)-\left(\eta_{\mu\nu}\eta_{\rho\sigma}\right)\right.\\
\\
\left.-{\displaystyle \frac{1}{2}\left(\eta_{\mu\rho}\omega_{\nu\sigma}+\eta_{\mu\sigma}\omega_{\nu\rho}+\eta_{\nu\sigma}\omega_{\mu\rho}+\eta_{\nu\rho}\omega_{\mu\sigma}\right)}\right]h^{\rho\sigma} & ={\displaystyle \frac{\kappa}{k^{2}}}\tau_{\mu\nu},
\end{array}\label{eq:58}
\end{equation}
and since, as shown in Appendix $B.2,$ the following relations hold
\begin{equation}
\begin{array}{l}
{\displaystyle \frac{1}{2}}\left(\eta_{\mu\rho}\eta_{\nu\sigma}+\eta_{\mu\sigma}\eta_{\nu\rho}\right)=\left(\mathcal{P}^{2}+\mathcal{P}^{1}+\mathcal{P}_{s}^{0}+\mathcal{P}_{w}^{0}\right)_{\mu\nu\rho\sigma},\\
\\
\eta_{\mu\nu}\omega_{\sigma\rho}+\eta_{\rho\sigma}\omega_{\mu\nu}=\left(\sqrt{3}\left(\mathcal{P}_{sw}^{0}+\mathcal{P}_{ws}^{0}\right)+2\mathcal{P}_{w}^{0}\right)_{\mu\nu\rho\sigma},\\
\\
{\displaystyle \frac{1}{2}\left(\eta_{\mu\rho}\omega_{\nu\sigma}+\eta_{\mu\sigma}\omega_{\nu\rho}+\eta_{\nu\sigma}\omega_{\mu\rho}+\eta_{\nu\rho}\omega_{\mu\sigma}\right)}=\left(\mathcal{P}^{1}+2\mathcal{P}_{w}^{0}\right)_{\mu\nu\rho\sigma},\\
\\
\eta_{\mu\nu}\eta_{\rho\sigma}=\left(3\mathcal{P}_{s}^{0}+\mathcal{P}_{w}^{0}+\sqrt{3}\left(\mathcal{P}_{sw}^{0}+\mathcal{P}_{ws}^{0}\right)\right)_{\mu\nu\rho\sigma},
\end{array}\label{eq:59}
\end{equation}
the equations (\ref{eq:58}) become 
\begin{equation}
\begin{array}{rl}
\left[\left(\mathcal{P}^{2}+\mathcal{P}^{1}+\mathcal{P}_{s}^{0}+\mathcal{P}_{w}^{0}\right)+\left(\sqrt{3}\left(\mathcal{P}_{sw}^{0}+\mathcal{P}_{ws}^{0}\right)+2\mathcal{P}_{w}^{0}\right)\right.\\
\\
\left.-\left(3\mathcal{P}_{s}^{0}+\mathcal{P}_{w}^{0}+\sqrt{3}\left(\mathcal{P}_{sw}^{0}+\mathcal{P}_{ws}^{0}\right)\right)-\left(\mathcal{P}^{1}+2\mathcal{P}_{w}^{0}\right)\right]_{\mu\nu\rho\sigma}h^{\rho\sigma} & {\displaystyle =\frac{\kappa}{k^{2}}}\tau_{\mu\nu}.
\end{array}\label{eq:60}
\end{equation}
In terms of the spin projector operators $\mathcal{P}^{2}$ and $\mathcal{P}_{s}^{0}$
(\ref{eq:60}) read 
\begin{equation}
\left(\mathcal{P}^{2}-2\mathcal{P}_{s}^{0}\right)_{\mu\nu\rho\sigma}h^{\rho\sigma}=\frac{\kappa}{k^{2}}\tau_{\mu\nu}.\label{eq:61}
\end{equation}
For reasons of simplicity, often we will write the equations, involving
the spin projector operators, without writing the indices, so (\ref{eq:61})
can be also written as
\begin{equation}
\left(\mathcal{P}^{2}-2\mathcal{P}_{s}^{0}\right)h=\frac{\kappa}{k^{2}}\tau.
\end{equation}
Note that to rewrite the field equations in terms of the spin projector
operators, we have also rewritten the operator $\mathcal{O}$ in (\ref{eq:42})
in terms of them. Indeed, from the Lagrangian (\ref{eq:41}), the
associated field equations in momentum space turn out to be
\begin{equation}
\mathcal{O}_{\mu\nu\rho\sigma}h^{\rho\sigma}=\kappa\tau_{\mu\nu}.\label{eq:125 field equation}
\end{equation}

By comparing the equation (\ref{eq:125 field equation}) with (\ref{eq:61})
we notice that 
\begin{equation}
\mathcal{O}_{\mu\nu\rho\sigma}=k^{2}\left(\mathcal{P}^{2}-2\mathcal{P}_{s}^{0}\right)_{\mu\nu\rho\sigma}.\label{eq:126operator}
\end{equation}

\begin{rem}
Let us observe that not only the spin-$2$ component is present but
also a spin-$0$ component, while the spin-$1$ component is absent.
Thus, in total we have six degree of freedom: five spin-$2$ and one
spin-$0$ components. The other four degrees of freedom drop out because
of the presence of gauge invariance, as we shall see below. Studying
the graviton propagator we can appreciate the importance of the spin-$0$
component; we shall discuss on it in the subsection $2.4.2.$ 
\end{rem}
$ $

\textbf{Gauge invariance of graviton Lagrangian}

$ $\\
At the end of Chapter $1$ we have seen that $\delta A_{\mu}(x)=\partial_{\mu}\alpha(x)$
is a gauge symmetry for the Lagrangian $\mathcal{L}_{V}$ and for
the associated field equations. We could ask what is the gauge symmetry
for $\mathcal{L}_{HE}$ and its field equations. Let us consider the
one-parameter group of diffeomorphisms $x'^{\mu}\equiv x'^{\mu}(x)$
and in particular its infinitesimal form
\begin{equation}
x'^{\mu}=x^{\mu}+\xi^{\mu}(x)\Leftrightarrow x^{\mu}=x'^{\mu}-\xi^{\mu}(x)\label{eq:62}
\end{equation}
Now we want to study how the metric tensor $g_{\mu\nu}(x)$ transforms
under (\ref{eq:62}):
\begin{equation}
\begin{array}{rl}
g'_{\mu\nu}(x')= & {\displaystyle \frac{\partial x^{\alpha}}{\partial x'^{\mu}}\frac{\partial x^{\beta}}{\partial x'^{\nu}}g_{\alpha\beta}}(x)=\left(\delta_{\mu}^{\alpha}-\partial_{\mu}\xi^{\alpha}\right)\left(\delta_{\nu}^{\beta}-\partial_{\nu}\xi^{\beta}\right)g_{\alpha\beta}(x)\\
\\
= & g_{\mu\nu}(x)-\left(\partial_{\mu}\xi^{\alpha}\right)g_{\alpha\nu}-\left(\partial_{\nu}\xi^{\alpha}\right)g_{\alpha\mu}.
\end{array}\label{eq:63}
\end{equation}
Moreover, by expanding in Taylor series $g'_{\mu\nu}(x')$ one gets%
\footnote{Note that we are using the fact that the transformation is infinitesimal,
namely $\partial_{\alpha}g'{}_{\mu\nu}(x)=\partial_{\alpha}g{}_{\mu\nu}(x)+\mathcal{O}(\xi^{2})$
and $\partial'_{\mu}\xi^{\alpha}(x')=\partial_{\mu}\xi^{\alpha}(x)+\mathcal{O}(\xi^{2}).$%
}
\begin{equation}
g'_{\mu\nu}(x')\simeq g'_{\mu\nu}(x)+\xi^{\alpha}\partial_{\alpha}g{}_{\mu\nu}(x).
\end{equation}
Substituting the latter equation in (\ref{eq:63}), we obtain
\[
\begin{array}{rl}
\delta g_{\mu\nu}(x)\equiv g'_{\mu\nu}(x)-g_{\mu\nu}(x)= & -\left(\partial_{\mu}\xi^{\alpha}\right)g_{\alpha\nu}-\left(\partial_{\nu}\xi^{\alpha}\right)g_{\mu\alpha}-\xi^{\alpha}\partial_{\alpha}g_{\mu\nu}\\
\\
= & -\partial_{\mu}\xi_{\nu}-\partial_{\nu}\xi_{\mu}+\xi^{\alpha}\left(\partial_{\mu}g_{\alpha\nu}+\partial_{\nu}g_{\mu\alpha}-\partial_{\alpha}g_{\mu\nu}\right)\\
\\
= & -\partial_{\mu}\xi_{\nu}-\partial_{\nu}\xi_{\mu}+2\Gamma_{\mu\nu}^{\alpha}\xi_{\alpha}
\end{array}
\]
\begin{equation}
\Leftrightarrow\delta g_{\mu\nu}(x)=-\nabla_{\mu}\xi_{\nu}-\nabla_{\nu}\xi_{\mu}.\label{eq:64}
\end{equation}
For Minkowski background we have 
\[
\nabla_{\mu}\rightarrow\partial_{\mu}\Rightarrow\delta h_{\mu\nu}=-\partial_{\mu}\xi_{\nu}-\partial_{\nu}\xi_{\mu};
\]
and finally, if we redefine $-\xi_{\mu}\rightarrow\xi_{\mu}$ the
variation of $h_{\mu\nu}$ reads as 
\begin{equation}
\delta h_{\mu\nu}=\partial_{\mu}\xi_{\nu}+\partial_{\nu}\xi_{\mu}.\label{eq:65}
\end{equation}
We can easily verify that the transformation (\ref{eq:65}) is a gauge
symmetry for the Lagrangian $\mathcal{L}_{HE}$ and for the associated
field equations. Indeed, substituting (\ref{eq:65}) in the equations
(\ref{eq:50}) one has
\[
\boxempty\partial_{\mu}\xi_{\nu}+\boxempty\partial_{\nu}\xi_{\mu}+2\partial_{\mu}\partial_{\nu}\partial_{\alpha}\xi^{\alpha}-\partial_{\mu}\partial_{\alpha}\partial^{\alpha}\xi_{\nu}-\partial_{\nu}\partial_{\alpha}\partial^{\alpha}\xi_{\mu}-2\partial_{\nu}\partial_{\mu}\partial_{\alpha}\xi^{\alpha}=0,
\]
i.e. the field equations (\ref{eq:50}) are invariant under gauge
transformation. \\
By means the gauge symmetry (\ref{eq:65}) we can transform the tensor
field $h_{\mu\nu},$ so that we are able to choose a special gauge
in which, for example, the field equations simplify. In the case of
vector field we considered the Lorenz gauge $\partial_{\mu}A^{\mu}=0.$
Analogously for the symmetric tensor field a possible gauge is the
$\mathit{De}$ $\mathit{Donder}$ $\mathit{gauge}$ 
\begin{equation}
\partial_{\alpha}h_{\mu}^{\alpha}-\frac{1}{2}\partial_{\mu}h=0\Leftrightarrow\partial_{\alpha}\left(h_{\mu}^{\alpha}-\frac{1}{2}\delta_{\mu}^{\alpha}h\right)=0.\label{eq:66}
\end{equation}
By choosing this gauge the field equations in the vacuum reduce to
the famous wave equation (i.e. the same equations of ED that we obtain
in the Lorenz gauge):
\[
\begin{array}{rl}
 & \boxempty h_{\mu\nu}+\partial_{\mu}\partial_{\nu}h-\partial_{\mu}\partial_{\alpha}h_{\nu}^{\alpha}-\partial_{\nu}\partial_{\alpha}h_{\mu}^{\alpha}=0\\
\\
\Leftrightarrow & \boxempty h_{\mu\nu}-\partial_{\mu}\left(\partial_{\alpha}h_{\nu}^{\alpha}-\frac{1}{2}\partial_{\nu}h\right)-\partial_{\nu}\left(\partial_{\alpha}h_{\mu}^{\alpha}-\frac{1}{2}\partial_{\mu}h\right)=0,
\end{array}
\]
\begin{equation}
\eqref{eq:66}\Rightarrow\boxempty h_{\mu\nu}=0.\label{eq:67}
\end{equation}
Going to the momentum space (\ref{eq:67}) becomes
\begin{equation}
-k^{2}h_{\mu\nu}=0\Rightarrow k^{2}=0,
\end{equation}
namely $h_{\mu\nu}$ describes a massless graviton.

\section{Graviton degrees of freedom}

As we have done for the vector field case, now we want to understand
how many physical degrees of freedom the graviton has, and we shall
see that by imposing the field equations and the gauge symmetry we
can get rid of the spurious degrees of freedom. Since the two-rank
tensor $h_{\mu\nu}$ is symmetric it has only 10 independent components;
our aim is to find and eliminate the unphysical ones. Also for the
graviton case we shall distinguish the on-shell and off-shell case.
We are going to start studying the on-shell case {[}\ref{-J.-Helayel}{]}.

\subsection{On-shell graviton}

Let us consider the field equations and the gauge symmetry in  momentum
space
\begin{equation}
\begin{cases}
k^{2}h_{\mu\nu}+k_{\mu}k_{\nu}h-k_{\mu}k_{\alpha}h_{\nu}^{\alpha}-k_{\nu}k_{\alpha}h_{\mu}^{\alpha}=0\\
\\
\delta h_{\mu\nu}=i\left(k_{\mu}\xi_{\nu}+k_{\nu}\xi_{\mu}\right)
\end{cases}\label{eq:68}
\end{equation}
and the basis of four-vectors $\left\{ k^{\mu},\tilde{k}^{\mu},\varepsilon_{1}^{\mu},\varepsilon_{2}^{\mu}\right\} ,$
such that the relations (\ref{eq:8}) hold 
\[
\begin{array}{l}
k^{\mu}\equiv(k^{0},\bar{k}),\,\,\,\,\,\tilde{k}^{\mu}\equiv(\tilde{k}^{0},-\bar{k}),\,\,\,\,\,\varepsilon_{i}^{\mu}\equiv(0,\bar{\varepsilon}_{i}),\\
\\
k^{\mu}\varepsilon_{i,\mu}=0=\tilde{k}^{\mu}\varepsilon_{i,\mu},\,\,\,\,\,\,\,\varepsilon_{i}^{\mu}\varepsilon_{j,\mu}=-\bar{\varepsilon}_{i}\cdot\bar{\varepsilon}_{j}=-\delta_{ij},
\end{array}\,\,\,\,\, i=1,2.
\]
The tensor field $h_{\mu\nu}$ can be expanded in this basis as follow
\begin{equation}
\begin{array}{rl}
h_{\mu\nu}(k)= & a(k)k_{\mu}k_{\nu}+b(k)k_{(\mu}\tilde{k}_{\nu)}+c_{i}(k)k_{(\mu}\varepsilon_{\nu)}^{i}\\
\\
 & +d(k)\tilde{k}_{\mu}\tilde{k}_{\nu}+e_{i}(k)\tilde{k}_{(\mu}\varepsilon_{\nu)}^{i}+f_{ij}(k)\varepsilon_{(\mu}^{i}\varepsilon_{\nu)}^{j},
\end{array}\label{eq:69}
\end{equation}
where the coefficients $\left\{ a,b,c_{1},c_{2},d,e_{1},e_{2},f_{11},f_{12},f_{21},f_{22}\right\} $%
\footnote{Note that we have only $11$ coefficients and not $16$ because the
expansion (\ref{eq:69}) takes already into account the symmetry of
$h_{\mu\nu},$ in fact only symmetrized products are present. %
} take into account the graviton degrees of freedom. By substituting
(\ref{eq:69}) in the first of (\ref{eq:68}) $a,$ $b$ and $c$
cancel each other disappearing from the equation and what remains
is
\begin{equation}
\begin{array}{rl}
d(k)k^{2}\tilde{k}_{\mu}\tilde{k}_{\nu}+e_{i}(k)k^{2}\tilde{k}_{(\mu}\varepsilon_{\nu)}^{i}+f_{ij}(k)k^{2}\varepsilon_{(\mu}^{i}\varepsilon_{\nu)}^{j}+d(k)\tilde{k}^{2}k_{\mu}k_{\nu}+e_{i}(k)k_{\mu}k_{\nu}\tilde{k}\cdot\varepsilon^{i}\\
\\
+f_{ij}(k)k_{\mu}k_{\nu}\left(\varepsilon^{i}\cdot\varepsilon^{j}\right)-d(k)k_{\mu}\left(k\cdot\tilde{k}\right)\tilde{k}_{\nu}-e_{i}(k)k_{\mu}k^{\alpha}\tilde{k}_{(\alpha}\varepsilon_{\nu)}^{i}-f_{ij}(k)k_{\mu}k_{\alpha}\varepsilon^{i,\alpha}\varepsilon_{\nu}^{j}\\
\\
-d(k)k_{\nu}k^{\alpha}\tilde{k}_{\mu}\tilde{k}_{\alpha}-e_{i}(k)k_{\nu}k^{\alpha}\tilde{k}_{(\mu}\varepsilon_{\alpha)}^{i}-f_{ij}(k)k_{\nu}k^{\alpha}\varepsilon_{(\mu}^{i}\varepsilon_{\alpha)}^{j} & =0
\end{array}\label{eq:70}
\end{equation}
Because of the (\ref{eq:8}) it turns out that
\begin{equation}
\begin{array}{l}
e_{i}(k)k_{\mu}k_{\nu}\tilde{k}\cdot\varepsilon^{i}=f_{ij}(k)k_{\mu}k_{\alpha}\varepsilon^{i,\alpha}\varepsilon_{\nu}^{j}=f_{ij}(k)k_{\nu}k^{\alpha}\varepsilon_{(\mu}^{i}\varepsilon_{\alpha)}^{j}=0,\\
\\
e_{i}(k)k_{\nu}k^{\alpha}\tilde{k}_{(\mu}\varepsilon_{\alpha)}^{i}=\frac{1}{2}e_{i}(k)k_{\nu}\left(k\cdot\tilde{k}\right)\varepsilon_{\mu}^{i},\\
\\
f_{ij}(k)k_{\mu}k_{\nu}\left(\varepsilon^{i}\cdot\varepsilon^{j}\right)=-f_{ij}(k)k_{\mu}k_{\nu}\delta^{ij}=f_{ii}k_{\mu}k_{\nu},
\end{array}\label{eq:71}
\end{equation}
where $f_{ii}\equiv\sum_{i=1,2}f_{ii}=f_{11}+f_{22}$ is the trace
of $f.$ Hence, the relations (\ref{eq:71}) reduce the expansion
(\ref{eq:70}) as
\begin{equation}
\begin{array}{rl}
d(k)k^{2}\tilde{k}_{\mu}\tilde{k}_{\nu}+e_{i}(k)k^{2}\tilde{k}_{(\mu}\varepsilon_{\nu)}^{i}+f_{ij}(k)k^{2}\varepsilon_{(\mu}^{i}\varepsilon_{\nu)}^{j}+d(k)\tilde{k}^{2}k_{\mu}k_{\nu}-f_{i}^{i}k_{\mu}k_{\nu}\\
\\
-d(k)k_{\mu}\left(k\cdot\tilde{k}\right)\tilde{k}_{\nu}-d(k)k_{\nu}\left(k\cdot\tilde{k}\right)\tilde{k}_{\mu}-\frac{1}{2}e_{i}(k)k_{\nu}\left(k\cdot\tilde{k}\right)\varepsilon_{\mu}^{i}-\frac{1}{2}e_{i}(k)k_{\mu}\left(k\cdot\tilde{k}\right)\varepsilon_{\nu}^{i} & =0
\end{array}\label{eq:72}
\end{equation}
Let us consider the $\mu=0=\nu$ component of (\ref{eq:72}):
\begin{equation}
\begin{array}{rl}
 & dk^{2}(k^{0})^{2}+d\tilde{k}^{2}(k^{0})^{2}-f_{ii}(k^{0})^{2}-2d(k^{0})^{2}\left(k\cdot\tilde{k}\right)=0\\
\\
\Leftrightarrow & 2d\left(k^{2}-k\cdot\tilde{k}\right)=f_{ii}\Leftrightarrow f_{ii}=-2d\bar{k}^{2}.
\end{array}\label{eq:73}
\end{equation}
Now contract (\ref{eq:72}) with $\eta^{\mu\nu}$ and use (\ref{eq:73}):
\[
2dk^{2}\tilde{k}^{2}-2f_{ii}k^{2}-2d\left(k\cdot\tilde{k}\right)^{2}=0\Leftrightarrow d\left(k^{2}\tilde{k}^{2}-\left(k\cdot\tilde{k}\right)^{2}+2\bar{k}^{2}k^{2}\right)=0
\]
\begin{equation}
\Rightarrow-2d\bar{k}^{2}\left(k\cdot\tilde{k}\right)=0\underset{{\scriptscriptstyle \bar{k}^{2}(k\cdot\tilde{k})\neq0}}{\Rightarrow}d=0,\label{eq:74}
\end{equation}
namely, since (\ref{eq:73}) holds, $f_{ij}$ is traceless, $f_{ii}=0;$
thus (\ref{eq:72}) becomes
\begin{equation}
e_{i}k^{2}\tilde{k}_{(\mu}\varepsilon_{\nu)}^{i}+f_{ij}k^{2}\varepsilon_{(\mu}^{i}\varepsilon_{\nu)}^{j}-\frac{1}{2}e_{i}k_{\mu}\left(k\cdot\tilde{k}\right)\varepsilon_{\nu}^{i}-\frac{1}{2}e_{i}k_{\nu}\left(k\cdot\tilde{k}\right)\varepsilon_{\mu}^{i}=0.\label{eq:75}
\end{equation}
Noting that $\varepsilon_{0}^{i}=0,$ if now we consider the components
$\mu=0$ and a generic $\nu$ in (\ref{eq:75}), we obtain 
\begin{equation}
\begin{array}{rl}
 & \frac{1}{2}e_{i}k^{2}k^{0}\varepsilon_{\nu}^{i}-\frac{1}{2}e_{i}k^{0}\left(k\cdot\tilde{k}\right)\varepsilon_{\nu}^{i}=0\\
\\
\Leftrightarrow & e_{i}k_{0}\left(k^{2}-k\cdot\tilde{k}\right)\varepsilon_{\nu}^{i}=0\Leftrightarrow e_{i}=0,\,\,\, i=1,2,
\end{array}
\end{equation}
since the four-vectors $\varepsilon_{\nu}^{i}$ are linearly independent
and $k^{0}\left(k^{2}-k\cdot\tilde{k}\right)k^{0}\left(-2\bar{k}^{2}\right)\neq0.$\\
Hence the field equations get rid of five coefficients: $d,f_{ii},e_{1},e_{2}$
and the expansion (\ref{eq:69}) reads as
\begin{equation}
h_{\mu\nu}(k)=a(k)k_{\mu}k_{\nu}+b(k)k_{(\mu}\tilde{k}_{\nu)}+c_{i}(k)k_{(\mu}\varepsilon_{\nu)}^{i}+f_{ij}\varepsilon_{(\mu}^{i}\varepsilon_{\nu)}^{j},\label{eq:76}
\end{equation}
with condition $f_{ii}=0.$ \\
Now we want to verify that the coefficients $a,b,c_{1},c_{2}$ can
be eliminated by using the gauge symmetry $\delta h_{\mu\nu}=i\left(k_{\mu}\xi_{\nu}+k_{\nu}\xi_{\mu}\right).$
First of all let us expand $\xi_{\mu}$ in the basis (\ref{eq:8})
\begin{equation}
\xi_{\mu}(k)=\alpha(k)k_{\mu}+\beta(k)\tilde{k}_{\mu}+\gamma_{i}(k)\varepsilon_{\mu}^{i},\label{eq:77}
\end{equation}
thus the gauge symmetry relation becomes
\begin{equation}
\delta h_{\mu\nu}=i2\left(\alpha k_{\mu}k_{\nu}+\beta k_{(\mu}\tilde{k}_{\nu)}+\gamma_{i}k_{(\mu}\varepsilon_{\nu)}^{i}\right).\label{eq:78}
\end{equation}
 By the gauge transformation corresponding to (\ref{eq:78}) we can
go from the tensor field in (\ref{eq:76}) to a new one $h'_{\mu\nu}:$

\begin{equation}
\begin{array}{rrl}
 & h'_{\mu\nu}=h_{\mu\nu}+\delta h_{\mu\nu}\\
\\
\Leftrightarrow & a'k_{\mu}k_{\nu}+b'k_{(\mu}\tilde{k}_{\nu)}+c_{i}'k_{(\mu}\varepsilon_{\nu)}^{i}+f_{ij}'\varepsilon_{(\mu}^{i}\varepsilon_{\nu)}^{j}= & (a+i2\alpha)k_{\mu}k_{\nu}+(b+i2\beta)k_{(\mu}\tilde{k}_{\nu)}\\
\\
 &  & +(c_{i}+i2\gamma_{i})k_{(\mu}\varepsilon_{\nu)}^{i}+f_{ij}'\varepsilon_{(\mu}^{i}\varepsilon_{\nu)}^{j}
\end{array}\label{eq:79}
\end{equation}
The last equation implies that the following relations hold%
\footnote{Don't get confused! We are indicating with the letter ``$i"$ both
the imaginary unit and the index component.%
}
\begin{equation}
a'=a+i2\alpha,\,\,\, b'=b+i2\beta,\,\,\, c'_{i}=c_{i}+i2\gamma_{i},\,\,\, f'_{ij}=f_{ij}.\label{eq:80}
\end{equation}
We can immediately notice that by choosing 
\begin{equation}
\alpha=i\frac{a}{2},\,\,\,\beta=i\frac{b}{2},\,\,\,\gamma_{i}=i\frac{c_{i}}{2},\label{eq:81}
\end{equation}
we get rid of the coefficients $a,$ $b$ and $c_{i}.$ \\
Hence the only remaining coefficients are $f_{12}=f_{21}$ and $f_{11}=-f_{22}.$
Finally we can state that the on-shell massless graviton has only
$\mathit{two}$ physical degrees of freedom.

\subsection{Off-shell graviton}

If we consider the off-shell case, since we cannot use the constraint
of the field equations, we won't be able to eliminate the coefficients
$d,e_{1},e_{2},f_{ii}$ and so an off-shell graviton will have $\mathit{six}$
degrees of freedom. \\
\\
We can summarize the counting of degrees of freedom for on-shell and
off-shell photon in the following expression%
\footnote{In general, if we are in $D\textrm{-}$dimensions the following rules
hold: 
\begin{equation}
D\textrm{-}\mathrm{dimension}:\,\,\,\begin{cases}
\mathrm{off\textrm{-}shell}: & \frac{D(D-1)}{2}\,\,\mathrm{d.o.f.}\\
\mathrm{on\textrm{-}shell}: & \frac{D(D-3)}{2}\,\,\mathrm{d.o.f.}
\end{cases};\label{eq:82-1}
\end{equation}
as particular case we can see that in $D=2$
\[
2\textrm{-}\mathrm{dimension}:\,\,\,\begin{cases}
\mathrm{off\textrm{-}shell}: & 1\,\,\mathrm{d.o.f.}\\
\mathrm{on\textrm{-}shell}: & -1\,\,\mathrm{d.o.f.}
\end{cases},
\]
but we can notice that for the real (on-shell) graviton we have a
negative number of degrees of freedom, so this case is not considerable;
then in $D=3$ \\
\[
3\textrm{-}\mathrm{dimension}:\,\,\,\begin{cases}
\mathrm{off\textrm{-}shell}: & 3\,\,\mathrm{d.o.f.}\\
\mathrm{on\textrm{-}shell}: & 0\,\,\mathrm{d.o.f.}
\end{cases},
\]
namely the physical graviton doesn't propagate.%
}:

\begin{equation}
4\textrm{-}\mathrm{dimension}:\,\,\,\begin{cases}
\mathrm{off\textrm{-}shell}: & 6\,\,\mathrm{d.o.f.}\\
\mathrm{on\textrm{-}shell}: & 2\,\,\mathrm{d.o.f.}
\end{cases}.\label{eq:82}
\end{equation}

\section{Graviton propagator}

Now our aim is to obtain the propagator for the graviton; we will
use the formalism of the spin projector operators. We shall proceed
in the same way we did for the vector field case, but since we are
dealing with two-rank tensor the propagator will have four indices.
The graviton propagator is defined as the inverse operator of the
operator $\mathcal{O}$ in (\ref{eq:42}). We will present two equivalent
methods:
\begin{enumerate}
\item To obtain the propagator one proceed straightforwardly with the inversion
of the operator $\mathcal{O},$ finding the operator $\mathcal{O}^{-1}$
such that $\mathcal{O}\mathcal{O}^{-1}=\mathbb{I};$ 
\item To obtain the propagator one always inverts the operator $\mathcal{O},$
but by acting with the spin projector operators on the field equations.
\end{enumerate}
First we are going to consider the general case by studying the symmetric
operator $\mathcal{O}$ associated to any symmetric two-rank field
Lagrangian. Then, we will specialize to the case of GR where the operator
$\mathcal{O}$ is defined in (\ref{eq:42}). 

$ $

$ $

\textbf{Method 1}\\
\textbf{$ $}\\
Given a symmetric two-rank tensor field Lagrangian, its associated
operator $\mathcal{O}$ can be expanded in terms of the basis of spin
projector operators%
\footnote{We have to note that we are doing an abuse of nomenclature, in fact
we are using the word ``projector'' also for $\mathcal{P}_{sw}^{0}$
and $\mathcal{P}_{ws}^{0}$ that are not projectors.%
}(\ref{eq:52})%
\footnote{Note that we are still suppressing the indices for simplicity.%
}:
\begin{equation}
\mathcal{O}=A\mathcal{P}^{2}+B\mathcal{P}^{1}+C\mathcal{P}_{s}^{0}+D\mathcal{P}_{w}^{0}+E\mathcal{P}_{sw}^{0}+F\mathcal{P}_{ws}^{0},\label{eq:83}
\end{equation}
and, similarly, for its inverse we have
\begin{equation}
\mathcal{O}^{-1}=X\mathcal{P}+Y\mathcal{P}^{1}+Z\mathcal{P}_{s}^{0}+W\mathcal{P}_{w}^{0}+R\mathcal{P}_{sw}^{0}+S\mathcal{P}_{ws}^{0}.\label{eq:84}
\end{equation}
By imposing $\mathcal{O}\mathcal{O}^{-1}=\mathbb{I}$ and using the
orthogonality relations (\ref{eq:53}) we can find the relations among
the two sets of coefficients $\left\{ A,B,C,D,E,F\right\} $ and $\left\{ X,Y,Z,W,R,S\right\} :$

\begin{equation}
\begin{array}{rl}
\mathcal{O}\mathcal{O}^{-1}= & AX\mathcal{P}^{2}+BY\mathcal{P}^{1}+\left(CZ+ES\right)\mathcal{P}_{s}^{0}+\left(DW+FR\right)\mathcal{P}_{w}^{0}\\
\\
 & +\left(CR+EW\right)\mathcal{P}_{sw}^{0}+\left(DS+FZ\right)\mathcal{P}_{ws}^{0}\\
\\
= & \mathbb{I}
\end{array}\label{eq:85}
\end{equation}
The equation is satisfied if, and only if, the following relations
hold:
\[
\begin{array}{ccc}
AX=1, & BY=1, & CZ+ES=1,\\
\\
DW+FR=1, & CR+EW=0, & DS+FZ=0;
\end{array}
\]
namely
\begin{equation}
\begin{array}{ccc}
X={\displaystyle \frac{1}{A},} & Y={\displaystyle \frac{1}{B},} & Z={\displaystyle \frac{D}{CD-EF},}\\
\\
W={\displaystyle \frac{C}{CD-EF},} & R={\displaystyle \frac{F}{EF-CD},} & S={\displaystyle \frac{E}{EF-CD}.}
\end{array}\label{eq:86}
\end{equation}
The inverse operator $\mathcal{O}^{-1}$ turns out to be
\begin{equation}
\begin{array}{rl}
\mathcal{O}^{-1}= & {\displaystyle \frac{1}{A}\mathcal{P}^{2}+\frac{1}{B}\mathcal{P}^{1}+\frac{D}{CD-EF}\mathcal{P}_{s}^{0}+\frac{C}{CD-EF}\mathcal{P}_{w}^{0}}\\
\\
 & +{\displaystyle \frac{F}{EF-CD}\mathcal{P}_{sw}^{0}+\frac{E}{EF-CD}\mathcal{P}_{ws}^{0}.}
\end{array}\label{eq:87}
\end{equation}
\\
Now let us come back to GR case, specializing Method $1$ to the H-E
Lagrangian (\ref{eq:41}).\\
First of all we need the operator $\mathcal{O}$ in (\ref{eq:42})
expressed in terms of the spin projector operators. We have already
gotten its expression in (\ref{eq:126operator}) and, by suppressing
the indices for simplicity, is given by 
\begin{equation}
\mathcal{O}=k^{2}\left(\mathcal{P}^{2}-2\mathcal{P}_{s}^{0}\right).\label{eq:88}
\end{equation}
Comparing with (\ref{eq:83}) we notice that only two coefficients
are not zero: 
\begin{equation}
\begin{array}{ccc}
A=k^{2}, & B=0, & C=-2k^{2},\\
\\
D=0, & E=0, & F=0.
\end{array}
\end{equation}
We notice that, as we have already seen for vector field, the operator
$\mathcal{O}$ defined in (\ref{eq:42}) is not invertible, in fact
we cannot invert something equal to zero.%
\footnote{Physically we can interpret this result saying that the spin-$1$
and spin-$0$ $w$ components don't propagate, so they won't appear
in the physical part of the propagator.%
} Hence we have the same mathematical obstacle already met with the
vector field case, and we learned that it can be got over by adding
a gauge fixing term to the Lagrangian (\ref{eq:40}). We have also
learned that the saturated propagator in which appears only the physical
(gauge-independent) part of the propagator is invertible. \\

Before introducing the gauge fixing term and inverting the propagator
let us introduce the second method to determine the propagator.

$ $

\textbf{Method 2}

$ $\\
Also with this second method, first we shall consider a general case
and then specialize to GR case. The field equations, in terms of the
operator $\mathcal{O},$ in the presence of a matter source $\tau_{\mu\nu}$
read as
\begin{equation}
\mathcal{O}^{\mu\nu\rho\sigma}h_{\rho\sigma}=\kappa\tau^{\mu\nu},\label{eq:90}
\end{equation}
or, without specifying the indices,
\begin{equation}
\mathcal{O}h=\kappa\tau.\label{eq:91}
\end{equation}
To derive the propagator the prescription is the following:
\begin{itemize}
\item We go to the momentum space and express the field equations (\ref{eq:91})
in terms of the spin projector operators%
\footnote{Let us note again that we use the basis of $\mathit{six}$ operators
to expand the operator $\mathcal{O};$ while we need just the four
projectors to decompose the symmetric two-rank tensor $\tau.$%
}:
\begin{equation}
\begin{array}{l}
\left(A\mathcal{P}^{2}+B\mathcal{P}^{1}+C\mathcal{P}_{s}^{0}+D\mathcal{P}_{w}^{0}+E\mathcal{P}_{sw}^{0}+F\mathcal{P}_{ws}^{0}\right)h=\kappa\left(\mathcal{P}^{2}+\mathcal{P}^{1}+\mathcal{P}_{s}^{0}+\mathcal{P}_{w}^{0}\right)\tau;\end{array}\label{eq:92}
\end{equation}

\item By acting with each spin projector operators on (\ref{eq:92}) one can decompose the field equations into a decoupled set
of new equations corresponding to the relevant degrees of freedom;
\item Then these equations will be invertible and we can obtain the propagator. 
\end{itemize}
$ $\\
Hence, following the above prescription, let us act with each spin
projector operators on (\ref{eq:92}):
\begin{equation}
\mathcal{P}^{2}\rightarrow\mathcal{P}^{2}h=\frac{1}{A}\kappa\mathcal{P}^{2}\tau;\label{eq:93}
\end{equation}
\begin{equation}
\mathcal{P}^{1}\rightarrow\mathcal{P}^{1}h=\frac{1}{B}\kappa\mathcal{P}^{1}\tau;\label{eq:94}
\end{equation}
\begin{equation}
\mathcal{P}_{s}^{0}\rightarrow\left(C\mathcal{P}_{s}^{0}+E\mathcal{P}_{sw}^{0}\right)h=\kappa\mathcal{P}_{s}^{0}\tau;\label{eq:95}
\end{equation}
\begin{equation}
\mathcal{P}_{w}^{0}\rightarrow\left(D\mathcal{P}_{w}^{0}+F\mathcal{P}_{ws}^{0}\right)h=\kappa\mathcal{P}_{w}^{0}\tau;\label{eq:96}
\end{equation}
\begin{equation}
\mathcal{P}_{sw}^{0}\rightarrow\left(D\mathcal{P}_{sw}^{0}+F\mathcal{P}_{s}^{0}\right)h=\kappa\mathcal{P}_{sw}^{0}\tau;\label{eq:97}
\end{equation}
\begin{equation}
\mathcal{P}_{ws}^{0}\rightarrow\left(C\mathcal{P}_{ws}^{0}+E\mathcal{P}_{w}^{0}\right)h=\kappa\mathcal{P}_{ws}^{0}\tau.\label{eq:98}
\end{equation}
We can see that the equations for $\mathcal{P}^{2}$ and $\mathcal{P}^{1}$
are decoupled, but for the scalar components seems to be coupled.
Consider the system composed of the equations (\ref{eq:95}) and (\ref{eq:97}):
\begin{equation}
\begin{cases}
\left(C\mathcal{P}_{s}^{0}+E\mathcal{P}_{sw}^{0}\right)h=\kappa\mathcal{P}_{s}^{0}\tau\\
\\
\left(D\mathcal{P}_{sw}^{0}+F\mathcal{P}_{s}^{0}\right)h=\kappa\mathcal{P}_{sw}^{0}\tau
\end{cases},\label{eq:99}
\end{equation}
or, equivalently, 
\begin{equation}
\begin{pmatrix}C & E\\
F & D
\end{pmatrix}\begin{pmatrix}\mathcal{P}_{s}^{0}h\\
\mathcal{P}_{sw}^{0}h
\end{pmatrix}=\kappa\begin{pmatrix}\mathcal{P}_{s}^{0}\tau\\
\mathcal{P}_{sw}^{0}\tau
\end{pmatrix}.\label{eq:100}
\end{equation}
By noting that the inverse of the matrix 
\begin{equation}
M=\begin{pmatrix}C & E\\
F & D
\end{pmatrix}
\end{equation}
is 
\begin{equation}
M^{-1}=\frac{1}{CD-EF}\begin{pmatrix}D & -F\\
-E & C
\end{pmatrix},\label{eq:169M}
\end{equation}
we are able to invert the system (\ref{eq:100}) by multiplying with
(\ref{eq:169M})\\
 
\begin{equation}
\begin{pmatrix}\mathcal{P}_{s}^{0}h\\
\mathcal{P}_{sw}^{0}h
\end{pmatrix}={\displaystyle \kappa\frac{1}{CD-EF}\begin{pmatrix}D & -F\\
-E & C
\end{pmatrix}\begin{pmatrix}\mathcal{P}_{s}^{0}\tau\\
\mathcal{P}_{sw}^{0}\tau
\end{pmatrix}},\label{eq:101}
\end{equation}
or, equivalently,
\begin{equation}
\begin{cases}
\mathcal{P}_{s}^{0}h={\displaystyle \kappa\frac{1}{CD-EF}}\left(D\mathcal{P}_{s}^{0}-F\mathcal{P}_{sw}^{0}\right)\tau\\
\\
\mathcal{P}_{sw}^{0}h=\kappa{\displaystyle \frac{1}{CD-EF}}\left(-E\mathcal{P}_{s}^{0}+C\mathcal{P}_{sw}^{0}\right)\tau
\end{cases}.\label{eq:102}
\end{equation}
We can do the same for the equations (\ref{eq:96}) and (\ref{eq:98})
and obtain
\begin{equation}
\begin{cases}
\mathcal{P}_{w}^{0}h=\kappa\frac{1}{CD-EF}\left(C\mathcal{P}_{w}^{0}-E\mathcal{P}_{ws}^{0}\right)\tau\\
\\
\mathcal{P}_{ws}^{0}h=\kappa{\displaystyle \frac{1}{CD-EF}\left(-F\mathcal{P}_{w}^{0}+D\mathcal{P}_{sw}^{0}\right)}\tau
\end{cases}.\label{eq:103}
\end{equation}
We have seen that also the scalar components decouple. Hence, using
(\ref{eq:93}), (\ref{eq:94}) and the firsts of (\ref{eq:102}) and
(\ref{eq:103}) we are able to obtain the inverse operator $\mathcal{O}^{-1}:$
\begin{equation}
\begin{array}{rl}
\left(\mathcal{P}^{2}+\mathcal{P}^{1}+\mathcal{P}_{s}^{0}+\mathcal{P}_{w}^{0}\right)h= & \kappa{\displaystyle \left[{\displaystyle \frac{1}{A}\mathcal{P}^{2}+\frac{1}{B}\mathcal{P}^{1}+\frac{D}{CD-EF}\mathcal{P}_{s}^{0}}\right.}\\
\\
 & {\displaystyle {\displaystyle \left.+\frac{C}{CD-EF}\mathcal{P}_{w}^{0}+\frac{F}{EF-CD}\mathcal{\mathcal{P}}_{sw}^{0}+\frac{E}{EF-CD}\mathcal{P}_{ws}^{0}\right]}\tau}.
\end{array}\label{eq:104}
\end{equation}
We can observe that the expression in brackets in the equation (\ref{eq:104})
is the inverse operator $\mathcal{O}^{-1},$ i.e. the propagator.
Note also that, as we expected, the result in (\ref{eq:104}) coincides
with the expression in the equation (\ref{eq:87}) obtained with the
first method.\\
$ $

Now let us come back to GR case. In (\ref{eq:61}) we have already
obtained the field equations in terms of the spin projector operators:
\begin{equation}
k^{2}\left(\mathcal{P}^{2}-2\mathcal{P}_{s}^{0}\right)h=\kappa\left(\mathcal{P}^{2}+\mathcal{P}^{1}+\mathcal{P}_{s}^{0}+\mathcal{P}_{w}^{0}\right)\tau.\label{eq:105}
\end{equation}
By acting with $\mathcal{P}^{2}$ we have
\begin{equation}
\frac{k^{2}}{2}\mathcal{P}^{2}h=\kappa\mathcal{P}^{2}\tau\,\Rightarrow\,\mathcal{P}^{2}h=\kappa\left(\frac{\mathcal{P}^{2}}{k^{2}}\right)\tau;\label{eq:106}
\end{equation}
 and by acting with $\mathcal{P}_{s}^{0}$ 
\begin{equation}
-k^{2}2\mathcal{P}_{s}^{0}h=\kappa\mathcal{P}_{s}^{0}\tau\,\Rightarrow\,\mathcal{P}_{s}^{0}h=\kappa\left(-\frac{\mathcal{P}_{s}^{0}}{2k^{2}}\right)\tau.\label{eq:107}
\end{equation}
By acting with $\mathcal{P}^{1}$ and $\mathcal{P}_{w}^{0}$ we have:
\begin{equation}
0h=\kappa\mathcal{P}^{1}\tau\,\Rightarrow\,\mathcal{P}^{1}\tau=0,\label{eq:108}
\end{equation}
\begin{equation}
0h=\kappa\mathcal{P}_{w}^{0}\tau\,\Rightarrow\,\mathcal{P}_{w}^{0}\tau=0,\label{eq:109}
\end{equation}
so it's impossible to obtain the components $\mathcal{P}^{1}h$ and
$\mathcal{P}_{w}^{0}h$ because we have zero on the left side. Recall
that in ED case we had the spin-$0$ component undetermined, instead
in GR case we have the spin-$1$ and spin-$0$ $w$ components that
are undetermined.%
\footnote{The equations (\ref{eq:108}) and (\ref{eq:109}) impose the constraints
on the matter source and correspond to a gauge freedom. The concept
is the same that we have mentioned in the previous chapter for the
vector field $A.$ In fact, if we analyze the GR Lagrangian:
\begin{equation}
\mathcal{L}_{HE}=-\frac{1}{2}h_{\mu\nu}\boxempty\left(\mathcal{P}^{2}-2\mathcal{P}_{s}^{0}\right)^{\mu\nu\rho\sigma}h_{\rho\sigma},\label{eq:111}
\end{equation}
it is invariant under spin-$1$ transformation as
\begin{equation}
\delta h_{\mu\nu}\sim\mathcal{P}_{\mu\nu\rho\sigma}^{1}h^{\rho\sigma},
\end{equation}
because of the orthogonality relations, $\mathcal{P}^{1}\mathcal{P}^{2}=\mathcal{P}^{1}\mathcal{P}_{s}^{0}=0,$
and there are also the restrictions (\ref{eq:108}) and (\ref{eq:109})
on the source. It means that for the Lagrangian (\ref{eq:111}) there
is a gauge symmetry that corresponds to the gauge invariance under
transformations $\delta h_{\mu\nu}=\partial_{\mu}\xi_{\nu}+\partial_{\nu}\xi_{\mu}.$
The arbitrary four-vector $\xi_{\mu}$ is the vector field associated
to the spin-$1$ symmetry. %
}

$ $\\
Now, we are going to proceed as already done in the previous Chapter
for photon propagator: firstly we want to introduce a gauge fixing
term to invert the operator $\mathcal{O},$ secondly we want to determine
the saturated propagator without the choice of any gauge fixing term.

\subsection{Gauge fixing term for graviton Lagrangian}

We are going to introduce the following gauge fixing term, called
$\mathit{De}$ $\mathit{Donder}$ $\mathit{gauge}$ (see (\ref{eq:66})):
\begin{equation}
\mathcal{L}_{gf}\coloneqq-\frac{1}{2\alpha}\left(\partial_{\rho}h_{\mu}^{\rho}-\frac{1}{2}\partial_{\mu}h\right)\left(\partial_{\sigma}h^{\mu\sigma}-\frac{1}{2}\partial^{\mu}h\right);\label{eq:117}
\end{equation}
where $\alpha$ is called $\mathit{gauge}$ $\mathit{parameter}.$
The total GR Lagrangian becomes
\begin{equation}
\tilde{\mathcal{L}}_{HE}=\mathcal{L}_{HE}+\mathcal{L}_{gf}.\label{eq:118}
\end{equation}
As we have done for the Lagrangian $\mathcal{L}_{HE}$ in (\ref{eq:41})
raising and lowering the indices with the metric tensor $\eta_{\mu\nu},$
we can easily rewrite the gauge fixing term (\ref{eq:117}) as
\begin{equation}
\mathcal{L}_{gf}=\frac{1}{2}h_{\mu\nu}\mathcal{O}_{gf}^{\mu\nu\rho\sigma}h_{\rho\sigma},\label{eq:119}
\end{equation}
where the operator $\mathcal{O}_{gf}^{\mu\nu\rho\sigma}$ is defined
as
\begin{equation}
\mathcal{O}_{gf}^{\mu\nu\rho\sigma}\coloneqq-\frac{2}{\alpha}\eta^{\mu\rho}\partial^{\nu}\partial^{\sigma}+\frac{1}{\alpha}\eta^{\rho\sigma}\partial^{\mu}\partial^{\nu}+\frac{1}{\alpha}\eta^{\mu\nu}\partial^{\rho}\partial^{\sigma}-\frac{1}{2\alpha}\eta^{\mu\nu}\eta^{\rho\sigma}\boxempty.\label{eq:120}
\end{equation}
Now we can rewrite also the total Lagrangian (\ref{eq:118}) as a
quadratic form
\begin{equation}
\mathcal{\tilde{L}}_{HE}=\frac{1}{2}h_{\mu\nu}\left(\mathcal{O}+\mathcal{O}_{gf}\right)^{\mu\nu\rho\sigma}h_{\rho\sigma},\label{eq:121}
\end{equation}
or, defining $\tilde{\mathcal{O}}\coloneqq\mathcal{O}+\mathcal{O}_{gf},$
in a more compact form we read 
\begin{equation}
\mathcal{\tilde{L}}_{HE}=\frac{1}{2}h_{\mu\nu}\tilde{\mathcal{O}}^{\mu\nu\rho\sigma}h_{\rho\sigma},\label{eq:122}
\end{equation}
 with 
\begin{equation}
\begin{array}{rl}
\tilde{\mathcal{O}}^{\mu\nu\rho\sigma}= & -{\displaystyle \left(\frac{1}{2}\eta^{\mu\rho}\eta^{\nu\sigma}+\frac{1}{2}\eta^{\mu\sigma}\eta^{\nu\rho}-\left(1-\frac{1}{2\alpha}\right)\eta^{\mu\nu}\eta^{\rho\sigma}\right)\boxempty}\\
\\
 & +{\displaystyle \left(\frac{1}{\alpha}-1\right)\left(\eta^{\mu\nu}\partial^{\rho}\partial^{\sigma}+\eta^{\rho\sigma}\partial^{\mu}\partial^{\nu}\right)}\\
\\
 & +{\displaystyle \frac{1}{2}\left(1-\frac{1}{\alpha}\right)\left(\eta^{\nu\rho}\partial^{\mu}\partial^{\sigma}+\eta^{\nu\sigma}\partial^{\mu}\partial^{\rho}+\eta^{\mu\rho}\partial^{\nu}\partial^{\sigma}+\eta^{\mu\sigma}\partial^{\nu}\partial^{\rho}\right).}
\end{array}\label{eq:123}
\end{equation}
Now we want to rewrite the operator $\tilde{\mathcal{O}}$ in terms
of the spin projectors operators and, again, since it is symmetric
we need only the six operator introduced above. By proceeding in the
same way we have done without gauge fixing term, going into momentum
space, we obtain:
\begin{equation}
\tilde{\mathcal{O}}=k^{2}\left[\mathcal{P}^{2}+\frac{1}{\alpha}\mathcal{P}^{1}+\left(\frac{3}{2\alpha}-2\right)\mathcal{P}_{s}^{0}+\frac{1}{2\alpha}\mathcal{P}_{w}^{0}-\frac{\sqrt{3}}{2\alpha}\mathcal{P}_{sw}^{0}-\frac{\sqrt{3}}{2\alpha}\mathcal{P}_{ws}^{0}\right].\label{eq:124}
\end{equation}
 We notice that this operator is invertible, in fact no coefficients
is equal to zero (see (\ref{eq:83})):
\begin{equation}
\begin{array}{ccc}
A=k^{2}, & B{\displaystyle =\frac{k^{2}}{\alpha},} & C={\displaystyle k^{2}\left(\frac{3}{2\alpha}-2\right),}\\
\\
D={\displaystyle \frac{k^{2}}{2\alpha},} & E=-{\displaystyle k^{2}\frac{\sqrt{3}}{2\alpha},} & F=-{\displaystyle k^{2}\frac{\sqrt{3}}{2\alpha}.}
\end{array}\label{eq:125}
\end{equation}
The coefficients in (\ref{eq:86}) specialized to the operator (\ref{eq:124})
are 
\begin{equation}
\begin{array}{ccc}
X={\displaystyle \frac{1}{k^{2}},} & Y={\displaystyle \frac{\alpha}{k^{2}},} & Z=-{\displaystyle \frac{1}{2k^{2}},}\\
\\
W={\displaystyle \frac{4\alpha-3}{2k^{2}},} & R=-{\displaystyle \frac{\sqrt{3}}{2k^{2}},} & S={\displaystyle -\frac{\sqrt{3}}{2k^{2}},}
\end{array}\label{eq:126}
\end{equation}
and the general form of the propagator with arbitrary coefficients
(\ref{eq:87}) in this case reads as%
\footnote{Note that often the propagator is also defined as the vacuum expectation
value of the time ordered product:
\[
\begin{array}{rl}
\left\langle T\left\{ h_{\mu\nu}(-k)h_{\rho\sigma}(k)\right\} \right\rangle = & i\Pi_{GR,\mu\nu\rho\sigma}(k)\\
\\
= & {\displaystyle \frac{i}{k^{2}}\left[\mathcal{P}^{2}+\alpha\mathcal{P}^{1}-\frac{1}{2}\mathcal{P}_{s}^{0}+\frac{4\alpha-3}{2}\mathcal{P}_{w}^{0}-\frac{\sqrt{3}}{2}\mathcal{P}_{sw}^{0}-\frac{\sqrt{3}}{2}\mathcal{P}_{ws}^{0}\right]_{\mu\nu\rho\sigma}}.
\end{array}
\]
The expression $T\left\{ h_{\mu\nu}(-k)h_{\rho\sigma}(k)\right\} $
is the Fourier transform of the time ordered product that in coordinate
space is defined as
\[
T\left\{ T\left\{ h_{\mu\nu}(x)h_{\rho\sigma}(y)\right\} \right\} \coloneqq h_{\mu\nu}(x)h_{\rho\sigma}(y)\Theta(x_{0}-y_{0})+h_{\rho\sigma}(y)h_{\mu\nu}(x)\Theta(y_{0}-x_{0}),
\]
where the function $\Theta(x_{0}-y_{0})$ is equal to $1$ if $x_{0}>y_{0},$
and to $0$ otherwise.%
}
\begin{equation}
\Pi_{GR}\equiv\tilde{\mathcal{O}}^{-1}=\frac{1}{k^{2}}\left[\mathcal{P}^{2}+\alpha\mathcal{P}^{1}-\frac{1}{2}\mathcal{P}_{s}^{0}+\frac{4\alpha-3}{2}\mathcal{P}_{w}^{0}-\frac{\sqrt{3}}{2}\mathcal{P}_{sw}^{0}-\frac{\sqrt{3}}{2}\mathcal{P}_{ws}^{0}\right].\label{eq:127}
\end{equation}
As we have already seen for the vector field case, a special gauge
is called $\mathit{Feynman}$ $\mathit{gauge},$ corresponding to
the choice $\alpha=1:$
\[
\begin{array}{rl}
\Pi_{GR,\mu\nu\rho\sigma}= & {\displaystyle \frac{1}{k^{2}}\left[\mathcal{P}^{2}+\mathcal{P}^{1}-\frac{1}{2}\mathcal{P}_{s}^{0}+\frac{1}{2}\mathcal{P}_{w}^{0}-\frac{\sqrt{3}}{2}\mathcal{P}_{sw}^{0}-\frac{\sqrt{3}}{2}\mathcal{P}_{ws}^{0}\right]_{\mu\nu\rho\sigma}}\\
\\
= & {\displaystyle \frac{1}{k^{2}}\left[\left(\mathcal{P}^{2}+\mathcal{P}^{1}+\mathcal{P}_{s}^{0}+\mathcal{P}_{w}^{0}\right)\right.}\\
\\
 & \left.-{\displaystyle \left(\frac{3}{2}\mathcal{P}_{s}^{0}+\frac{1}{2}\mathcal{P}_{w}^{0}\frac{\sqrt{3}}{2}\left(\mathcal{P}_{sw}^{0}+\mathcal{P}_{ws}^{0}\right)\right)}\right]_{\mu\nu\rho\sigma}
\end{array}
\]
 and since the following relations hold
\begin{equation}
\begin{array}{l}
\left(\mathcal{P}^{2}+\mathcal{P}^{1}+\mathcal{P}_{s}^{0}+\mathcal{P}_{w}^{0}\right)_{\mu\nu\rho\sigma}={\displaystyle \frac{1}{2}}\left(\eta_{\mu\rho}\eta_{\nu\sigma}+\eta_{\nu\rho}\eta_{\mu\sigma}\right),\\
\\
{\displaystyle \left(\frac{3}{2}\mathcal{P}_{s}^{0}+\frac{1}{2}\mathcal{P}_{w}^{0}-\frac{\sqrt{3}}{2}\left(\mathcal{P}_{sw}^{0}+\mathcal{P}_{ws}^{0}\right)\right)_{\mu\nu\rho\sigma}}={\displaystyle \frac{1}{2}\eta_{\mu\nu}\eta_{\rho\sigma}},
\end{array}\label{eq:191realations}
\end{equation}
we obtain a very simple form for the propagator in the Feynman gauge:
\begin{equation}
\Pi_{GR,\mu\nu\rho\sigma}=\frac{1}{2k^{2}}\left(\eta_{\mu\rho}\eta_{\nu\sigma}+\eta_{\nu\rho}\eta_{\mu\sigma}-\eta_{\mu\nu}\eta_{\rho\sigma}\right).\label{eq:128}
\end{equation}

\subsection{Saturated graviton propagator}

Our starting point are the field equations (\ref{eq:61}) that we
write again for convenience
\begin{equation}
k^{2}\left(\mathcal{P}^{2}-2\mathcal{P}_{s}^{0}\right)h(k)=\kappa\left(\mathcal{P}^{2}+\mathcal{P}^{1}+\mathcal{P}_{s}^{0}+\mathcal{P}_{w}^{0}\right)\tau(k).
\end{equation}
Add to the both members of the last equations the terms $k^{2}\left(\mathcal{P}^{1}+3\mathcal{P}_{s}^{0}+\mathcal{P}_{w}^{0}\right)h:$
\begin{equation}
\begin{array}{rl}
k^{2}\left(\mathcal{P}^{2}+\mathcal{P}^{1}+\mathcal{P}_{s}^{0}+\mathcal{P}_{w}^{0}\right)h(k)= & \kappa\left(\mathcal{P}^{2}+\mathcal{P}^{1}+\mathcal{P}_{s}^{0}+\mathcal{P}_{w}^{0}\right)\tau(k)\\
\\
 & +k^{2}\left(\mathcal{P}^{1}+3\mathcal{P}_{s}^{0}+\mathcal{P}_{w}^{0}\right)h(k);
\end{array}\label{eq:112}
\end{equation}
then, multiply for $\tau(-k)$ on the left side 
\begin{equation}
\begin{array}{rl}
k^{2}\tau(-k)\left(\mathcal{P}^{2}+\mathcal{P}^{1}+\mathcal{P}_{s}^{0}+\mathcal{P}_{w}^{0}\right)h(k)= & \kappa\tau(-k)\left(\mathcal{P}^{2}+\mathcal{P}^{1}+\mathcal{P}_{s}^{0}+\mathcal{P}_{w}^{0}\right)\tau(k)\\
\\
 & +k^{2}\tau(-k)\left(\mathcal{P}^{1}+3\mathcal{P}_{s}^{0}+\mathcal{P}_{w}^{0}\right)h(k).
\end{array}\label{eq:113}
\end{equation}
Because of the equations (\ref{eq:106})-(\ref{eq:109}) the last
equation reduces to
\begin{equation}
\tau(-k)\left(\mathcal{P}^{2}+\mathcal{P}^{1}+\mathcal{P}_{s}^{0}+\mathcal{P}_{w}^{0}\right)h(k)=\kappa\tau(-k)\frac{1}{k^{2}}\left(\mathcal{P}^{2}-\frac{1}{2}\mathcal{P}_{s}^{0}\right)\tau(k),\label{eq:114}
\end{equation}
namely we have derived the saturated propagator for the massless graviton:
\begin{equation}
\tau(-k)\Pi_{GR}(k)\tau(k)\equiv\tau(-k)\frac{1}{k^{2}}\left(\mathcal{P}^{2}-\frac{1}{2}\mathcal{P}_{s}^{0}\right)\tau(k),\label{eq:115}
\end{equation}
or, if we want to write the equations with the indices, one has
\begin{equation}
\tau^{\mu\nu}(-k)\Pi_{GR,\mu\nu\rho\sigma}(k)\tau^{\rho\sigma}(k)\equiv\tau^{\mu\nu}(-k)\frac{1}{k^{2}}\left(\mathcal{P}^{2}-\frac{1}{2}\mathcal{P}_{s}^{0}\right)_{\mu\nu\rho\sigma}\tau^{\rho\sigma}(k).\label{eq:116}
\end{equation}

Now we want to rewrite (\ref{eq:116}) in the following by making
explicit the form of the projectors $\mathcal{P}^{2}$ and $\mathcal{P}_{s}^{0}.$
By keeping in mind that $k_{\mu}\tau^{\mu\nu}=0,$ one has 
\begin{equation}
\begin{array}{rl}
{\displaystyle \tau^{\mu\nu}(-k)\frac{1}{k^{2}}\left(\mathcal{P}^{2}-\frac{1}{2}\mathcal{P}_{s}^{0}\right)_{\mu\nu\rho\sigma}\tau^{\rho\sigma}(k)}= & {\displaystyle \tau^{\mu\nu}(-k)\frac{1}{k^{2}}\left[\frac{1}{2}\left(\theta_{\mu\rho}\theta_{\nu\sigma}+\theta_{\mu\sigma}\theta_{\nu\rho}\right)-\frac{1}{2}\theta_{\mu\nu}\theta_{\rho\sigma}\right]\tau^{\rho\sigma}(}k)\\
\\
= & {\displaystyle \tau^{\mu\nu}(-k)\frac{1}{k^{2}}\left[\theta_{\mu\rho}\theta_{\nu\sigma}-\frac{1}{2}\theta_{\mu\nu}\theta_{\rho\sigma}\right]\tau^{\rho\sigma}(k})\\
\\
= & {\displaystyle \tau^{\mu\nu}(-k)\frac{1}{k^{2}}\left[\left(\eta_{\mu\rho}-\omega_{\mu\rho}\right)\left(\eta_{\nu\sigma}-\omega_{\nu\sigma}\right)\right.}\\
\\
 & {\displaystyle \left.-\frac{1}{2}\left(\eta_{\mu\nu}-\omega_{\mu\nu}\right)\left(\eta_{\rho\sigma}-\omega_{\rho\sigma}\right)\right]\tau^{\rho\sigma}(k)}\\
\\
= & {\displaystyle \tau^{\mu\nu}(-k)\frac{1}{k^{2}}}\left[\left(\eta_{\mu\rho}\eta_{\nu\sigma}-\eta_{\mu\rho}\omega_{\nu\sigma}-\omega_{\mu\rho}\eta_{\nu\sigma}+\omega_{\mu\rho}\omega_{\nu\sigma}\right)\right.\\
\\
 & \left.-{\displaystyle \frac{1}{2}}\left(\eta_{\mu\rho}\eta_{\nu\sigma}-\eta_{\mu\rho}\omega_{\nu\sigma}-\omega_{\mu\rho}\eta_{\nu\sigma}+\omega_{\mu\rho}\omega_{\nu\sigma}\right)\right]\tau^{\rho\sigma}(k)\\
\\
= & \tau^{\mu\nu}(-k){\displaystyle \frac{1}{2k^{2}}}\left(\eta_{\mu\rho}\eta_{\nu\sigma}+\eta_{\nu\rho}\eta_{\mu\sigma}-\eta_{\mu\nu}\eta_{\rho\sigma}\right)\tau^{\rho\sigma}(k).
\end{array}\label{eq:203saturated}
\end{equation}
The equation (\ref{eq:203saturated}) tells us that the physical (gauge-independent)
part of the propagator can be easily express as a product of metric
tensors. Note also that the physical part of the propagator coincides
with the propagator in the Feynman gauge $(\alpha=1)$ we have determined
in (\ref{eq:128}).\\
Moreover, it is worth observing that, since in the saturated propagator
just the physical part appears (see (\ref{eq:116})), if we consider
the sandwich between two conserved currents of the propagator in De
Donder gauge (\ref{eq:127}) that has both physical and gauge dependent
parts, the latter should vanishes. To show this, first notice that
all the spin projector operators proportional to the momentum $k^{\mu},$
i.e. $\mathcal{P}^{1},$ $\mathcal{P}_{w}^{0},$ $\mathcal{P}_{sw}^{0}$
and $\mathcal{P}_{ws}^{0},$ acting on the source give us a null contribution
because of the conservation law of the source (see (\ref{eq:52})
for the expression of the projectors): 
\begin{equation}
\begin{array}{rl}
\mathcal{P}_{\mu\nu\rho\sigma}^{1}\tau^{\rho\sigma}= & {\displaystyle \frac{1}{2}\left(\theta_{\mu\rho}\omega_{\nu\sigma}+\theta_{\mu\sigma}\omega_{\nu\rho}+\theta_{\nu\rho}\omega_{\mu\sigma}+\theta_{\nu\sigma}\omega_{\mu\rho}\right)\tau^{\rho\sigma},}\\
\\
= & {\displaystyle \frac{1}{2}\left(\theta_{\mu\rho}\frac{k_{\nu}k_{\sigma}}{k^{2}}+\theta_{\mu\sigma}\frac{k_{\nu}k_{\rho}}{k^{2}}+\theta_{\nu\rho}\frac{k_{\mu}k_{\sigma}}{k^{2}}+\theta_{\nu\sigma}\frac{k_{\mu}k_{\rho}}{k^{2}}\right)}\tau^{\rho\sigma}=0,\\
\\
\mathcal{P}_{w,\,\mu\nu\rho\sigma}^{0}\tau^{\rho\sigma}= & \omega_{\mu\nu}\omega_{\rho\sigma}\tau^{\rho\sigma}{\displaystyle =\frac{k_{\nu}k_{\sigma}}{k^{2}}\frac{k_{\nu}k_{\sigma}}{k^{2}}\tau^{\rho\sigma}=0,}\\
\\
\mathcal{P}_{sw,\,\mu\nu\rho\sigma}^{0}\tau^{\rho\sigma}= & {\displaystyle \frac{1}{\sqrt{3}}\theta_{\mu\nu}\omega_{\rho\sigma}\tau^{\rho\sigma}=\frac{1}{\sqrt{3}}\theta_{\mu\nu}\frac{k_{\rho}k_{\sigma}}{k^{2}}\tau^{\rho\sigma}=0,}\\
\\
\tau^{\mu\nu}\mathcal{P}_{ws,\,\mu\nu\rho\sigma}^{0}= & {\displaystyle \frac{1}{\sqrt{3}}\tau^{\mu\nu}\omega_{\mu\nu}\theta_{\rho\sigma}=\frac{1}{\sqrt{3}}\tau^{\mu\nu}\frac{k_{\mu}k_{\nu}}{k^{2}}\theta_{\rho\sigma}=0.}
\end{array}\label{eq:131}
\end{equation}
While for $\mathcal{P}^{2}$ and $\mathcal{P}_{s}^{0}$ the contribution
are not zero, 
\begin{equation}
\tau^{\mu\nu}(-k)\mathcal{P}_{\mu\nu\rho\sigma}^{2}\tau^{\rho\sigma}(k)\neq0,\,\,\,\,\,\,\tau^{\mu\nu}(-k)\mathcal{P}_{s,\,\mu\nu\rho\sigma}^{0}\tau^{\rho\sigma}(k)\neq0.\label{eq:132}
\end{equation}

We are now able to calculate the sandwich between two conserved currents
of (\ref{eq:127}):
\[
\begin{array}{l}
{\displaystyle \tau{}^{\mu\nu}(-k)\frac{1}{k^{2}}\left[\mathcal{P}^{2}+\alpha\mathcal{P}^{1}-\frac{1}{2}\mathcal{P}_{s}^{0}+\frac{4\alpha-3}{2}\mathcal{P}_{w}^{0}-\frac{\sqrt{3}}{2}\mathcal{P}_{sw}^{0}-\frac{\sqrt{3}}{2}\mathcal{P}_{ws}^{0}\right]_{\mu\nu\rho\sigma}}\tau^{\rho\sigma}(k)=\end{array}
\]

\begin{equation}
=\tau^{\mu\nu}(-k)\frac{1}{k^{2}}\left(\mathcal{P}^{2}-\frac{1}{2}\mathcal{P}_{s}^{0}\right)_{\mu\nu\rho\sigma}\tau^{\rho\sigma}(k).\label{eq:133}
\end{equation}
What we found is that also starting from a generic gauge, we have
just had the confirmation that the physical part of the graviton propagator
is%
\footnote{We have to observe that (\ref{eq:134}) doesn't represent the propagator
but its physical part that remain in the saturated propagator, i.e.
in (\ref{eq:133}). Sometime it can happen that we make an abuse of
nomenclature calling it $\mathit{just}$ with word propagator.%
} 
\begin{equation}
\Pi_{GR}=\frac{1}{k^{2}}\left(\mathcal{P}^{2}-\frac{1}{2}\mathcal{P}_{s}^{0}\right).\label{eq:134}
\end{equation}

In the next chapters, especially in Chapter $3$ and $4,$ our discussions
will concern the physical part of the propagator, i.e. (\ref{eq:134})
and its modification in the framework of special theories of modified
gravity.

\section{Graviton propagator and polarization sums}

\subsection{Polarization tensors}

In the subsection $1.4.1$ we constructed a set of spin-$1$ polarization
vector for the photon, $\epsilon_{{\scriptscriptstyle (j=1,j_{z})}}^{\mu}.$
Our aim now is to do the same with graviton, i.e. to introduce a set
of $\mathit{polarization}$ $\mathit{tensors}.$ In particular we
want the spin-$2$ polarization tensors, $\epsilon_{{\scriptscriptstyle (j,j_{z})}}^{\mu\nu}.$
An easy way to construct $j=2$ polarization tensors is to take products
of the $j=1$ polarization vectors given in (\ref{eq:31-1}), that
we write down again for convenience: 
\[
\epsilon_{{\scriptscriptstyle (1,+1)}}\equiv\frac{1}{\sqrt{2}}\begin{pmatrix}0\\
1\\
i\\
0
\end{pmatrix},\,\,\,\,\epsilon_{{\scriptscriptstyle (1,-1)}}\equiv\frac{1}{\sqrt{2}}\begin{pmatrix}0\\
1\\
-i\\
0
\end{pmatrix},\,\,\,\,\epsilon_{{\scriptscriptstyle (1,0)}}\equiv\frac{1}{k}\begin{pmatrix}k^{3}\\
0\\
0\\
k^{0}
\end{pmatrix}.
\]
The right products can be obtained using Clebsch-Gordan coefficients
(Appendix $E)$ corresponding to the product of two spin-$1:$ $1\otimes1.$
In fact, looking at  Appendix $E,$ the product of two $j=1$ gives
$j=2,$ $j=1$ and $j=0$ polarization tensors.\\
$j=2$ gives us five polarization tensors, $j_{z}=+2,+1,0,-1,-2:$
\begin{equation}
\epsilon_{{\scriptscriptstyle (2,+2)}}^{\mu\nu}=\epsilon_{{\scriptscriptstyle (1,+1)}}^{\mu}\otimes\epsilon_{{\scriptscriptstyle (1,+1)}}^{\nu}=\frac{1}{2}\begin{pmatrix}0 & 0 & 0 & 0\\
0 & 1 & i & 0\\
0 & i & -1 & 0\\
0 & 0 & 0 & 0
\end{pmatrix},\label{eq:142}
\end{equation}

$ $

\begin{equation}
\epsilon_{{\scriptscriptstyle (2,-2)}}^{\mu\nu}=\epsilon_{{\scriptscriptstyle (1,-1)}}^{\mu}\otimes\epsilon_{{\scriptscriptstyle (1,-1)}}^{\nu}=\frac{1}{2}\begin{pmatrix}0 & 0 & 0 & 0\\
0 & 1 & -i & 0\\
0 & -i & -1 & 0\\
0 & 0 & 0 & 0
\end{pmatrix},\label{eq:143}
\end{equation}

$ $

\begin{equation}
\begin{array}{rl}
\epsilon_{{\scriptscriptstyle (2,+1)}}^{\mu\nu}= & \frac{1}{\sqrt{2}}\left(\epsilon_{{\scriptscriptstyle (1,+1)}}^{\mu}\otimes\epsilon_{{\scriptscriptstyle (1,0)}}^{\nu}+\epsilon_{{\scriptscriptstyle (1,0)}}^{\mu}\otimes\epsilon_{{\scriptscriptstyle (1,+1)}}^{\nu}\right)\\
\\
= & \frac{1}{2k}\begin{pmatrix}0 & k^{3} & ik^{3} & 0\\
k^{3} & 0 & 0 & k^{0}\\
ik^{3} & 0 & 0 & ik^{0}\\
0 & k^{0} & ik^{0} & 0
\end{pmatrix},
\end{array}\label{eq:144}
\end{equation}

$ $

\begin{equation}
\begin{array}{rl}
\epsilon_{{\scriptscriptstyle (2,-1)}}^{\mu\nu}= & \frac{1}{\sqrt{2}}\left(\epsilon_{{\scriptscriptstyle (1,-1)}}^{\mu}\otimes\epsilon_{{\scriptscriptstyle (1,0)}}^{\nu}+\epsilon_{{\scriptscriptstyle (1,0)}}^{\mu}\otimes\epsilon_{{\scriptscriptstyle (1,-1)}}^{\nu}\right)\\
\\
= & \frac{1}{2k}\begin{pmatrix}0 & k^{3} & -ik^{3} & 0\\
k^{3} & 0 & 0 & k^{0}\\
-ik^{3} & 0 & 0 & -ik^{0}\\
0 & k^{0} & -ik^{0} & 0
\end{pmatrix},
\end{array}\label{eq:145}
\end{equation}

$ $

\begin{equation}
\begin{array}{rl}
\epsilon_{{\scriptscriptstyle (2,0)}}^{\mu\nu}= & \frac{1}{\sqrt{6}}\left(\epsilon_{{\scriptscriptstyle (1,+1)}}^{\mu}\otimes\epsilon_{{\scriptscriptstyle (1,-1)}}^{\nu}+\epsilon_{{\scriptscriptstyle (1,-1)}}^{\mu}\otimes\epsilon_{{\scriptscriptstyle (1,+1)}}^{\nu}+2\epsilon_{{\scriptscriptstyle (1,0)}}^{\mu}\otimes\epsilon_{{\scriptscriptstyle (1,0)}}^{\nu}\right)\\
\\
= & \frac{1}{\sqrt{6}}\begin{pmatrix}2\frac{(k^{3})^{2}}{k^{2}} & 0 & 0 & 2\frac{k^{3}k^{0}}{k^{2}}\\
0 & 1 & 0 & 0\\
0 & 0 & 1 & 0\\
2\frac{k^{3}k^{0}}{k^{2}} & 0 & 0 & 2\frac{(k^{0})^{2}}{k^{2}}
\end{pmatrix}.
\end{array}\label{eq:146}
\end{equation}

$ $\\
Then the polarization tensor corresponding to $j=0$ is
\begin{equation}
\begin{array}{rl}
\epsilon_{{\scriptscriptstyle (0,0)}}^{\mu\nu}= & \frac{1}{\sqrt{3}}\left(\epsilon_{{\scriptscriptstyle (1,+1)}}^{\mu}\otimes\epsilon_{{\scriptscriptstyle (1,-1)}}^{\nu}+\epsilon_{{\scriptscriptstyle (1,-1)}}^{\mu}\otimes\epsilon_{{\scriptscriptstyle (1,+1)}}^{\nu}-\epsilon_{{\scriptscriptstyle (1,0)}}^{\mu}\otimes\epsilon_{{\scriptscriptstyle (1,0)}}^{\nu}\right)\\
\\
= & \frac{1}{\sqrt{3}}\begin{pmatrix}-\frac{(k^{3})^{2}}{k^{2}} & 0 & 0 & -\frac{k^{3}k^{0}}{k^{2}}\\
0 & 1 & 0 & 0\\
0 & 0 & 1 & 0\\
-\frac{k^{3}k^{0}}{k^{2}} & 0 & 0 & -\frac{(k^{0})^{2}}{k^{2}}
\end{pmatrix}.
\end{array}\label{eq:147}
\end{equation}
In principle we also have the spin-$1$ polarization tensor $\epsilon_{{\scriptscriptstyle (1,+1)}}^{\mu\nu},$
$\epsilon_{{\scriptscriptstyle (1,0)}}^{\mu\nu}$ and $\epsilon_{{\scriptscriptstyle (1,-1)}}^{\mu\nu}:$
\begin{equation}
\epsilon_{{\scriptscriptstyle (1,+1)}}^{\mu\nu}=\frac{1}{\sqrt{2}}\left(\epsilon_{{\scriptscriptstyle (1,+1)}}^{\mu}\otimes\epsilon_{{\scriptscriptstyle (1,0)}}^{\nu}-\epsilon_{{\scriptscriptstyle (1,0)}}^{\mu}\otimes\epsilon_{{\scriptscriptstyle (1,+1)}}^{\nu}\right),\label{eq:148}
\end{equation}
\begin{equation}
\epsilon_{{\scriptscriptstyle (1,-1)}}^{\mu\nu}=\frac{1}{\sqrt{2}}\left(\epsilon_{{\scriptscriptstyle (1,0)}}^{\mu}\otimes\epsilon_{{\scriptscriptstyle (1,-1)}}^{\nu}-\epsilon_{{\scriptscriptstyle (1,-1)}}^{\mu}\otimes\epsilon_{{\scriptscriptstyle (1,0)}}^{\nu}\right),\label{eq:149}
\end{equation}
\begin{equation}
\epsilon_{{\scriptscriptstyle (1,0)}}^{\mu\nu}=\frac{1}{\sqrt{2}}\left(\epsilon_{{\scriptscriptstyle (1,+1)}}^{\mu}\otimes\epsilon_{{\scriptscriptstyle (1,-1)}}^{\nu}-\epsilon_{{\scriptscriptstyle (1,-1)}}^{\mu}\otimes\epsilon_{{\scriptscriptstyle (1,+1)}}^{\nu}\right).\label{eq:150}
\end{equation}
But (\ref{eq:148}), (\ref{eq:149}) and (\ref{eq:150}) say that
the polarization tensor with $j=1$ are antisymmetric, so we shall
not consider them.%
\footnote{Even if we consider them, since they appear sandwiched between two
conserved currents, they give a null contribution when are multiplied
with symmetric tensors, $\tau^{\mu\nu}\epsilon_{{\scriptscriptstyle (1,j_{z})},\mu\nu}=0.$ %
}\\
One can easily compute the value of the helicity of each polarization
tensor by acting twice with the rotation matrix around the third axis
(see (\ref{eq:75 rotation}) for its definition). The following relations
hold:
\begin{equation}
\begin{array}{ll}
R_{\rho}^{(z)\mu}(\vartheta)R_{\sigma}^{(z)\nu}(\vartheta)\epsilon_{{\scriptscriptstyle (2,+2)}}^{\rho\sigma}=e^{i2\vartheta}\epsilon_{{\scriptscriptstyle (2,+2)}}^{\mu\nu}, & R_{\rho}^{(z)\mu}(\vartheta)R_{\sigma}^{(z)\nu}(\vartheta)\epsilon_{{\scriptscriptstyle (2,-2)}}^{\rho\sigma}=e^{-i2\vartheta}\epsilon_{{\scriptscriptstyle (2,-2)}}^{\mu\nu},\\
\\
R_{\rho}^{(z)\mu}(\vartheta)R_{\sigma}^{(z)\nu}(\vartheta)\epsilon_{{\scriptscriptstyle (2,+1)}}^{\rho\sigma}=e^{i\vartheta}\epsilon_{{\scriptscriptstyle (2,+1)}}^{\mu\nu}, & R_{\rho}^{(z)\mu}(\vartheta)R_{\sigma}^{(z)\nu}(\vartheta)\epsilon_{{\scriptscriptstyle (2,-1)}}^{\rho\sigma}=e^{-i\vartheta}\epsilon_{{\scriptscriptstyle (2,-1)}}^{\mu\nu},\\
\\
R_{\rho}^{(z)\mu}(\vartheta)R_{\sigma}^{(z)\nu}(\vartheta)\epsilon_{{\scriptscriptstyle (2,0)}}^{\rho\sigma}=\epsilon_{{\scriptscriptstyle (2,0)}}^{\mu\nu}, & R_{\rho}^{(z)\mu}(\vartheta)R_{\sigma}^{(z)\nu}(\vartheta)\epsilon_{{\scriptscriptstyle (0,0)}}^{\rho\sigma}=\epsilon_{{\scriptscriptstyle (0,0)}}^{\mu\nu}.
\end{array}
\end{equation}
Hence we have seen that we have six polarization tensors, (\ref{eq:142})-(\ref{eq:147}),
one per each degree of freedom of a virtual graviton. For a real graviton
we expect that only two polarization tensors are physical; indeed
in the next subsection we shall see that the physical ones are those
corresponding to the helicity states $j_{z}=+2,-2.$

\subsection{Graviton propagator in terms of polarization tensors}

As we have done for the photon propagator, we want to rewrite the
saturated propagator in terms of the polarization tensors and see
which components are present for either on-shell and off-shell graviton.
Unlike the case of the virtual photon, which presented only a spin-$1$
component, we will have confirmation that a virtual graviton in addition
to the spin-$2$ component has also a spin-$0$ components. We will
be able to write these two different components in terms of the polarization
tensors. The fact that the graviton has a $j=0$ component, as we
have verified in the section $2.3,$ doesn't violate any fundamental
principle, but its presence is very important because cancels the
$j=2,$ $j_{z}=0$ component when we consider a on-shell graviton.
We are also taking inspiration from {[}\ref{-Dicus}{]}.

$ $

\textbf{Off-shell graviton}

$ $\\
Our starting point is the saturated propagator in (\ref{eq:116}).
By expliciting the spin projector operators in terms of $\eta_{\mu\nu}$
and $k_{\mu},$ using the relations (\ref{eq:191realations}) and
implementing the conservation law of the source, $k_{\mu}\tau^{\mu\nu}=0,$
one gets
\begin{equation}
\tau^{\mu\nu}\left(-k\right)\Pi_{GR,\mu\nu\rho\sigma}(k)\tau^{\rho\sigma}(k)=\tau^{\mu\nu}\left(-k\right)\frac{1}{2k^{2}}\left(\eta_{\mu\rho}\eta_{\nu\sigma}+\eta_{\mu\sigma}\eta_{\nu\sigma}-\eta_{\mu\nu}\eta_{\rho\sigma}\right)\tau^{\rho\sigma}(k).\label{eq:151}
\end{equation}
The explicit expression of (\ref{eq:151}), once we compute the products
between $\tau^{\mu\nu}$ and the metric tensor $\eta_{\mu\nu},$ is
\begin{equation}
\begin{array}{rl}
\tau^{\mu\nu}\left(-k\right)\Pi_{GR,\mu\nu\rho\sigma}(k)\tau^{\rho\sigma}(k)= & {\displaystyle \frac{1}{k^{2}}\left\{ \frac{1}{2}\tau^{00}(-k)\left[\tau^{00}(k)+\tau^{11}(k)+\tau^{22}(k)+\tau^{33}(k)\right]\right.}\\
\\
 & +{\displaystyle \frac{1}{2}\tau^{11}(-k})\left[\tau^{00}(k)+\tau^{11}(k)-\tau^{22}(k)-\tau^{33}(k)\right]\\
\\
 & +{\displaystyle \frac{1}{2}\tau^{22}}(-k)\left[\tau^{00}(k)-\tau^{11}(k)+\tau^{22}(k)-\tau^{33}(k)\right]\\
\\
 & +{\displaystyle \frac{1}{2}}\tau^{33}(-k)\left[\tau^{00}(k)-\tau^{11}(k)-\tau^{22}(k)+\tau^{33}(k)\right]\\
\\
 & -2\tau^{01}(-k)\tau^{01}(k)-2\tau^{02}(-k)\tau^{02}(k)-2\tau^{03}(-k)\tau^{03}(k)\\
\\
 & \left.+2\tau^{12}(-k)\tau^{12}(k)+2\tau^{13}(-k)\tau^{13}(k)+2\tau^{23}(-k)\tau^{23}(k)\right\} .
\end{array}\label{eq:152}
\end{equation}
The symmetry of the source, $\tau^{\mu\nu}=\tau^{\nu\mu},$ has been
used to obtain the last expression.\\
Let us suppose that the virtual graviton has the four-momentum $k^{\mu}\equiv\left(k^{0},0,0,k^{3}\right).$
In the off-shell case the conservation law $k_{\mu}\tau^{\mu\nu}=0$
assumes the following form:
\begin{equation}
k^{0}\tau^{0\nu}=k^{3}\tau^{3\nu}.\label{eq:153}
\end{equation}
By performing easy calculations, using symmetry and (\ref{eq:153}),
we can rewrite (\ref{eq:152}) in a more convenient form:
\begin{equation}
\tau^{\mu\nu}\left(-k\right)\Pi_{GR,\mu\nu\rho\sigma}(k)\tau^{\rho\sigma}(k)=\Pi_{1}+\Pi_{2}+\Pi_{3}+\Pi_{4},\label{eq:154}
\end{equation}
where
\begin{equation}
\begin{array}{rl}
\Pi_{1}= & {\displaystyle \frac{1}{k^{2}}\left\{ \frac{1}{2}\left[\left(\tau^{11}(-k)-\tau^{22}(-k)\right)\left(\tau^{11}(k)-\tau^{22}(k)\right)\right]+2\tau^{12}(-k)\tau^{12}(k)\right\} ,}\\
\\
\Pi_{2}= & {\displaystyle \frac{2}{k^{2}}\left[\tau^{13}(-k)\tau^{13}(k)+\tau^{23}(-k)\tau^{23}(k)-\tau^{01}(-k)\tau^{01}(k)-\tau^{02}(-k)\tau^{02}(k)\right]},\\
\\
\Pi_{3}= & {\displaystyle \frac{1}{6k^{2}}}\left[2\left(\tau^{00}(-k)-\tau^{33}(-k)\right)+\tau^{11}(-k)+\tau^{22}(-k)\right]\\
\\
 & \times\left[2\left(\tau^{00}(k)-\tau^{33}(k)\right)+\tau^{11}(k)+\tau^{22}(k)\right],\\
\\
\Pi_{4}= & {\displaystyle \frac{1}{6k^{2}}}\left[-\tau^{00}(-k)+\tau^{33}(-k)+\tau^{11}(-k)+\tau^{22}(-k)\right]\\
\\
 & \times\left[-\tau^{00}(k)+\tau^{33}(k)+\tau^{11}(k)+\tau^{22}(k))\right].
\end{array}
\end{equation}

Now we want to rewrite each $\Pi_{i}$ in terms of the polarization
tensors. Always using symmetry and conservation law, we notice that:
\[
\tau^{\mu\nu}\left(-k\right)\left(\sum_{{\scriptscriptstyle j_{z}=+2,-2}}\epsilon_{{\scriptscriptstyle (2,j_{z})},\mu\nu}\epsilon_{{\scriptscriptstyle (2,j_{z})},\rho\sigma}^{*}\right)\tau^{\rho\sigma}(k)=
\]
\[
=\tau^{\mu\nu}\left(-k\right)\left[\frac{1}{2}\begin{pmatrix}0 & 0 & 0 & 0\\
0 & 1 & i & 0\\
0 & i & -1 & 0\\
0 & 0 & 0 & 0
\end{pmatrix}_{\mu\nu}\frac{1}{2}\begin{pmatrix}0 & 0 & 0 & 0\\
0 & 1 & -i & 0\\
0 & -i & -1 & 0\\
0 & 0 & 0 & 0
\end{pmatrix}_{\rho\sigma}\right]\tau^{\rho\sigma}(k)
\]
\[
+\tau^{\mu\nu}(-k)\left[\frac{1}{2}\begin{pmatrix}0 & 0 & 0 & 0\\
0 & 1 & -i & 0\\
0 & -i & -1 & 0\\
0 & 0 & 0 & 0
\end{pmatrix}_{\mu\nu}\frac{1}{2}\begin{pmatrix}0 & 0 & 0 & 0\\
0 & 1 & i & 0\\
0 & i & -1 & 0\\
0 & 0 & 0 & 0
\end{pmatrix}_{\rho\sigma}\right]\tau^{\rho\sigma}(k)=
\]
\[
=\frac{1}{4}\left[\tau^{11}(-k)+2i\tau^{12}(-k)-\tau^{22}(-k)\right]\left[\tau^{11}(k)-2i\tau^{12}(k)-\tau^{22}(k)\right]
\]
\[
+\frac{1}{4}\left[\tau^{11}(-k)-2i\tau^{12}(-k)-\tau^{22}(-k)\right]\left[\tau^{11}(k)+2i\tau^{12}(k)-\tau^{22}(k)\right]
\]
\begin{equation}
=\frac{1}{2}\left[\left(\tau^{11}(-k)-\tau^{22}(-k)\right)\left(\tau^{11}(k)-\tau^{22}(k)\right)\right]+2\tau^{12}(-k)\tau^{12}(k)=\Pi_{1}.\label{eq:155}
\end{equation}
So the first line of (\ref{eq:154}) corresponds to $j_{z}=+2,-2.$
In the same way we can connect $\Pi_{2},\Pi_{3},\Pi_{4}$ to the other
polarization tensors. Although we are not going to do all the tedious
calculations, one can see that:
\begin{equation}
\begin{array}{rl}
\tau^{\mu\nu}\left(-k\right)\left(\sum_{{\scriptscriptstyle j_{z}=+1,-1}}\epsilon_{{\scriptscriptstyle (2,j_{z})},\mu\nu}\epsilon_{{\scriptscriptstyle (2,j_{z})},\rho\sigma}^{*}\right)\tau^{\rho\sigma}(k)= & 2\left[\tau^{13}(-k)\tau^{13}(k)+\tau^{23}(-k)\tau^{23}(k)\right.\\
\\
 & \left.-\tau^{01}(-k)\tau^{01}(k)-\tau^{02}(-k)\tau^{02}(k)\right]=\Pi_{2};
\end{array}\label{eq:156}
\end{equation}
\begin{equation}
\begin{array}{rl}
\tau^{\mu\nu}\left(-k\right)\epsilon_{{\scriptscriptstyle (2,0)},\mu\nu}\epsilon_{{\scriptscriptstyle (2,0)},\rho\sigma}^{*}\tau^{\rho\sigma}(k)= & {\displaystyle \frac{1}{6}}\left[2\left(\tau^{00}(-k)-\tau^{33}(-k)\right)+\tau^{11}(-k)+\tau^{22}(-k)\right]\\
\\
 & \times\left[2\left(\tau^{00}(k)-\tau^{33}(k)\right)+\tau^{11}(k)+\tau^{22}(k)\right]=\Pi_{3};
\end{array}\label{eq:157}
\end{equation}
\begin{equation}
\begin{array}{rl}
\tau^{\mu\nu}\left(-k\right)\epsilon_{{\scriptscriptstyle (0,0)},\mu\nu}\epsilon_{{\scriptscriptstyle (0,0)},\rho\sigma}^{*}\tau^{\rho\sigma}(k)= & {\displaystyle \frac{1}{3}}\left[-\tau^{00}(-k)+\tau^{33}(-k)+\tau^{11}(-k)+\tau^{22}(-k)\right]\\
\\
 & \times\left[-\tau^{00}(k)+\tau^{33}(k)+\tau^{11}(k)+\tau^{22}(k))\right]=\Pi_{4}.
\end{array}\label{eq:158}
\end{equation}
So $\Pi_{2}$ corresponds to $j=2,j_{z}=\pm1$ component, $\Pi_{3}$
to $j=2,$ $j_{z}=0$ and $\Pi_{4}$ to the spin-$0$ component, $j=0.$
\\
Now we need to rewrite the saturated propagator in terms of the polarization
tensors; in fact from (\ref{eq:155})-(\ref{eq:158}) we obtain
\begin{equation}
\begin{array}{rl}
\tau^{\mu\nu}\left(-k\right)\Pi_{GR,\mu\nu\rho\sigma}(k)\tau^{\rho\sigma}(k)=\tau^{\mu\nu}{\displaystyle \left(-k\right)\frac{1}{k^{2}}} & \biggl({\displaystyle \sum_{{\scriptscriptstyle j_{z}=-2}}^{{\scriptscriptstyle +2}}\epsilon_{{\scriptscriptstyle (2,j_{z})},\mu\nu}\epsilon_{{\scriptscriptstyle (2,j_{z})},\rho\sigma}^{*}}\\
\\
 & \left.-{\displaystyle \frac{1}{2}}\epsilon_{{\scriptscriptstyle (0,0)},\mu\nu}\epsilon_{{\scriptscriptstyle (0,0)},\rho\sigma}^{*}\right)\tau^{\rho\sigma}(k),
\end{array}\label{eq:159}
\end{equation}
where the sum runs on over the five $j=2$ polarization tensors. Equation
(\ref{eq:159}) confirms that a virtual graviton has also a scalar
component, 
\[
\tau^{\mu\nu}(-k)\frac{1}{2}\epsilon_{{\scriptscriptstyle (0,0)},\mu\nu}\epsilon_{{\scriptscriptstyle (0,0)},\rho\sigma}^{*}\tau^{\rho\sigma}(k).
\]
 By comparing the saturated propagator in (\ref{eq:159}) with its
form in (\ref{eq:116}) we can read the spin projector operators $\mathcal{P}^{2}$
and $\mathcal{P}_{s}^{0}$ in terms of the polarization tensors:
\begin{equation}
\begin{array}{rl}
\mathcal{P}_{\mu\nu\rho\sigma}^{2}\equiv & {\displaystyle \sum_{{\scriptscriptstyle j_{z}=-2}}^{{\scriptscriptstyle +2}}\epsilon_{{\scriptscriptstyle (2,j_{z})},\mu\nu}\epsilon_{{\scriptscriptstyle (2,j_{z})},\rho\sigma}^{*}}\\
\\
\mathcal{P}_{s,\,\mu\nu\rho\sigma}^{0}\equiv & \epsilon_{{\scriptscriptstyle (0,0)},\mu\nu}\epsilon_{{\scriptscriptstyle (0,0)},\rho\sigma}^{*}.
\end{array}\label{eq:160}
\end{equation}

$ $

\textbf{On-shell graviton}

$ $\\
As for the real graviton the saturated propagator (\ref{eq:154})
reduces in a more simple form when on-shell condition is imposed.
In fact, in the on-shell case $(k^{0}=k^{3})$ the conservation law
\ref{eq:153} becomes
\begin{equation}
\tau^{0\nu}=\tau^{3\nu}.\label{eq:161}
\end{equation}
But (\ref{eq:161}) implies also $\tau^{00}=\tau^{03}=\tau^{33}.$
Hence because of this last relation and (\ref{eq:151}) the saturated
propagator for a real graviton becomes
\begin{equation}
\begin{array}{rl}
\tau^{\mu\nu}\left(-k\right)\Pi_{GR,\mu\nu\rho\sigma}(k)\tau^{\rho\sigma}(k)= & {\displaystyle \frac{1}{k^{2}}}\left\{ 2\tau^{12}(-k)\tau^{12}(k)\right.\\
\\
 & \left.+{\displaystyle \frac{1}{2}\left[\left(\tau^{11}(-k)-\tau^{22}(-k)\right)\left(\tau^{11}(k)-\tau^{22}(k)\right)\right]}\right\} \\
\\
= & \tau^{\mu\nu}{\displaystyle \left(-k\right)\frac{1}{k^{2}}}{\displaystyle \left(\sum_{{\scriptscriptstyle j_{z}=+2,-2}}\epsilon_{{\scriptscriptstyle (2,j_{z})},\mu\nu}\epsilon_{{\scriptscriptstyle (2,j_{z})},\rho\sigma}^{*}\right)\tau^{\rho\sigma}}(k).
\end{array}\label{eq:162}
\end{equation}
The last equation confirms that an on-shell graviton has only two
degrees of freedom, indeed only the two polarization tensors with
helicity $j_{z}=+2$ and $j_{z}=-2$ are present.\\
\\
We must also notice that when we go on-shell, it happens that in equation
(\ref{eq:154}) the component $\Pi_{4}$ cancels with $\Pi_{3},$
namely the scalar component cancels with the $j=2,$ $j_{z}=0$ component
(longitudinal component). Thus we have shown what we have anticipated
at the beginning of this subsection, i.e. that the $j=0$ component
is necessary in order to ensure that a real graviton has no $j=2,$
$j_{z}=0$ component.

\section{Ghosts and unitarity analysis in General Relativity}

In this section we want to check whether ghosts are absent, so, whether
the unitarity is preserved in GR {[}\ref{-J.-Helayel}{]},{[}\ref{-A.-Accioly}{]}.
As we have done for the vector field case we shall follow the method
discussed in Appendix $C,$ i.e. we will verify the positivity of
the imaginary part of the amplitude residue at the pole $k^{2}=0.$\\
Let us consider the propagator in (\ref{eq:127}) with a generic parameter
$\alpha.$ As we have already seen above, the choice of De Donder
gauge, corresponding to the symmetry $\delta h_{\mu\nu}=\partial_{\mu}\xi_{\nu}+\partial_{\nu}\xi_{\mu},$
implies the conservation of the energy-momentum tensor $\partial_{\mu}\tau^{\mu\nu}=0,$
or equivalently in momentum space $k_{\mu}\tau^{\mu\nu}=0.$ The amplitude
of a process in which a graviton is created at a point of the space-time
and is annihilated at another point is described by an amplitude of
the following type (see (\ref{eq:116}) or (\ref{eq:133})):
\begin{equation}
\begin{array}{rl}
\mathcal{A}= & \tau^{*\mu\nu}(k)\left\langle T\left(h_{\mu\nu}(-k)h_{\rho\sigma}(k)\right)\right\rangle \tau^{\rho\sigma}(k)\\
\\
= & i\tau^{*\mu\nu}(k)\Pi_{GR,\mu\nu\rho\sigma}\tau^{\rho\sigma}(k),\\
\\
= & i\tau^{*\mu\nu}(k)\frac{1}{k^{2}}\left(\mathcal{P}^{2}-\frac{1}{2}\mathcal{P}_{s}^{0}\right)_{\mu\nu\rho\sigma}\tau^{\rho\sigma}(k).
\end{array}\label{eq:129}
\end{equation}
where $k^{\mu}$ is the four-momentum of the graviton. By following
the calculations in (\ref{eq:203saturated}) we can also recast the
equation (\ref{eq:129}) just in terms of the Minkowski metric tensor:
\begin{equation}
\begin{array}{rl}
\mathcal{A}= & i\tau^{*\mu\nu}(k{\displaystyle )\frac{1}{2k^{2}}\left(\eta_{\mu\rho}\eta_{\nu\sigma}+\eta_{\nu\rho}\eta_{\mu\sigma}-\eta_{\mu\nu}\eta_{\rho\sigma}\right)}\tau^{\rho\sigma}(k)\\
\\
= & {\displaystyle \frac{i}{k^{2}}\left(\tau^{*\mu\nu}(k)\tau_{\mu\nu}(k)-\frac{1}{2}\tau_{\rho}^{*\rho}(k)\tau_{\sigma}^{\sigma}(k)\right)}\\
\\
= & {\displaystyle \frac{i}{k^{2}}\left(|\tau^{\mu\nu}|^{2}-\frac{1}{2}|\tau_{\sigma}^{\sigma}|^{2}\right)}.
\end{array}\label{eq:135}
\end{equation}
The residue of the amplitude (\ref{eq:135}) in $k^{2}=0$ is
\begin{equation}
Res_{k^{2}=0}\mathcal{A}=\lim_{k^{2}\rightarrow0}k^{2}\frac{i}{k^{2}}\left(|\tau^{\mu\nu}|^{2}-\frac{1}{2}|\tau_{\sigma}^{\sigma}|^{2}\right)=i\left(|\tau^{\mu\nu}|^{2}-\frac{1}{2}|\tau_{\sigma}^{\sigma}|^{2}\right)
\end{equation}
\begin{equation}
\Rightarrow\mathrm{Im}\left\{ Res_{k^{2}=0}\mathcal{A}\right\} =|\tau^{\mu\nu}|^{2}-\frac{1}{2}|\tau|^{2}.\label{eq:136}
\end{equation}
To understand if the imaginary part of the residue is either positive
or negative we have to analyze the structure of the current $\tau^{\mu\nu}.$
Let us consider again the basis (\ref{eq:8}), $\left\{ k^{\mu},\tilde{k}^{\mu},\varepsilon_{1}^{\mu},\varepsilon_{2}^{\mu}\right\} ,$
in the space of the four-vectors, and expand the current $\tau^{\mu\nu}$
as
\begin{equation}
\begin{array}{rl}
\tau_{\mu\nu}(k)= & a(k)k_{\mu}k_{\nu}+b(k)k_{(\mu}\tilde{k}_{\nu)}+c_{i}(k)k_{(\mu}\varepsilon_{\nu)}^{i}\\
\\
 & +d(k)\tilde{k}_{\mu}\tilde{k}_{\nu}+e_{i}(k)\tilde{k}_{(\mu}\varepsilon_{\nu)}^{i}+f_{ij}(k)\varepsilon_{(\mu}^{i}\varepsilon_{\nu)}^{j}.
\end{array}\label{eq:137}
\end{equation}
Impose the conservation law for $\tau^{\mu\nu}:$ 
\begin{equation}
\left[k^{\mu}\tau_{\mu\nu}(k)\right]_{k^{2}=0}=\frac{1}{2}b(k)\left(k^{\mu}\tilde{k}_{\mu}\right)k_{\nu}+d(k)\left(k^{\mu}\tilde{k}_{\mu}\right)\tilde{k}_{\nu}+\frac{1}{2}e_{i}(k)\left(k^{\mu}\tilde{k}_{\mu}\right)\varepsilon_{\nu}^{i}=0.\label{eq:138}
\end{equation}
Note that we have valuated the last equation in $k^{2}=0$ and we
shall also do the same with the following formulas, because we are
interested in the residue of the amplitude that is calculated in $k^{2}=0.$\\
Since $k\cdot\tilde{k}=k^{\mu}\tilde{k}_{\mu}=(k^{0})^{2}+(\bar{k})^{2}=2(\bar{k})^{2}\neq0,$
(\ref{eq:138}) reads as 
\begin{equation}
\frac{1}{2}b(k)k_{\nu}+d(k)\tilde{k}_{\nu}+\frac{1}{2}e_{i}(k)\varepsilon_{\nu}^{i}=0.\label{eq:139}
\end{equation}
If we consider the component $\nu=0$ of the the last equation we
obtain
\begin{equation}
\frac{1}{2}b(k)k_{0}+d(k)k_{0}=0,\,\,\, k_{0}\neq0\Rightarrow b(k)=-2d(k),
\end{equation}
so (\ref{eq:139}) becomes 
\begin{equation}
b(k)\left(k_{\nu}-\tilde{k}_{\nu}\right)+\frac{1}{2}e_{i}(k)\varepsilon_{\nu}^{i}=0,
\end{equation}
and multiplying for $\varepsilon^{j,\nu}$ we have
\begin{equation}
\frac{1}{2}e_{i}(k)\varepsilon^{j,\nu}\varepsilon_{\nu}^{i}=\frac{1}{2}e_{i}(k)\delta^{ji}=0\Leftrightarrow e_{i}(k)=0,\,\,\, i=1,2.
\end{equation}
Hence $k^{2}=0$ implies $d(k)=0$ and $e_{1}(k)=0=e_{2}(k),$ and
the expansion for $\tau^{\mu\nu}$ simplifies as
\[
\left[\tau_{\mu\nu}(k)\right]_{k^{2}=0}=a(k)k_{\mu}k_{\nu}+c_{i}(k)k_{(\mu}\varepsilon_{\nu)}^{i}+f_{ij}(k)\varepsilon_{(\mu}^{i}\varepsilon_{\nu)}^{j};
\]
instead for the trace $\tau_{\sigma}^{\sigma}(k)$ valuated in $k^{2}=0$
one has%
\footnote{Note that we are always following the Einstein convention. So, with
$f=f_{ii}$ we mean the trace $f=\sum_{i}f_{ii},$ with $|f|^{2}$
we mean $|f|^{2}=\left(\sum_{i}f_{ii}^{*}\right)\bigl(\sum_{j}f_{jj}\bigr)$
and with $|f_{ij}|^{2}$ we mean $|f_{ij}|^{2}=\sum_{ij}f_{ij}^{*}f_{ij}.$ %
}
\begin{equation}
\left[\tau_{\sigma}^{\sigma}(k)\right]_{k^{2}=0}=-\delta^{ij}f_{ij}=-f_{ii}
\end{equation}
Let us now calculate the quantities present in the imaginary part
of the residue.\\
We can start from $\tau^{*\mu\nu}(k)\tau_{\mu\nu}(k):$ 
\begin{equation}
\begin{array}{rl}
\left[\tau^{*\mu\nu}(k)\tau_{\mu\nu}(k)\right]_{k^{2}=0}= & {\displaystyle \frac{1}{4}}f_{ij}^{*}(k)f_{kl}(k)\left(\varepsilon^{i,\mu}\varepsilon^{j,\nu}+\varepsilon^{i,\mu}\varepsilon^{j,\nu}\right)\left(\varepsilon_{\mu}^{k}\varepsilon_{\nu}^{l}+\varepsilon_{\mu}^{k}\varepsilon_{\nu}^{l}\right)\\
\\
= & {\displaystyle \frac{1}{2}}f_{ij}^{*}(k)f_{kl}(k)\left(\delta^{ik}\delta^{jl}+\delta^{il}\delta^{jk}\right)=f_{ij}^{*}(k)f_{kl}(k)\\
\\
= & |f_{ij}|^{2}
\end{array}\label{eq:140}
\end{equation}
instead for the trace part $\tau^{*}(k)\tau(k)$ we obtain: 
\begin{equation}
\left[\tau^{*}(k)\tau(k)\right]_{k^{2}=0}=f_{ii}^{*}f_{jj}=|f_{ii}|^{2}.\label{eq:141}
\end{equation}
Now, by using the formulas (\ref{eq:140}) and (\ref{eq:141}) we
are able to rewrite the imaginary part of the residue in terms of
the coefficients $f_{ij}:$
\begin{equation}
\mathrm{Im}\left\{ Res_{k^{2}=0}\mathcal{A}\right\} =|f_{ij}|^{2}-\frac{1}{2}|f|^{2}.
\end{equation}
Going ahead with the calculations we obtain: 
\begin{equation}
\begin{array}{rl}
{\displaystyle |f_{ij}|^{2}-\frac{1}{2}|f|^{2}}= & f_{ij}^{*}f_{ij}-f_{ii}^{*}f_{jj}\\
\\
= & f_{11}^{*}f_{11}+2f_{12}^{*}f_{12}+f_{22}^{*}f_{22}\\
\\
 & -{\displaystyle \frac{1}{2}}\left(f_{11}^{*}f_{11}+f_{11}^{*}f_{22}+f_{22}^{*}f_{11}+f_{22}^{*}f_{22}\right)\\
\\
= & {\displaystyle \frac{1}{2}}\left(f_{11}-f_{22}\right)^{*}\left(f_{11}-f_{22}\right)+2f_{12}^{*}f_{12}\\
\\
= & \frac{1}{2}|f_{11}-f_{22}|^{2}+2|f_{12}|^{2}>0
\end{array}
\end{equation}
\begin{equation}
\Rightarrow\mathrm{Im}\left\{ Res_{k^{2}=0}\mathcal{A}\right\} >0,
\end{equation}
i.e. in GR the unitarity is not violated.  \\
$ $
\begin{rem}
The spin-$0$ part of the propagator, having a minus sign, can let
us suspect the presence of a ghost that could violate the unitarity
of the theory. We have just showed that the unitarity is preserved,
so it means that the spin-$0$ component corresponds to a ``good''
ghost that is fundamental to ensure the unitarity. In the previous
chapter we have seen why its presence is so important, in fact we
pointed out that the spin-$0$ component is equal and opposite to
the helicity-$0$ component of the spin-$2$ part of the propagator,
so they cancel out.
\end{rem}

\chapter{Quadratic gravity}

In the Introduction, we have emphasized that Einstein's General Relativity
is the best theory we have to describe the gravitational interaction,
although we need to face serious questions when one considers the
behavior at short distances (or high energy). Indeed, problems of
divergence emerge because of the presence of black hole and cosmological
singularities at the classical level and because of the UV incompleteness
at the quantum level. Because of these problems modification of Hilbert-Einstein action is demanded, but one must take care of preserving the well
known and valid behavior in the infrared regime (or large distances) that is consistent with the experiments.\\
In this chapter we shall consider one of the most intuitive modification
of GR, i.e. in addition to the Einstein-Hilbert term we are going
to consider all the possible terms quadratic in the curvatures {[}\ref{-Biswas-T,non local}{]}%
\footnote{See also Ref. {[}\ref{-G.-Lambiase,}{]},{[}\ref{-S.-Capozziello,}{]},{[}\ref{-G.-Lambiase,-1}{]},{[}\ref{-S.Capozziello,-G.}{]},{[}\ref{-S.-Calchi}{]}.%
}. Once we have the action, or in other words the Lagrangian, for this
kind of modified gravity, we will linearize the theory and then determine
equations of motion and propagator. We will take inspiration from
Ref. {[}\ref{-Biswas-T,non local}{]}.

\section{Most general quadratic curvature action of gravity}

The most general form for the quadratic parity-invariant and torsion free action of gravity is given by
\begin{equation}
S_{q}=\frac{1}{2}\int d^{4}x\sqrt{-g}\mathcal{R}_{\mu_{1}\nu_{1}\lambda_{1}\sigma_{1}}\mathcal{D}_{\mu_{2}\nu_{2}\lambda_{2}\sigma_{2}}^{\mu_{1}\nu_{1}\lambda_{1}\sigma_{1}}\mathcal{R}^{\mu_{2}\nu_{2}\lambda_{2}\sigma_{2}},\label{eq:163}
\end{equation}
where the operator $\mathcal{D}$ is made in such a way to preserve the general covariance, so it must be a differential operator containing only
covariant derivatives and the metric tensor $g_{\mu\nu}.$ Note that it can happen
that a differential operator acts on the left of the Riemann
tensor as well, but one can always recast that into the above expression (\ref{eq:163}) integrating by part. \\
By considering all possible independent combinations of covariant derivatives
and metric tensors in the operator $\mathcal{D},$ one can write down the most general quadratic action $S_{q}$ (parity-invariant and torsion free) explicitly: 
\begin{equation}
\begin{array}{rl}
S_{q}= & \int d^{4}x\sqrt{-g}\left[\mathcal{R}F_{1}(\boxempty)\mathcal{R}+\mathcal{R}F_{2}(\boxempty)\nabla_{\mu}\nabla_{\nu}\mathcal{R}^{\mu\nu}\right.\\
\\
 & +\mathcal{R}_{\mu\nu}F_{3}(\boxempty)\mathcal{R}^{\mu\nu}+\mathcal{R}_{\mu}^{\nu}F_{4}(\boxempty)\nabla_{\nu}\nabla_{\lambda}\mathcal{R}^{\mu\lambda}\\
\\
 & +\mathcal{R}^{\lambda\sigma}F_{5}(\boxempty)\nabla_{\mu}\nabla_{\sigma}\nabla_{\nu}\nabla_{\lambda}\mathcal{R}^{\mu\nu}+\mathcal{R}F_{6}(\boxempty)\nabla_{\mu}\nabla_{\lambda}\nabla_{\nu}\nabla_{\sigma}\mathcal{R}^{\mu\nu\lambda\sigma}\\
\\
 & +\mathcal{R}_{\mu\lambda}F_{7}(\boxempty)\nabla_{\nu}\nabla_{\sigma}\mathcal{R}^{\mu\nu\lambda\sigma}+\mathcal{R}_{\lambda}^{\rho}F_{8}(\boxempty)\nabla_{\mu}\nabla_{\sigma}\nabla_{\nu}\nabla_{\rho}\mathcal{R}^{\mu\nu\lambda\sigma}\\
\\
 & +\mathcal{R}^{\mu_{1}\nu_{1}}F_{9}(\boxempty)\nabla_{\mu_{1}}\nabla_{\nu_{1}}\nabla_{\mu}\nabla_{\nu}\nabla_{\lambda}\nabla_{\sigma}\mathcal{R}^{\mu\nu\lambda\sigma}+\mathcal{R}_{\mu\nu\lambda\sigma}F_{10}(\boxempty)\mathcal{R}^{\mu\nu\lambda\sigma}\\
\\
 & +\mathcal{R}_{\,\mu\nu\lambda}^{\rho}F_{11}(\boxempty)\nabla_{\rho}\nabla_{\sigma}\mathcal{R}^{\mu\nu\lambda\sigma}+\mathcal{R}_{\mu\lambda_{1}\nu\sigma_{1}}F_{12}(\boxempty)\nabla^{\lambda_{1}}\nabla^{\sigma_{1}}\nabla_{\lambda}\nabla_{\sigma}\mathcal{R}^{\mu\lambda\nu\sigma}\\
\\
 & \left.+\mathcal{R}^{\mu_{1}\nu_{1}\rho_{1}\sigma_{1}}F_{14}(\boxempty)\nabla_{\rho1}\nabla_{\sigma_{1}}\nabla_{\nu_{1}}\nabla_{\mu_{1}}\nabla_{\mu}\nabla_{\nu}\nabla_{\lambda}\nabla_{\sigma}\mathcal{R}^{\mu\nu\lambda\sigma}\right]\\
\\
\equiv & \int d^{4}x\sqrt{-g}\mathcal{L}_{q}
\end{array}\label{eq:164}
\end{equation}
Note that the order of the covariant derivatives does not matter
because their commutator is proportional to another curvature leading
to $\mathcal{O}(\mathcal{R}^{3})$ modifications. The most general
quadratic action is captured by $14$ arbitrary functions, the $F_{i}(\boxempty)'$s,
that are functions of the D'Alambertian operator, $\boxempty=g^{\mu\nu}\nabla_{\mu}\nabla_{\nu}.$
For now we do not make other assumptions on the functions $F_{i}(\boxempty)'s,$
but in the next chapter their form will be fundamental for our results.
The Lagrangian introduced in the last line of (\ref{eq:164}) is the
quadratic curvature Lagrangian of the modified theory of gravity we
are considering.\\
\\
We notice that not all the $14$ terms are independent. Indeed by
using the antisymmetry properties of the Riemann tensor,
\[
\mathcal{R}_{(\mu\nu)\lambda\sigma}=\mathcal{R}_{\mu\nu(\lambda\sigma)}=0,
\]
and
\[
\nabla_{\alpha}\mathcal{R}_{\mu\nu\lambda\sigma}+\nabla_{\sigma}\mathcal{R}_{\mu\nu\alpha\lambda}+\nabla_{\lambda}\mathcal{R}_{\mu\nu\sigma\alpha}=0,
\]
the action (\ref{eq:164}) can be reduced to the following simpler
form: 
\begin{equation}
\begin{array}{rl}
S_{q}= & \int d^{4}x\sqrt{-g}\left[\mathcal{R}\mathcal{F}_{1}(\boxempty)\mathcal{R}+\mathcal{R}_{\mu\nu}\mathcal{F}_{2}(\boxempty)\mathcal{R}^{\mu\nu}\right.\\
\\
 & +\mathcal{R}_{\mu\nu\lambda\sigma}\mathcal{F}_{3}(\boxempty)\mathcal{R}^{\mu\nu\lambda\sigma}+\mathcal{R}\mathcal{F}_{4}(\boxempty)\nabla_{\mu}\nabla_{\lambda}\nabla_{\nu}\nabla_{\sigma}\mathcal{R}^{\mu\nu\lambda\sigma}\\
\\
 & +\mathcal{R}_{\mu}^{\,\nu_{1}\lambda_{1}\sigma_{1}}\mathcal{F}_{5}(\boxempty)\nabla_{\lambda_{1}}\nabla_{\sigma_{1}}\nabla_{\nu_{1}}\nabla_{\nu}\nabla_{\lambda}\nabla_{\sigma}\mathcal{R}^{\mu\nu\lambda\sigma}\\
\\
 & \left.+\mathcal{R}^{\mu_{1}\nu_{1}\rho_{1}\sigma_{1}}\mathcal{F}_{6}(\boxempty)\nabla_{\rho1}\nabla_{\sigma_{1}}\nabla_{\nu_{1}}\nabla_{\mu_{1}}\nabla_{\mu}\nabla_{\nu}\nabla_{\lambda}\nabla_{\sigma}\mathcal{R}^{\mu\nu\lambda\sigma}\right],
\end{array}\label{eq:165}
\end{equation}
where the functions $\mathcal{F}_{i}(\boxempty)'$s are new functions
depending on the $F_{i}(\boxempty)'$s. So we got rid of $8$ of the
$14$ terms already in a curved background. \\
Remember that for a complete and consistent description, we also need to include the contribution from the well known Einstein-Hilbert term. Thus, the full action we need to consider is
\begin{equation}
S=-\int d^{4}x\sqrt{-g}\mathcal{R}+S_{q}.\label{eq:166}
\end{equation}
$ $

\subsection{Linearized quadratic action}

We shall follow by following the same steps and prescriptions we have already seen for GR in the previous Chapter. Hence, our first task is to obtain the quadratic (in $h_{\mu\nu})$
free part of the above action (\ref{eq:166}). We can note that the
covariant derivatives must be taken on the Minkowski space-time, and we can
commute them freely as they are simple partial derivatives. Thus, the terms with $\mathcal{F}_{4}(\boxempty),\mathcal{F}_{5}(\boxempty)$
and $\mathcal{F}_{6}(\boxempty)$ in the action (\ref{eq:165}) do not
contribute in the limit of flat background because of the vanishing value of the symmetric-antisymmetric products between partial derivatives and Riemann tensor, whose index pairs are antisymmetric. For example, one has
\begin{equation}
\begin{array}{rl}
\mathcal{R}\mathcal{F}_{4}(\boxempty)\nabla_{\mu}\nabla_{\lambda}\nabla_{\nu}\nabla_{\sigma}\mathcal{R}^{\mu\nu\lambda\sigma}= & \mathcal{R}\mathcal{F}_{4}(\boxempty)\partial_{\mu}\partial_{\lambda}\partial_{\nu}\partial_{\sigma}\mathcal{R}^{\mu\nu\lambda\sigma}\\
\\
= & \mathcal{R}\mathcal{F}_{4}(\boxempty)\partial_{\mu}\partial_{\nu}\partial_{\lambda}\partial_{\sigma}\mathcal{R}^{\mu\nu\lambda\sigma}\\
\\
= & 0
\end{array}
\end{equation}
and one can do the same with $\mathcal{F}_{4}$ and $\mathcal{F}_{5}.$
At the end the linearized form of the action (\ref{eq:166}) reads
as
\begin{equation}
S=\int d^{4}x\left[-\mathcal{R}+\mathcal{R}\mathcal{F}_{1}(\boxempty)\mathcal{R}+\mathcal{R}_{\mu\nu}\mathcal{F}_{2}(\boxempty)\mathcal{R}^{\mu\nu}+\mathcal{R}_{\mu\nu\lambda\sigma}\mathcal{F}_{3}(\boxempty)\mathcal{R}^{\mu\nu\lambda\sigma}\right].\label{eq:167}
\end{equation}
We recall for convenience the linearized forms of the Riemann tensor,
the Ricci tensor and the scalar curvature, already introduced in the
previous chapter: 
\begin{equation}
\mathcal{R}_{\mu\nu\lambda\sigma}=\frac{1}{2}\left(\partial_{\nu}\partial_{\lambda}h_{\mu\sigma}+\mathcal{\partial_{\mu}\partial_{\sigma}}h_{\nu\lambda}-\mathcal{\partial_{\sigma}\partial_{\nu}}h_{\mu\lambda}-\partial_{\mu}\partial_{\lambda}h_{\nu\sigma}\right),\label{eq:168}
\end{equation}
\begin{equation}
\mathcal{R_{\mu\nu}}=\frac{1}{2}\left(\partial_{\rho}\partial_{\nu}h_{\mu}^{\rho}+\partial_{\rho}\partial_{\mu}h_{\nu}^{\rho}-\partial_{\mu}\partial_{\nu}h-\boxempty h_{\mu\nu}\right),\label{eq:169}
\end{equation}
\begin{equation}
\mathcal{R}=\partial_{\mu}\partial_{\nu}h^{\mu\nu}-\boxempty h.\label{eq:170}
\end{equation}
Substitute now the linearized curvatures (\ref{eq:168}),(\ref{eq:169})
and (\ref{eq:170}) in the action (\ref{eq:167}). Then, ignoring surface
terms one has:
\begin{equation}
\begin{array}{rl}
\mathcal{R}\mathcal{F}_{1}(\boxempty)\mathcal{R}= & \left(\partial_{\mu}\partial_{\nu}h^{\mu\nu}-\boxempty h\right)\mathcal{F}_{1}(\boxempty)\left(\partial_{\alpha}\partial_{\beta}h^{\alpha\beta}-\boxempty h\right)\\
\\
= & h\mathcal{F}_{1}(\boxempty)\boxempty^{2}h+h^{\mu\nu}\partial_{\nu}\partial_{\mu}\partial_{\alpha}\partial_{\beta}h^{\alpha\beta}\\
\\
 & -h\mathcal{F}_{1}(\boxempty)\boxempty\partial_{\mu}\partial_{\nu}h^{\mu\nu}-h^{\mu\nu}\mathcal{F}_{1}(\boxempty)\boxempty\partial_{\nu}\partial_{\mu}h.
\end{array}
\end{equation}
Since $\mathcal{F}_{i}(\boxempty)'$s are functions of covariant D'Alambertian
(on Minkowski space) we can always use the integration by parts and
so, ignoring surface terms again, we have:
\begin{equation}
\mathcal{R}\mathcal{F}_{1}(\boxempty)\mathcal{R}=\mathcal{F}_{1}(\boxempty)\left[h\boxempty^{2}h+h^{\mu\nu}\partial_{\nu}\partial_{\mu}\partial_{\alpha}\partial_{\beta}h^{\alpha\beta}-2h\boxempty\partial_{\mu}\partial_{\nu}h^{\mu\nu}\right].\label{eq:171}
\end{equation}
For the two other terms we have:

\begin{equation}
\begin{array}{rl}
\mathcal{R}_{\mu\nu}\mathcal{F}_{2}(\boxempty)\mathcal{R}^{\mu\nu}= & \frac{1}{2}\left(\partial_{\rho}\partial_{\nu}h_{\mu}^{\rho}+\partial_{\rho}\partial_{\mu}h_{\nu}^{\rho}-\partial_{\mu}\partial_{\nu}h-\boxempty h_{\mu\nu}\right)\mathcal{F}_{3}(\boxempty)\\
\\
 & \times\frac{1}{2}\left(\partial_{\alpha}\partial^{\nu}h^{\alpha\mu}+\partial_{\alpha}\partial^{\mu}h^{\alpha\nu}-\partial^{\mu}\partial^{\nu}h-\boxempty h^{\mu\nu}\right)\\
\\
= & \frac{1}{4}\mathcal{F}_{2}(\boxempty)\left[\left(\partial_{\mu}\partial_{\nu}h\partial^{\mu}\partial^{\nu}h\right)+\left(\boxempty h_{\mu\nu}\boxempty h^{\mu\nu}\right)\right.\\
\\
 & +\left(-\partial_{\rho}\partial_{\nu}h_{\mu}^{\rho}\partial^{\mu}\partial^{\nu}h-\partial_{\rho}\partial_{\mu}h_{\nu}^{\rho}\partial^{\mu}\partial^{\nu}h-\partial_{\mu}\partial_{\nu}h\partial_{\alpha}\partial^{\nu}h^{\alpha\mu}\right.\\
\\
 & \left.-\partial_{\mu}\partial_{\nu}h\partial_{\alpha}\partial^{\mu}h^{\alpha\nu}+\partial_{\mu}\partial_{\nu}h\boxempty h^{\mu\nu}+\boxempty h_{\mu\nu}\partial^{\mu}\partial^{\nu}h\right)\\
\\
 & +\left(\partial_{\rho}\partial_{\nu}h_{\mu}^{\rho}\partial_{\alpha}\partial^{\nu}h^{\alpha\mu}-\partial_{\rho}\partial_{\nu}h_{\mu}^{\rho}\boxempty h^{\mu\nu}+\partial_{\rho}\partial_{\mu}h_{\nu}^{\rho}\partial_{\alpha}\partial^{\mu}h^{\alpha\nu}\right.\\
\\
 & \left.-\partial_{\rho}\partial_{\mu}h_{\nu}^{\rho}\boxempty h^{\mu\nu}-\boxempty h_{\mu\nu}\partial_{\alpha}\partial^{\nu}h^{\alpha\mu}-\boxempty h_{\mu\nu}\partial_{\alpha}\partial^{\mu}h^{\alpha\nu}\right)\\
\\
 & \left.+\left(\partial_{\rho}\partial_{\nu}h_{\mu}^{\rho}\partial_{\alpha}\partial^{\mu}h^{\alpha\nu}+\partial_{\rho}\partial_{\mu}h_{\nu}^{\rho}\partial_{\alpha}\partial^{\nu}h^{\alpha\mu}\right)\right]\\
\\
= & \mathcal{F}_{2}(\boxempty)\left[\frac{1}{4}h\boxempty^{2}h+\frac{1}{4}h_{\mu\nu}\boxempty^{2}h^{\mu\nu}+\frac{1}{4}\left(-2h\boxempty\partial_{\mu}\partial_{\nu}h^{\mu\nu}\right)\right.\\
\\
 & \left.+\frac{1}{4}\left(-2h_{\mu}^{\rho}\boxempty\partial_{\rho}\partial_{\nu}h^{\mu\nu}\right)+\frac{1}{4}\left(2h^{\lambda\sigma}\partial_{\sigma}\partial_{\lambda}\partial_{\mu}\partial_{\nu}h^{\mu\nu}\right)\right]\\
\\
= & \mathcal{F}_{2}(\boxempty)\left[\frac{1}{4}h\boxempty^{2}h+\frac{1}{4}h_{\mu\nu}\boxempty^{2}h^{\mu\nu}-\frac{1}{2}h\boxempty\partial_{\mu}\partial_{\nu}h^{\mu\nu}\right.\\
\\
 & \left.-\frac{1}{2}h_{\mu}^{\rho}\boxempty\partial_{\rho}\partial_{\nu}h^{\mu\nu}+\frac{1}{2}h^{\lambda\sigma}\partial_{\sigma}\partial_{\lambda}\partial_{\mu}\partial_{\nu}h^{\mu\nu}\right].
\end{array}\label{eq:172}
\end{equation}
Finally, following the same steps the last term becomes:
\begin{equation}
\begin{array}{rl}
\mathcal{R}_{\mu\nu\lambda\sigma}\mathcal{F}_{3}(\boxempty)\mathcal{R}^{\mu\nu\lambda\sigma}= & \frac{1}{2}\left(\partial_{\nu}\partial_{\lambda}h_{\mu\sigma}+\mathcal{\partial_{\mu}\partial_{\sigma}}h_{\nu\lambda}-\mathcal{\partial_{\sigma}\partial_{\nu}}h_{\mu\lambda}-\partial_{\mu}\partial_{\lambda}h_{\nu\sigma}\right)\mathcal{F}_{3}(\boxempty)\\
\\
 & \times\frac{1}{2}\left(\partial^{\nu}\partial^{\lambda}h^{\mu\sigma}+\mathcal{\partial^{\mu}\partial^{\sigma}}h^{\nu\lambda}-\mathcal{\partial^{\sigma}\partial^{\nu}}h^{\mu\lambda}-\partial^{\mu}\partial^{\lambda}h^{\nu\sigma}\right)\\
\\
= & \mathcal{F}_{3}(\boxempty)\left[h_{\mu\nu}\boxempty^{2}h^{\mu\nu}+h^{\lambda\sigma}\partial_{\sigma}\partial_{\lambda}\partial_{\mu}\partial_{\nu}h^{\mu\nu}-2h_{\mu}^{\rho}\boxempty\partial_{\rho}\partial_{\nu}h^{\mu\nu}\right].
\end{array}\label{eq:173}
\end{equation}
Furthermore we have to consider the quadratic $h\textrm{-}$terms
due to the Hilbert-Einstein action that we introduced in $(2.2.1):$
\begin{equation}
S_{HE}=\int d^{4}x\left(h_{\sigma}^{\mu}\partial^{\sigma}\partial^{\nu}h_{\mu}^{\nu}-h\partial^{\mu}\partial^{\nu}h_{\mu\nu}-\frac{1}{2}h_{\mu\nu}\boxempty h^{\mu\nu}+\frac{1}{2}h\boxempty h\right).\label{eq:174}
\end{equation}
It remains to write the whole $h$-quadratic action with all the terms
due to the Hilbert-Einstein action and the quadratic terms in the
curvatures; we can quickly see that we have only $5$ different combination
terms:
\begin{equation}
\begin{array}{rl}
S_{q}= & \displaystyle{-\int d^{4}x}\left\{ \frac{1}{2}h_{\mu\nu}\boxempty\left[1-\frac{1}{2}\mathcal{F}_{2}(\boxempty)\boxempty-2\mathcal{F}_{3}(\boxempty)\boxempty\right]h^{\mu\nu}\right.\\
\\
 & +\displaystyle{h_{\mu}^{\sigma}\left[-1+\frac{1}{2}\mathcal{F}_{2}(\boxempty)\boxempty+2\mathcal{F}_{3}(\boxempty)\boxempty\right]\partial_{\sigma}\partial_{\nu}h^{\mu\nu}}\\
\\
 & +\displaystyle{\frac{1}{2}h\left[1+2\mathcal{F}_{1}(\boxempty)\boxempty+\frac{1}{2}\mathcal{F}_{2}(\boxempty)\boxempty\right]\partial_{\mu}\partial_{\nu}h^{\mu\nu}}\\
\\
 & +\displaystyle{\frac{1}{2}h\boxempty\left[-2\mathcal{F}_{1}(\boxempty)\boxempty-\frac{1}{2}\mathcal{F}_{2}(\boxempty)\boxempty-1\right]h}\\
\\
 & \left.+\displaystyle{\frac{1}{2}h^{\lambda\sigma}\frac{1}{\boxempty}}\left[-2\mathcal{F}_{1}(\boxempty)\boxempty-\mathcal{F}_{2}(\boxempty)\boxempty-2\mathcal{F}_{3}(\boxempty)\boxempty\right]\partial_{\sigma}\partial_{\lambda}\partial_{\mu}\partial_{\nu}h^{\mu\nu}\right\} .
\end{array}\label{eq:175}
\end{equation}

\begin{rem*}
It is important to observe that this last action $S_{q}$ is not the
starting action (\ref{eq:163}). Indeed (\ref{eq:163}) the letter
$"q"$ means that the action is quadratic in the curvatures, instead
now it means that the action is quadratic in $h.$ In the last action
we have written there is also the Hilbert-Einstein contribution, linear
in the curvature.
\end{rem*}
$ $\\
We can write the action in a more compact form defining the following
coefficients:
\begin{equation}
\begin{array}{l}
a(\boxempty)\coloneqq1-\frac{1}{2}\mathcal{F}_{2}(\boxempty)\boxempty-2\mathcal{F}_{3}(\boxempty)\boxempty,\\
\\
b(\boxempty)\coloneqq-1+\frac{1}{2}\mathcal{F}_{2}(\boxempty)\boxempty+2\mathcal{F}_{3}(\boxempty)\boxempty,\\
\\
c(\boxempty)\coloneqq1+2\mathcal{F}_{1}(\boxempty)\boxempty+\frac{1}{2}\mathcal{F}_{2}(\boxempty)\boxempty,\\
\\
d(\boxempty)\coloneqq-1-2\mathcal{F}_{1}(\boxempty)\boxempty-\frac{1}{2}\mathcal{F}_{2}(\boxempty)\boxempty,\\
\\
f(\boxempty)\coloneqq-2\mathcal{F}_{1}(\boxempty)\boxempty-\mathcal{F}_{2}(\boxempty)\boxempty-2\mathcal{F}_{3}(\boxempty)\boxempty,
\end{array}\label{eq:176}
\end{equation}
so we obtain
\begin{equation}
\begin{array}{rl}
S_{q}= & -{\displaystyle \int d^{4}x\left[\frac{1}{2}h_{\mu\nu}\boxempty a(\boxempty)h^{\mu\nu}+h_{\mu}^{\sigma}b(\boxempty)\partial_{\sigma}\partial_{\nu}h^{\mu\nu}\right.}\\
\\
 & \left.{\displaystyle +hc(\boxempty)\partial_{\mu}\partial_{\nu}h^{\mu\nu}+\frac{1}{2}h\boxempty d(\boxempty)h+\frac{1}{2}h^{\lambda\sigma}\frac{f(\boxempty)}{\boxempty}\partial_{\sigma}\partial_{\lambda}\partial_{\mu}\partial_{\nu}h^{\mu\nu}}\right],
\end{array}\label{eq:177}
\end{equation}
or, equivalently the Linearized quadratic Lagrangian is
\begin{equation}
\begin{array}{rl}
\mathcal{L}_{q}= & {\displaystyle -\frac{1}{2}h_{\mu\nu}\boxempty a(\boxempty)h^{\mu\nu}-h_{\mu}^{\sigma}b(\boxempty)\partial_{\sigma}\partial_{\nu}h^{\mu\nu}-hc(\boxempty)\partial_{\mu}\partial_{\nu}h^{\mu\nu}}\\
\\
 & {\displaystyle -\frac{1}{2}h\boxempty d(\boxempty)h-\frac{1}{2}h^{\lambda\sigma}\frac{f(\boxempty)}{\boxempty}\partial_{\sigma}\partial_{\lambda}\partial_{\mu}\partial_{\nu}h^{\mu\nu}}
\end{array}\label{eq:178}
\end{equation}
The function $f(\boxempty)$ appears only in higher order theories.\\
From the above expressions (\ref{eq:176}) of the coefficients we
easily deduce the following interesting relations
\begin{equation}
\begin{cases}
a(\boxempty)+b(\boxempty)=0\\
c(\boxempty)+d(\boxempty)=0\\
b(\boxempty)+c(\boxempty)+f(\boxempty)=0
\end{cases}\label{eq:179}
\end{equation}
so that we are really left with only two independent arbitrary functions:
a number much smaller than the beginning one, $14.$ Below we shall see that this can be better understood as a consequence of the Bianchi identities.\\
As we have already done with ED and GR cases, by raising and lowering
the space-time indices with the metric tensor $\eta_{\mu\nu},$ we can rewrite the Lagrangian (\ref{eq:178}) in the following form
\begin{equation}
\mathcal{L}_{q}=\frac{1}{2}h_{\mu\nu}\mathcal{O}_{q}^{\mu\nu\rho\sigma}h_{\rho\sigma},\label{eq:180}
\end{equation}
 where the operator $\mathcal{O}_{q}^{\mu\nu\rho\sigma}$ is defined
as
\[
\mathcal{O}_{q}^{\mu\nu\rho\sigma}\coloneqq-\left(\frac{a(\boxempty)}{2}\eta^{\mu\rho}\eta^{\nu\sigma}+\frac{a(\boxempty)}{2}\eta^{\mu\sigma}\eta^{\nu\rho}+d(\boxempty)\eta^{\mu\nu}\eta^{\rho\sigma}\right)-b(\boxempty)\left(\eta^{\mu\nu}\partial^{\rho}\partial^{\sigma}+\eta^{\rho\sigma}\partial^{\mu}\partial^{\nu}\right)
\]
\begin{equation}
-\frac{c(\boxempty)}{2}\left(\eta^{\mu\rho}\partial^{\nu}\partial^{\sigma}+\eta^{\mu\sigma}\partial^{\nu}\partial^{\sigma}+\eta^{\nu\rho}\partial^{\mu}\partial^{\sigma}+\eta^{\nu\sigma}\partial^{\mu}\partial^{\rho}\right)-\frac{f(\boxempty)}{\boxempty}\partial^{\mu}\partial^{\nu}\partial^{\rho}\partial^{\sigma}.\label{eq:181}
\end{equation}

\subsection{Field equations}

We want to derive the field equations associated to the action (\ref{eq:177}).
In this case, since the Lagrangian contains higher orders, we
also need to consider the functional derivatives with respect to the second
derivatives of the field $h_{\mu\nu}$ in the Euler-Lagrange equations.
Let us determine the Euler-Lagrange equations for the linearized Lagrangian
$\mathcal{L}_{q},$
\begin{equation}
\partial_{\alpha}\frac{\partial\mathcal{L}_{q}}{\partial(\partial_{\alpha}h^{\mu\nu})}-\partial_{\alpha}\partial_{\beta}\frac{\partial\mathcal{L}_{q}}{\partial(\partial_{\alpha}\partial_{\beta}h^{\mu\nu})}=\frac{\partial\mathcal{L}_{q}}{\partial h^{\mu\nu}}.\label{eq:182}
\end{equation}
To apply (\ref{eq:182}) we have to rewrite the term with $f(\boxempty)$
in (\ref{eq:178}) in the following more convenient form:
\begin{equation}
\begin{array}{rl}
\mathcal{L}_{q}= & {\displaystyle -\frac{1}{2}h_{\mu\nu}\boxempty a(\boxempty)h^{\mu\nu}-h_{\mu}^{\sigma}b(\boxempty)\partial_{\sigma}\partial_{\nu}h^{\mu\nu}-hc(\boxempty)\partial_{\mu}\partial_{\nu}h^{\mu\nu}}\\
\\
 & {\displaystyle -\frac{1}{2}h\boxempty d(\boxempty)h-\frac{1}{2}\partial_{\sigma}\partial_{\lambda}h^{\lambda\sigma}\frac{f(\boxempty)}{\boxempty}\partial_{\mu}\partial_{\nu}h^{\mu\nu}.}
\end{array}\label{eq:183}
\end{equation}
Now we can proceed with the computation of the derivatives. First
of all we notice that in (\ref{eq:183}) there are not terms containing
first derivatives of the field $h_{\mu\nu}:$
\begin{equation}
\partial_{\sigma}\frac{\partial\mathcal{L}_{q}}{\partial(\partial_{\sigma}h^{\mu\nu})}=0.\label{eq:184}
\end{equation}
As for the the second term on the left side of (\ref{eq:182}) we
have: 
\[
\begin{array}{rl}
{\displaystyle \frac{\partial\mathcal{L}_{q}}{\partial(\partial_{\alpha}\partial_{\beta}h^{\mu\nu})}}= & {\displaystyle -\frac{1}{2}a(\boxempty)\eta^{\alpha\beta}h_{\mu\nu}-b(\boxempty)h_{\mu}^{\alpha}\delta_{\nu}^{\beta}-c(\boxempty)h_{\rho\sigma}\eta^{\rho\sigma}\delta_{\mu}^{\alpha}\delta_{\nu}^{\beta}}\\
\\
 & {\displaystyle -\frac{1}{2}d(\boxempty)h_{\rho\sigma}\eta^{\rho\sigma}\eta_{\mu\nu}\eta^{\alpha\beta}-f(\boxempty)\boxempty^{-1}\partial_{\mu}\partial_{\nu}h^{\alpha\beta}}
\end{array}
\]
\begin{equation}
\begin{array}{rl}
\Rightarrow\partial_{\alpha}{\displaystyle \partial_{\beta}\frac{\partial\mathcal{L}_{q}}{\partial(\partial_{\alpha}\partial_{\beta}h^{\mu\nu})}}= & {\displaystyle -\frac{1}{2}a(\boxempty)\boxempty h_{\mu\nu}-b(\boxempty)\partial_{\nu}\partial_{\alpha}h_{\mu}^{\alpha}-c(\boxempty)\partial_{\mu}\partial_{\nu}h_{\rho\sigma}\eta^{\rho\sigma}}\\
\\
 & {\displaystyle -\frac{1}{2}d(\boxempty)\eta_{\mu\nu}\eta^{\rho\sigma}\boxempty h_{\rho\sigma}-f(\boxempty)\boxempty^{-1}\partial_{\mu}\partial_{\nu}\partial_{\alpha}\partial_{\beta}h^{\alpha\beta}.}
\end{array}\label{eq:185}
\end{equation}
Instead the derivative with respect to the field $h^{\mu\nu}$ gives
us
\begin{equation}
\begin{array}{rl}
{\displaystyle \frac{\partial\mathcal{L}_{q}}{\partial h^{\mu\nu}}}= & {\displaystyle -\frac{1}{2}a(\boxempty)\boxempty h_{\mu\nu}-b(\boxempty)\partial_{\mu}\partial_{\alpha}h_{\nu}^{\alpha}}\\
\\
 & {\displaystyle -c(\boxempty)\eta_{\mu\nu}\partial_{\alpha}\partial_{\beta}h^{\alpha\beta}-\frac{1}{2}d(\boxempty)\eta_{\mu\nu}\eta^{\rho\sigma}\boxempty h_{\rho\sigma}.}
\end{array}\label{eq:186}
\end{equation}
Hence from (\ref{eq:184}), (\ref{eq:185}) and (\ref{eq:186}) we
deduce that the field equations for the quadratic Lagrangian (\ref{eq:183})
are
\begin{equation}
\begin{array}{rl}
a(\boxempty)\boxempty h_{\mu\nu}+b(\boxempty)\left(\partial_{\mu}\partial_{\alpha}h_{\nu}^{\alpha}+\partial_{\alpha}\partial_{\nu}h_{\mu}^{\alpha}\right)+c(\boxempty)\left(\eta_{\mu\nu}\partial_{\alpha}\partial_{\beta}h^{\alpha\beta}+\partial_{\mu}\partial_{\nu}h\right)\\
\\
+\eta_{\mu\nu}d(\boxempty)\boxempty h+f(\boxempty)\boxempty^{-1}\partial_{\beta}\partial_{\alpha}\partial_{\mu}\partial_{\nu}h^{\alpha\beta} & =0
\end{array}\label{eq:187}
\end{equation}
If there is also the matter contribute in the action, described by
energy-momentum tensor of matter $\tau_{\mu\nu},$ we have to add
the term $-\kappa\tau_{\rho\sigma}h^{\rho\sigma}$ to the Lagrangian
(\ref{eq:183}) and when we compute the derivative $\frac{\partial\mathcal{L}_{q}}{\partial h^{\mu\nu}}$
we have also to consider the contribution $-\kappa\tau_{\mu\nu}.$
Thus the field equations in presence of matter will read as
\begin{equation}
\begin{array}{rl}
a(\boxempty)\boxempty h_{\mu\nu}+b(\boxempty)\left(\partial_{\mu}\partial_{\alpha}h_{\nu}^{\alpha}+\partial_{\alpha}\partial_{\nu}h_{\mu}^{\alpha}\right)+c(\boxempty)\left(\eta_{\mu\nu}\partial_{\alpha}\partial_{\beta}h^{\alpha\beta}+\partial_{\mu}\partial_{\nu}h\right)\\
\\
+\eta_{\mu\nu}d(\boxempty)\boxempty h+f(\boxempty)\boxempty^{-1}\partial_{\beta}\partial_{\alpha}\partial_{\mu}\partial_{\nu}h^{\alpha\beta} & =-\kappa\tau_{\mu\nu}.
\end{array}\label{eq:188}
\end{equation}
One can demonstrate that the energy-momentum tensor is conserved because of the generalized Bianchi identity {[}\ref{-T.-Koivisto,}{]} due to diffeomorphism invariance: $\nabla_{\mu}\tau_{\nu}^{\mu}=0.$
Thus, by acting with the covariant derivative on the field equation 
\[
\begin{array}{rl}
a(\boxempty)\boxempty\partial_{\mu}h_{\nu}^{\mu}+b(\boxempty)\left(\boxempty\partial_{\alpha}h_{\nu}^{\alpha}+\partial_{\alpha}\partial_{\nu}\partial_{\mu}h^{\alpha\mu}\right)+c(\boxempty)\left(\partial_{\nu}\partial_{\alpha}\partial_{\beta}h^{\alpha\beta}+\boxempty\partial_{\nu}h\right)\\
\\
+d(\boxempty)\boxempty\partial_{\nu}h+f(\boxempty)\boxempty^{-1}\partial_{\beta}\partial_{\alpha}\boxempty\partial_{\nu}h^{\alpha\beta} & =0
\end{array}
\]
\begin{equation}
\Leftrightarrow\left(c+d\right)\boxempty\partial_{\nu}h+\left(a+b\right)\boxempty\partial_{\mu}h_{\nu}^{\mu}+\left(b+c+f\right)\partial_{\nu}\partial_{\alpha}\partial_{\mu}h^{\alpha\mu}=0\label{eq:189}
\end{equation}
Now it is clearer why the relations (\ref{eq:179}) must be valid. In fact the
last equation (\ref{eq:189}) holds if, and only if, the quantities
in brackets are zero, i.e. if the relations (\ref{eq:179}) are satisfied.

\section{Propagator for quadratic Lagrangian}

Now we want to derive the form of the physical (gauge independent) part of the propagator for the quadratic Lagrangian (\ref{eq:178}). Remember that we have introduced two different
methods to calculate the propagator in the previous Chapter. In this
section we shall refer to the second one.\\
\\
The first step is to rewrite the field equations (\ref{eq:188}) in
terms of the spin projector operators (\ref{eq:52}). The field equations
(\ref{eq:188}) in  momentum space read as
\begin{equation}
\begin{array}{rl}
{\displaystyle a(-k^{2})h_{\mu\nu}+b(-k^{2})\left(k_{\mu}k_{\alpha}h_{\nu}^{\alpha}+k_{\alpha}k_{\nu}h_{\mu}^{\alpha}\right)+\frac{c(-k^{2})}{k^{2}}}\left(\eta_{\mu\nu}k_{\alpha}k_{\beta}h^{\alpha\beta}+k_{\mu}k_{\nu}h\right)\\
\\
+{\displaystyle \eta_{\mu\nu}\frac{d(-k^{2})}{k^{2}}h+\frac{f(-k^{2})}{k^{4}}k_{\beta}k_{\alpha}k_{\mu}k_{\nu}h^{\alpha\beta}} & ={\displaystyle \kappa\frac{\tau_{\mu\nu}}{k^{2}}.}
\end{array}\label{eq:190}
\end{equation}
Now we can write every term of the field equations (\ref{eq:190})
in the following way (use the relations (\ref{eq:b384rel})):

\begin{equation}
a(-k^{2})h_{\mu\nu}=a(-k^{2})\left(\mathcal{P}^{2}+\mathcal{P}^{1}+\mathcal{P}_{s}^{0}+\mathcal{P}_{w}^{0}\right)h;\label{eq:191}
\end{equation}
$ $
\begin{equation}
\begin{array}{rl}
b(-k^{2})\left(k_{\nu}k_{\alpha}h_{\mu}^{\alpha}+k_{\alpha}k_{\mu}h_{\nu}^{\alpha}\right)= & b(-k^{2})k^{2}\left(\omega_{\sigma\nu}h_{\mu}^{\sigma}+\omega_{\sigma\mu}h_{\nu}^{\sigma}\right)\\
\\
= & b(-k^{2})k^{2}\left(\eta_{\mu\rho}\omega_{\nu\sigma}+\eta_{\rho\nu}\omega_{\mu\sigma}\right)h^{\rho\sigma}\\
\\
= & b(-k^{2}){\displaystyle k^{2}\frac{1}{2}}\left(\eta_{\mu\rho}\omega_{\nu\sigma}+\eta_{\mu\sigma}\omega_{\nu\rho}+\eta_{\nu\rho}\omega_{\mu\sigma}+\eta_{\nu\sigma}\omega_{\mu\rho}\right)h^{\rho\sigma}\\
\\
= & b(-k^{2})k^{2}\left(\mathcal{P}^{1}+2\mathcal{P}_{w}^{0}\right)h;
\end{array}\label{eq:192}
\end{equation}
$ $
\begin{equation}
\begin{array}{rl}
c(-k^{2})\left(\eta_{\mu\nu}k_{\rho}k_{\sigma}h^{\rho\sigma}+k_{\mu}k_{\nu}h\right)= & c(-k^{2})k^{2}\left(\eta_{\mu\nu}\omega_{\rho\sigma}h^{\rho\sigma}+\omega_{\mu\nu}\eta_{\rho\sigma}h^{\rho\sigma}\right)\\
\\
= & c(-k^{2})k^{2}\left(\theta_{\mu\nu}\omega_{\rho\sigma}+\omega_{\mu\nu}\omega_{\rho\sigma}+\omega_{\mu\nu}\omega_{\rho\sigma}+\omega_{\mu\nu}\theta_{\rho\sigma}\right)h^{\rho\sigma}\\
\\
= & c(-k^{2})k^{2}\left(2\mathcal{P}_{w}^{0}+\sqrt{3}\left(\mathcal{P}_{sw}^{0}+\mathcal{P}_{ws}^{0}\right)\right)h;
\end{array}\label{eq:193}
\end{equation}

$ $
\begin{equation}
\begin{array}{rl}
d(-k^{2})\eta_{\mu\nu}\eta^{\rho\sigma}h_{\rho\sigma}= & d(-k^{2})(\theta_{\mu\nu}+\omega_{\mu\nu})\left(\theta_{\rho\sigma}+\omega_{\rho\sigma}\right)h^{\rho\sigma}\\
\\
= & d(-k^{2})\left(\theta_{\mu\nu}\theta_{\rho\sigma}+\theta_{\mu\nu}\omega_{\rho\sigma}+\omega_{\mu\nu}\theta_{\rho\sigma}+\omega_{\mu\nu}\omega_{\rho\sigma}\right)h^{\rho\sigma}\\
\\
= & d(-k^{2})\left(3\mathcal{P}_{s}^{0}+\mathcal{P}_{w}^{0}+\sqrt{3}\left(\mathcal{P}_{sw}^{0}+\mathcal{P}_{ws}^{0}\right)\right)h;
\end{array}\label{eq:194}
\end{equation}
$ $

\begin{equation}
f(-k^{2})k^{\rho}k^{\sigma}k_{\mu}k_{\nu}h_{\rho\sigma}=f(-k^{2})k^{4}\omega_{\mu\nu}\omega^{\rho\sigma}h_{\rho\sigma}=f(-k^{2})\mathcal{P}_{w}^{0}h.\label{eq:195}
\end{equation}
For the above calculations we have used a lot the relations (\ref{eq:59})
(see in Appendix $B.2)$ already used for GR. Furthermore, we are
still suppressing the space-time indices for convenience. \\
Hence, from the relations (\ref{eq:191})-(\ref{eq:195}) the field
equations in terms of the spin projector operators read as:
\[
\left[a(-k^{2})\left(\mathcal{P}^{2}+\mathcal{P}^{1}+\mathcal{P}_{s}^{0}+\mathcal{P}_{w}^{0}\right)+b(-k^{2})\left(\mathcal{P}^{1}+2\mathcal{P}_{w}^{0}\right)+c(-k^{2})\left(2\mathcal{P}_{w}^{0}+\sqrt{3}\left(\mathcal{P}_{sw}^{0}+\mathcal{P}_{ws}^{0}\right)\right)\right.
\]
\begin{equation}
\left.+d(-k^{2})\left(3\mathcal{P}_{s}^{0}+\mathcal{P}_{w}^{0}+\sqrt{3}\left(\mathcal{P}_{sw}^{0}+\mathcal{P}_{ws}^{0}\right)\right)+f(-k^{2})\mathcal{P}_{w}^{0}\right]h=\kappa\frac{\left(\mathcal{P}^{2}+\mathcal{P}^{1}+\mathcal{P}_{s}^{0}+\mathcal{P}_{w}^{0}\right)}{k^{2}}\tau.\label{eq:196}
\end{equation}
Now we are ready to invert the field equations and then obtain the
corresponding propagator: we shall proceed following the prescription
introduced in the section $2.3-$Method $2.$ \\
By acting with the projector $\mathcal{P}^{2}$ on (\ref{eq:196})
and using the orthogonality relations (\ref{eq:53}) we find
\begin{equation}
\mathcal{P}^{2}h=\kappa\left(\frac{\mathcal{P}^{2}}{a(-k^{2})k^{2}}\right)\tau;\label{eq:197}
\end{equation}
by acting with $\mathcal{P}^{1},$ one finds
\begin{equation}
\left(a(-k^{2})+b(-k^{2})\right)\mathcal{P}^{1}h=\kappa\frac{\mathcal{P}^{1}}{k^{2}}\tau,\label{eq:198}
\end{equation}
but since $a+b=0,$ then there are no vector degrees of freedom, and
accordingly the stress-energy tensor must have no vector part: 
\begin{equation}
0\mathcal{P}^{1}h=\kappa\frac{\mathcal{P}^{1}}{k^{2}}\tau\Rightarrow\mathcal{P}^{1}\tau=0.\label{eq:199}
\end{equation}
\\
Next let us look at the scalar multiplets. By acting with $\mathcal{P}_{s}^{0}$
we find
\[
\left((a+3d)\mathcal{P}_{s}^{0}+\sqrt{3}(c+d)\mathcal{P}_{sw}^{0}\right)h=\kappa\frac{\mathcal{P}_{s}^{0}}{k^{2}}\tau,
\]
but since $c+d=0,$ then we obtain
\begin{equation}
\mathcal{P}_{s}^{0}h=\kappa\left(\frac{\mathcal{P}_{s}^{0}}{(a-3c)k^{2}}\right)\tau;\label{eq:200}
\end{equation}
instead by acting with $\mathcal{P}_{w}^{0}$ 
\begin{equation}
(a+2b+2c+d+f)\mathcal{P}_{w}^{0}h=\kappa\frac{\mathcal{P}_{w}^{0}}{k^{2}}\tau,\label{eq:201}
\end{equation}
but
\[
\begin{array}{rl}
a+2b+2c+d+f= & (a+b)+b+c+(c+d)+f\\
\\
= & b+c+f=0
\end{array}
\]
\begin{equation}
\Rightarrow0\mathcal{P}_{w}^{0}h=\kappa\frac{\mathcal{P}_{w}^{0}}{k^{2}}\tau\Rightarrow\mathcal{P}_{w}^{0}\tau=0,\label{eq:202}
\end{equation}
so there is no $w-$multiplet that contributes to the degrees of freedom
in the propagator. We can also note that, in principle, the scalar
multiplets are coupled, but by acting with the spin projector operators
the scalars decouple.\\
As it happens in GR the spin components $\mathcal{P}^{1}h$ and $\mathcal{P}_{w}^{0}h$
of the tensor field $h_{\mu\nu}$ are undetermined and two restrictions
of the source $\tau_{\mu\nu}$ hold, i.e. (\ref{eq:199}) and (\ref{eq:202}).
Remember that these two source constraints are associated to some
gauge freedom which in turn is associated to an invariance of the
Lagrangian. Indeed also the quadratic Lagrangian (\ref{eq:183})  is
invariant under spin-$1$ transformation: $\delta h\sim\mathcal{P}^{1}h.$
\\
Because of (\ref{eq:197}) and (\ref{eq:200}) we expect that the
physical part of the propagator contains only the spin-$2$ and spin-$0$
components, $\mathcal{P}^{2}$ and $\mathcal{P}_{s}^{0}.$ We can
easily show this last statement by proceeding as we did for the saturated
GR propagator following the steps from (\ref{eq:112}) to (\ref{eq:115}).
Doing this we obtain the saturated propagator for the quadratic Lagrangian
(\ref{eq:178}): 
\begin{equation}
\tau(-k)\Pi(k)\tau(k)\equiv\tau(-k)\left(\frac{\mathcal{P}^{2}}{ak^{2}}+\frac{\mathcal{P}_{s}^{0}}{(a-3c)k^{2}}\right)\tau(k),\label{eq:203}
\end{equation}
and the physical (gauge independent) part of the propagator reads
as
\begin{equation}
\Pi(k)=\frac{\mathcal{P}^{2}}{ak^{2}}+\frac{\mathcal{P}_{s}^{0}}{(a-3c)k^{2}}.\label{eq:204}
\end{equation}
$ $

\section{Ghosts and unitarity analysis in quadratic gravity}

Once we have determined the propagator associated to the Lagrangian
of the theory we can study whether ghosts and tachyons are absent
and whether the unitarity condition is preserved. Hence we have to
study the positivity of the imaginary part of the current-current
amplitude residue. The amplitude is given by
\begin{equation}
\mathcal{A}=i\tau^{*\mu\nu}(k)\frac{1}{k^{2}}\left(\frac{\mathcal{P}^{2}}{ak^{2}}+\frac{\mathcal{P}_{s}^{0}}{(a-3c)k^{2}}\right)_{\mu\nu\rho\sigma}\tau^{\rho\sigma}(k).
\end{equation}
We know that the coefficients $a\equiv a(-k^{2})$ and $c\equiv c(-k^{2})$
depend on the square momentum $k^{2},$ thus their special form could
bring other poles in the propagator and so new states in addition
to the graviton one. For example, if the coefficients $a$ and $b$
are polynomial in $k^{2}$ we will have new poles for sure. Hence
without specifying the particular dependence of the coefficients on
$k^{2}$ we are not able to state anything neither about the presence
of ghosts and tachyon, nor about unitarity condition. \\
In the next section and in the next Chapter we shall make special
choices for the coefficients $a$ and $c.$

\section{Applications to special cases}

\subsection{General Relativity}

In this chapter we have been studying a case of modified, or extended,
theories of gravity because we want to try to solve problems that
GR is not able to explain. A good new theory must give us GR when
we consider some limits. In fact, since we want to recover the right
infrared behavior of GR, we require from any viable theory that for
$k^{2}\rightarrow0$
\begin{equation}
a(0)=c(0)=-b(0)=-d(0)=1,\label{eq:205}
\end{equation}
corresponding to the GR values (in GR these functions are the same
constants for any Fourier mode). The condition (\ref{eq:205}) ensures
that as $k^{2}\rightarrow0,$ we have only the physical graviton propagator
\begin{equation}
\lim_{k^{2}\rightarrow0}\Pi=\frac{\mathcal{P}^{2}}{k^{2}}-\frac{\mathcal{P}_{s}^{0}}{2k^{2}}\equiv\Pi_{GR}.\label{eq:206}
\end{equation}
We have already seen that the GR is a free-ghost theory, and so the
choice (\ref{eq:205}) gives us a theory that preserves the unitarity,
without ghosts and tachyons.\\
Thus, we conclude that $k^{2}=0$ pole just describes the physical
graviton state, i.e. there are no new states, but just the state predicted
by GR. Secondly, we can also note that we are left with only a single
arbitrary function, $a=1.$

\subsection{$f(\mathcal{R})$ theory}

The $f(\mathcal{R})$ gravity is actually a family of theories, each
one defined by a different function of the Ricci scalar. The general
action of the this theory is
\begin{equation}
S_{f}=\int d^{4}x\sqrt{-g}f(\mathcal{R}).
\end{equation}
For our aims here, one can just consider the expansion of the
Lagrangian around flat space ($\mathcal{R}=0):$
\begin{equation}
f(\mathcal{R})=f(0)+f'(0)\mathcal{R}+\frac{1}{2}f''(0)\mathcal{R}^{2}+\cdots.
\end{equation}
The zero order term identifies with the cosmological constant, $f(0)=-\kappa\Lambda,$
and the first order term should reduce to the Einstein-Hilbert term in
a healthy theory, $f'(0)=-1.$ The relevant modification of the theory
is contained in the quadratic part, i.e. the second order term in the expansion. Since now only the function $F_{1}(\boxempty)$
is nonzero in (\ref{eq:164}), the relations (\ref{eq:176}) become
\footnote{Here we must be careful because we are using the letter $f$ both
for the coefficient and for the Lagrangian function term in $f(\mathcal{R})$theory.
So don't get confused!%
}
\begin{equation}
\begin{array}{l}
a(\boxempty)=1,\\
\\
b(\boxempty)=-1,\\
\\
c(\boxempty)=1+2\mathcal{F}_{1}(\boxempty)\boxempty=1+f''(0)\boxempty,\\
\\
d(\boxempty)=-1-2\mathcal{F}_{1}(\boxempty)\boxempty=-1-f''(0)\boxempty,\\
\\
f(\boxempty)=-2\mathcal{F}_{1}(\boxempty)\boxempty=+f''(0)\boxempty.
\end{array}
\end{equation}
The physical part of the propagator (\ref{eq:204}) specialized to
$f(\mathcal{R})$ is given by
\begin{equation}
\begin{array}{rl}
\Pi_{f}(k)= & {\displaystyle \frac{\mathcal{P}^{2}}{k^{2}}}+{\displaystyle \frac{\mathcal{P}_{s}^{0}}{k^{2}(1-3+3f''(0)k^{2})}}\\
\\
= & \Pi_{GR}+{\displaystyle \frac{1}{2}\frac{\mathcal{P}_{s}^{0}}{k^{2}-m^{2}},}
\end{array}
\end{equation}
where $m^{2}\coloneqq\frac{2}{3}\frac{1}{f''(0)}.$ The scalar part
of the propagator is modified as we expected. In fact, since these theories are a
particular class of scalar-tensor theories, an extra scalar
degree of freedom must be taken into account. Hence, the $f(\mathcal{R})$ modification of GR introduces an additional spin-$0$ particle which is not a ghost and moreover is non-tachyonic as long as $f''(0)>0.$

\subsection{Conformally invariant gravity}

We are now going to consider the Weyl squared gravity as an example of theory with presence of ghost. The Weyl tensor is defined as
\[
C_{\mu\nu\rho\sigma}\coloneqq\mathcal{R}_{\mu\nu\rho\sigma}+\frac{\mathcal{R}}{6}\left(g_{\mu\rho}g_{\nu\sigma}-g_{\mu\sigma}g_{\nu\rho}\right)-\frac{1}{2}\left(g_{\mu\rho}\mathcal{R}_{\nu\sigma}-g_{\mu\sigma}\mathcal{R}_{\nu\rho}-g_{\nu\rho}\mathcal{R}_{\mu\sigma}+g_{\nu\sigma}\mathcal{R}_{\mu\rho}\right).
\]
The theory is then specified by the conformally invariant Weyl-squared
term, 
\begin{equation}
\mathcal{L}_{C}=-\left(\mathcal{R}+\frac{1}{m^{2}}C^{2}\right),
\end{equation}
where
\begin{equation}
C^{2}\equiv C_{\mu\nu\rho\sigma}C^{\mu\nu\rho\sigma}=\frac{1}{3}\mathcal{R}^{2}-2\mathcal{R}_{\mu\nu}\mathcal{R}^{\mu\nu}+\mathcal{R}_{\mu\nu\rho\sigma}\mathcal{R}^{\mu\nu\rho\sigma};
\end{equation}
hence the action is 
\begin{equation}
S_{C}=-\int d^{4}x\sqrt{-g}\left[\mathcal{R}+\frac{1}{m^{2}}\left(\frac{1}{3}\mathcal{R}^{2}-2\mathcal{R}_{\mu\nu}\mathcal{R}^{\mu\nu}+\mathcal{R}_{\mu\nu\rho\sigma}\mathcal{R}^{\mu\nu\rho\sigma}\right)\right].\label{eq:207}
\end{equation}
From (\ref{eq:207}) we can easily see that the relations (\ref{eq:176})
become
\begin{equation}
\begin{array}{l}
a(\boxempty)=1+{\displaystyle \frac{1}{m^{2}}\boxempty},\\
\\
b(\boxempty)=-1-{\displaystyle \frac{1}{m^{2}}}\boxempty,\\
\\
c(\boxempty)=1+{\displaystyle \frac{1}{3m^{2}}}\boxempty,\\
\\
d(\boxempty)=-1-{\displaystyle \frac{1}{3m^{2}}}\boxempty,\\
\\
f(\boxempty)=+{\displaystyle \frac{2}{3m^{2}}}\boxempty.
\end{array}\label{eq:208}
\end{equation}
Going into the momentum space and using the relations (\ref{eq:208}),
from (\ref{eq:204}) we obtain a propagator with a double pole in
the spin-$2$ component:
\begin{equation}
\begin{array}{rl}
\Pi_{C}= & {\displaystyle \frac{\mathcal{P}^{2}}{\left(1-\frac{1}{m^{2}}k^{2}\right)k^{2}}}+{\displaystyle \frac{\mathcal{P}_{s}^{0}}{\left(1-\frac{1}{m^{2}}k^{2}-3+\frac{1}{m^{2}}k^{2}\right)k^{2}}}\\
\\
= & {\displaystyle \frac{\mathcal{P}^{2}}{\left(1-\frac{1}{m^{2}}k^{2}\right)k^{2}}}{\displaystyle -\frac{\mathcal{P}_{s}^{0}}{2k^{2}}}=\Pi_{GR}-{\displaystyle \frac{\mathcal{P}^{2}}{k^{2}-m^{2}}}.
\end{array}\label{eq:304weyl}
\end{equation}
From the latter form of the propagator one can notice the presence of an extra spin-$2$ degree of freedom with respect to GR.
Moreover, the new contribution comes with the wrong sign:
this is the Weyl ghost. In particular we can show that this is a ``bad''
ghost that violates the unitarity. As we have already done for GR,
we can follow the same prescription calculating the imaginary part
of the current-current amplitude residue. Indeed:
\begin{equation}
\begin{array}{rl}
\mathcal{A}= & i\tau^{*\mu\nu}(k)\Pi_{C,\mu\nu\rho\sigma}(k)\tau^{\rho\sigma}(k)\\
\\
= & i\tau^{*\mu\nu}(k)\Pi_{GR,\mu\nu\rho\sigma}(k)\tau^{\rho\sigma}(k)-i\tau^{*\mu\nu}(k){\displaystyle \frac{\mathcal{P}^{2}}{k^{2}-m^{2}}}\tau^{\rho\sigma}(k)\\
\\
= & \mathcal{A}_{GR}+\mathcal{A}_{2},
\end{array}
\end{equation}
where
\begin{equation}
\begin{array}{rl}
\mathcal{A}_{GR}= & i\tau^{*\mu\nu}(k)\Pi_{GR,\mu\nu\rho\sigma}(k)\tau^{\rho\sigma}(k),\\
\\
\mathcal{A}_{2}= & -i\tau^{*\mu\nu}(k){\displaystyle \frac{\mathcal{P}^{2}}{k^{2}-m^{2}}}\tau^{\rho\sigma}(k).
\end{array}\label{eq:306weyl}
\end{equation}
One can show that $\mathrm{Im}Res_{k^{2}=0}\{\mathcal{A}_{GR}\}>0$
(see for the GR part eq. (\ref{eq:136})), and that $\mathrm{Im}Res_{k^{2}=m^{2}}\{\mathcal{A}_{2}\}<0.$
So the last inequality says that the presence of the spin-$2$ massive
ghost violates the unitarity condition (see also Appendix $C.4.).$
\\
$ $

In the next chapter we shall consider a special choice for the coefficients
$a,b,c,d$ and $f$ that will give us a particular theory in which
the form of the graviton propagator is modified without adding new
physical states other than the spin-$2$ massless, traceless and transverse
graviton of GR. \\

\chapter{Infinite derivative theories of gravity}

\section{Consistency conditions on $a(\boxempty),$ $c(\boxempty)$ }

In the previous chapter we have taken the most general covariant quadratic
free-torsion action for gravity (\ref{eq:164}); we have considered
the linearized form (\ref{eq:177}) and calculated the physical propagator
(\ref{eq:204}). At the end we examined different choices of the coefficients
$a,b,c,d,$ and $f$ that gave us different (sub-)theories: GR, $f(\mathcal{R})$
gravity and conformally invariant gravity. The situation is the following:
$f(\mathcal{R})$ theories can be ghost-free but they are not able to improve the
UV behavior, while modifications involving $\mathcal{R}_{\mu\nu\rho\sigma}$
can improve the UV behavior but, on the other hands, they suffer from the presence of the Weyl ghost.
Before seeing how to overcome this problem, it is worth listing the
conditions needed to have a stable and ghost free theory around
the Minkowski background {[}\ref{-T.-Biswas,}{]}:
\begin{itemize}
\item As we would expect from a healthy modified theory of gravity, we want to recover GR at large distances (infrared limit); this request
implies that $a(\boxempty),$ $c(\boxempty)$ must be analytic around
$\boxempty=0$ $(k^2=0);$ 
\item To avoid the presence of the Weyl ghost must we need to require that $a(\boxempty)$ cannot have any zeroes, so that the only part of the spin-$2$ component is the GR graviton one;
\item We also demand that the scalar mode does not have any ghosts other than the the scalar component of the GR propagator (i.e. the pole $k^2=0$). So, $a-3c$ in the propagator (\ref{eq:204}) can at most
have one zero. To satisfy these conditions we should be able to express $c(\boxempty)$ in the following general form
\begin{equation}
c(\boxempty)=\frac{a(\boxempty)}{3}\left[1+2\left(1-\frac{\boxempty}{m^{2}}\right)\tilde{c}(\boxempty)\right],\label{eq:209}
\end{equation}
where $\tilde{c}(\boxempty)$ must be analytic around $\boxempty=0$ and cannot have any zeros;
\item Moreover, the fact that we do not want any tachyonic modes implies
$m^{2}>0$ in (\ref{eq:209}).
$ $
\end{itemize}
From the above conditions, by making different choices for the coefficients $a(\boxempty)$ and $c(\boxempty),$ different types of theories emerge. The difference relies on the number of degrees of freedom included in the physical part of the propagator, and whether they are either massive or massless.
We will make the choice $m^2\rightarrow0$ and $\tilde{c}(\boxempty)=0$, so that the resulting theory only contains the GR graviton as a propagating degree of freedom. Furthermore, we have only one independent function, $a(\boxempty)$=$c(\boxempty)$, that controls the modification in the UV regime. We shall see that this type
of theories is a good candidate for a ghost free and renormalizable
theory of gravity.

\section{Ghost and singularity free theories of gravity}

In the previous section we have seen that by making the choice of having only the GR graviton pole in the modified propagator, the expression (\ref{eq:209}) reduces to $a(\boxempty)=c(\boxempty).$ Moreover, because of (\ref{eq:179}) the following relations hold
\begin{equation}
a(\boxempty)=-b(\boxempty)=c(\boxempty)=-d(\boxempty)\Rightarrow f(\boxempty)=0.\label{eq:210}
\end{equation}
This means that we are essentially left with just a single free function
\begin{equation}
a(\boxempty)\coloneqq1-\frac{1}{2}\mathcal{F}_{2}(\boxempty)\boxempty-2\mathcal{F}_{3}(\boxempty)\boxempty,
\end{equation}
and from the relations (\ref{eq:176}) we are able to write $\mathcal{F}_{3}(\boxempty)$
as
\begin{equation}
\mathcal{F}_{3}(\boxempty)=-\left(\mathcal{F}_{1}(\boxempty)+\frac{\mathcal{F}_{2}}{2}(\boxempty)\right).\label{eq:211}
\end{equation}
Now, since the function $\mathcal{F}_{3}$ satisfies (\ref{eq:211}),
the action (\ref{eq:167}) becomes
\begin{equation}
\begin{array}{rl}
S= & \int d^{4}x\biggl[-\mathcal{R}+\mathcal{R}\mathcal{F}_{1}(\boxempty)\mathcal{R}+\mathcal{R}_{\mu\nu}\mathcal{F}_{2}(\boxempty)\mathcal{R}^{\mu\nu}\\
\\
 & \left.-\mathcal{R}_{\mu\nu\lambda\sigma}\left(\mathcal{F}_{1}(\boxempty)+{\displaystyle \frac{\mathcal{F}_{2}(\boxempty)}{2}}\right)\mathcal{R}^{\mu\nu\lambda\sigma}\right].
\end{array}\label{eq:306}
\end{equation}

By working around Minkowski space-time, since the covariant derivatives
become simple partial derivatives, we can move the function $\mathcal{F}_{i}(\boxempty)$
on the left of the curvatures by integrating by parts. In this way,
we are able to get rid of the product of two Riemann tensor by implementing
the Euler topological invariant relation:
\begin{equation}
\mathcal{R}^{\mu\nu\rho\sigma}\mathcal{R}_{\mu\nu\rho\sigma}-4\mathcal{R}^{\mu\nu}\mathcal{R}_{\mu\nu}+\mathcal{R}^{2}=\nabla_{\mu}K^{\mu},
\end{equation}
where $\nabla_{\mu}K^{\mu}$ is a four-divergence (surface term) that
doesn't contribute to the action variation. Thus, the action (\ref{eq:306})
becomes
\begin{equation}
S=\int d^{4}x\left[-\mathcal{R}+\mathcal{R}\mathcal{F}_{1}(\boxempty)\mathcal{R}+\mathcal{R}_{\mu\nu}\mathcal{F}_{2}(\boxempty)\mathcal{R}^{\mu\nu}\right];
\end{equation}
while in terms of the coefficient $a(\boxempty),$ by assuming $\mathcal{F}_{3}(\boxempty)=0,$
we have 
\begin{equation}
S=\int d^{4}x\left[-\mathcal{R}-\mathcal{R}\frac{1}{2}\frac{1-a(\boxempty)}{\boxempty}\mathcal{R}+\mathcal{R}_{\mu\nu}2\frac{1-a(\boxempty)}{\boxempty}\mathcal{R}^{\mu\nu}\right].
\end{equation}

Then, since $a(\boxempty)$ is the only function remaining, by using
the linearized form of the curvatures (see (\ref{eq:97linearized curv}))
the linearized action and the linearized field equations can be obtained:
\begin{equation}
S_{q}=-\int d^{4}x\left[\frac{1}{2}h_{\mu\nu}a(\boxempty)\boxempty h^{\mu\nu}-h_{\mu}^{\sigma}a(\boxempty)\partial_{\sigma}\partial_{\nu}h^{\mu\nu}+ha(\boxempty)\partial_{\mu}\partial_{\nu}h^{\mu\nu}-\frac{1}{2}ha(\boxempty)\boxempty h\right]
\end{equation}
and
\begin{equation}
a(\boxempty)\left[\boxempty h_{\mu\nu}-\left(\partial_{\mu}\partial_{\alpha}h_{\nu}^{\alpha}+\partial_{\alpha}\partial_{\nu}h_{\mu}^{\alpha}\right)+\left(\eta_{\mu\nu}\partial_{\alpha}\partial_{\beta}h^{\alpha\beta}+\partial_{\mu}\partial_{\nu}h\right)-\eta_{\mu\nu}\boxempty h\right]=-\kappa\tau_{\mu\nu}.\label{eq:211eq}
\end{equation}
Our theory has to contain only the graviton pole, so the spin-$2$
and spin-$0$ component cannot have any other pole (zeroes). The physical
propagator (\ref{eq:204}) because of the choice (\ref{eq:210}) becomes
\begin{equation}
\Pi(k)=\frac{1}{a(-k^{2})}\left(\frac{\mathcal{P}^{2}}{k^{2}}-\frac{\mathcal{P}_{s}^{0}}{2k^{2}}\right)=\frac{1}{a(-k^{2})}\Pi_{GR}(k),\label{eq:212}
\end{equation}
i.e. we obtain the the GR propagator modified by the factor $\frac{1}{a(-k^{2})}.$
The next step is choosing a special form for the coefficient $a(\boxempty)$
that suits our aims. We require that there are no gauge-invariant
poles other than the transverse and traceless massless physical graviton
pole, thus the coefficient $a(\boxempty)$ cannot vanish in the complex
plane. Special functions that satisfies these characteristics are
the exponentials of entire functions: indeed they do not have
poles in the complex plane and vanish only at infinity. \\
Hence, for the coefficient $a(\boxempty)$ can be done the following
choice:
\begin{equation}
a(\boxempty)=e^{-\frac{\boxempty}{M^{2}}},\label{eq:213}
\end{equation}
where $M$ is a parameter that makes dimensionless the exponent of
the exponential and, physically, corresponds to the scale at which
the modification to GR made by this theory should appear. We can note
that expanding in Taylor series the exponential in (\ref{eq:213})
one has
\begin{equation}
a(\boxempty)=e^{-\frac{\boxempty}{M^{2}}}=\sum_{n=0}^{\infty}\frac{1}{n!}\left(-\frac{\boxempty}{M^{2}}\right)^{n}.\label{eq:214}
\end{equation}
The meaning of (\ref{eq:214}) is that we are dealing with a theory
containing an infinite set of derivatives expressed in the form of an exponential function%
\footnote{The most general quadratic action for IDG theories is given by 
\[
S=\int\sqrt{-g}\left(-\mathcal{R}+\mathcal{R}\mathcal{F}_{1}(\boxempty)\mathcal{R}+\mathcal{R}^{\mu\nu}\mathcal{F}_{2}(\boxempty)\mathcal{R}_{\mu\nu}+\mathcal{R}^{\mu\nu\rho\sigma}\mathcal{F}_{3}(\boxempty)\mathcal{R}_{\mu\nu\rho\sigma}\right),
\]
where the $\mathcal{F}_{i}(\boxempty)'$s are functions of the D'Alambertian
operator, $\boxempty=g^{\mu\nu}\nabla_{\mu}\nabla_{\nu},$ and contain
an infinite set derivatives:
\[
\mathcal{F}_{i}(\boxempty)=\sum_{n=0}^{\infty}f_{i,n}\boxempty^{n},\,\,\,\,\, i=1,2,3.
\]
One requires that the $\mathcal{F}_{i}(\boxempty)'$s are analytic
at $\boxempty=0$ so that one can recover GR in the infrared regime.%
}. For this reason these particular theories of gravity are called
$\mathit{Infinite}$ $\mathit{Derivative}$ $\mathit{Theories}$ $\mathit{of}$
$\mathit{Gravity}$ (IDG). In  momentum space one has
\begin{equation}
a(-k^{2})=e^{\frac{k^{2}}{M^{2}}}=\sum_{n=0}^{\infty}\frac{1}{n!}\left(\frac{k^{2}}{M^{2}}\right)^{n},\label{eq:215}
\end{equation}
so the physical propagator read as
\begin{equation}
\Pi(k)=e^{-\frac{k^{2}}{M^{2}}}\left(\frac{\mathcal{P}^{2}}{k^{2}}-\frac{\mathcal{P}_{s}^{0}}{2k^{2}}\right)=e^{-\frac{k^{2}}{M^{2}}}\Pi_{GR}(k),\label{eq:216}
\end{equation}
and we notice that the only difference from GR case is that the graviton
propagator turns out to be modified by a multiplicative trascendental function.

\subsection{Non-singular Newtonian potential}

In this subsection we shall see how the potential modifies in this
special theory of gravity when one considers the Newtonian approximation
(or weak field approximation). We will see that the resulting potential
is non-singular for values of the radius equal to zero, i.e. is not
divergent like Newtonian potential $\frac{1}{r}$ but it is finite.
\\
\\
Hence, we are going to focus particularly on the classical short-distance
behavior. As is usual, we want to solve the linearized modified field
equations (\ref{eq:211eq}) for a point source:
\begin{equation}
\tau_{\mu\nu}=\rho\delta_{\mu}^{0}\delta_{\nu}^{0}=m_{g}\delta^{3}(\bar{x})\delta_{\mu}^{0}\delta_{\nu}^{0},\label{eq:217}
\end{equation}
where $m_{g}$ is the mass of the object that is generating the gravitational
potential. In Newtonian approximation the metric, besides to have
small perturbation, is stationary and the sources are static (or,
anyway, with neglecting velocities). Hence the metric reduces to
\begin{equation}
ds^{2}=\left(1+2\Phi\right)dt^{2}-\left(1-2\Psi\right)|d\bar{x}|^{2},\label{eq:218}
\end{equation}
with $|\Phi|,|\Psi|<<1,$ since in Newtonian approximation the potentials
are weak.\\
Now our aim is to determine a form for the potentials $\Phi$ and
$\Psi,$ i.e. to find the metric. Let us consider the trace and the
$00$ component of the equations (\ref{eq:211eq}) keeping in mind
that in Newtonian approximation $\partial_{0}h_{\mu\nu}=0:$ 
\begin{lyxlist}{00.00.0000}
\item [{Trace:}] 
\begin{equation}
\begin{array}{rl}
 & a(\boxempty)\left[\boxempty h-\left(2\partial_{\alpha}\partial_{\beta}h^{\alpha\beta}\right)+\left(4\partial_{\alpha}\partial_{\beta}h^{\alpha\beta}+\boxempty h\right)-4\boxempty h\right]=-\kappa\tau\\
\\
\Leftrightarrow & 2a(\boxempty)\left[-\boxempty h+\partial_{\alpha}\partial_{\beta}h^{\alpha\beta}\right]=-\kappa\rho;
\end{array}
\end{equation}

\item [{$00\,\,$component:}] 
\begin{equation}
a(\boxempty)\left[\boxempty h_{00}+\partial_{\alpha}\partial_{\beta}h^{\alpha\beta}-\boxempty h\right]=-\kappa\rho.
\end{equation}

\end{lyxlist}
$ $\\
Note that the metric (\ref{eq:218}) can be rewritten making explicit
the perturbation $h_{\mu\nu}:$
\begin{equation}
\begin{array}{rl}
ds^{2}= & dt^{2}-|d\bar{x}|^{2}+\left(2\Phi dt^{2}+2\Psi|d\bar{x}|^{2}\right)\\
\\
= & \eta_{\mu\nu}+h_{\mu\nu},
\end{array}\label{eq:219}
\end{equation}
with
\begin{equation}
h_{\mu\nu}\equiv\begin{pmatrix}2\Phi & 0 & 0 & 0\\
0 & 2\Psi & 0 & 0\\
0 & 0 & 2\Psi & 0\\
0 & 0 & 0 & 2\Psi
\end{pmatrix}.\label{eq:220}
\end{equation}
From (\ref{eq:219}) and the assumption of static potential we have:
\begin{equation}
h=\eta^{\mu\nu}h_{\mu\nu}=2\left(\Phi-3\Psi\right),\label{eq:221}
\end{equation}
\begin{equation}
h_{00}=2\Phi,
\end{equation}
\begin{equation}
\partial^{\mu}\partial^{\nu}h_{\mu\nu}=\partial^{i}\partial^{j}h_{ij}=2\nabla^{2}\Psi,
\end{equation}
\begin{equation}
\boxempty\rightarrow-\nabla^{2}=-\delta^{ij}\partial_{i}\partial_{j}.\label{eq:222}
\end{equation}
Applying the (\ref{eq:221})-(\ref{eq:222}), the trace and $00$
component equations become respectively:
\begin{equation}
2a(\nabla^{2})\left[2\nabla^{2}\left(\Phi-3\Psi\right)+2\nabla^{2}\Psi\right]=-\kappa\rho
\end{equation}
and
\begin{equation}
a(\nabla^{2})\left[4\nabla^{2}\Psi\right]=\kappa\rho.
\end{equation}
By comparing these last equations, at the end, we notice that the
two potentials $\Phi$ and $\Psi$ satisfy the same equation:
\begin{equation}
4a(\nabla^{2})\nabla^{2}\Phi=4a(\nabla^{2})\nabla^{2}\Psi=\kappa\rho=\kappa m_{g}\delta^{3}(\bar{x}).\label{eq:223}
\end{equation}
We can find the solutions of the last equations by going into the
momentum space and then going back to the coordinate space. Since
the two potentials satisfy the same equations, let us consider the
potential $\Phi:$
\begin{equation}
4a(\nabla^{2})\nabla^{2}\Phi=\kappa m_{g}\delta^{3}(\bar{x})\longrightarrow-4a(\bar{k}^{2})\bar{k}^{2}\Phi(k)=\kappa m_{g}
\end{equation}
\begin{equation}
{\displaystyle \begin{array}{rl}
\Rightarrow\Phi(r)= & -{\displaystyle \frac{\kappa m_{g}}{4}\int\frac{d^{3}k}{(2\pi)^{3}}\frac{e^{i\bar{k}\cdot\bar{r}}}{\bar{k}^{2}a(-\bar{k}^{2})}}\\
\\
= & {\displaystyle -\frac{\kappa m_{g}}{4(2\pi)^{3}}\int_{0}^{\infty}d|\bar{k}||\bar{k}|\int_{-1}^{+1}d(\cos\vartheta){\displaystyle \int_{0}^{2\pi}}d\varphi\frac{e^{i\bar{k}\cdot\bar{r}}}{\bar{k}^{2}a(-\bar{k}^{2})}}\\
\\
= & {\displaystyle -\frac{\kappa m_{g}}{8\pi^{2}}\frac{1}{r}\int_{0}^{\infty}d|\bar{k}|\frac{e^{-\frac{\bar{k}^{2}}{M^{2}}}\sin|\bar{k}|r}{|\bar{k}|}.}
\end{array}}
\end{equation}
From the last equation we can already notice that the potential is
modified, in fact apart from the Newtonian contribution $\frac{1}{r}$
we also have a new factor given by the integral. This integral corresponds
to one of the special functions, i.e. the $Error$ $Function:$ 
\begin{equation}
\int_{0}^{\infty}d|\bar{k}|\frac{e^{-\frac{\bar{k}^{2}}{M^{2}}}\sin|\bar{k}|r}{|\bar{k}|}=\frac{\pi}{2}\mathrm{Erf}\left(\frac{rM}{2}\right).
\end{equation}
Hence the solutions of the equations (\ref{eq:223}) are given by ({[}\ref{-T.-Biswas,prl}{]}, {[}\ref{-modesto}{]}, {[}\ref{-shapmod1}{]})
\begin{equation}
\Phi(r)=\Psi(r)=-\frac{1}{16\pi}\frac{m_{g}}{M_{p}^{2}}\frac{1}{r}\mathrm{Erf}\left(\frac{rM}{2}\right),\label{eq:330potential}
\end{equation}
where we have used the fact that $\kappa=\frac{1}{M_{p}^{2}},$ with
$M_{p}\simeq1.2\times10^{18}\,\mathrm{GeV}$ reduced Planck mass.
\\
We observe that as $r\rightarrow\infty,$ $\mathrm{Erf}\left(\frac{rM}{2}\right)\rightarrow1,$
and we recover the GR limit (infrared limit), i.e. the usual Newtonian
potential $\Phi(r)=\Psi(r)\sim-\frac{1}{r}.$ On the other hand, as
$r\rightarrow0,$ $\mathrm{Erf}\left(\frac{rM}{2}\right)\rightarrow\frac{rM}{2},$
namely
\begin{equation}
r\rightarrow0\Rightarrow\Phi(r)=\Psi(r)\sim-\frac{m_{g}M}{M_{p}^{2}}.\label{eq:224}
\end{equation}
We have found that there are no divergences. The potential, in fact,
converges to a finite value as shown in (\ref{eq:224}). Thus, although
the matter source has a delta function singularity, the potentials
remain finite. Further, since we are working in the approximation
of weak potentials, the whole discussion holds as long as 
\begin{equation}
\frac{m_{g}M}{M_{p}^{2}}<<1\Leftrightarrow m_{g}M<<M_{p}^{2}.\label{eq:225}
\end{equation}
It is clear that for small masses our theory provides a very different
description of space-time as compared to GR. In fact, according to
our model there are no black-hole like solutions (no horizon and no
singularity) as long as the mass source satisfies the condition (\ref{eq:225}).
Unfortunately, our analysis cannot say anything about large mass astrophysical
black holes because the Newtonian potentials become too large for
us to be able to trust the perturbative calculations.\\
Below we have made more explicit how Newton potential is modified
in this model by plotting the both singular and non-singular functions:

\begin{figure}[H]
\includegraphics[scale=0.85]{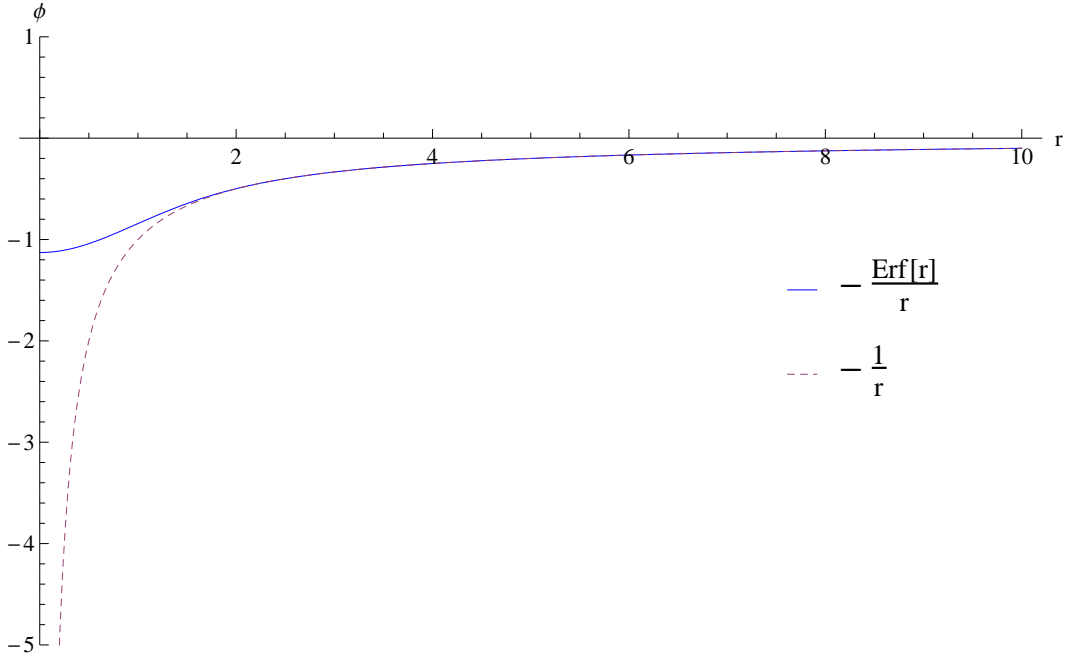}

\protect\caption{In this plot both singular and non-singular functions are drawn: the
blue line represents the non-singular potential, the dashed line the
singular Newton potential. The plot was obtained by using Wolfram
Mathematica 9.0.}

\end{figure}

We can also study what is the gravitational force $\bar{F}$ that
a test mass $m$ undergoes in the gravitational potential (\ref{eq:330potential})
generated by the point source $m_{g}.$ O nce we have the potential
(\ref{eq:330potential}) we can obtain the force by calculating its
derivative with respect to $r:$
\begin{equation}
\begin{array}{rl}
F=m\ddot{r}= & {\displaystyle -m\frac{\partial\phi(r)}{\partial r}}\\
\\
= & {\displaystyle \frac{mm_{g}}{M_{P}^{2}}\left[\frac{M}{\sqrt{\pi}}\frac{e^{-(\frac{Mr}{2})^{2}}}{r}-\frac{Erf(\frac{Mr}{2})}{r^{2}}\right].}
\end{array}\label{eq:333force}
\end{equation}
In the following plot we have plotted the gravitational force (\ref{eq:333force})
as a function of $\frac{Mr}{2}.$ We can notice that there is a minimum
at%
\footnote{The minimum was calculated by using Wolfram Mathematica 9.0.%
} 
\begin{equation}
\frac{Mr_{min}}{2}=0.9678\simeq1.\label{eq:334 minimum}
\end{equation}
It means that for for value of the coordinate $r<r_{min}\simeq\frac{2}{M}$
the gravitational force starts decreasing, until it vanishes for $r=0.$
The scale at which this happens is dictated by the parameter $M$
(see next section).

We can state that this model describe a $\mathit{classic}$ $\mathit{asymptotic}$
$\mathit{free}$ theory of gravity%
\footnote{A theory is ``asymptotic free'' when the perturbation approach gets
better and better at higher energies, and in the infinite momentum
limit, the coupling constant vanishes. It means that particles become
asymptotically weaker as energy increase and distance decrease. An
example is the theory of Quantum Chromodynamics (QCD). The author
in Ref. {[}\ref{-T.-Biswas,prl}{]},{[}\ref{-S.-Talaganis,}{]} and
{[}\ref{-T.-Biswas,}{]} believe that in the framework of IDG theories
the gravitational interaction could behave like QCD interaction.Since
at the classical level we obtain a gravitational force the decrease
with the distance, we can use the expression ``classical asymptotic
freedom'' and so the non-singular potential behavior is a $\mathit{strong}$
clue in favour of an ``asymptotically free quantum theory of gravity''.%
}. 

\begin{figure}[H]
\includegraphics[scale=0.85]{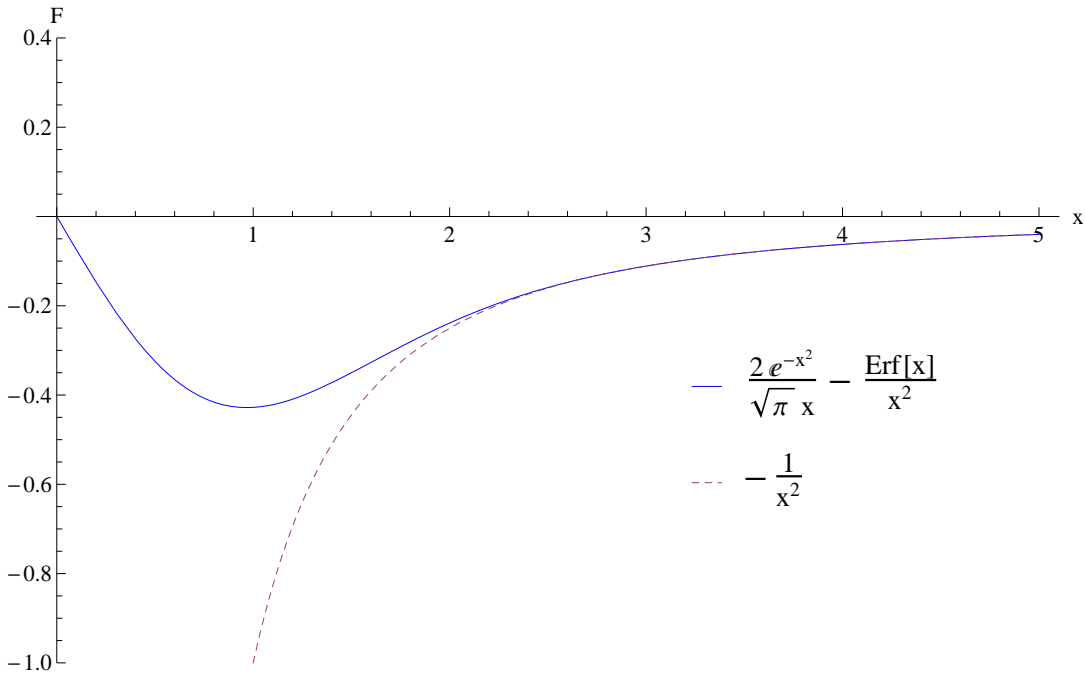}

\protect\caption{In this plot both gravitational force deriving from the non-singular
potential (\ref{eq:330potential}) and Newton force are drawn: the
blue line represents the non-singular gravitational force, the dashed
line the singular Newton force. The axis $x$ defines the variable
$x=\frac{Mr}{2}.$ The plot was obtained by using Wolfram Mathematica
9.0.}

\end{figure}

\begin{rem}
We have seen that an UV modification involving the exponential function
$e^{-\frac{\boxempty}{M^{2}}}$ is able to eliminate the Newtonian
singularity. This is also a promising clue that let us believe in
the realization of asymptotic freedom within IDG models {[}\ref{-T.-Biswas,prl}{]},{[}\ref{-S.-Talaganis,}{]},{[}\ref{-T.-Biswas,}{]}.
It could be possible that a non-singular behavior (``classical''
asymptotic freedom) can be connected to a ``quantum'' asymptotic
freedom behavior. \\
In the above discussion we have made use of one special transcendental function
given by (\ref{eq:213}). Note that a more general choice could be
\begin{equation}
a(\boxempty)=e^{-\gamma(\boxempty),}
\end{equation}
where $\gamma(\boxempty)$ is an analytic function of $\boxempty.$
It is then easy to see that for any polynomial $\gamma(\boxempty),$
as long as the highest power has positive coefficient, we will have
a $\mathit{potentially}$ asymptotic free theory and the propagator
will be even more convergent than the exponential case (\ref{eq:213}). 
\end{rem}

\section{Parameter $\mathbf{M}$ }

In the previous section we have seen that by making an appropriate choice for the only independent coefficient
$a(\boxempty),$ we are able to obtain a non-singular gravitational
potential that in the UV limit gives us the GR behavior. So far we
haven't talked yet about the kind of modification we are considering
at the physical level, but we have just done mathematical consideration.
\\
We know that GR is recovered in the low energy regime (infrared limit), so it
means that modifications of GR should be visible for higher energy (short distances). The energy scale at which the modification becomes noticeable is given
by the mass $M$ appearing in the exponent of the exponential $a(\boxempty).$
It would be interesting to constrain the parameter by putting some
bounds. We know for sure that we have an upper bound given by the
reduced mass Planck, $M<M_{P}\simeq2.4\times10^{18}\, GeV,$ but it
is too high to get interesting physical results.\\
\\
It is worth noting that $M$ corresponds to a scale of $\mathit{non}\textrm{-}\mathit{locality},$
and this is due to the fact that IDG theories describe a non-local%
\footnote{Informally, locality means that physics over here is independent of
physics over there; we don't have to have the wavefunction of the
universe to see what happens in our lab {[}\ref{-M.-Schwartz,}{]}.%
} gravitational interactions. Let us clarify why infinitive derivative
theories of gravity are non-local.\\
From Cauchy theorem on differential equation we know that the higher
the number of derivatives is the more initial data you have to provide
to find a solution. If you have some Lagrangian that contains an infinite
number of derivatives (or derivatives appearing non-polynomially,
such as one over derivative) then you have to provide an infinite
amount of initial data which amounts to non-local info, in the sense
we now explain. If we have a theory with second derivatives it means
that we have just to provide the field and its first derivative as
initial values, at a specific point. So we don't need to know the
whole function, but just its value in the neighborhood of a point
of its domain, i.e. we just need to know the function $\mathit{locally}.$
Instead, if we have infinite derivative, for example if we think in
terms of Taylor expansions around an initial value, then you have
to provide the full function (and thus non-local information). \\
\\
In Ref. {[}\ref{-J.-Edholm,}{]} the authors put a bottom bound on
the parameter $M$ by using the results obtained by the tests of $\frac{1}{r}$
fall of Newtonian gravity, that has been tested in the laboratory
up to $5.6\times10^{-5}\, m$ {[}\ref{-D.-J. alderbrger}{]}%
\footnote{ The experiment discussed in Ref. {[}\ref{-D.-J. alderbrger}{]} was
conducted with torsion-balance and the inverse-square law was tested
with $95\%$ confidence. %
}, which implies that the scale of non-locality should be bigger than
$M>2\times10^{4}\, m^{-1}=0.01\, eV.$ This gives us also an upper
limit for the minimum radius in equation (\ref{eq:334 minimum}):
\begin{equation}
r_{min}<\frac{2}{2\times10^{4}\, m^{-1}}=10^{-4}\, m.
\end{equation}
\\
Hence the current experimental limits put the bound on non-locality
to be around greater than $0.01\, eV$ and smaller than $10^{18}\, GeV.$
It means that any modification from Newtonian potential can occur
in the gulf of scales spanning some $30$ orders of magnitude. Our
knowledge about the gravitational interaction at short distances is
really limited!

\chapter*{Conclusions}\addcontentsline{toc}{chapter}{Conclusions}
\markboth{Conclusions}{CONCLUSIONS}

In the last ten years, a lot of work was made by several groups. Today there are promising clues and a lot of confidence to continue working
on infinitive derivative theories of gravity. In particular we have
seen that they can be free from instabilities and ghosts around Minkowski
background, but at the same time resolve the singularity problem for
static mini-black holes. IDG theories could be also able to resolve the
cosmological singularity problem, but so far only partially. In Ref. {[}\ref{-T.-Biswas,prl}{]}
and {[}\ref{-T.-Biswas,}{]} the authors resolve the problem in the
linearized regime, obtaining sinusoidal (periodic) solutions for the scale
factor $a(t),$ i.e. bounce solutions. These solutions hold only for
vanishing energy-momentum tensors. They also notice that the singularity
can be avoided only if one introduces an
additional massive scalar degree of freedom, whose corresponding mass will be connected to the frequency of expansion and contraction of the universe. Furthermore,
in Ref. {[}\ref{-T.-Biswas,bouncing}{]}, {[}\ref{-T.-Biswas,}{]}
and {[}\ref{-Biswas-T, cosmological pert}{]} the authors also face
the cosmological singularity problem in the full (not linearized)
theory by using the following action
\[
S=\int d^{4}x\left[-\mathcal{R}+\mathcal{R}\mathcal{F}_{1}(\boxempty)\mathcal{R}\right];
\]
i.e. a sub-case of the general action (\ref{eq:167}),
$\mathcal{F}_{2}(\boxempty)=0=\mathcal{F}_{3}(\boxempty).$ Thus the
information about the non-locality, and so about the presence of an
infinite set of higher derivatives, is contained in the entire function
$\mathcal{F}_{1}(\boxempty)=\sum_{n=0}^{\infty}f_{1,n}\boxempty^{n}.$
\\
These classical results are in contrast with extended theories of gravity with finite higher derivatives, which show be either ghost-free or singularity-free,
but not both.\\
In this thesis we haven't faced the quantum aspects of IDG theories
but we have given a little overview in the previous chapter. Several
groups have addressed the question of renormalizability and obtained
good and promising results so far. The fundamental role is played
by the exponential function $e^{-\frac{\boxempty}{M^{2}}}$
that makes the theory softened in the UV regime and gives a first
strong clue that the theory could describe an asymptotically free
gravity. \\
\\
As we can notice, there are still several questions that remain to be faced and better understood. Let us conclude listing some of these remaining challenges by following the authors in Ref. {[}\ref{-T.-Biswas,}{]}:
\begin{itemize}
\item \textbf{Black hole singularity:} we have seen that IDG theories solve the singularity problem in the weak field approximation as shown above by the form of the modified non-singular Newtonian potential. At the classical level, one of the remaining aim is to understand whether the singularity can be also avoided in the strong field regime, i.e. for astrophysical black holes.  	
\item \textbf{Cosmological singularity:} as we have already mentioned, in the linear regime, the cosmological
singularity problem was solved in absence of matter source. To conduct
a perturbative study for generic matter sources, the quadratic terms in the perturbation are not sufficient and we need to include cubic interactions. Furthermore, the exact cosmological solutions were only obtained in the presence
of a cosmological constant {[}\ref{-T.-Biswas,bouncing}{]}. A realistic cosmological scenario must also give an explanation for the inflationary phase. People are still trying to find a way to include and describe this primordial transition.
\item \textbf{Extension to different backgrounds:} while IDG models have been investigated around
the Minkowski space-time, one needs to extend this way of proceeding to
de Sitter and FLRW backgrounds. Some progress has already been done in
Ref. {[}\ref{de sitt mazum}{]} and {[}\ref{de sitt and anti mazum}{]}.
\item \textbf{Unitarity:} The absence of "bad" ghost in the bare propagator implies that the unitarity condition is preserved at the tree level but it says nothing when loops are involved. While unitarity and causality of non-local theories have been formally argued for example in Ref. {[}\ref{-G.-V. efimov}{]},
{[}\ref{-V.-A.}{]} and {[}\ref{-E.-T.}{]}, one should check that the several techniques used to calculate the loop integrals do not violate the unitarity condition once one or more loops are considered.
\item \textbf{UV behavior:} as we have already emphasized, for these models the quantum UV behavior
seems to be improved, but so far computations have only been carried out
up to $2$ loops. So far just a toy scalar model {[}\ref{-S.-Talaganis,}{]} has been considered, but obviously as a next step one has to deal with gravitational interaction in the framework of IDG theories. It could also happen that the exponential choice we have considered is not the most suitable to get a renormalizable and unitary theory. If this is the case, there might be different entire functions (satisfying the same properties we mentioned several times) that could solve this diatribe.
\end{itemize}
\renewcommand{\theequation}
{\thechapter.\arabic{equation}}
\setcounter{equation}{0}

\appendix

\chapter{Off-shell and on-shell particles}

In quantum field theory, off-shell (virtual) particles don't satisfy
the Einstein dispersion relation; instead, on-shell (real) particles
do satisfy this relation. In formula it means that:
\[
\begin{array}{rl}
\mathrm{off\textrm{-}shell:} & E^{2}-|\bar{p}|^{2}\neq m^{2}\\
\\
\mathrm{on\textrm{-}shell:} & E^{2}-|\bar{p}|^{2}=m^{2}
\end{array}.
\]
Naturally, if we deal with massless particles the last relations have
to be considered with $m=0.$ \\
\\
Let us consider a massive scalar field for simplicity. We know that
the Lagrangian for a massive scalar field is given by
\[
\mathcal{L}=-\frac{1}{2}\phi\left(\boxempty+m^{2}\right)\phi,
\]
and the Euler-Lagrange equations are given by the Klein-Gordon equations:
\[
\left(\boxempty+m^{2}\right)\phi=0,
\]
that in  momentum space becomes 
\[
\left(-p^{2}+m^{2}\right)\phi=0\Rightarrow p^{2}-m^{2}=0,
\]
where $p^{2}=p^{\mu}p_{\mu}=E^{2}-|\bar{p}|^{2}.$ Hence a scalar
field satisfying the field equations satisfies also the Einstein dispersion
relation and so it will describe on-shell particles. Note that the
Noether theorem is an on-shell theorem, in fact in the demonstration
we have to impose the validity of the field equations to obtain the
conservation law of the current, $\partial_{\mu}J^{\mu}=0,$ that,
in turn, is an on-shell equation. \\
\\
Note that the virtual particles corresponding to internal propagators
in a Feynman diagram are in general allowed to be off-shell. In fact,
given a virtual particle with $4-$momentum $q,$ if the Einstein
dispersion relation held ($q^{2}=m^{2})$ we would have a singularity
in the propagator:
\[
\mathcal{P}(q)=\frac{1}{q^{2}-m^{2}}.
\]
\\
Note also that, since the field equations are a constraint for the
field, it means that off-shell particles turn out to be ``more free''
than on-shell particles in terms of degrees of freedom (see sections
$1.2$ and $2.2).$

\chapter{Spin projector operators decomposition}

\section{Tensor decomposition}

We have used a lot the formalism of the spin projector operators to
derive most of the results of this thesis. Especially we have found
them very useful to understand which are the spin components of photon
and graviton, and for the calculation of the propagators. In this
appendix our aim is to understand why we can decompose either a vector
or a two-rank tensor in terms of the spin projector operators. First
of all we shall study the tensor representations of the Lorentz Group,
especially the irreducible tensor representations under $SO(3);$
then we shall introduce special projector operators, called $\mathit{spin}$
$\mathit{projector}$ $\mathit{operators},$ by which we can decompose
our tensor in two scalar, one transverse vector and one transverse
and traceless tensor components. As for the part on tensor representation
we shall also take inspiration from Ref. {[}\ref{-M.-Maggiore,}{]}.

\subsection{Lorentz tensor representation}

Let us consider a two-rank tensor $\varphi^{\mu\nu}$ with two contravariant
indices in Minkowski space. By definition $\varphi^{\mu\nu}$ is an
object that under Lorentz transformations transforms as
\begin{equation}
\varphi'^{\mu\nu}=\Lambda_{\rho}^{\mu}\Lambda_{\sigma}^{\nu}\varphi^{\rho\sigma}.\label{eq:b1}
\end{equation}
Tensors are examples of representations of the Lorentz group. For
instance a generic two-rank tensor $\varphi^{\mu\nu}$ has $16$ components
and (\ref{eq:b1}) shows that these components transform among themselves,
i.e. they form a basis for a $16$-dimensional representation of the
Lorentz group. \\
In Group Theory the irreducible representations of any group turn
out to be very important; for example, they are very useful when we
want to decompose a tensor object in its several spin components.
We can notice that the $16$-dimensional representation, we have just
introduced, is reducible in different irreducible parts. First of
all we easily understand that if $\varphi$ is symmetric (antisymmetric)
then also $\varphi'$ will be symmetric (antisymmetric), so the symmetric
and antisymmetric parts of a tensor $\varphi^{\mu\nu}$ don't mix,
and the $16$-dimensional representation is for sure reducible into
a $6$-dimensional antisymmetric representation $\psi^{\mu\nu}$ and
a $10$-dimensional symmetric representation $h^{\mu\nu}.$ One can
explicitly see this decomposition in symmetric and antisymmetric parts
in the following way: 
\begin{equation}
\varphi^{\mu\nu}=h^{\mu\nu}+\psi^{\mu\nu},\,\,\,\,\begin{cases}
h^{\mu\nu}\coloneqq\frac{1}{2}\left(\varphi^{\mu\nu}+\varphi^{\nu\mu}\right)\\
\psi^{\mu\nu}\coloneqq\frac{1}{2}\left(\varphi^{\mu\nu}-\varphi^{\nu\mu}\right)
\end{cases}.\label{eq:b2}
\end{equation}
Furthermore, the trace of a symmetric tensor can be also isolated.
Indeed, it is invariant under Lorentz transformation:
\[
h'=\eta_{\mu\nu}h'^{\mu\nu}=\eta_{\mu\nu}\Lambda_{\rho}^{\mu}\Lambda_{\sigma}^{\nu}h^{\rho\sigma}=\eta_{\rho\sigma}h^{\rho\sigma}=h;
\]
so a traceless tensor remains traceless after a Lorentz transformation,
and thus the $10$-dimensional symmetric representation decomposes
into a $9$-dimensional irreducible symmetric traceless representation
and a $1$-dimensional scalar representation. In formula this means
that 
\begin{equation}
h^{T\mu\nu}\coloneqq h^{\mu\nu}-\frac{1}{4}\eta_{\mu\nu}h,\,\,\,\, h=\eta_{\mu\nu}h^{\mu\nu},\label{eq:b3}
\end{equation}
where the apex ``$T"$ means ``traceless'', in fact $\eta_{\mu\nu}h^{T\mu\nu}=h-\frac{4}{4}h=0.$\\
In representation theory the following notation is commonly used:
an irreducible representation is denoted by its dimensionality (the
number of components), written in boldface. Thus the scalar representation
is denoted as $\mathbf{1},$ the four-vector representation as $\mathbf{4},$
the antisymmetric representation as $\mathbf{6}$ and the traceless
symmetric representation as $\mathbf{9}.$ \\
The tensor representation (\ref{eq:b1}) sees the action of two Lorentz
matrices. It means that the representation (\ref{eq:b1}) is a tensor
product of two four-vector representations, namely each of two contravariant
indices of $\varphi^{\mu\nu}$ transforms separately as a four-vector
index. The tensor product of two representation is denoted by the
symbol $\otimes.$ Since we have found that a two-rank tensor can
be decomposed into the direct sum of three irreducible representations,
denoting the direct sum by $\oplus,$ we can express the irreducible
representation in terms of the dimensionality introduced above:
\begin{equation}
\mathbf{4}\otimes\mathbf{4}=\mathbf{1}\oplus\mathbf{6}\oplus\mathbf{9}.\label{eq:b4}
\end{equation}
Analogously one can obtain the tensor decomposition into irreducible
parts when more than two indices are present. The most general irreducible
representation of the Lorentz group are found starting from a generic
tensor with an arbitrary number of indices, removing first all traces,
and then symmetrizing or antisymmetrizing over all pairs of indices.
Note also that, by raising and lowering the indices with the Minkowski
metric tensor $\eta_{\mu\nu},$ we can always restrict to contravariant
tensors. Thus, for instance, $V^{\mu}$ and $V_{\mu}$ are equivalent
representations.\\
All tensor representations are in a sense derived from the four-vector
representation, since the transformation law of a tensor is obtained
applying separately on each Lorentz index the matrix $\Lambda_{\nu}^{\mu}$
that defines the transformation law of a four-vector. This means that
tensor representations are tensor product of four-vector representations
and for this reason, the four-vector plays a fundamental role.

\subsection{Decomposition of Lorentz tensors under $SO(3)$ }

We know how a tensor behaves under a generic Lorentz transformation.
Now, we are going to focus particularly on the transformation properties
of a tensor under the $SO(3)$ rotation subgroup, and we can therefore
ask what is the angular momentum $j$ of the various tensor representations.
We will be able to decompose a generic two-rank tensor in terms of
its spin components.\\

Let us recall that the representations of $SO(3)$ are labeled by
an index $j$ which assumes integer values $j=0,1,2\dots;$ while
the dimension of the representation, labeled by $j,$ is defined by
$2j+1.$ Then within each representation, there are $2j+1$ states
labeled by $j_{z}=-j,\dots,j.$ Note that for $SO(3)$ it is more
common to denote the representation as $\mathbf{j},$ i.e. to label
it with the associated angular momentum rather than with the dimension
of the representation, $2j+1.$ Hence in this notation, $\mathbf{0}$
is the scalar (singlet, spin-$0$), $\mathbf{1}$ is a triplet (spin-$1)$
with components $j_{z}=-1,0,1,$ while $\mathbf{2}$ is a representation
of dimension $5$ (spin-$2),$ and so on with higher dimensionality.\\
A Lorentz scalar is of course also scalar under rotations, so it has
$j=0.$ A four-vector $V^{\mu}=(V_{0},\bar{V})$ is an irreducible
representation of the Lorentz group, since a generic Lorentz transformation
mixes all four components, but under $SO(3)$ it is reducible. Indeed,
spatial rotations do not mix $V^{0}$ with $\bar{V}:$ $V^{0}$ is
invariant under spatial rotations, so it has $j=0,$ while the three
spatial components $V^{i}$ form an irreducible $3$-dimensional representations
of $SO(3),$ with $j=1.$ \\
By adopting the above convention according to which the representations
are indicated by the associated angular momentum, the decomposition
of a four-vector into the direct sum of a scalar and a spin-$1$ representation
under $SO(3)$ can be written as
\begin{equation}
V^{\mu}\in\mathbf{0}\oplus\mathbf{1}.\label{eq:b5}
\end{equation}
While in terms of their dimensions we should write
\begin{equation}
V^{\mu}\in\mathbf{4}=\mathbf{1}\oplus\mathbf{3}.\label{eq:b6}
\end{equation}
Now we would like to understand which are the spin components of a
two-rank tensor $\varphi^{\mu\nu},$ i.e. what angular momenta appear.
By definition we know that $\varphi^{\mu\nu}$ transforms as the tensor
product of two four-vector representations. Since, from a point of
view of $SO(3),$ a four-vector decomposes as $\mathbf{0}\oplus\mathbf{1},$
a generic two-rank tensor has the following decomposition in angular
momenta
\begin{equation}
\begin{array}{rl}
\varphi^{\mu\nu}\in(\mathbf{0}\oplus\mathbf{1})\otimes(\mathbf{0}\oplus\mathbf{1})= & (\mathbf{0}\otimes\mathbf{0})\oplus(\mathbf{0}\otimes\mathbf{1})\oplus(\mathbf{1}\otimes\mathbf{0})\oplus(\mathbf{1}\otimes\mathbf{1})\\
\\
= & \mathbf{0}\oplus\mathbf{1}\oplus\mathbf{1}\oplus(\mathbf{0}\oplus\mathbf{1}\oplus\mathbf{2}).
\end{array}\label{eq:b7}
\end{equation}
In (\ref{eq:b7}) we have used the usual rule to compose angular momenta,
according to which the composition of two angular momentum $j_{1}$
and $j_{2}$ is given by all angular momentum between $|j_{1}-j_{2}|$
and $j_{1}+j_{2}:$
\begin{equation}
\mathbf{0}\otimes\mathbf{0}=\mathbf{0},\,\,\,\mathbf{0}\otimes\mathbf{1}=\mathbf{1}\otimes\mathbf{0}=\mathbf{1},\,\,\,\mathbf{1}\otimes\mathbf{1}=\mathbf{0}\oplus\mathbf{1}\oplus\mathbf{2}.\label{eq:b8}
\end{equation}
Thus, under the rotation group $SO(3),$ $\varphi^{\mu\nu}$ decomposes
as two spin-$0$ representations, three spin-$1$ representations
and one spin-$2$ representation.\\
It would be interesting to see how these representations are shared
between the symmetric traceless, the trace and the antisymmetric part
of the tensor $\varphi^{\mu\nu},$ since these are irreducible Lorentz
representations. So, let us see how these two different irreducible
decompositions match to each other. \\
The trace is a Lorentz scalar, so it is in particular scalar under
$SO(3)$ and therefore is a $\mathbf{0}$ representation. An antisymmetric
tensor $\psi^{\mu\nu}$ has six components, which can be written as
the direct sum of the two three-vectors $\psi^{0i}$ and $\frac{1}{2}\epsilon^{ijk}\psi^{jk}.$
These are two spatial vectors (two triplets) and so%
\footnote{In Electrodynamics one has an important example of antisymmetric tensor,
i.e the tensor field $F_{\mu\nu}=\partial_{\mu}A_{\nu}-\partial_{\nu}A_{\mu}.$
In this case the two vectors are $E^{i}=-F^{0i}$ and $B^{i}=-\frac{1}{2}\epsilon^{ijk}F^{jk},$
i.e. the electric and magnetic fields.%
}
\begin{equation}
\psi^{\mu\nu}\in\mathbf{1}\oplus\mathbf{1}.\label{eq:b9}
\end{equation}
Since we have identified the trace $h$ with $\mathbf{0}$ and $\psi^{\mu\nu}$
with $\mathbf{1}\oplus\mathbf{1},$ by comparing (\ref{eq:b4}) and
(\ref{eq:b7}) we can see that the nine components of a symmetric
traceless tensor $h^{T\mu\nu}$ decompose, under the subgroup $SO(3),$
as
\begin{equation}
h^{T\mu\nu}\in\mathbf{0}\oplus\mathbf{1}\oplus\mathbf{2}.\label{eq:b10b10}
\end{equation}

\begin{rem}
We have seen that a generic two-rank tensor can be written as a tensor
product of two four-vectors. So let us observe that when we write
$\varphi^{\mu\nu}$ as $(\mathbf{0}\oplus\mathbf{1})\otimes(\mathbf{0}\oplus\mathbf{1}),$
the first $\mathbf{0}$ corresponds to taking the index $\mu=0,$
the first $\mathbf{1}$ corresponds to taking the index $\mu=i,$
and similarly for the second factor $(\mathbf{0}\oplus\mathbf{1})$
and the index $\nu.$ Therefore in equation (\ref{eq:b7}) we have
the following correspondence:
\[
\mathbf{0}\otimes\mathbf{0}\rightarrow\varphi^{00},\,\,\mathbf{0}\otimes\mathbf{1}\rightarrow\varphi^{0i},\,\,\mathbf{1}\otimes\mathbf{0}\rightarrow\varphi^{i0},\,\,\mathbf{1}\otimes\mathbf{1}\rightarrow\varphi^{ij}.
\]
It is clear that, under spatial rotations $SO(3),$ $\varphi^{00}$
behaves like a scalar, while $\varphi^{0i}$ and $\varphi^{i0}$ like
spatial vectors. As for the spatial components$\varphi^{ij},$ its
antisymmetric part $\psi^{ij}=\frac{1}{2}\left(\varphi^{ij}-\varphi^{ji}\right)$
has only three independent component and so it turns out to be a spatial
vector, giving a $3$-dimensional representation $\mathbf{1};$ while
its symmetric part can be separated into its trace, which gives the
second $\mathbf{0}$ representation, and the traceless symmetric part,
which must have%
\footnote{In General Relativity an important example is given by the physical
graviton. It can be described by a traceless symmetric spatial tensor
(transverse to the propagation direction) corresponding to spin-$2.$%
} $j=2.$ \\

In general, a symmetric tensor with $N$ indices contains angular
momentum up to $j=N.$ In four dimensions, higher antisymmetric tensors
with four indices, $\varphi^{\mu\nu\rho\sigma},$ has only one independent
component $\varphi^{0123},$ so it must be a Lorentz scalar. An antisymmetric
tensor with three indices, $\varphi^{\mu\nu\rho},$ has $\frac{4\times3\times2}{3!}=4$
components and it has the same transformation properties of a four-vector.
\end{rem}

\subsection{Tensor decomposition in curved space-time}

Up to now we have only considered the behavior of a Lorentz tensor
and we managed to obtain its decomposition in terms of spin-$0,$
spin-$1$ and spin-$2$ components. We can also do the same with a
generic two-rank tensor (or any rank) defined by means diffeomorphism
group transformations. More generally a two-rank tensor $\varphi^{\mu\nu}$
is an object that transforms in the following way
\begin{equation}
x'^{\mu}\equiv x'^{\mu}(x)\Rightarrow\varphi'^{\mu\nu}(x')=\frac{\partial x'^{\mu}}{\partial x^{\rho}}\frac{\partial x'^{\nu}}{\partial x^{\sigma}}\varphi^{\rho\sigma}(x).\label{eq:b11}
\end{equation}
Also here we can easily see that symmetric and antisymmetric parts
don't mix between them, so first of all $\varphi^{\mu\nu}$ decomposes
in symmetric part $h^{\mu\nu}$ and antisymmetric part $\psi^{\mu\nu}$
as in (\ref{eq:b2}). Then we can isolate the trace component in the
symmetric part $h^{\mu\nu}$ because of $h$ being invariant under
diffeomorphism group, in fact
\begin{equation}
\begin{array}{rrl}
h(x)\equiv g_{\mu\nu}(x)h^{\mu\nu}(x)\Rightarrow & h'(x')= & g_{\mu\nu}(x')h^{\mu\nu}(x')\\
\\
 & = & g_{\mu\nu}(x'){\displaystyle \frac{\partial x'^{\mu}}{\partial x^{\rho}}\frac{\partial x'^{\nu}}{\partial x^{\sigma}}}h^{\rho\sigma}(x)\\
\\
 & = & h(x),
\end{array}=g_{\rho\sigma}(x)h^{\rho\sigma}(x)
\end{equation}
so $h$ turns out to be an invariant also under diffeomorphism transformations.
We have learned that also for a generic tensor, that transforms as
in (\ref{eq:b11}), we can have a decomposition in trace, antisymmetric
and symmetric traceless components.\\
Now we want to study the behavior of $\varphi^{\mu\nu}(x)$ from the
point of view of $SO(3),$ trying to obtain a kind of decomposition
as in (\ref{eq:b7}). We know that a generic rotation transformation
doesn't mix time and space components, i.e. we can define a rotation
in the following way
\begin{equation}
\begin{cases}
x^{0}\rightarrow x'^{0}=x^{0}\\
\bar{x}\rightarrow\bar{x}'=f(\bar{x})
\end{cases}.\label{eq:b13}
\end{equation}
Let us study separately the antisymmetric and the symmetric parts
(see (\ref{eq:b2})): 
\[
\varphi^{\mu\nu}(x)=h^{\mu\nu}(x)+\psi^{\mu\nu}(x).
\]
Recall that $\psi^{\mu\nu}(x)$ has $6$-independent component and
$h^{\mu\nu}(x)$ has $10$-independent components because of their
antisymmetric and symmetric nature. First we can notice that $\psi^{0i}$
and $\psi^{ij}$ are two three-vectors. In fact, by implementing the
transformation (\ref{eq:b13}) one gets
\begin{equation}
\begin{array}{rl}
\psi^{0i}(x')= & {\displaystyle \frac{\partial x'^{0}}{\partial x^{\rho}}\frac{\partial x'^{i}}{\partial x^{\sigma}}\psi^{\rho\sigma}}(x)\\
\\
= & {\displaystyle \frac{\partial x'^{0}}{\partial x^{0}}\frac{\partial x'^{i}}{\partial x^{\sigma}}}\psi^{0\sigma}(x)={\displaystyle \frac{\partial x'^{i}}{\partial x^{j}}}\psi^{0j}(x);
\end{array}\label{eq:b14}
\end{equation}
then we can introduce the other vector in the following way
\begin{equation}
\psi^{k}\coloneqq\frac{1}{2}\varepsilon^{ijk}\psi^{ij}.\label{eq:b15}
\end{equation}
As for the symmetric part we can notice that it decomposes into the
scalar trace component, the scalar component $h^{00},$ the three-vector
component $h^{0i}$ and the three-traceless tensor component $h^{ij}.$
In fact
\begin{equation}
\begin{array}{ll}
h'^{00}(x')={\displaystyle \frac{\partial x'^{0}}{\partial x^{\alpha}}\frac{\partial x'^{0}}{\partial x^{\beta}}}h^{\alpha\beta}(x)={\displaystyle \frac{\partial x'^{0}}{\partial x^{0}}\frac{\partial x'^{0}}{\partial x^{0}}}h^{00}(x)=h^{00}(x) & (\mathrm{scalar}),\\
\\
h'^{0i}(x')={\displaystyle \frac{\partial x'^{0}}{\partial x^{\alpha}}\frac{\partial x'^{i}}{\partial x^{\beta}}}h^{\alpha\beta}(x)={\displaystyle \frac{\partial x'^{0}}{\partial x^{0}}\frac{\partial x'^{i}}{\partial x^{j}}}h^{0j}(x)={\displaystyle \frac{\partial x'^{i}}{\partial x^{j}}}h^{0j}(x) & (3\textrm{-}\mathrm{vector}),\\
\\
h'^{ij}(x')={\displaystyle \frac{\partial x'^{i}}{\partial x^{\alpha}}\frac{\partial x'^{j}}{\partial x^{\beta}}}h^{\alpha\beta}(x)={\displaystyle \frac{\partial x'^{i}}{\partial x^{k}}\frac{\partial x'^{j}}{\partial x^{l}}}h^{kl}(x) & (3\textrm{-}\mathrm{vector}).
\end{array}\label{eq:b16}
\end{equation}
Then we can easily see that the trace $\tilde{h}$%
\footnote{We are using the symbol $\tilde{h}$ to not create confusion with
trace $h$ in four dimensions. Recall that, how we can see in {[}\ref{-Landau-book}{]},
in $3$-dimensions we have to consider the spatial metric tensor defined
as $\gamma_{ij}=-g_{ij}+\frac{g_{0i}g_{0j}}{g_{00}}$ and $\gamma^{ij}=-g^{ij},$
to define and calculate the trace $\tilde{h}$ in $3$-dimensions.
Hence $\tilde{h}=\gamma_{ij}h^{ij}=\gamma^{ij}h_{ij}.$%
} of the three-tensor $h^{ij}$ is an invariant from the point of view
of $SO(3),$
\begin{equation}
\begin{array}{rl}
\tilde{h}'\equiv & -g^{ij}(x')h'_{ij}(x')=-g^{ij}(x'){\displaystyle \frac{\partial x{}^{k}}{\partial x'^{i}}\frac{\partial x{}^{l}}{\partial x'^{j}}}h_{kl}(x)\\
\\
= & -g^{kl}(x)h_{kl}(x)=\tilde{h};
\end{array}
\end{equation}
 where in the last step, since the metric tensor $g_{\mu\nu}(x)$
is symmetric, we have considered $g_{ij}(x)$ as a three-tensor. Hence
we can define a three-traceless tensor as
\begin{equation}
h^{Tij}\coloneqq h^{ij}-\frac{g^{ij}}{g_{kl}g^{kl}}\tilde{h};
\end{equation}
we can also verify that it is traceless:
\begin{equation}
\tilde{h}^{T}=-g^{ij}h_{ij}^{T}=-g^{ij}h_{ij}+\frac{g_{ij}g^{ij}}{g_{kl}g^{kl}}\tilde{h}=\tilde{h}-\tilde{h}=0.\label{eq:b21}
\end{equation}
Hence, finally, because of the equations (\ref{eq:b14}) and (\ref{eq:b15})
for the antisymmetric part, and the equations (\ref{eq:b16})-(\ref{eq:b21})
for the symmetric part, the two-rank tensor $\varphi^{\mu\nu}$ decomposes
into two scalar components, three three-vector components and one
three-traceless tensor component.\\
This kind of decomposition is very important in the theory of cosmological
perturbations where the two-rank tensor that decomposes is the metric
tensor perturbation that, being symmetric, has two scalar, one vector
and one tensor components.

\section{Spin projector operators}

So far we have seen how to decompose a generic two-rank tensor (or
more generally a $N-$rank tensor) into scalar, vector and tensor
components. At this point one question that we can ask could be: can
we define a complete set of projector operators by which we are able
to project the tensor $\varphi^{\mu\nu}$ along its scalar, vector
and tensor components? The answer is ``yes'' and in this section
we shall introduce these useful operators. \\
Furthermore, we are going to introduce also a basis in terms of which
one can decompose any four-rank tensor operator%
\footnote{In this Appendix, when we say ``four-rank tensor operator'' we refer
to the operator $\mathcal{O}^{\mu\nu\rho\sigma}$ that appears in
a given parity-invariant Lagrangian for a two-rank tensor field. %
} $\mathcal{O}^{\mu\nu\rho\sigma}$ appearing in a given parity-invariant
Lagrangian 
\begin{equation}
\mathcal{\mathcal{L}}=\frac{1}{2}\varphi_{\mu\nu}\mathcal{O}^{\mu\nu\rho\sigma}\varphi_{\rho\sigma},\label{eq:b361}
\end{equation}
or in the associated field equations once we consider the presence
of a source $J^{\mu\nu},$ 
\begin{equation}
\mathcal{O}^{\mu\nu\rho\sigma}\varphi_{\rho\sigma}=\lambda J^{\mu\nu},\label{eq:b362}
\end{equation}
where $\lambda$ is the coupling constant. In other words we can say
that, the operator space of the field equations (\ref{eq:b362}) can
be spanned in the basis mentioned above.

Note that in the case of GR we have symmetric tensors $\varphi_{\mu\nu}\rightarrow h_{\mu\nu}$
and $J_{\mu\nu}\rightarrow\tau_{\mu\nu},$ the coupling constant is
given by $\lambda\rightarrow\kappa$ and the operator $\mathcal{O}^{\mu\nu\rho\sigma}$
is symmetric.

\subsection{Four-vector decomposition}

A generic $4\textrm{-}$vector $V^{\mu}$ can be projected along transverse
and longitudinal components, $V^{\mu}\in\mathbf{0}\oplus\mathbf{1},$
and we can perform the projection by introducing a set of two projectors,
$\left\{ \theta,\omega\right\} ,$  in the following way:
\begin{equation}
V_{\mu}=\theta_{\mu\nu}V^{\nu}+\omega_{\mu\nu}V^{\nu},\label{eq:b345}
\end{equation}
where the operators $\theta_{\mu\nu}$ and $\omega_{\mu\nu}$ are
defined as

\begin{equation}
{\displaystyle \begin{cases}
{\displaystyle \theta_{\mu\nu}\coloneqq\eta_{\mu\nu}-\frac{\partial_{\mu}\partial_{\nu}}{\boxempty}}\\
{\displaystyle \omega_{\mu\nu}\coloneqq\frac{\partial_{\mu}\partial_{\nu}}{\boxempty}}
\end{cases}.}\label{eq:4-1}
\end{equation}

These projectors in momentum space are given by

\begin{equation}
\theta_{\mu\nu}=\eta_{\mu\nu}-\frac{k_{\mu}k_{\nu}}{k^{2}},\,\,\,\,\,\omega_{\mu\nu}=\frac{k_{\mu}k_{\nu}}{k^{2}}.\label{eq:4momentum-1}
\end{equation}
It is easy to show that the following properties hold%
\footnote{As we have already said more times, we shall often write the projectors
suppressing the indices.%
}
\begin{equation}
\begin{array}{rl}
 & \theta+\omega=\mathbb{I}\Leftrightarrow\theta_{\mu\nu}+\omega_{\mu\nu}=\eta_{\mu\nu}\\
\\
 & \theta^{2}=\theta,\,\,\,\omega^{2}=\omega,\,\,\,\theta\omega=0\\
\\
\Leftrightarrow & \theta_{\mu\nu}\theta_{\rho}^{\nu}=\theta_{\mu\rho},\,\,\,\omega_{\mu\nu}\omega_{\rho}^{\nu}=\omega_{\mu\rho},\,\,\,\theta_{\mu\nu}\omega_{\rho}^{\nu}=0,
\end{array}\label{eq:5-1}
\end{equation}

namely the set $\{\theta,\omega\}$ turns out to be complete and orthogonal. 

One can also verify that this special decomposition corresponds to
that in which $V^{\mu}$ decomposes in transverse and longitudinal
components. In fact, if $k^{\mu}$ is the $4\textrm{-}$momentum associated
to the electromagnetic wave (or photon) we can immediately see that
\begin{equation}
k^{\mu}\theta_{\mu\nu}=0,\,\,\,\,\, k^{\mu}\omega_{\mu\nu}=k_{\nu};
\end{equation}
hence $\theta$ and $\omega$ project along the transverse and longitudinal
components respectively. \\
Furthermore, we notice that the transverse component has spin-$1$
and the longitudinal one spin-$0$ by calculating the trace of the
two projectors:
\begin{equation}
\begin{array}{rl}
\eta^{\mu\nu}\theta_{\mu\nu}=3=2(1)+1\,\,(\mathrm{spin}\textrm{-}1),\\
\\
\eta^{\mu\nu}\omega_{\mu\nu}=1=2(0)+1\,\,(\mathrm{spin}\textrm{-}0).
\end{array}\label{eq:b350-1}
\end{equation}

The relations (\ref{eq:b350-1}) tell us that (\ref{eq:b345}) corresponds
to the decomposition of a four-vector in terms of spin-$1$ and spin-$0$
components under the rotation group $SO(3),$ i.e. $V^{\mu}\in\mathbf{0}\oplus\mathbf{1}$
(see the Subsection $B.1.2).$

\subsection{Two-rank tensor decomposition}

We know that a two-rank tensor behaves like the tensor product of
two four-vector, so we can find the projector operators for $\varphi^{\mu\nu}$
by decomposing each index separately:
\begin{equation}
\varphi^{\mu\nu}\equiv V^{\mu}\otimes U^{\nu}.\label{eq:b349}
\end{equation}
Moreover, we know that we can decompose $\varphi^{\mu\nu}$ in its
symmetric and antisymmetric parts as done in (\ref{eq:b2}), so one
can study the two parts separately.

$ $

\textbf{Symmetric decomposition}

$ $\\
Let us start with the symmetric part $h_{\mu\nu}\in\mathbf{0}\oplus\mathbf{0}\oplus\mathbf{1}\oplus\mathbf{2}.$
By seeing $h_{\mu\nu}$ as a symmetric tensor product of two four-vectors,
the decomposition can be performed as follow: 
\begin{equation}
\begin{array}{rl}
h_{\mu\nu}= & \left(\theta_{\mu\rho}+\omega_{\mu\rho}\right)\left(\theta_{\nu\sigma}+\omega_{\nu\sigma}\right)h^{\rho\sigma}\\
\\
= & \left(\theta_{\mu\rho}\theta_{\nu\sigma}+\theta_{\mu\rho}\omega_{\nu\sigma}+\omega_{\mu\rho}\theta_{\nu\sigma}+\omega_{\mu\rho}\omega_{\nu\sigma}\right)h^{\rho\sigma}\\
\\
= & \frac{1}{2}\left(\theta_{\mu\rho}\theta_{\nu\sigma}+\theta_{\mu\sigma}\theta_{\nu\rho}\right)h^{\rho\sigma}-\frac{1}{3}\theta_{\mu\nu}\theta_{\rho\sigma}h^{\rho\sigma}\\
\\
 & +\frac{1}{3}\theta_{\mu\nu}\theta_{\rho\sigma}h^{\rho\sigma}+\omega_{\mu\nu}\omega_{\rho\sigma}h^{\rho\sigma}\\
\\
 & +\frac{1}{2}\left(\theta_{\mu\rho}\omega_{\nu\sigma}+\theta_{\mu\sigma}\omega_{\nu\rho}+\theta_{\nu\rho}\omega_{\mu\sigma}+\theta_{\nu\sigma}\omega_{\mu\rho}\right)h^{\rho\sigma}.
\end{array}\label{eq:b350}
\end{equation}
Now we can define the $\mathit{spin}$ $\mathit{projector}$ $\mathit{operators}:$
\footnote{We are labeling the spin-$1$ projector operator also with the letter
$m.$ When we consider only symmetric tensors we can avoid it and
write directly $\mathcal{P}^{1},$ as it has been done in this thesis
(see Chapter $2).$%
}
\begin{equation}
\begin{array}{rl}
\mathcal{P}_{\mu\nu\rho\sigma}^{2}= & \frac{1}{2}\left(\theta_{\mu\rho}\theta_{\nu\sigma}+\theta_{\mu\sigma}\theta_{\nu\rho}\right)-\frac{1}{3}\theta_{\mu\nu}\theta_{\rho\sigma},\\
\\
\mathcal{P}_{m,\,\mu\nu\rho\sigma}^{1}= & \frac{1}{2}\left(\theta_{\mu\rho}\omega_{\nu\sigma}+\theta_{\mu\sigma}\omega_{\nu\rho}+\theta_{\nu\rho}\omega_{\mu\sigma}+\theta_{\nu\sigma}\omega_{\mu\rho}\right),\\
\\
\mathcal{P}_{s,\,\mu\nu\rho\sigma}^{0}= & \frac{1}{3}\theta_{\mu\nu}\theta_{\rho\sigma},\,\,\,\,\,\,\,\mathcal{P}_{w,\,\mu\nu\rho\sigma}^{0}=\omega_{\mu\nu}\omega_{\rho\sigma}.
\end{array}\label{eq:b2.6}
\end{equation}
The set 
\begin{equation}
\mathcal{O}^{i}\equiv\left\{ \mathcal{P}^{2},\mathcal{P}_{m}^{1},\mathcal{P}_{s}^{0},\mathcal{P}_{w}^{0}\right\} ,\,\,\,\,\, i=1,2,3,4,
\end{equation}
forms a complete set of spin projector operators in terms of which
a symmetric two-rank tensor can be decomposed. In fact, one can easily
verify that%
\footnote{We are suppressing the indices, but if we want to be more precise
we should write $\mathcal{O}_{\mu\nu\alpha\beta}^{i}\mathcal{O}_{\alpha\beta\rho\sigma}^{j}=\delta_{ij}\mathcal{O}_{\mu\nu\rho\sigma}^{i}$
or $\mathcal{P}_{a,\,\mu\nu\alpha\beta}^{i}\mathcal{P}_{b,\,\alpha\beta\rho\sigma}^{i}=\delta_{ij}\delta_{ab}\mathcal{P}_{a,\,\mu\nu\rho\sigma}^{i}.$%
} 
\begin{equation}
\mathcal{O}^{i}\mathcal{O}^{j}=\delta_{ij}\mathcal{O}^{i},\,\,\,\,\,\mathcal{O}^{1}+\mathcal{O}^{2}+\mathcal{O}^{3}+\mathcal{O}^{4}=\mathbb{I},\label{eq:b2.7}
\end{equation}
or in terms of $\mathcal{P}'$s 
\begin{equation}
\mathcal{P}_{a}^{i}\mathcal{P}_{b}^{j}=\delta_{ij}\delta_{ab}\mathcal{P}_{a}^{i},\,\,\,\,\,\mathcal{P}^{2}+\mathcal{P}^{1}+\mathcal{P}_{s}^{0}+\mathcal{P}_{w}^{0}=\mathbb{I}.\label{eq:b2.8}
\end{equation}
The second property of (\ref{eq:b2.7}) (or (\ref{eq:b2.8})) has
been already showed when we constructed and defined the set of operators
in (\ref{eq:b350}), but we can also show it explicitly:
\[
\begin{array}{rl}
\mathcal{P}^{2}+\mathcal{P}_{m}^{1}+\mathcal{P}_{s}^{0}+\mathcal{P}_{w}^{0}= & \frac{1}{2}\left(\theta_{\mu\rho}\theta_{\nu\sigma}+\theta_{\mu\sigma}\theta_{\nu\rho}\right)+\omega_{\mu\nu}\omega_{\rho\sigma}\\
\\
 & +\frac{1}{2}\left(\theta_{\mu\rho}\omega_{\nu\sigma}+\theta_{\mu\sigma}\omega_{\nu\rho}+\theta_{\nu\rho}\omega_{\mu\sigma}+\theta_{\nu\sigma}\omega_{\mu\rho}\right)\\
\\
= & \frac{1}{2}\eta_{\nu\sigma}\theta_{\mu\rho}+\frac{1}{2}\eta_{\nu\rho}\theta_{\mu\sigma}+\frac{1}{2}\theta_{\nu\rho}\omega_{\mu\sigma}+\frac{1}{2}\theta_{\nu\sigma}\omega_{\mu\rho}+\omega_{\mu\nu}\omega_{\rho\sigma}\\
\\
= & \frac{1}{2}\left(\eta_{\mu\rho}\eta_{\nu\sigma}+\eta_{\nu\rho}\eta_{\mu\sigma}\right)+\frac{1}{2}\eta_{\nu\rho}\omega_{\mu\sigma}+\frac{1}{2}\eta_{\nu\sigma}\omega_{\mu\rho}\\
\\
 & -\frac{1}{2}\eta_{\nu\sigma}\omega_{\mu\rho}-\frac{1}{2}\eta_{\nu\rho}\omega_{\mu\sigma}\\
\\
= & \frac{1}{2}\left(\eta_{\mu\rho}\eta_{\nu\sigma}+\eta_{\nu\rho}\eta_{\mu\sigma}\right)=\mathbb{I}.
\end{array}
\]
 Hence, we have found a complete set of projector operators to decompose
$h^{\mu\nu}:$ 
\begin{equation}
\begin{array}{rl}
h_{\mu\nu}= & \mathcal{P}_{\mu\nu\rho\sigma}^{2}h^{\rho\sigma}+\mathcal{P}_{m,\,\mu\nu\rho\sigma}^{1}h^{\rho\sigma}+\mathcal{P}_{s,\,\mu\nu\rho\sigma}^{0}h^{\rho\sigma}+\mathcal{P}_{w,\,\mu\nu\rho\sigma}^{0}h^{\rho\sigma}\\
\\
= & \left(\mathcal{P}^{2}+\mathcal{P}_{m}^{1}+\mathcal{P}_{s}^{0}+\mathcal{P}_{w}^{0}\right)_{\mu\nu\rho\sigma}h^{\rho\sigma}.
\end{array}\label{eq:b.356sym}
\end{equation}
Note that to form a basis in terms of which any symmetric four-rank
tensor can be expanded we also need to introduce other two operators
that mix the two scalar components $s$ and $w.$ They are required
to close the algebra of the spin projector operators %
\footnote{Note that we are doing an abuse because we are calling $\mathcal{P}_{sw}^{0}$
and $\mathcal{P}_{sd}^{0}$ projectors, but they are not like that.
In fact this becomes very clear by looking at the orthogonality relations
below, (\ref{eq:b356}). Often we will make this abuse of nomenclature. %
}. These two new operators are defined as follow
\begin{equation}
\mathcal{P}_{sw,\,\mu\nu\rho\sigma}^{0}=\frac{1}{\sqrt{3}}\theta_{\mu\nu}\omega_{\rho\sigma},\,\,\,\,\,\mathcal{P}_{ws,\,\mu\nu\rho\sigma}^{0}=\frac{1}{\sqrt{3}}\omega_{\mu\nu}\theta_{\rho\sigma}.\label{eq:b2.10}
\end{equation}

Now, the orthogonality relations in (\ref{eq:b2.8}) can be extended
to the operators $\mathcal{P}_{sw}^{0}$ and $\mathcal{P}_{ws}^{0},$
so that we obtain (when $a\neq b$ and $c\neq d)$ 
\begin{equation}
\begin{array}{lll}
\mathcal{P}_{a}^{i}\mathcal{P}_{b}^{j}=\delta_{ij}\delta_{ab}\mathcal{P}_{a}^{j}, &  & \mathcal{P}_{ab}^{0}\mathcal{P}_{c}^{i}=\delta_{i0}\delta_{bc}\mathcal{P}_{ab}^{i},\\
\\
\mathcal{P}_{c}^{i}\mathcal{P}_{ab}^{0}=\delta_{i0}\delta_{ac}\mathcal{P}_{ab}^{0}, &  & \mathcal{P}_{ab}^{0}\mathcal{P}_{cd}^{0}=\delta_{ad}\delta_{bc}\mathcal{P}_{a}^{0},
\end{array}\label{eq:b356}
\end{equation}
where $i,j=2,1,0$ and $a,b,c,d=m,s,w,\mathrm{absent}.$%
\footnote{Note that the projector $\mathcal{P}^{2}$ does not have any lower
index, so it can happen that $a,b,c,d$ are absent.%
}\\
Hence, the set $\left\{ \mathcal{P}^{2},\mathcal{P}_{m}^{1},\mathcal{P}_{s}^{0},\mathcal{P}_{w}^{0},\mathcal{P}_{sw}^{0},\mathcal{P}_{ws}^{0}\right\} $
forms a basis of symmetric four-rank tensors.
\begin{rem}
Since we have applied the formalism of the spin projector operators
to gravity theories, it is worth observing that the basis of projectors represents six field degrees of freedom. The other four fields in a symmetric tensor field, as
usual, represent the gauge (unphysical) degrees of freedom. $\mathcal{P}^{2}$
and $\mathcal{P}^{1}$ represent transverse and traceless spin-$2$
and spin-$1$ degrees, accounting for four degrees of freedom, while
$\mathcal{P}_{s}^{0}$ and $\mathcal{P}_{w}^{0}$ represent the spin-$0$
scalar multiplets. In addition to the four projectors we also need to introduce the operators $\mathcal{P}_{sw}^{0}$ and
$\mathcal{P}_{ws}^{0}$ which are necessary to close the algebra and form a basis of operators acting in the space of the symmetric two-rank tensors.
From the relations (\ref{eq:b356}) we notice that $\mathcal{P}_{sw}^{0}$
and $\mathcal{P}_{ws}^{0}$ are not projector operators, but transition
operators that mix the two spin-$0$ projector operators, $s$ and
$w.$
\end{rem}
We can also verify that $\mathcal{P}^{2}$ is traceless and transverse,
in fact:
\begin{equation}
\begin{array}{rl}
\eta^{\mu\nu}\mathcal{P}_{\mu\nu\rho\sigma}^{2}h^{\rho\sigma}= & \left[\frac{1}{2}\left(\eta^{\mu\nu}\theta_{\mu\rho}\theta_{\nu\sigma}+\eta^{\mu\nu}\theta_{\mu\sigma}\theta_{\nu\rho}\right)-\frac{1}{3}\eta^{\mu\nu}\theta_{\mu\nu}\theta_{\rho\sigma}\right]h^{\rho\sigma}\\
\\
= & \left[\frac{1}{2}\left(\theta_{\rho}^{\nu}\theta_{\nu\sigma}+\theta_{\sigma}^{\nu}\theta_{\nu\rho}\right)-\frac{1}{3}\left(4-1\right)\theta_{\rho\sigma}\right]h^{\rho\sigma}\\
\\
= & \left[\frac{1}{2}\left(\theta_{\rho\sigma}+\theta_{\rho\sigma}\right)-\theta_{\rho\sigma}\right]h^{\rho\sigma}=0
\end{array}
\end{equation}
Then
\begin{equation}
k^{\mu}\mathcal{P}_{\mu\nu\rho\sigma}^{2}h^{\rho\sigma}=\left[\frac{1}{2}\left(k^{\mu}\theta_{\mu\rho}\theta_{\nu\sigma}+k^{\mu}\theta_{\mu\sigma}\theta_{\nu\rho}\right)-\frac{1}{3}k^{\mu}\theta_{\mu\nu}\theta_{\rho\sigma}\right]h^{\rho\sigma};\label{eq:b2.13}
\end{equation}
since $k^{\mu}\theta_{\mu\rho}=k^{\mu}\theta_{\mu\sigma}=k^{\mu}\theta_{\mu\nu}=0,$
(\ref{eq:b2.13}) becomes $k^{\mu}\mathcal{P}_{\mu\nu\rho\sigma}^{2}h^{\rho\sigma}=0.$
\\
With this choice, none among the operators $\mathcal{P}_{s}^{0},\mathcal{P}_{w}^{0},\mathcal{P}_{sw}^{0}$
and $\mathcal{P}_{ws}^{0}$ corresponds to the trace operator: we
shall show how to construct an operator that acting on the tensor
field $h_{\mu\nu}$ gives us the trace $h$ at the end of the section
$B.3.$ 

$ $\\
\textbf{Antisymmetric decomposition}

$ $\\
Let us now work on the antisymmetric part $\psi^{\mu\nu}\in\mathbf{1}\oplus\mathbf{1}.$
By proceeding as we have already done for the symmetric part, we get:

\begin{equation}
\begin{array}{rl}
\psi_{\mu\nu}= & \left(\theta_{\mu\rho}+\omega_{\mu\rho}\right)\left(\theta_{\nu\sigma}+\omega_{\nu\sigma}\right)\psi^{\rho\sigma}\\
\\
= & \left(\theta_{\mu\rho}\theta_{\nu\sigma}+\theta_{\mu\rho}\omega_{\nu\sigma}+\omega_{\mu\rho}\theta_{\nu\sigma}+\omega_{\mu\rho}\omega_{\nu\sigma}\right)\psi^{\rho\sigma}\\
\\
= & \frac{1}{2}\left(\theta_{\mu\rho}\theta_{\nu\sigma}-\theta_{\mu\sigma}\theta_{\nu\rho}\right)\psi^{\rho\sigma}\\
\\
 & +\frac{1}{2}\left(\theta_{\mu\rho}\omega_{\nu\sigma}-\theta_{\mu\sigma}\omega_{\nu\rho}-\theta_{\nu\rho}\omega_{\mu\sigma}+\theta_{\nu\sigma}\omega_{\mu\rho}\right)\psi^{\rho\sigma}
\end{array}
\end{equation}
We can define the spin projector operators for the antisymmetric part
as follow:
\begin{equation}
\begin{array}{rl}
\mathcal{P}_{b,\,\mu\nu\rho\sigma}^{1}= & \frac{1}{2}\left(\theta_{\mu\rho}\theta_{\nu\sigma}-\theta_{\mu\sigma}\theta_{\nu\rho}\right),\\
\\
\mathcal{P}_{e,\,\mu\nu\rho\sigma}^{1}= & \frac{1}{2}\left(\theta_{\mu\rho}\omega_{\nu\sigma}-\theta_{\mu\sigma}\omega_{\nu\rho}-\theta_{\nu\rho}\omega_{\mu\sigma}+\theta_{\nu\sigma}\omega_{\mu\rho}\right).
\end{array}
\end{equation}
 Thus, we obtain

\begin{equation}
\begin{array}{rl}
\psi_{\mu\nu}= & \mathcal{P}_{b,\,\mu\nu\rho\sigma}^{1}\psi^{\rho\sigma}+\mathcal{P}_{e,\,\mu\nu\rho\sigma}^{1}\psi^{\rho\sigma}\\
\\
= & \left(\mathcal{P}_{b}^{1}+\mathcal{P}_{e}^{1}\right)_{\mu\nu\rho\sigma}\psi^{\rho\sigma}
\end{array}\label{eq:b363}
\end{equation}
The set $\left\{ \mathcal{P}_{b}^{1},\mathcal{P}_{e}^{1}\right\} $
is complete and allow us to project every antisymmetric tensor along
its two vector components. Observe that the letters $b$ and $e$
refer, respectively, to $\mathit{magnetic}$ spin-$1$ and $\mathit{electric}$
spin-$1,$ due to the fact that in electrodynamics the same happens
with the antisymmetric tensor $F^{\mu\nu}=\partial^{\mu}A^{\nu}-\partial^{\nu}A^{\mu}.$
\\
Note that, in the antisymmetric case, the completeness relation is
given by
\begin{equation}
\left(\mathcal{P}_{b}^{1}+\mathcal{P}_{e}^{1}\right)_{\mu\nu\rho\sigma}=\frac{1}{2}\left(\eta_{\mu\rho}\eta_{\nu\sigma}-\eta_{\mu\sigma}\eta_{\nu\rho}\right).
\end{equation}

$ $

\textbf{Full decomposition}

$ $

We are now able to decompose any two-rank tensor $\varphi^{\mu\nu}$
along the spin components corresponding to the irreducible representations
of the group $SO(3),$
\[
\varphi^{\mu\nu}\in\mathbf{1}\oplus\mathbf{1}\oplus\mathbf{0}\oplus\mathbf{0}\oplus\mathbf{1}\oplus\mathbf{2},
\]
in terms of the spin projector operators. Indeed, we can extend the
symmetric set $\left\{ \mathcal{P}^{2},\mathcal{P}_{m}^{1},\mathcal{P}_{s}^{0},\mathcal{P}_{w}^{0}\right\} $
by including the antisymmetric part $\left\{ \mathcal{P}_{b}^{1},\mathcal{P}_{e}^{1}\right\} .$
Thus any two-rank tensor can be decomposed in terms of the complete
set of spin projectors operators, 
\begin{equation}
\mathcal{O}^{i}\equiv\left\{ \mathcal{P}^{2},\mathcal{P}_{m}^{1},\mathcal{P}_{s}^{0},\mathcal{P}_{w}^{0},\mathcal{P}_{b}^{1},\mathcal{P}_{e}^{1}\right\} ,\,\,\,\, i=1,2,3,4,5,6,
\end{equation}
in the following way:
\begin{equation}
\begin{array}{rl}
\varphi^{\mu\nu}= & \mathcal{P}_{\mu\nu\rho\sigma}^{2}\varphi^{\rho\sigma}+\mathcal{P}_{m,\,\mu\nu\rho\sigma}^{1}\varphi^{\rho\sigma}+\mathcal{P}_{s,\,\mu\nu\rho\sigma}^{0}\varphi^{\rho\sigma}\\
\\
 & +\mathcal{P}_{w,\,\mu\nu\rho\sigma}^{0}\varphi^{\rho\sigma}+\mathcal{P}_{b,\,\mu\nu\rho\sigma}^{1}\varphi^{\rho\sigma}+\mathcal{P}_{e,\,\mu\nu\rho\sigma}^{1}\varphi^{\rho\sigma}\\
\\
= & \left(\mathcal{P}^{2}+\mathcal{P}_{m}^{1}+\mathcal{P}_{s}^{0}+\mathcal{P}_{w}^{0}+\mathcal{P}_{b}^{1}+\mathcal{P}_{e}^{1}\right)_{\mu\nu\rho\sigma}\varphi^{\rho\sigma}.
\end{array}
\end{equation}

We are also interested to form a basis in terms of which any four-rank
tensor can be expanded. We have already seen that for the symmetric
part we needed to define two operators that mix the scalar components.
To complete the full basis we need to introduce other two spin-$1$
operator that mix the spin-$1$ components. They are defined as  
\begin{equation}
\begin{array}{rl}
\mathcal{P}_{me,\,\mu\nu\rho\sigma}^{1}= & \frac{1}{2}\left(\theta_{\mu\rho}\omega_{\nu\sigma}-\theta_{\mu\sigma}\omega_{\nu\rho}+\theta_{\nu\rho}\omega_{\mu\sigma}-\theta_{\nu\sigma}\omega_{\mu\rho}\right),\\
\\
\mathcal{P}_{em,\,\mu\nu\rho\sigma}^{1}= & \frac{1}{2}\left(\theta_{\mu\rho}\omega_{\nu\sigma}+\theta_{\mu\sigma}\omega_{\nu\rho}-\theta_{\nu\rho}\omega_{\mu\sigma}-\theta_{\nu\sigma}\omega_{\mu\rho}\right).
\end{array}
\end{equation}

In this way we have closed the algebra and formed the basis of four-rank
tensors 
\begin{equation}
\left\{ \mathcal{P}^{2},\mathcal{P}_{m}^{1},\mathcal{P}_{s}^{0},\mathcal{P}_{w}^{0},\mathcal{P}_{e}^{1},\mathcal{P}_{b}^{1},\mathcal{P}_{sw}^{0},\mathcal{P}_{ws}^{0},\mathcal{P}_{em}^{1},\mathcal{P}_{me}^{1}\right\} .\label{eq:b2.18}
\end{equation}
It easy to show that the following orthogonal relations hold (when
$a\neq b$ and $c\neq d)$ 
\begin{equation}
\begin{array}{lll}
\mathcal{P}_{a}^{i}\mathcal{P}_{b}^{j}=\delta_{ij}\delta_{ab}\mathcal{P}_{a}^{j}, &  & \mathcal{P}_{ab}^{0}\mathcal{P}_{c}^{i}=\delta_{i0}\delta_{bc}\mathcal{P}_{ab}^{i},\\
\\
\mathcal{P}_{c}^{i}\mathcal{P}_{ab}^{0}=\delta_{i0}\delta_{ac}\mathcal{P}_{ab}^{0}, &  & \mathcal{P}_{ab}^{0}\mathcal{P}_{cd}^{0}=\delta_{ad}\delta_{bc}\mathcal{P}_{a}^{0},
\end{array}\label{eq:b2.19}
\end{equation}
where $i,j=2,1,0$ and $a,b,c,d=m,s,w,b,e,\mathrm{absent}.$ Hence,
the introduction of the four operators%
\footnote{No operators which connect electric and magnetic spin-$1$ spaces
$(\mathcal{P}_{eb}^{1}$ and $\mathcal{P}_{be}^{1}),$ nor operators
which connect the third pair of spin-$1$ spaces $(\mathcal{P}_{bm}^{1}$
and $\mathcal{P}_{mb}^{1}).$ To understand why these operators are
not needed we have to observe that the four-rank tensor operators
we want to expand in the full basis is present in the Lagrangians
and so in the associated field equations. For instance, given the
following free Lagrangian 
\[
\mathcal{L}=\frac{1}{2}\varphi_{\mu\nu}\mathcal{O}^{\mu\nu\rho\sigma}\varphi_{\rho\sigma},
\]
we need to expand the operator $\mathcal{O}^{\mu\nu\rho\sigma}$ in
terms of the full basis (\ref{eq:b2.18}). Now, if the Lagrangian
is invariant under parity transformations the presence of such transition
operators is excluded. While, a parity-violation case would bring
to the presence of terms like $\epsilon_{\mu\nu\rho\sigma}\varphi^{\mu\nu}\varphi^{\rho\sigma}$
or $\epsilon_{\mu\nu\rho\sigma}\varphi^{\mu\nu}\partial^{\rho}\psi^{\sigma}$
in the Lagrangian and so the operators $\mathcal{P}_{eb}^{1},\mathcal{P}_{be}^{1},\mathcal{P}_{mb}^{1}$
and $\mathcal{P}_{bm}^{1}$ would appear. See Ref. {[}\ref{-P.-Van}{]}
for more details.%
} $\mathcal{P}_{ws}^{0},\mathcal{P}_{sw}^{0},\mathcal{P}_{em}^{1},\mathcal{P}_{me}^{1}$
is important to satisfy the relations (\ref{eq:b2.19}) that define
the algebra of the operators.\\
Note that for the full decomposition the completeness relations takes
into account both symmetric and antisymmetric part, and it is given
by
\begin{equation}
\begin{array}{rl}
{\displaystyle \left(\mathcal{P}^{2}+\mathcal{P}_{m}^{1}+\mathcal{P}_{s}^{0}+\mathcal{P}_{w}^{0}+\mathcal{P}_{b}^{1}+\mathcal{P}_{e}^{1}\right)_{\mu\nu\rho\sigma}}= & {\displaystyle \frac{1}{2}\left(\eta_{\mu\rho}\eta_{\nu\sigma}+\eta_{\mu\sigma}\eta_{\nu\rho}\right)}\\
\\
 & +{\displaystyle \frac{1}{2}}\left(\eta_{\mu\rho}\eta_{\nu\sigma}-\eta_{\mu\sigma}\eta_{\nu\rho}\right)\\
\\
= & \eta_{\mu\rho}\eta_{\nu\sigma}.
\end{array}
\end{equation}

$ $

We can also find the spin value of each spin projector operators by
contracting with the identity matrix $\eta_{\mu\rho}\eta_{\nu\sigma}.$%
\footnote{As we have already seen considering symmetric and antisymmetric decompositions,
the identity matrix in the symmetric (antisymmetric) case can be rewritten
as ${\displaystyle \frac{1}{2}\left(\eta_{\mu\rho}\eta_{\nu\sigma}+\eta_{\mu\sigma}\eta_{\nu\rho}\right)}$
$({\displaystyle \frac{1}{2}}\left(\eta_{\mu\rho}\eta_{\nu\sigma}-\eta_{\mu\sigma}\eta_{\nu\rho}\right)).$ %
} Indeed the following relation holds:
\begin{equation}
\left(\eta^{\mu\rho}\eta^{\nu\sigma}\right)\mathcal{P}_{\mu\nu\rho\sigma}^{j}=2(j)+1,
\end{equation}
where $j$ is the spin associated to the spin projector operator $\mathcal{P}^{j}.$
Note that, because of the symmetry, the product can also read as $\eta^{\mu\rho}\eta^{\nu\sigma}\mathcal{P}^{j}.$
Hence we can easily verify that:
\begin{equation}
\begin{array}{ll}
\eta^{\mu\rho}\eta^{\nu\sigma}\mathcal{P}_{\mu\nu\rho\sigma}^{2}=5=2(2)+1 & (\mathrm{spin\textrm{-}2),}\\
\\
\eta^{\mu\rho}\eta^{\nu\sigma}\mathcal{P}_{m,\,\mu\nu\rho\sigma}^{1}=3=2(1)+1 & (\mathrm{spin\textrm{-}1),}\\
\\
\eta^{\mu\rho}\eta^{\nu\sigma}\mathcal{P}_{s,\,\mu\nu\rho\sigma}^{0}=1=2(0)+1 & (\mathrm{spin\textrm{-}0),}\\
\\
\eta^{\mu\rho}\eta^{\nu\sigma}\mathcal{P}_{w,\,\mu\nu\rho\sigma}^{0}=1=2(0)+1 & (\mathrm{spin\textrm{-}0),}\\
\\
\eta^{\mu\rho}\eta^{\nu\sigma}\mathcal{P}_{b,\,\mu\nu\rho\sigma}^{1}=3=2(1)+1 & (\mathrm{spin\textrm{-}1),}\\
\\
\eta^{\mu\rho}\eta^{\nu\sigma}\mathcal{P}_{e,\,\mu\nu\rho\sigma}^{1}=3=2(1)+1 & (\mathrm{spin\textrm{-}1).}
\end{array}\label{eq:b2.21}
\end{equation}
Note that the relations (\ref{eq:b2.21}) don't hold for the operators
$\mathcal{P}_{sw,}^{0}$ $\mathcal{P}_{ws,}^{0}$ $\mathcal{P}_{me,}^{1}$
$\mathcal{P}_{em}^{1}$ because they are not projectors as we have
already pointed out.

The relations (\ref{eq:b2.21}) tell us that (\ref{eq:b.356sym})
corresponds to the decomposition of a symmetric two-rank tensor in
terms of one spin-$2,$ one spin-$1$ and two spin-$0$ components
under the rotation group $SO(3),$ i.e. $h^{\mu\nu}\in\mathbf{0}\oplus\mathbf{0}\oplus\mathbf{1}\oplus\mathbf{2};$
while (\ref{eq:b363}) corresponds to the decomposition of an antisymmetric
tensor in terms of two spin-$1$ components, i.e. $h^{\mu\nu}\in\mathbf{1}\oplus\mathbf{1}$
(see the Subsection $B.1.2).$ 
\begin{rem}
The basis (\ref{eq:b2.18}) is important, for instance, when we want
to determine the propagator associated to any Lagrangian. We have
used only the symmetric space in this thesis, but in general we can
have a general Lagrangian that required the use of the complete basis
containing also the antisymmetric operators {[}\ref{-P.-Van}{]}.
Thus in the general case, to invert the operators $\mathcal{O}$ of
any Lagrangian (see Chapter $2)$ we need to expand it in terms of
the symmetric and the antisymmetric spin projector operators, and
to invert the field equations we have to act with both kind of operators,
i.e. with the full basis (\ref{eq:b2.18}). In formula, the equations
(\ref{eq:b361}) and (\ref{eq:b362}) can be recast in terms of the
spin projector operators as follow. First of all let us recall the
full set of the projectors as
\begin{equation}
\mathcal{O}^{i}\equiv\left\{ \mathcal{P}^{2},\mathcal{P}_{m}^{1},\mathcal{P}_{s}^{0},\mathcal{P}_{w}^{0},\mathcal{P}_{b}^{1},\mathcal{P}_{e}^{1},\mathcal{P}_{ws}^{0},\mathcal{P}_{sw}^{0},\mathcal{P}_{em}^{1},\mathcal{P}_{me}^{1}\right\} ,\,\,\,\, i=1,\ldots,10.
\end{equation}

We are now able to expand the operator $\mathcal{O}$ in a compact
way:

\begin{equation}
\begin{array}{rl}
\mathcal{L}= & {\displaystyle \frac{1}{2}}\varphi_{\mu\nu}\mathcal{O}^{\mu\nu\rho\sigma}\varphi_{\rho\sigma}\\
\\
= & {\displaystyle \frac{1}{2}}\varphi_{\mu\nu}{\displaystyle \left(\sum_{i=1}^{10}C_{i}\mathcal{O}^{i,\,\mu\nu\rho\sigma}\right)}\varphi_{\rho\sigma},
\end{array}\label{eq:b391}
\end{equation}

or, in other words,we can say that the operator space of the field
equations can be spanned as

\begin{equation}
\begin{array}{rl}
 & \mathcal{O}^{\mu\nu\rho\sigma}\varphi_{\rho\sigma}=\lambda J^{\mu\nu}\\
\\
\Leftrightarrow & {\displaystyle \left(\sum_{i=1}^{10}C_{i}\mathcal{O}^{i,\,\mu\nu\rho\sigma}\right)}\varphi_{\rho\sigma}=\lambda{\displaystyle \left(\sum_{i=1}^{6}C_{i}\mathcal{O}^{i,\,\mu\nu\rho\sigma}\right)}J_{\rho\sigma},
\end{array}\label{eq:b392}
\end{equation}
where the coefficients $C_{i}$ are defined by the specific Lagrangian
we are considering. \\
Now, for instance, one could determine the propagator for the generic
Lagrangian (\ref{eq:b391}) by following the prescription we introduced
in Chapter $2.$ This case would is more general because the Lagrangian
is neither symmetric or antisymmetric, but it has both parts. See
Ref. {[}\ref{-P.-Van}{]} for more details. 
\end{rem}
$ $

\textbf{Useful relations}

$ $\\
Now we want to list some important relations between spin projector
operators, among which some of them turned out to be very useful to
rewrite the Lagrangians and the field equations in terms of the spin
projector operators. These relations are:
\begin{equation}
\begin{array}{l}
\frac{1}{2}\left(\eta_{\mu\rho}\eta_{\nu\sigma}+\eta_{\mu\sigma}\eta_{\nu\rho}\right)=\left(\mathcal{P}^{2}+\mathcal{P}^{1}+\mathcal{P}_{s}^{0}+\mathcal{P}_{w}^{0}\right)_{\mu\nu\rho\sigma},\\
\\
\eta_{\mu\nu}\omega_{\sigma\rho}+\eta_{\rho\sigma}\omega_{\mu\nu}=\left(\sqrt{3}\left(\mathcal{P}_{sw}^{0}+\mathcal{P}_{ws}^{0}\right)+2\mathcal{P}_{w}^{0}\right)_{\mu\nu\rho\sigma},\\
\\
\frac{1}{2}\left(\eta_{\mu\rho}\omega_{\nu\sigma}+\eta_{\mu\sigma}\omega_{\nu\rho}+\eta_{\nu\sigma}\omega_{\mu\rho}+\eta_{\nu\rho}\omega_{\mu\sigma}\right)=\left(\mathcal{P}^{1}+2\mathcal{P}_{w}^{0}\right)_{\mu\nu\rho\sigma},\\
\\
\eta_{\mu\nu}\eta_{\rho\sigma}=\left(3\mathcal{P}_{s}^{0}+\mathcal{P}_{w}^{0}+\sqrt{3}\left(\mathcal{P}_{sw}^{0}+\mathcal{P}_{ws}^{0}\right)\right)_{\mu\nu\rho\sigma},\\
\\
\mathcal{P}_{\mu\nu\rho\sigma}^{2}=\frac{1}{2}\left(\eta_{\mu\rho}\eta_{\nu\sigma}+\eta_{\mu\sigma}\eta_{\nu\rho}\right)-\frac{1}{3}\eta_{\mu\nu}\eta_{\rho\sigma}-\left[\mathcal{P}_{m}^{1}+\frac{2}{3}\mathcal{P}_{w}^{0}-\frac{1}{\sqrt{3}}\left(\mathcal{P}_{sw}^{0}+\mathcal{P}_{ws}^{0}\right)\right]_{\mu\nu\rho\sigma},\\
\\
\mathcal{P}_{s,\,\mu\nu\rho\sigma}^{0}=\frac{1}{3}\eta_{\mu\nu}\eta_{\rho\sigma}-\frac{1}{3}\left[\mathcal{P}_{w}^{0}+\sqrt{3}\left(\mathcal{P}_{sw}^{0}+\mathcal{P}_{ws}^{0}\right)\right]_{\mu\nu\rho\sigma}.
\end{array}\label{eq:b384rel}
\end{equation}

\section{Examples of metric decompositions}

In this section we shall see examples of tensor decomposition, in
particular the metric perturbation decomposition, so we shall deal
only with symmetric tensor.\\
Let us find the decomposition for the choice, (\ref{eq:b2.6}) and
(\ref{eq:b2.10}), we have made for the complete set of spin projector
operators. We can note that the following metric decomposition can
be written in terms of our set of spin projector operators matching
metric components with spin projector components. This metric decomposition
is
\begin{equation}
h_{\mu\nu}=h_{\mu\nu}^{TT}+\frac{1}{2}\left(\partial_{\mu}\xi_{\nu}^{T}+\partial_{\nu}\xi_{\mu}^{T}\right)+\left(\boxempty\eta_{\mu\nu}-\partial_{\mu}\partial_{\nu}\right)s+\partial_{\mu}\partial_{\nu}w,\label{eq:b3.1}
\end{equation}
with $h_{\mu\nu}^{TT}$ transverse and traceless with respect to the
Lorentz indices,
\begin{equation}
\partial^{\mu}h_{\mu\nu}^{TT}=0,\,\,\,\,\,\eta^{\mu\nu}h_{\mu\nu}^{TT}=0;
\end{equation}
and $\partial^{\mu}\xi_{\mu}^{T}=0.$\\
To match with our choice of spin projector operators (\ref{eq:b2.6})
and (\ref{eq:b2.10}) we have to choose:
\begin{equation}
\begin{array}{rl}
\xi_{\mu}^{T}\coloneqq & \displaystyle{\frac{2}{\boxempty}}\left(\eta_{\mu\nu}-{\displaystyle \frac{\partial_{\mu}\partial_{\nu}}{\boxempty}}\right)\partial_{\sigma}h^{\nu\sigma},\\
\\
s\coloneqq & \displaystyle{\frac{1}{3}\frac{1}{\boxempty}}\left(\eta_{\mu\nu}-{\displaystyle \frac{\partial_{\mu}\partial_{\nu}}{\boxempty}}\right)h^{\mu\nu},\\
\\
w\coloneqq & {\displaystyle \frac{\partial_{\mu}\partial_{\nu}}{\boxempty}}h^{\mu\nu};\\
\\
h_{\mu\nu}^{TT}\coloneqq & h_{\mu\nu}-{\displaystyle \frac{\partial_{\mu}}{\boxempty}}\left({\displaystyle \eta_{\nu\rho}}-{\displaystyle \frac{\partial_{\nu}\partial_{\rho}}{\boxempty}}\right)\partial_{\sigma}h^{\rho\sigma}-{\displaystyle \frac{\partial_{\nu}}{\boxempty}\left(\eta_{\mu\rho}-{\displaystyle \frac{\partial_{\mu}\partial_{\rho}}{\boxempty}}\right)}\partial_{\sigma}h^{\rho\sigma}\\
\\
 & -{\displaystyle \frac{1}{3}\left(\eta_{\mu\nu}-\frac{\partial_{\mu}\partial_{\nu}}{\boxempty}\right)\left(\eta_{\rho\sigma}-\frac{\partial_{\rho}\partial_{\sigma}}{\boxempty}\right)h^{\rho\sigma}-\partial_{\mu}\partial_{\nu}\frac{\partial_{\rho}\partial_{\sigma}}{\boxempty}h^{\rho\sigma}.}
\end{array}\label{eq:b3.3}
\end{equation}
Let us rewrite the equations (\ref{eq:b3.1})-(\ref{eq:b3.3}) in
 momentum space:
\begin{equation}
h_{\mu\nu}=h_{\mu\nu}^{TT}+\frac{i}{2}\left(k_{\mu}\xi_{\nu}^{T}+k_{\nu}\xi_{\mu}^{T}\right)+\left(-k^{2}\eta_{\mu\nu}+k_{\mu}k_{\nu}\right)s-k_{\mu}k_{\nu}w;
\end{equation}
\begin{equation}
\begin{array}{rl}
\xi_{\mu}^{T}= & {\displaystyle -\frac{2i}{k^{2}}\left(\eta_{\mu\nu}-\frac{k_{\mu}k_{\nu}}{k^{2}}\right)k_{\sigma}h^{\nu\sigma},}\\
\\
s= & -{\displaystyle \frac{1}{3}\frac{1}{k^{2}}\left(\eta_{\mu\nu}-\frac{k_{\mu}k_{\nu}}{k^{2}}\right)h^{\mu\nu},}\\
\\
w= & {\displaystyle \frac{k_{\mu}k_{\nu}}{k^{2}}}h^{\mu\nu},\\
\\
h_{\mu\nu}^{TT}= & h_{\mu\nu}{\displaystyle -\frac{k_{\mu}}{k^{2}}\left(\eta_{\nu\rho}-\frac{k_{\nu}k_{\rho}}{k^{2}}\right)k_{\sigma}h^{\rho\sigma}-\frac{k_{\nu}}{k^{2}}\left(\eta_{\mu\rho}-\frac{k_{\mu}k_{\rho}}{k^{2}}\right)k_{\sigma}h^{\rho\sigma}}\\
\\
 & +{\displaystyle \frac{1}{3}\frac{\left(-k^{2}\eta_{\mu\nu}+k_{\mu}k_{\nu}\right)}{k^{2}}\left(\eta_{\rho\sigma}-\frac{k_{\rho}k_{\sigma}}{k^{2}}\right)h^{\rho\sigma}-\frac{k_{\mu}k_{\nu}k_{\rho}k_{\sigma}}{k^{2}}h^{\rho\sigma}}
\end{array}
\end{equation}
Now we shall identify each term in $h_{\mu\nu}$ with the corresponding
spin projector operator. 
\begin{equation}
\begin{array}{rl}
\frac{1}{2}\left(\partial_{\mu}\xi_{\nu}^{T}+\partial_{\nu}\xi_{\mu}^{T}\right)\rightarrow & {\displaystyle \frac{k_{\mu}}{k^{2}}\left(\eta_{\nu\rho}-\frac{k_{\nu}k_{\rho}}{k^{2}}\right)k_{\sigma}h^{\rho\sigma}+\frac{k_{\nu}}{k^{2}}\left(\eta_{\mu\rho}-\frac{k_{\mu}k_{\rho}}{k^{2}}\right)k_{\sigma}h^{\rho\sigma}}\\
\\
= & \left(\theta_{\mu\rho}\theta_{\nu\sigma}+\theta_{\mu\sigma}\theta_{\nu\rho}\right)h^{\rho\sigma}\\
\\
= & \frac{1}{2}\left(\theta_{\mu\rho}\omega_{\nu\sigma}+\theta_{\mu\sigma}\omega_{\nu\rho}+\theta_{\nu\rho}\omega_{\mu\sigma}+\theta_{\nu\sigma}\omega_{\mu\rho}\right)h^{\rho\sigma}\\
\\
= & \mathcal{P}_{\mu\nu\rho\sigma}^{1}h^{\rho\sigma};
\end{array}
\end{equation}
\begin{equation}
\begin{array}{rl}
\left(\boxempty\eta_{\mu\nu}-\partial_{\mu}\partial_{\nu}\right)s\rightarrow & {\displaystyle -\frac{1}{3}\frac{\left(-k^{2}\eta_{\mu\nu}+k_{\mu}k_{\nu}\right)}{k^{2}}\left(\eta_{\rho\sigma}-\frac{k_{\rho}k_{\sigma}}{k^{2}}\right)h^{\rho\sigma}}\\
\\
= & \frac{1}{3}\theta_{\mu\nu}\theta_{\rho\sigma}h^{\rho\sigma}\\
\\
= & \mathcal{P}_{s,\,\mu\nu\rho\sigma}^{0}h^{\rho\sigma};
\end{array}
\end{equation}

$ $

\[
\partial_{\mu}\partial_{\nu}r\rightarrow\frac{k_{\mu}k_{\nu}k_{\rho}k_{\sigma}}{k^{2}}h^{\rho\sigma}=\omega_{\mu\nu}\omega_{\rho\sigma}h^{\rho\sigma}=\mathcal{P}_{w,\,\mu\nu\rho\sigma}h^{\rho\sigma};
\]
and as for $h_{\mu\nu}^{TT},$ of course, we have 
\begin{equation}
h_{\mu\nu}^{TT}\rightarrow\mathcal{P}_{\mu\nu\rho\sigma}^{2}h^{\rho\sigma}.
\end{equation}
Hence we have decomposed the metric perturbation (\ref{eq:b3.1})
in terms of the spin projection operators and we matched each of them
with the corresponding scalar, vector and tensor components:
\begin{equation}
\mathcal{P}^{2}\leftrightarrow h_{\mu\nu}^{TT},\,\,\,\,\mathcal{P}^{1}\leftrightarrow\xi_{\mu}^{T},\,\,\,\,\mathcal{P}_{s}^{0}\leftrightarrow s,\,\,\,\,\mathcal{P}_{w}^{0}\leftrightarrow w.
\end{equation}
$ $\\
Let us consider another example of decomposition:
\begin{equation}
h_{\mu\nu}=h_{\mu\nu}^{TT}+\frac{1}{2}\left(\partial_{\mu}\xi_{\nu}^{T}+\partial_{\nu}\xi_{\mu}^{T}\right)+\partial_{\mu}\partial_{\nu}\alpha+\frac{1}{3}\eta_{\mu\nu}\beta,
\end{equation}
where $h_{\mu\nu}^{TT}$ is always traceless and transverse and $\xi_{\mu}^{T}$
transverse. The definition of $\xi_{\mu}^{T}$ is the same of the
previous example as in (\ref{eq:b3.3}), instead the two scalar components
are defined as 
\begin{equation}
\begin{array}{rl}
\alpha\coloneqq & {\displaystyle \left(\eta_{\mu\nu}-\frac{\partial_{\mu}\partial_{\nu}}{\boxempty}\right)h^{\mu\nu},}\\
\\
\beta\coloneqq & -{\displaystyle \frac{1}{3}\frac{1}{\boxempty}\left(\eta_{\mu\nu}-4\frac{\partial_{\mu}\partial_{\nu}}{\boxempty}\right)}h^{\mu\nu}.
\end{array}
\end{equation}
By following the procedure showed in the first example we can obtain
\begin{equation}
\begin{array}{lll}
h_{\mu\nu}^{TT}\rightarrow\mathcal{P}_{\mu\nu\rho\sigma}^{2}h^{\rho\sigma}, &  & \frac{1}{2}\left(\partial_{\mu}\xi_{\nu}^{T}+\partial_{\nu}\xi_{\mu}^{T}\right)\rightarrow\mathcal{P}_{\mu\nu\rho\sigma}^{1}h^{\rho\sigma},\\
\\
\partial_{\mu}\partial_{\nu}\alpha\rightarrow\left(\mathcal{P}_{w}^{0}-{\displaystyle \frac{\sqrt{3}}{3}}{\displaystyle \mathcal{P}_{ws}^{0}}\right)_{\mu\nu\rho\sigma}h^{\rho\sigma}, &  & \frac{1}{3}\eta_{\mu\nu}\beta\rightarrow\left(\mathcal{P}_{s}^{0}+{\displaystyle \frac{\sqrt{3}}{3}\mathcal{P}_{ws}^{0}}\right)_{\mu\nu\rho\sigma}h^{\rho\sigma}.
\end{array}
\end{equation}
\\
We can also have a metric perturbation decomposition in which one
of the two scalar components corresponds to the trace $h$ of $h_{\mu\nu},$
and in this case we need to know how to write the trace in terms of
the spin projector operators. One can easily check that the trace
assume the following form
\begin{equation}
h=\eta^{\mu\nu}\left(\mathcal{P}_{s}^{0}+\mathcal{P}_{w}^{0}\right)_{\mu\nu\rho\sigma}h^{\rho\sigma}.
\end{equation}

\chapter{Ghosts and unitarity}

In this appendix we want to define and show the validity of some tools
we used in this thesis to verify whether the unitarity is preserved.
We saw that a necessary condition for a theory to be unitary is that
(bad) ghosts are absent. To check that ghosts field doesn't violate
the unitarity condition we made use of a prescription by which we
verified the positivity of the imaginary part of the residue current-current
amplitude.\\
First of all we are going to define the concept of unitarity, then
to define ghost fields and at the end to show why the prescription
we used is valid.

\section{Unitarity condition}

In a quantum theory, we expect that the sum of all probabilities is
equal to $1.$ This mean that the norm of a state $\left|s\right\rangle $
at time $t=0$ should be the same at a later time $t:$
\begin{equation}
\left\langle s,t=0|s,t=0\right\rangle =\left\langle s,t|s,t\right\rangle .
\end{equation}
This means that the Hamiltonian should be Hermitian, $H^{\dagger}=H,$
because
\begin{equation}
\left|s,t\right\rangle =e^{iHt}\left|s,t=0\right\rangle ,
\end{equation}
but it also means that $S-$matrix should be unitary, since by definition
one has $S=e^{iHt}.$ Thus
\[
\left\langle s,t|s,t\right\rangle =\left\langle s,t=0\left|e^{-iHt}e^{iHt}\right|s,t=0\right\rangle =\bigl\langle s,t=0\bigl|S^{\dagger}S\bigr|s,t=0\bigr\rangle
\]
\begin{equation}
\Rightarrow S^{\dagger}S=1.\label{eq:c1}
\end{equation}
The unitarity of the $S-$matrix is equivalent to conservation of
probability, which seems to be a property of our universe as far as
anyone can tell.

$ $

$ $

$ $

\textbf{Optical theorem}%
\footnote{See Ref. {[}\ref{-M.-Schwartz,}{]} for more details.%
}

$ $\\
One of the most important implications of unitarity is the optical
theorem that we are going to discuss below. \\
Let us write the $S-$matrix as
\begin{equation}
S=1+iT,\label{eq:c2}
\end{equation}
where $T$ is called transfer matrix and its elements are defined
as
\[
\left\langle f\left|T\right|i\right\rangle =(2\pi)^{4}\delta^{4}(p_{f}-p_{i})\mathcal{M}(i\rightarrow f),
\]
with $\mathcal{M}(i\rightarrow f)$ scattering amplitude. The matrix
$T$ is not hermitian, in fact from (\ref{eq:c2}) we have
\[
1=S^{\dagger}S=\left(1-iT^{\dagger}\right)\left(1+iT\right)
\]
\begin{equation}
\Rightarrow i\left(T^{\dagger}-T\right)=T^{\dagger}T,\label{eq:c3}
\end{equation}
that is an equivalent form to express the unitarity condition. \\
Let us now introduce the following orthonormal and complete set of
states $\left|n\right\rangle :$
\begin{equation}
\left\langle n|m\right\rangle =\delta_{nm},\,\,\,\,\sum_{n}\left|n\left\rangle \right\langle n\right|=1.\label{eq:c4}
\end{equation}
We can also write each state $\left|n\right\rangle $ in terms of
the momenta $k_{i}$ of the particles in it, so in this way the completeness
relation reads as
\begin{equation}
\begin{array}{rl}
1= & \sum_{n}\int d\Pi_{n}\left|n\left\rangle \right\langle n\right|\\
\\
= & {\displaystyle \sum_{n}\prod_{j\in n}\int\frac{dk_{j}}{(2\pi)^{3}}\frac{1}{2E_{j}}\left|k_{1},k_{2},\dots,k_{n}\left\rangle \right\langle k_{1},k_{2},\dots,k_{n}\right|.}
\end{array}\label{eq:c5}
\end{equation}
The left-side of (\ref{eq:c3}) is%
\footnote{Don't get confused because of the presence of two ``$i":$ one is
the imaginary unit and the other one represents a generic initial
state!%
} 
\begin{equation}
\bigl\langle f\bigl|i\bigl(T^{\dagger}-T\bigr)\bigr|i\bigr\rangle=i(2\pi)^{4}\delta^{4}(p_{f}-p_{i})\left(\mathcal{M}^{\dagger}(i\rightarrow f)-\mathcal{M}(i-f)\right);
\end{equation}
instead, the left side, by using the relation (\ref{eq:c5}), reads
as
\[
\bigl\langle f\bigl|T^{\dagger}T\bigr|i\bigr\rangle=\sum_{n}\int d\Pi_{n}\bigl\langle f\bigl|T^{\dagger}\bigr|n\bigr\rangle\left\langle n\left|T\right|i\right\rangle =
\]
\begin{equation}
=\left(i(2\pi)^{4}\delta^{4}(p_{f}-p_{n})\right)\left(i(2\pi)^{4}\delta^{4}(p_{n}-p_{i})\right)\sum_{n}\int d\Pi_{n}\mathcal{M}^{\dagger}(n\rightarrow f)\mathcal{M}(i\rightarrow n).
\end{equation}
Then the unitarity condition (\ref{eq:c3}) implies 
\begin{equation}
\mathcal{M}^{\dagger}(i\rightarrow f)-\mathcal{M}(i-f)=-i\sum_{n}\int d\Pi_{n}(2\pi)^{4}\delta^{4}(p_{n}-p_{i})\mathcal{M}^{\dagger}(n\rightarrow f)\mathcal{M}(i\rightarrow n),\label{eq:c6}
\end{equation}
that represents the generalized optical theorem.\\
Note that the relation (\ref{eq:c6}) must work order by order in
perturbation theory. But while the left hand side is matrix elements,
the right hand side is matrix elements squared. This means that at
order $\lambda^{2}$ in some coupling the left hand side must be a
loop to match a tree-level calculation on the right hand side. Thus,
the imaginary parts of loop amplitudes are determined completely by
tree-level amplitudes. In particular, we must have the loops otherwise
without loops unitarity would be violated. \\
To the extent that trees represent classical physics and loops represent
quantum effects, the optical theorem implies that the quantum theory
is uniquely determined by the classical theory because of unitarity
{[}\ref{-M.-Schwartz,}{]}.\\
\\
The most important case is when $\left|f\right\rangle =\left|i\right\rangle =\left|X\right\rangle $
for some state $X.$ In particular, when $\left|X\right\rangle $
is a $1$-particle state, (\ref{eq:c6}) becomes
\begin{equation}
\mathrm{Im}\mathcal{M}(X\rightarrow X)=m_{X}\sum_{n}\Gamma(X\rightarrow n).\label{eq:c7}
\end{equation}
Here $\mathcal{M}(X\rightarrow X)$ is a $2$-point function, i.e.
a propagator. So (\ref{eq:c7}) says that the imaginary part of the
propagator is equal to the sum of the decay rates into every possible
particle.\\
If $\left|X\right\rangle $ is a $2$-particle state, then (\ref{eq:c6})
becomes
\begin{equation}
\mathrm{Im}\mathcal{M}(X\rightarrow X)=2E_{CM}|\bar{p}_{CM}|\sum_{n}\sigma(X\rightarrow n).
\end{equation}
This says that the imaginary part of the forward scattering amplitude
is proportional to the total cross section, which is the optical theorem
from optics.

\section{Ghost fields}

In this section we define a ghost field and we shall see what its
presence implies both at the classical level and at the quantum level.
We will only consider a scalar field for simplicity but the results
also hold for vector and tensor field. We shall follow Ref. {[}\ref{-F.-Sbisa,}{]}
in the first subsection. \\
\\
Let us consider the following Lagrangian:
\begin{equation}
\mathcal{L}=\frac{a}{2}\partial_{\mu}\phi\partial^{\mu}\phi-\frac{b}{2}m^{2}\phi^{2},\label{eq:c8}
\end{equation}
where $a=\pm1$ and $b=\pm.$ The momentum conjugate to $\phi$ is
defined by
\begin{equation}
\pi\coloneqq\frac{\partial\mathcal{L}}{\partial(\partial_{0}\phi)}=a\partial_{0}\phi\equiv a\dot{\phi},
\end{equation}
We can obtain the Hamiltonian density by performing the following Legendre transformation
\begin{equation}
\begin{array}{rl}
\mathcal{H}= & \pi\dot{\phi}-\mathcal{L}\\
\\
= & {\displaystyle \pi\dot{\phi}-\frac{a}{2}\partial_{\mu}\phi\partial^{\mu}\phi+\frac{b}{2}m^{2}\phi^{2}}\\
\\
= & {\displaystyle \frac{a}{2}\left(\dot{\phi}^{2}+(\nabla\phi)^{2}\right)+\frac{b}{2}m^{2}\phi^{2}}
\end{array}
\end{equation}
in terms of which the Hamiltonian is defined as 
\begin{equation}
H=\int d^{3}x\left[\frac{a}{2}\left(\dot{\phi}^{2}+(\nabla\phi)^{2}\right)+\frac{b}{2}m^{2}\phi^{2}\right].\label{eq:c9}
\end{equation}
Note that 
\begin{itemize}
\item $a=b=+1:\,\,$ the Hamiltonian is positive semi-definite and therefore
bounded from below; 
\item $a=b=-1:\,\,$ the Hamiltonian is negative semi-definite and therefore
bounded from above;
\item $a=-b:\,\,$ the Hamiltonian is indefinite and so it is not bounded
either from below or from above. 
\end{itemize}
If $a=b=-1,$ the field $\phi$ is called $\mathit{ghost}$ $\mathit{field}$, if
$a=+1$ and $b=-1$ one has a $\mathit{tachyon}$ $\mathit{field}$, finally if $a=-1$ and $b=+1.$ it is called $\mathit{tachyonic}$ $\mathit{ghost}$
$\mathit{field}.$ More generally one gives the following definition:

$ $

$\hspace{2cm}$``A ghost field is defined as a field which has negative
kinetic energy''.

$ $\\
\\
In the next two subsections we shall see what happens if ghosts are
present in the Lagrangian. We shall see that the presence of ghost has implications either if we perform a classical study or if we study the quantum aspects of the theory. Indeed, at the classical level the presence of ghost could cause instabilities of the vacuum since the energy is not bounded from below, while at the quantum level ghosts correspond to states with negative norm and they
could violate the unitarity condition. However, the presence of ghost
does not always violate fundamental principles. In fact, we have to
distinguish $\mathit{good}$ ghosts from $\mathit{bad}$ ghosts. A
good ghost doesn't violate any fundamental principles since they never
appear as observable physical state; while a bad ghost does violate
fundamental principles because it is associated with physical particles.

\subsection{Ghosts at the classical level}

If a Hamiltonian is unbounded from below (like in the cases $a=b=+1$
and $a=-b)$ instabilities can emerge in the system.
However, if a ghost field $\phi$ is free, namely if interactions are absent, one can easily see that the system will be still stable as the energy is conserved, independently of its sign. In fact, any constant that mupltiplies
a (classical) Lagrangian does not change the physics, since it does not appear
in the equations of motion. Thus, the choices
$a=b=+1$ and $a=b=-1$ are completely equivalent at the classical level. Translating the previous words in formula, the
field equation for the Lagrangian (\ref{eq:c1}) is given by 
\begin{equation}
\left(a\boxempty+bm^{2}\right)\phi(x)=0.\label{eq:c10}
\end{equation}
So, if we take the cases $a=b=+1$ and $a=b=-1$ we notice that the
field equations are the same.\\
In momentum space the last equation becomes
\begin{equation}
\left(-ak^{2}+bm^{2}\right)\phi(k)=0\Rightarrow a(k^{0})^{2}-a\bar{k}^{2}=bm^{2}>0.\label{eq:c11}
\end{equation}
We can easily verify what we have already said above, namely that
for $a=-b$ we have also a tachyonic solution. In fact (\ref{eq:c11})
gives $(k^{0})^{2}-\bar{k}^{2}=-m^{2}>0$ that implies $m^{2}<0,$
i.e. complex masses.\\
If we consider the following Fourier decomposition
\begin{equation}
\phi(\bar{x},t)=\int\frac{d^{3}k}{(2\pi)^{3}}\phi_{\bar{k}}(t)e^{i\bar{k}\cdot\bar{x}},
\end{equation}
we obtain that, for $a=b=\pm1,$ every mode $\phi_{\bar{k}}(t)$ evolves independently from the others and satisfies the equation
\begin{equation}
\ddot{\phi}_{\bar{k}}(t)+\left(m^{2}+\bar{k}^{2}\right)\phi_{\bar{k}}(t)=0,
\end{equation}
which exhibits oscillatory solutions of frequency given by $\omega(\bar{k})=\sqrt{m^{2}+\bar{k}^{2}}.$
A small perturbation at $t=t_{0}$ from the configuration $\phi=0$ is described by small Fourier coefficients $\phi_{\bar{k}}(t_{0}),$ and the
oscillatory behavior ensures that there is no exponential enhancement, i.e. the perturbation remains small for $t>t_{0}.$ Instead, if $a=-b,$
the frequency $\omega(\bar{k})=\sqrt{\bar{k}^{2}-m^{2}}$ turns out to be
imaginary when $\bar{k}^{2}<m^{2}$ holds and so the Fourier
modes suffer from an exponential growth, implying the presence of an instability in the theory.
However, the situation changes if there is interaction, and so energy exchanges, between a ghost field and a normal (non-ghost) field.\\
Let us, in fact, consider the following interacting Lagrangian
\begin{equation}
\mathcal{L}=\frac{a}{2}\partial_{\mu}\phi\partial^{\mu}\phi-\frac{a}{2}m_{\phi}^{2}\phi^{2}+\frac{1}{2}\partial_{\mu}\psi\partial^{\mu}\psi-\frac{1}{2}m_{\psi}^{2}\psi^{2}-V_{int}(\phi,\psi),\label{eq:432}
\end{equation}
where for hypothesis the potential $V_{int}$ does not contain derivative
interaction terms, but only depends on the two fields, and admits the solution
$\phi=\psi=0$ as a local minimum. Performing the
Legendre transformation we obtain the Hamiltonian
\begin{equation}
\mathcal{H}=\frac{a}{2}\left(\dot{\phi}^{2}+(\nabla\phi)^{2}\right)+\frac{a}{2}m^{2}\phi^{2}+\frac{1}{2}\left(\dot{\phi}^{2}+(\nabla\phi)^{2}\right)+\frac{1}{2}m^{2}\phi^{2}+V_{int}(\phi,\psi).
\end{equation}
First, note that, if $V_{int}(\phi,\psi)=0,$ the state $\phi=\psi=0$
is still stable independently from the sign of $a.$ The stability is preserved as the conservation energy law can be applied separately for the two
non-interacting fields $\phi$ and $\psi.$ However, we have to point out that already at this level 
there is a difference between the cases $a=+1$ and $a=-1.$ Although the system is stable, the choice $a=-1$ corresponds
to an infinite number of different states with $E=0$ which
cannot be associated to small perturbations of the vacuum (minimum) $\phi=\psi=0.$ \\
Secondly, in the case $V_{int}(\phi,\psi)\neq0,$ the minimum configuration is
still a solution of the field equation and one can show that the Hamiltonian
can be bounded from below for constant values of the dynamical fields.
Note that, since now the interaction term is not vanishing, the configurations
cannot have zero energy anymore. However, by perturbing the
vacuum configuration, one can construct states with
energy values very close to zero. Therefore, if $a=-1,$ the available volume of the momentum space turns out to be infinite with an infinite number of excited states. Thus, since for entropy reasons the
total energy is redistributed into the largest possible class of states,
the decay towards these excited states is extremely favoured, and
this can be summarized saying that the system is unstable for small oscillations%
\footnote{See Ref. {[}\ref{-F.-Sbisa,}{]} for more details.%
}.

As we shall see in the section $C.3,$ a Lagrangians of the type (\ref{eq:432})
with $a=-1$ is equivalent to a Lagrangian containing higher derivatives.
In 1850, Ostrogradsky demonstrated a theorem in which he states that
the Hamiltonian of a non-degenerate higher derivative theory is unbounded
from below, and also from above {[}\ref{-R.-P.woodard}{]}, so instabilities
are present. For this reason, the classical instability due to the
presence of a negative kinetic term, that we have treated in this
subsection, is called $\mathit{Ostrogradskian}$ $\mathit{instability.}$

\subsection{Ghosts at the quantum level}

We have seen that instability problems are associated to the presence
of a ghost at the classical level. At the quantum level the presence
of a (bad) ghost is even more problematic. Let us just consider the
Lagrangian (\ref{eq:c1}) with the ghost choice for the coefficients,
$a=b=-1:$
\begin{equation}
\mathcal{L}=-\frac{1}{2}\partial_{\mu}\phi\partial^{\mu}\phi+\frac{1}{2}m^{2}\phi^{2}=\frac{1}{2}\phi\left(\boxempty+m^{2}\right)\phi.
\end{equation}
We can see that for a ghost field the propagator in momentum space
is given by
\begin{equation}
\mathcal{P}(k)=-\frac{1}{k^{2}-m^{2}},
\end{equation}
i.e. it turns out to have a minus sign of difference with respect
to an ordinary field, thus its residue is negative%
\footnote{Often the negativity of the propagator residue is taken to define
a ghost field. Furthermore, some authors define the propagator including
the imaginary unit ``$i"$ and so they define a ghost field in terms
of the imaginary part of its residue.%
}. \\
To quantize the (ghost) scalar field theory we need to impose the
following commutation relations:
\begin{equation}
\begin{array}{rl}
\left[\phi(\bar{x},t),\pi(\bar{x}',t)\right]= & i\delta^{3}(\bar{x}-\bar{x}'),\\
\\
\left[\phi(\bar{x},t),\phi(\bar{x}',t)\right]= & 0,\\
\\
\left[\pi(\bar{x},t),\pi(\bar{x}',t)\right]= & 0,
\end{array}\label{eq:c13}
\end{equation}
where 
\begin{equation}
\phi(\bar{x},t)=\int\frac{d^{3}k}{\sqrt{2\omega_{\dot{k}}(2\pi)^{3}}}\left(a_{\bar{k}}e^{i(\bar{k}\cdot\bar{x}-\omega_{\bar{k}}t)}+a_{\bar{k}}^{\dagger}e^{-i(\bar{k}\cdot\bar{x}-\omega_{\bar{k}}t)}\right)
\end{equation}
and
\begin{equation}
\pi(\bar{x},t)\coloneqq\frac{\partial\mathcal{L}}{\partial\dot{\phi}}=-\dot{\phi}.\label{eq:c12}
\end{equation}
The coefficients $a_{\bar{k}}$ and $a_{\bar{k}}^{\dagger}$ are the
usual annihilation and creation operators, respectively. Because of
the minus sign in the definition of the conjugate momentum (\ref{eq:c12})
and to have consistency with the commutation relations (\ref{eq:c13}),
the commutation relations for the annihilation and creation operators
must be
\begin{equation}
\begin{array}{rl}
\bigl[a_{\bar{k}},a_{\bar{k}'}^{\dagger}\bigr]= & -\delta^{3}(\bar{k}-\bar{k}'),\\
\\
\left[a_{\bar{k}},a_{\bar{k}'}\right]= & 0,\\
\\
\left[a_{\bar{k}},a_{\bar{k}'}\right]= & 0.
\end{array}\label{eq:c14}
\end{equation}
The problem of states with negative norms associated to the presence
of a ghost field becomes evident if we try to construct the Fock space
for the ghost scalar field. Let us assume that a normalized vacuum
state $\left|0\right\rangle $ exists which has the property
\begin{equation}
a_{\bar{k}}\left|0\right\rangle =0\,\,\,\,\forall\bar{k},\,\,\,\,\,\left\langle 0|0\right\rangle =1.
\end{equation}
As usual the eigenstates of the occupation number can be constructed
by applying the creation operators $a_{\bar{k}}^{\dagger}$ to the
vacuum. The state vector containing $n_{\bar{k}}$ ghost fields reads
\begin{equation}
\bigl|n_{\bar{k}}\bigr\rangle=\frac{1}{\sqrt{n_{\bar{k}}!}}\left(a_{\bar{k}}^{\dagger}\right)^{n_{\bar{k}}}\left|0\right\rangle .
\end{equation}
Now, if we calculate the norm of the one-particle state $1_{\bar{k}},$
we obtain
\begin{equation}
\begin{array}{rl}
\bigl\langle1_{\bar{k}}|1_{\bar{k}}\bigr\rangle= & \bigl\langle0\bigl|a_{\bar{k}}a_{\bar{k}}^{\dagger}\bigr|0\bigr\rangle=\bigl\langle0\bigl|-\delta^{3}(\bar{k}-\bar{k})+a_{\bar{k}}^{\dagger}a_{\bar{k}}\bigr|0\bigr\rangle\\
\\
= & -\delta^{3}(\bar{0})\left\langle 0|0\right\rangle <0,
\end{array}
\end{equation}
where we have used the commutation relation (\ref{eq:c14}). Hence,
we have just showed that the state with one ghost field has negative
norm.\\
Another example of ghost field one has in Electrodynamics where the
time component of the four-vector $A_{\mu}$ has a negative kinetic
termgmn. Also in this case one can show that state associated to the
scalar time component have negative norm caused by a minus sign in
the definition of the commutation relation for the $0\textrm{-}$component.
In Electrodynamics these ghost field states don't appear as physical
states, but they are very important because are necessary to cancel
out the longitudinal component of the vector field. Since it doesn't
violate the unitarity condition is a good ghost. We have also seen
an example of good ghost in GR, where we have a scalar graviton component
that behave as a ghost.\\
In gauge quantum field theory we have another important example of
good ghost, the Faddev-Popov ghost. It was firstly introduced to maintain
the consistency between gauge invariance and path integral formulation.
Secondly, as Feynman noticed, it is very necessary to preserve unitarity,
in fact its absence wouldn't satisfy the optical theorem that is an
implication of the unitarity condition. It is remarkable, doubtless
of profound significance, that good ghosts solve, simultaneously,
the problem of unitarity and gauge invariance {[}\ref{-L.-Ryder,}{]}.
\\
Instead, an example of bad ghost is given by the Pauli-Villars ghost
that they introduce to define a regulation scheme to solve the divergence
problem in quantum field theory. They add a particle with very large
mass $M$ whose propagator has a minus sign:
\[
\mathcal{P}_{P-V}(k)=-\frac{1}{k^{2}-M^{2}}.
\]
One can show that the presence of this propagator in quantum loop
doesn't preserve the unitarity, in fact the optical theorem is not
satisfied.\\
Then, we have also already encountered another example of bad ghost
in subsection $3.4.3,$ i.e the Weyl ghost (see eq. (\ref{eq:304weyl}),
(\ref{eq:306weyl}) and also Appendix $C.4.).$

\subsection{Ghost-unitarity analysis}

So far in this appendix we have introduced the concept of unitarity
and the optical theorem, see as one of the unitarity implications,
and defined the concept of ghost field analizying its classical and
quantum nature. We learned that good ghosts preserve the unitarity
condition and that bad ghosts violate it. Now, we are going to introduce
a method by which we can verify whether the presence of a ghost preserve
the unitarity or not, namely whether it is a good or a bad ghost {[}\ref{-A.-Accioly}{]},
{[}\ref{-F.-C.}{]},{[}\ref{-S.F.-Hassan,}{]}. To reach our aim we
need to work with the path integral formulation.\\
From the path integral formulation of the quantum field theory we
know that the vacuum-vacuum transition amplitude in the presence of
a source $J$ corresponds to the generating function $Z_{0}[J],$
where the subscript $0$ means that we are dealing with the free theory.
As we can see in many books of quantum field theory, like {[}\ref{-M.-Schwartz,}{]}
and {[}\ref{-L.-Ryder,}{]}, one can shows that the vacuum-vacuum
amplitude is given by
\begin{equation}
Z_{0}[J]=\bigl\langle0,\infty|0,-\infty\bigr\rangle^{J}=\exp\left\{ i\int d^{4}x\int d^{4}y\frac{1}{2}J(x)\mathcal{P}(x-y)J(y)\right\} ,\label{eq:c15}
\end{equation}

with the normalization choice $Z_{0}[0]=1.$ 

Now, we want to recast the integrals in (\ref{eq:c15}) in terms of
integrals on the momentum $k.$ To do this we need to rewrite the
integrand as Fourier transforms: 
\begin{equation}
\begin{array}{rl}
\mathcal{P}(x-y)= & {\displaystyle \int\frac{d^{4}k}{(2\pi)^{4}}\mathcal{P}(k)e^{ik\cdot(x-y)}=\int\frac{d^{4}k}{(2\pi)^{4}}\frac{1}{k^{2}-m^{2}}e^{ik\cdot(x-y)},}\\
\\
J(x)= & {\displaystyle \int\frac{d^{4}k_{1}}{(2\pi)^{4}}J(k_{1})e^{ik_{1}\cdot x},}\\
\\
J(y)= & {\displaystyle \int\frac{d^{4}k_{2}}{(2\pi)^{4}}J(k_{2})e^{ik_{2}\cdot y}}.
\end{array}
\end{equation}
Thus, the integral in (\ref{eq:c15}) becomes:
\[
{\displaystyle \int d^{4}x\int d^{4}yJ(x)\mathcal{P}(x-y)J(y)}=
\]
\begin{equation}
={\displaystyle \int\frac{d^{4}k}{(2\pi)^{4}}\int d^{4}k_{1}\int d^{4}k_{2}J(k_{1})\mathcal{P}(k)J(k_{2})\left[\int \frac{d^{4}x}{(2\pi)^{4}}e^{ix\cdot(k+k_{1})}\right]\left[\int \frac{d^{4}y}{(2\pi)^{4}}e^{-iy\cdot(k-k_{2})}\right].}
\end{equation}
By using the integral representation in momentum space of the delta
function in four dimensions 
\begin{equation}
\delta^{4}(k+k_1)=\int\frac{d^{4}x}{(2\pi)^{4}}e^{-i(k+k_1)\cdot x},\,\,\,\,\delta^{4}(k-k_2)=\int\frac{d^{4}y}{(2\pi)^{4}}e^{-i(k-k_2)\cdot x},
\end{equation}
we obtain%
\footnote{We are ignoring the $2\pi$ factors.%
} 
\begin{equation}
\int d^{4}x\int d^{4}yJ(x)\mathcal{P}(x-y)J(y)\sim\int d^{4}kJ(-k)\mathcal{P}(k)J(k).
\end{equation}
Thus, the vacuum-vacuum amplitude in the presence of a source $J$
becomes
\begin{equation}
\bigl\langle0,\infty|0,-\infty\bigr\rangle^{J}\sim\exp\left\{ i\int d^{4}kJ(-k)\mathcal{P}(k)J(k)\right\} .\label{eq:c16}
\end{equation}
Note that we can decompose the integral on the momentum $k$ as
\begin{equation}
i\int d^{4}kJ(-k)\mathcal{P}(k)J(k)=\int d^{3}k\left[\int dk_{0}iJ(-k)\mathcal{P}(k)J(k)\right]
\end{equation}
and we can calculate the integral on the time component $k_{0}$ by
using the residue theorem of Cauchy:
\begin{equation}
\int d^{4}kJ(-k)\mathcal{P}(k)J(k)=\int d^{3}k\left[2\pi iRes\left\{ iJ(-k)\mathcal{P}(k)J(k)\right\} _{k^{2}=0}\right].
\end{equation}
Hence, the equation (\ref{eq:c16}) can be recast in the following
way%
\footnote{In the equation (\ref{eq:c17}) we have written $k^{2}=m^{2}$ meaning
that the residue is calculate at the pole $m^{2},$ but in this thesis
we have considered massless photon and massless graviton with poles
$k^{2}=0.$ %
}
\begin{equation}
\bigl\langle0,\infty|0,-\infty\bigr\rangle^{J}\sim\exp\left\{ \int d^{3}k\left[2\pi iRes\left\{ iJ(-k)\mathcal{P}(k)J(k)\right\} _{k^{2}=m^{2}}\right]\right\} .\label{eq:c17}
\end{equation}
Let us note that the integrand represents the current-current amplitude
in momentum space: $\mathcal{A}(k)=iJ(-k)\mathcal{P}(k)J(k).$ We
can easily notice that the sign of the imaginary part of the residue
is crucial%
\footnote{Note that if the integrand $2\pi iRes\left\{ iJ(-k)\mathcal{P}(k)J(k)\right\} _{k^{2}=m^{2}}$
is positive (negative), the integral $\int d^{3}k\left[2\pi iRes\left\{ iJ(-k)\mathcal{P}(k)J(k)\right\} _{k^{2}=m^{2}}\right]$
will be positive (negative) too.%
}: if it is positive we obtain a negative exponent but if it is negative
the exponential is positive giving a vacuum-vacuum amplitude greater
than $1.$ The quantity $\bigl\langle0,\infty|0,-\infty\bigr\rangle^{J}$
is the transition amplitude to go from the initial state $|0,-\infty\bigr\rangle$
to the final state $|0,\infty\bigr\rangle.$ The probability to find
the system in the initial state is given by $P$ and it has to be
less than $1,$ $(P<1);$ while the probability to transit to the
final state is given by $1-P$ and it has to be less than one too.
We notice that if the imaginary part of the amplitude residue in (\ref{eq:c17})
gives a negative value we have that $1-P>1\Rightarrow P<0,$ i.e.
as result we obtain negative probabilities that makes the theory non-unitary.\\
In this thesis we have applied this analysis to photon and graviton
case and we verified that the unitarity is non violated.\\
\\
We can conclude this section saying that we have found a method to
verify whether the presence of ghosts violate the unitarity condition.
The only thing we need to do is to check the sign of the imaginary
part of the amplitude residue: if it is positive the unitarity condition
is preserved; if it is negative the unitarity condition is violated.

\section{Ghosts in higher derivative theories}

We have already said many times that when one consider theories with
higher derivatives ghost fields appear in the theory. The more important
example is the Fourth Derivative Gravity in which the propagator has
a double pole at $k^{2}=0$ suggesting the appearance of a ghost in
the Hilbert space and signifying that either unitarity or positivity
of the energy spectrum might be violated {[}\ref{-Stelle...renormaliz}{]}.
\\
In this section we want to show that a Lagrangian with higher derivatives
is equivalent to a Lagrangian with lower derivatives but with the
presence of ghost fields. We shall do this just for a quadratic fourth
derivative Lagrangian {[}\ref{-S.W-Hawking,}{]}, {[}\ref{-A.-Gavrielides,}{]}.\\
\\
Let us consider a scalar field with fourth derivative Lagrangian given
by
\begin{equation}
\mathcal{L}=-\frac{1}{2}\phi\left(\boxempty+m^{2}\right)\left(\boxempty+M^{2}\right)\phi,\label{eq:c18}
\end{equation}
where the masses $m$ and $M$ can be also equal to zero, but we are
going to consider the massive case%
\footnote{See Ref. {[}\ref{-A.-Gavrielides,}{]} for a more rigorous treatment
of both massless and massive cases.%
} as in Ref. {[}\ref{-S.W-Hawking,}{]}. So, let us suppose $M>m.$
Defining the following two new fields
\begin{equation}
\psi_{1}\coloneqq\frac{\left(\boxempty+M^{2}\right)\phi}{\sqrt{\left(M^{2}-m^{2}\right)}},\,\,\,\,\,\psi_{2}\coloneqq\frac{\left(\boxempty+m^{2}\right)\phi}{\sqrt{\left(M^{2}-m^{2}\right)}},
\end{equation}
the Lagrangian (\ref{eq:c18}) can be rewritten as
\begin{equation}
\mathcal{L}=-\frac{1}{2}\psi_{1}\left(\boxempty+m^{2}\right)\psi_{1}+\frac{1}{2}\psi_{2}\left(\boxempty+M^{2}\right)\psi_{2},
\end{equation}
where the the term corresponding to the field $\psi_{2}$ has the
wrong sign, i.e. it is a ghost field. We are considering a theory
without interaction but, naturally, the study can be extend to Lagrangians
with interaction potential $V_{int}(\phi).$ We know that at the classical
level if the potential is set to zero ghost fields don't violate any
fundamental principles. At the quantum level, on the other hand, one
could be in trouble even in the absence of interaction, as can be
seen by looking at the free field propagator for $\phi.$ In momentum
space, this is the inverse of a fourth order expression in $k,$ $\left[\left(-k^{2}+m^{2}\right)\left(-k^{2}+M^{2}\right)\right]^{-1},$
which can be expanded as
\begin{equation}
\mathcal{P}(k)=\frac{1}{\left(M^{2}-m^{2}\right)}\left(\frac{1}{k^{2}-m^{2}}-\frac{1}{k^{2}-M^{2}}\right).
\end{equation}
This is just the difference of the propagators for $\psi_{1}$ and
$\psi_{2}.$ The important point, that we have already mentioned other
times, is that the propagator for $\psi_{2}$ appears with a negative
sign and it could violate the unitarity condition.\\
\\
We can also consider other cases, for example the case in which one
has only one has a massless ordinary field and a massive ghost field
with mass $m.$ The starting Lagrangian, in this case, is given by
\begin{equation}
\mathcal{L}=-\frac{1}{2}\phi\boxempty\left(\boxempty+m^{2}\right)\phi\label{eq:c19}
\end{equation}
and it is equivalent to the following Lagrangian

\begin{equation}
\mathcal{L}=-\frac{1}{2}\psi_{1}\boxempty\psi_{1}+\frac{1}{2}\psi_{2}\left(\boxempty+m^{2}\right)\psi_{2},\label{eq:c20}
\end{equation}
where $\psi_{1}$ is an ordinary scalar field with positive kinetic
term, while $\psi_{2}$ a massive ghost field. The authors in Ref.
{[}\ref{-A.-Gavrielides,}{]} show that the Lagrangians (\ref{eq:c19})
and (\ref{eq:c20}) describe equivalent theories by showing that the
respective generating functional are the same apart from a multiplicative
factor if we impose that $J_{1}=-J_{2}=\frac{J}{m}.$ They also do
the same analysis for a gauge field theory.\\
\\
Although we have neither gone into details nor been very rigorous,
we have have seen why higher derivative terms in the Lagrangian correspond
to ghost fields. In the next section we shall consider Higher (or
Fourth) Derivative Gravity and shall show that it is affected by the
presence of bad ghost in the spin-$2$ sector.

\section{Fourth Derivative Gravity}

The action for higher derivative gravity is:
\begin{equation}
S=\int d^{4}x\sqrt{-g}\left[-\mathcal{R}+\alpha\mathcal{R}_{\mu\nu}\mathcal{R}^{\mu\nu}+\beta\mathcal{R}^{2}\right].\label{eq:457}
\end{equation}
The linearized action, quadratic in the perturbation $h_{\mu\nu}$
is given by (see eq. (\ref{eq:177})) 
\begin{equation}
\begin{array}{rl}
S_{q}= & -{\displaystyle \int d^{4}x\left[\frac{1}{2}h_{\mu\nu}\boxempty a(\boxempty)h^{\mu\nu}+h_{\mu}^{\sigma}b(\boxempty)\partial_{\sigma}\partial_{\nu}h^{\mu\nu}\right.}\\
\\
 & \left.{\displaystyle +hc(\boxempty)\partial_{\mu}\partial_{\nu}h^{\mu\nu}+\frac{1}{2}h\boxempty d(\boxempty)h+\frac{1}{2}h^{\lambda\sigma}f(\boxempty)\partial_{\sigma}\partial_{\lambda}\partial_{\mu}\partial_{\nu}h^{\mu\nu}}\right],
\end{array}\label{eq:458}
\end{equation}
where the coefficients (\ref{eq:176}) in this case are
\begin{equation}
\begin{array}{l}
a(\boxempty){\displaystyle \coloneqq1-\frac{1}{2}\alpha}\boxempty,\\
\\
b(\boxempty){\displaystyle \coloneqq-1+\frac{1}{2}\alpha\boxempty},\\
\\
c(\boxempty)\coloneqq{\displaystyle 1+2\beta\boxempty+\frac{1}{2}\alpha\boxempty},\\
\\
d(\boxempty)\coloneqq{\displaystyle -1-2\beta\boxempty-\frac{1}{2}\alpha\boxempty},\\
\\
f(\boxempty)\coloneqq-2\beta-\alpha.
\end{array}\label{eq:459}
\end{equation}
One can show that the physical part (gauge-independent) of the propagator
in momentum space obtained in (\ref{eq:204}), 
\begin{equation}
\Pi(k)=\frac{\mathcal{P}^{2}}{ak^{2}}+\frac{\mathcal{P}_{s}^{0}}{(a-3c)k^{2}},
\end{equation}
in the case of the action (\ref{eq:459}) (or (\ref{eq:458})) assumes
the following expression 
\begin{equation}
{\displaystyle \Pi(k)=\frac{\mathcal{P}^{2}}{\left[1+\frac{1}{2}\alpha k^{2}\right]k^{2}}+\frac{\mathcal{P}_{s}^{0}}{\left[-2+\left(2\alpha+6\beta\right)k^{2}\right]k^{2}}}.
\end{equation}
By playing with the fractions one can obtain the following form for
the propagator:
\begin{equation}
{\displaystyle \Pi(k)=\frac{1}{k^{2}}\left(\mathcal{P}^{2}-\frac{\mathcal{P}_{s}^{0}}{2}\right)-\frac{\mathcal{P}^{2}}{k^{2}-m_{2}^{2}}+\frac{1}{2}\frac{\mathcal{P}_{s}^{0}}{k^{2}-m_{0}^{2}},}\label{eq:462}
\end{equation}
where $m_{2}=-\left(\frac{1}{2}\alpha\right)^{-1}$ and $m_{0}=\left(\alpha+\beta\right)^{-1}.$
Note that 
\begin{equation}
\Pi_{GR}(k)=\frac{1}{k^{2}}\left(\mathcal{P}^{2}-\frac{\mathcal{P}_{s}^{0}}{2}\right)
\end{equation}
is the GR graviton propagator corresponding to the Hilbert-Einstein
linearized action (quadratic in $h_{\mu\nu})$ obtained in Chapter
$2.$ While the second and the third terms in (\ref{eq:462}) correspond
to a massive spin-$2$ ghost with mass $m_{2}$ and a normal massive
scalar with mass $m_{0},$ respectively. We want to understand whether
the presence of the massive spin\textbf{-$2$ }ghost violates the
unitarity.\textbf{}\\
Let us consider the amplitude
\begin{equation}
\mathcal{A}=i\tau^{*\mu\nu}(k)\Pi(k)_{\mu\nu\rho\sigma}\tau^{\rho\sigma}(k)=\mathcal{A}_{GR}+\mathcal{A}_{2}+\mathcal{A}_{0},\label{eq:464}
\end{equation}
 where
\begin{equation}
\begin{array}{rl}
\mathcal{A}_{GR}= & i\tau^{*\mu\nu}(k)\Pi_{GR}(k)_{\mu\nu\rho\sigma}\tau^{\rho\sigma}(k),\\
\\
\mathcal{A}_{2}= & -i\tau^{*\mu\nu}(k){\displaystyle \frac{\mathcal{P}_{\mu\nu\rho\sigma}^{2}}{k^{2}-m_{2}^{2}}}\tau^{\rho\sigma}(k),\\
\\
\mathcal{A}_{0}= & i\tau^{*\mu\nu}(k){\displaystyle \frac{1}{2}\frac{\mathcal{P}_{s,\,\mu\nu\rho\sigma}^{0}}{k^{2}-m_{0}^{2}}}\tau^{\rho\sigma}(k).
\end{array}\label{eq:465}
\end{equation}
We need to calculate the imaginary part of the residue of the full
amplitude in (\ref{eq:464}), that corresponds to the sum of the residue
of the three amplitudes in (\ref{eq:465}). By using the definition
of the spin projector operators $\mathcal{P}^{2}$ and $\mathcal{P}_{s}^{0}$
one can show that
\begin{equation}
\begin{array}{rl}
\mathrm{Im}Res_{k^{2}=0}\left\{ \mathcal{A}_{GR}\right\} = & {\displaystyle |\tau^{\mu\nu}(0)|^{2}-\frac{1}{2}|\tau(0)|^{2}}\\
\\
\mathrm{Im}Res_{k^{2}=m_{2}^{2}}\left\{ \mathcal{A}_{2}\right\} = & -\left(|\tau^{\mu\nu}(m_{2})|^{2}-\frac{1}{3}|\tau(m_{2})|^{2}\right)\\
\\
\mathrm{Im}Res_{k^{2}=m_{0}^{2}}\left\{ \mathcal{A}_{0}\right\} = & {\displaystyle \frac{1}{6}|\tau(m_{0})|^{2}}.
\end{array}
\end{equation}
Now, let us consider the following expansion of the source $\tau(k)$
already used in Chapter $2:$ 
\begin{equation}
\begin{array}{rl}
\tau_{\mu\nu}(k)= & a(k)k_{\mu}k_{\nu}+b(k)k_{(\mu}\tilde{k}_{\nu)}+c_{i}(k)k_{(\mu}\varepsilon_{\nu)}^{i}\\
\\
 & +d(k)\tilde{k}_{\mu}\tilde{k}_{\nu}+e_{i}(k)\tilde{k}_{(\mu}\varepsilon_{\nu)}^{i}+f_{ij}(k)\varepsilon_{(\mu}^{i}\varepsilon_{\nu)}^{j},
\end{array}\label{eq:467}
\end{equation}
where the basis of the expansion is $\left\{ k^{\mu},\tilde{k}^{\mu},\varepsilon_{1}^{\mu},\varepsilon_{2}^{\mu}\right\} ,$
such that 
\begin{equation}
\begin{array}{l}
k^{\mu}\equiv(k^{0},\bar{k}),\,\,\,\,\,\tilde{k}^{\mu}\equiv(\tilde{k}^{0},-\bar{k}),\,\,\,\,\,\varepsilon_{i}^{\mu}\equiv(0,\bar{\varepsilon}_{i}),\\
\\
k^{\mu}\varepsilon_{i,\mu}=0=\tilde{k}^{\mu}\varepsilon_{i,\mu},\,\,\,\,\,\,\,\varepsilon_{i}^{\mu}\varepsilon_{j,\mu}=-\bar{\varepsilon}_{i}\cdot\bar{\varepsilon}_{j}=-\delta_{ij},
\end{array}\,\,\,\,\, i=1,2.
\end{equation}
One can show that
\begin{equation}
\begin{array}{rl}
\mathrm{Im}Res_{k^{2}=0}\left\{ \mathcal{A}_{GR}\right\} = & |f_{ij}(0)|^{2}-\frac{1}{2}|f(0)|^{2}=\frac{1}{2}|f_{11}(0)-f_{22}(0)|^{2}+2|f_{12}(0)|^{2}>0\\
\\
\mathrm{Im}Res_{k^{2}=m_{2}^{2}}\left\{ \mathcal{A}_{2}\right\} = & -{\displaystyle \left[\frac{2}{3}\left(a(m_{2})-d(m_{2})\right)^{2}m_{2}^{4}+\frac{m_{2}^{2}}{2}\left(|c_{i}(m_{2})|^{2}-|e_{i}(m_{2})|^{2}\right)\right.}\\
\\
 & {\displaystyle \left.+|f_{ij}(m_{2})|^{2}-\frac{1}{2}|f(m_{2})|^{2}-\frac{2}{3}\left(a(m_{2})-d(m_{2})\right)m_{2}^{2}f_{ii}(m_{2})\right]}\\
\\
\mathrm{Im}Res_{k^{2}=m_{0}^{2}}\left\{ \mathcal{A}_{0}\right\} = & {\displaystyle \frac{1}{6}\left[\left(a(m_{0})-d(m_{0})\right)^{2}m_{0}^{4}+|f_{ii}(m_{0})|^{2}\right.}\\
\\
 & \left.-2\left(d(m_{0})-a(m_{0})\right)m_{0}^{2}f_{ii}(m_{0})\right]>0
\end{array}\label{eq:469}
\end{equation}
\\
As usual, we consider $\tau>0,$ thus $f_{ii}<0$ {[}\ref{-A.-Accioly}{]}.
Since the source $\tau_{\mu\nu}(k)$ is arbitrary, its Fourier modes
(i.e. the coefficients $a(k),$ $b(k),\ldots)$ can be freely chosen
and we can make choices such that only one of the three residues contributes at
a time. We quickly notice that the massless and the massive scalar
poles (first and second lines in (\ref{eq:469})) are well defined
physical state. While, if we choose the source $\tau_{\mu\nu}(k)$
such that only the pole $m_{2}^{2}$ contributes one can see that
the second line in eq. (\ref{eq:469}) is not positive defined. For
example, if $|c_{i}(m_{2})|^{2}-|e_{i}(m_{2})|^{2}>0$ and $a(m_{2})-d(m_{2})>0$
we get 
\begin{equation}
\mathrm{Im}Res_{k^{2}=m_{2}^{2}}\left\{ \mathcal{A}_{2}\right\} <0
\end{equation}
that violates the unitarity at the tree level. \\
We could imagine to make special choices for the coefficients in the
source expansion (\ref{eq:467}) to obtain a positive value for the
sum of the three residues and so a ghost-free theory%
\footnote{Keep in mind that with the nomenclature ``ghost-free'' we refer
to a theory free from ``bad'' ghost.%
}. But, in this way we would restrict $\tau_{\mu\nu}(k)$ by hand to
get a ghost-free sum and it does not mean that the theory is healthy
because interactions can always generate the $\tau_{\mu\nu}(k)\textrm{-}$configurations
that have not been considered {[}\ref{-Private-conversationHassan}{]}.
\\
\\
Hence, we have seen that Higher Derivative Gravity is not a healthy
theory because of the presence of the massive spin-$2$ (bad) ghost
that violates the unitarity at the tree level.

\chapter{Clebsch-Gordan coefficients}

In the subsection $2.4.1$ we calculate the graviton polarization
tensors by starting from the photon polarization vectors. We saw every
polarization tensor as the composition of $\mathit{two}$ spin-$1$
polarization. In this way we obtained $\mathit{five}$ polarization
tensors of spin-$2,$ $\mathit{three}$ polarization tensors of spin-$1$
that we didn't considered because of their antisymmetric nature, and
$\mathit{one}$ polarization tensor of spin-$0.$ Since we have constructed
the set of polarization tensors by composition of two spin-$1$ vectors,
we made use of the table of Clebsch-Gordan coefficients.\\
Below we report the table of some Clebsch-Gordan coefficients. The
composition $1\otimes1$ we have used is indicated with red arrows. 

\begin{figure}[H]
\includegraphics[scale=0.6]{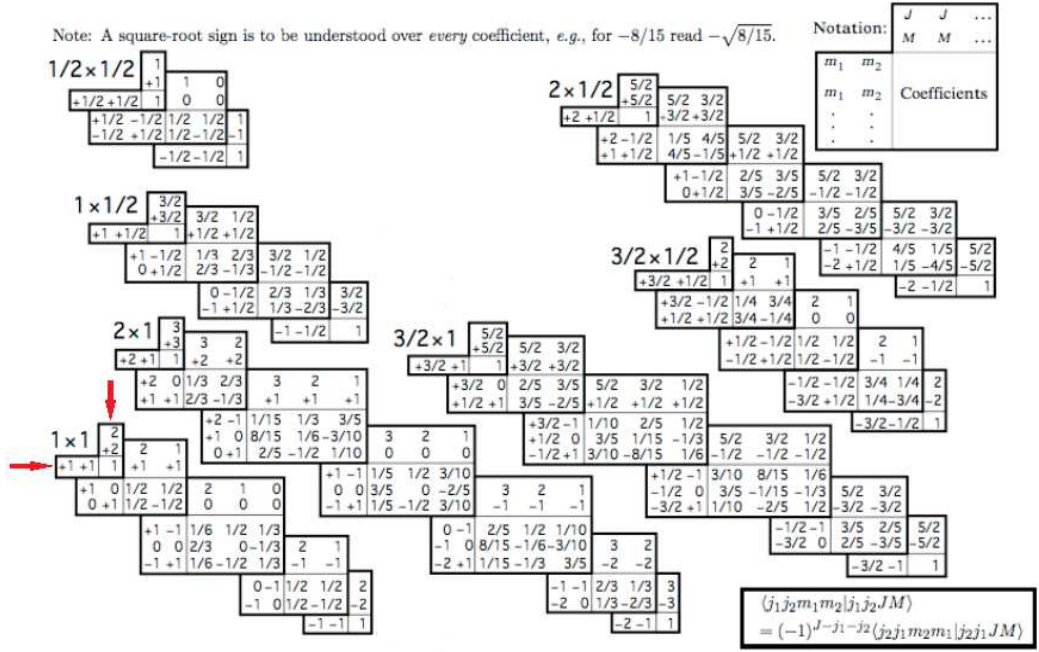}

\protect\caption{This Clebsch-Gordan coefficients table was taken from PDG (Particle
Data Group) listings. The product $1\otimes1$ is the composition
we are interested in and it is indicated by two red arrows.}
\end{figure}


\begin{thebibliography}{10}
\bibitem{key-1} \label{-Private-conversationHassan}Private conversation
with Prof. Fawad Hassan who was very kind, patient and clear in his
explanations.
\end{thebibliography}
\end{document}